\documentclass[11pt, a4paper, onecolumn, thu]{thuc3i}

\usepackage[authoryear, sort&compress, square]{natbib}
\usepackage{fontawesome5}
\usepackage[most]{tcolorbox}
\definecolor{AbstractBgColor}{HTML}{F4F7FB}
\definecolor{ProjectLinkBorder}{HTML}{39C4DC}
\definecolor{ProjectLinkBg}{HTML}{ECFAFD}
\newcommand{\projectlink}[2]{%
    \href{#1}{\tcbox[on line, colback=ProjectLinkBg,
        colframe=ProjectLinkBorder, boxrule=0.55pt, arc=3pt,
        boxsep=0pt, left=5pt, right=5pt, top=1.5pt, bottom=1.5pt]{%
        \makebox[64pt][c]{\color{black}\mdseries\fontsize{8.5}{10}\selectfont\rule[-2.5pt]{0pt}{10pt}#2}}}%
}
\newcommand{\githuburl}{https://github.com/AI45Lab/REX}
\newcommand{\huggingfaceurl}{https://huggingface.co/collections/AI45Research/rex}

\usepackage{mathtools}
\usepackage{booktabs}
\usepackage[capitalise,noabbrev]{cleveref}

\let\reportbibsection\bibsection
\renewcommand{\bibsection}{%
  \reportbibsection
  \phantomsection
  \addcontentsline{toc}{section}{\refname}%
}

\graphicspath{{figures/}}

\newcommand{\sysname}{REX}
\newcommand{\Ours}{\sysname}

\newcommand{\Sstd}{\ensuremath{S_{\mathrm{std}}}}

\expandafter\def\csname ev@historical:intern/bio/B/0M/mmlu\endcsname{85.1}
\expandafter\def\csname ev@historical:intern/bio/B/0M/wmdp\endcsname{83.7}
\expandafter\def\csname ev@historical:intern/bio/B/0M/ifeval\endcsname{81.3}
\expandafter\def\csname ev@historical:intern/bio/S_std(B)/250M/ifeval\endcsname{78.6}
\expandafter\def\csname ev@historical:intern/bio/F0/0M/mmlu\endcsname{84.9}
\expandafter\def\csname ev@historical:intern/bio/F0/0M/wmdp\endcsname{83.2}
\expandafter\def\csname ev@historical:intern/bio/A0/0M/mmlu\endcsname{83.4}
\expandafter\def\csname ev@historical:intern/bio/A0/0M/wmdp\endcsname{69.8}
\expandafter\def\csname ev@historical:intern/bio/R/0M/mmlu\endcsname{84.5}
\expandafter\def\csname ev@historical:intern/bio/R/0M/wmdp\endcsname{34.4}
\expandafter\def\csname ev@historical:intern/bio/R-SFT/250M/ifeval\endcsname{79.8}
\expandafter\def\csname ev@historical:intern/bio/CEU-SFT/250M/ifeval\endcsname{80.2}
\expandafter\def\csname ev@historical:intern/bio/GA-SFT/250M/ifeval\endcsname{78.9}
\expandafter\def\csname ev@historical:intern/bio/GD-SFT/250M/ifeval\endcsname{79.3}
\expandafter\def\csname ev@historical:intern/bio/PureNPO-SFT/250M/ifeval\endcsname{70.1}
\expandafter\def\csname ev@historical:intern/bio/ExternalRMU-SFT/250M/ifeval\endcsname{77.5}
\expandafter\def\csname ev@historical:intern/bio/SatImp-SFT/250M/ifeval\endcsname{80.0}
\expandafter\def\csname ev@historical:intern/bio/SimNPO-SFT/250M/ifeval\endcsname{66.0}
\expandafter\def\csname ev@historical:intern/bio/WGA-SFT/250M/ifeval\endcsname{78.7}
\expandafter\def\csname ev@historical:variant:intern/bio/A1-h0s1/0M/mmlu\endcsname{78.1}
\expandafter\def\csname ev@historical:variant:intern/bio/A1-h0s1/0M/wmdp\endcsname{57.0}
\expandafter\def\csname ev@historical:variant:intern/bio/F1-h0s1/0M/mmlu\endcsname{81.4}
\expandafter\def\csname ev@historical:variant:intern/bio/F1-h0s1/0M/wmdp\endcsname{81.3}
\expandafter\def\csname ev@historical:variant:intern/bio/F3-h0s1/250M/ifeval\endcsname{79.5}
\expandafter\def\csname ev@historical:variant:intern/bio/A3-h0s1/250M/ifeval\endcsname{78.0}
\expandafter\def\csname ev@historical:variant:intern/bio/A3-direct-h0s1/250M/ifeval\endcsname{78.4}
\expandafter\def\csname ev@historical:variant:intern/bio/F1-gradient_mask_only-h0s1/0M/mmlu\endcsname{85.1}
\expandafter\def\csname ev@historical:variant:intern/bio/F1-gradient_mask_only-h0s1/0M/wmdp\endcsname{84.1}
\expandafter\def\csname ev@historical:variant:intern/bio/A1-gradient_mask_only-h0s1/0M/mmlu\endcsname{83.2}
\expandafter\def\csname ev@historical:variant:intern/bio/A1-gradient_mask_only-h0s1/0M/wmdp\endcsname{68.1}
\expandafter\def\csname ev@historical:variant:intern/bio/F1-without_std_activation_mask-h0s1/0M/mmlu\endcsname{81.7}
\expandafter\def\csname ev@historical:variant:intern/bio/F1-without_std_activation_mask-h0s1/0M/wmdp\endcsname{82.1}
\expandafter\def\csname ev@historical:variant:intern/bio/A1-without_std_activation_mask-h0s1/0M/mmlu\endcsname{79.8}
\expandafter\def\csname ev@historical:variant:intern/bio/A1-without_std_activation_mask-h0s1/0M/wmdp\endcsname{62.7}
\expandafter\def\csname ev@historical:variant:intern/bio/F1-without_routing_loss-h0s1/0M/mmlu\endcsname{85.0}
\expandafter\def\csname ev@historical:variant:intern/bio/F1-without_routing_loss-h0s1/0M/wmdp\endcsname{83.5}
\expandafter\def\csname ev@historical:variant:intern/bio/A1-without_routing_loss-h0s1/0M/mmlu\endcsname{83.7}
\expandafter\def\csname ev@historical:variant:intern/bio/A1-without_routing_loss-h0s1/0M/wmdp\endcsname{70.2}
\expandafter\def\csname ev@historical:intern/bio/S_std(B)/10M/wmdp\endcsname{84.1}
\expandafter\def\csname ev@historical:intern/bio/S_std(B)/10M/mmlu\endcsname{85.1}
\expandafter\def\csname ev@historical:intern/bio/S_std(B)/50M/wmdp\endcsname{84.0}
\expandafter\def\csname ev@historical:intern/bio/S_std(B)/50M/mmlu\endcsname{85.1}
\expandafter\def\csname ev@historical:intern/bio/S_std(B)/100M/wmdp\endcsname{84.3}
\expandafter\def\csname ev@historical:intern/bio/S_std(B)/100M/mmlu\endcsname{84.9}
\expandafter\def\csname ev@historical:intern/bio/S_std(B)/250M/wmdp\endcsname{84.3}
\expandafter\def\csname ev@historical:intern/bio/S_std(B)/250M/mmlu\endcsname{84.8}
\expandafter\def\csname ev@historical:intern/bio/A2-masked/0M/wmdp\endcsname{29.5}
\expandafter\def\csname ev@historical:intern/bio/A2-masked/0M/mmlu\endcsname{80.9}
\expandafter\def\csname ev@historical:intern/bio/F3/10M/wmdp\endcsname{63.5}
\expandafter\def\csname ev@historical:intern/bio/F3/10M/mmlu\endcsname{82.3}
\expandafter\def\csname ev@historical:intern/bio/A3/10M/wmdp\endcsname{29.0}
\expandafter\def\csname ev@historical:intern/bio/A3/10M/mmlu\endcsname{80.9}
\expandafter\def\csname ev@historical:intern/bio/F3/50M/wmdp\endcsname{74.1}
\expandafter\def\csname ev@historical:intern/bio/F3/50M/mmlu\endcsname{83.4}
\expandafter\def\csname ev@historical:intern/bio/A3/50M/wmdp\endcsname{37.3}
\expandafter\def\csname ev@historical:intern/bio/A3/50M/mmlu\endcsname{82.1}
\expandafter\def\csname ev@historical:intern/bio/F3/100M/wmdp\endcsname{77.1}
\expandafter\def\csname ev@historical:intern/bio/F3/100M/mmlu\endcsname{84.0}
\expandafter\def\csname ev@historical:intern/bio/A3/100M/wmdp\endcsname{50.0}
\expandafter\def\csname ev@historical:intern/bio/A3/100M/mmlu\endcsname{82.6}
\expandafter\def\csname ev@historical:intern/bio/F3/250M/wmdp\endcsname{78.6}
\expandafter\def\csname ev@historical:intern/bio/F3/250M/mmlu\endcsname{84.3}
\expandafter\def\csname ev@historical:intern/bio/A3/250M/wmdp\endcsname{55.1}
\expandafter\def\csname ev@historical:intern/bio/A3/250M/mmlu\endcsname{82.9}
\expandafter\def\csname ev@historical:intern/bio/A3-direct/10M/wmdp\endcsname{29.8}
\expandafter\def\csname ev@historical:intern/bio/A3-direct/10M/mmlu\endcsname{81.1}
\expandafter\def\csname ev@historical:intern/bio/A3-direct/50M/wmdp\endcsname{42.3}
\expandafter\def\csname ev@historical:intern/bio/A3-direct/50M/mmlu\endcsname{82.0}
\expandafter\def\csname ev@historical:intern/bio/A3-direct/100M/wmdp\endcsname{58.3}
\expandafter\def\csname ev@historical:intern/bio/A3-direct/100M/mmlu\endcsname{83.0}
\expandafter\def\csname ev@historical:intern/bio/A3-direct/250M/wmdp\endcsname{64.2}
\expandafter\def\csname ev@historical:intern/bio/A3-direct/250M/mmlu\endcsname{83.1}
\expandafter\def\csname ev@historical:intern/bio/R-SFT/10M/wmdp\endcsname{33.7}
\expandafter\def\csname ev@historical:intern/bio/R-SFT/10M/mmlu\endcsname{84.5}
\expandafter\def\csname ev@historical:intern/bio/R-SFT/50M/wmdp\endcsname{52.7}
\expandafter\def\csname ev@historical:intern/bio/R-SFT/50M/mmlu\endcsname{85.0}
\expandafter\def\csname ev@historical:intern/bio/R-SFT/100M/wmdp\endcsname{76.2}
\expandafter\def\csname ev@historical:intern/bio/R-SFT/100M/mmlu\endcsname{84.9}
\expandafter\def\csname ev@historical:intern/bio/R-SFT/250M/wmdp\endcsname{80.4}
\expandafter\def\csname ev@historical:intern/bio/R-SFT/250M/mmlu\endcsname{84.9}
\expandafter\def\csname ev@historical:intern/bio/PDU-SFT/10M/wmdp\endcsname{55.5}
\expandafter\def\csname ev@historical:intern/bio/PDU-SFT/10M/mmlu\endcsname{83.6}
\expandafter\def\csname ev@historical:intern/bio/PDU-SFT/50M/wmdp\endcsname{77.0}
\expandafter\def\csname ev@historical:intern/bio/PDU-SFT/50M/mmlu\endcsname{84.8}
\expandafter\def\csname ev@historical:intern/bio/PDU-SFT/100M/wmdp\endcsname{81.8}
\expandafter\def\csname ev@historical:intern/bio/PDU-SFT/100M/mmlu\endcsname{84.9}
\expandafter\def\csname ev@historical:intern/bio/PDU-SFT/250M/wmdp\endcsname{83.0}
\expandafter\def\csname ev@historical:intern/bio/PDU-SFT/250M/mmlu\endcsname{84.7}
\expandafter\def\csname ev@historical:intern/bio/CEU-SFT/10M/wmdp\endcsname{70.4}
\expandafter\def\csname ev@historical:intern/bio/CEU-SFT/10M/mmlu\endcsname{75.9}
\expandafter\def\csname ev@historical:intern/bio/CEU-SFT/50M/wmdp\endcsname{81.4}
\expandafter\def\csname ev@historical:intern/bio/CEU-SFT/50M/mmlu\endcsname{83.6}
\expandafter\def\csname ev@historical:intern/bio/CEU-SFT/100M/wmdp\endcsname{83.3}
\expandafter\def\csname ev@historical:intern/bio/CEU-SFT/100M/mmlu\endcsname{84.5}
\expandafter\def\csname ev@historical:intern/bio/CEU-SFT/250M/wmdp\endcsname{84.3}
\expandafter\def\csname ev@historical:intern/bio/CEU-SFT/250M/mmlu\endcsname{84.7}
\expandafter\def\csname ev@historical:intern/bio/GA-SFT/10M/wmdp\endcsname{24.7}
\expandafter\def\csname ev@historical:intern/bio/GA-SFT/10M/mmlu\endcsname{22.9}
\expandafter\def\csname ev@historical:intern/bio/GA-SFT/50M/wmdp\endcsname{24.7}
\expandafter\def\csname ev@historical:intern/bio/GA-SFT/50M/mmlu\endcsname{24.0}
\expandafter\def\csname ev@historical:intern/bio/GA-SFT/100M/wmdp\endcsname{73.9}
\expandafter\def\csname ev@historical:intern/bio/GA-SFT/100M/mmlu\endcsname{48.3}
\expandafter\def\csname ev@historical:intern/bio/GA-SFT/250M/wmdp\endcsname{48.5}
\expandafter\def\csname ev@historical:intern/bio/GA-SFT/250M/mmlu\endcsname{46.3}
\expandafter\def\csname ev@historical:intern/bio/GD-SFT/10M/wmdp\endcsname{67.9}
\expandafter\def\csname ev@historical:intern/bio/GD-SFT/10M/mmlu\endcsname{82.4}
\expandafter\def\csname ev@historical:intern/bio/GD-SFT/50M/wmdp\endcsname{78.8}
\expandafter\def\csname ev@historical:intern/bio/GD-SFT/50M/mmlu\endcsname{84.1}
\expandafter\def\csname ev@historical:intern/bio/GD-SFT/100M/wmdp\endcsname{83.2}
\expandafter\def\csname ev@historical:intern/bio/GD-SFT/100M/mmlu\endcsname{84.5}
\expandafter\def\csname ev@historical:intern/bio/GD-SFT/250M/wmdp\endcsname{83.1}
\expandafter\def\csname ev@historical:intern/bio/GD-SFT/250M/mmlu\endcsname{84.1}
\expandafter\def\csname ev@historical:intern/bio/PureNPO-SFT/10M/wmdp\endcsname{76.7}
\expandafter\def\csname ev@historical:intern/bio/PureNPO-SFT/10M/mmlu\endcsname{78.6}
\expandafter\def\csname ev@historical:intern/bio/PureNPO-SFT/50M/wmdp\endcsname{77.1}
\expandafter\def\csname ev@historical:intern/bio/PureNPO-SFT/50M/mmlu\endcsname{79.2}
\expandafter\def\csname ev@historical:intern/bio/PureNPO-SFT/100M/wmdp\endcsname{76.9}
\expandafter\def\csname ev@historical:intern/bio/PureNPO-SFT/100M/mmlu\endcsname{80.0}
\expandafter\def\csname ev@historical:intern/bio/PureNPO-SFT/250M/wmdp\endcsname{77.9}
\expandafter\def\csname ev@historical:intern/bio/PureNPO-SFT/250M/mmlu\endcsname{80.5}
\expandafter\def\csname ev@historical:intern/bio/ExternalRMU-SFT/10M/wmdp\endcsname{36.6}
\expandafter\def\csname ev@historical:intern/bio/ExternalRMU-SFT/10M/mmlu\endcsname{84.3}
\expandafter\def\csname ev@historical:intern/bio/ExternalRMU-SFT/50M/wmdp\endcsname{62.2}
\expandafter\def\csname ev@historical:intern/bio/ExternalRMU-SFT/50M/mmlu\endcsname{84.6}
\expandafter\def\csname ev@historical:intern/bio/ExternalRMU-SFT/100M/wmdp\endcsname{80.3}
\expandafter\def\csname ev@historical:intern/bio/ExternalRMU-SFT/100M/mmlu\endcsname{84.7}
\expandafter\def\csname ev@historical:intern/bio/ExternalRMU-SFT/250M/wmdp\endcsname{81.9}
\expandafter\def\csname ev@historical:intern/bio/ExternalRMU-SFT/250M/mmlu\endcsname{84.6}
\expandafter\def\csname ev@historical:intern/bio/SatImp-SFT/10M/wmdp\endcsname{74.0}
\expandafter\def\csname ev@historical:intern/bio/SatImp-SFT/10M/mmlu\endcsname{77.9}
\expandafter\def\csname ev@historical:intern/bio/SatImp-SFT/50M/wmdp\endcsname{82.0}
\expandafter\def\csname ev@historical:intern/bio/SatImp-SFT/50M/mmlu\endcsname{84.0}
\expandafter\def\csname ev@historical:intern/bio/SatImp-SFT/100M/wmdp\endcsname{83.8}
\expandafter\def\csname ev@historical:intern/bio/SatImp-SFT/100M/mmlu\endcsname{84.3}
\expandafter\def\csname ev@historical:intern/bio/SatImp-SFT/250M/wmdp\endcsname{83.7}
\expandafter\def\csname ev@historical:intern/bio/SatImp-SFT/250M/mmlu\endcsname{84.3}
\expandafter\def\csname ev@historical:intern/bio/SimNPO-SFT/10M/wmdp\endcsname{74.7}
\expandafter\def\csname ev@historical:intern/bio/SimNPO-SFT/10M/mmlu\endcsname{77.7}
\expandafter\def\csname ev@historical:intern/bio/SimNPO-SFT/50M/wmdp\endcsname{74.9}
\expandafter\def\csname ev@historical:intern/bio/SimNPO-SFT/50M/mmlu\endcsname{77.9}
\expandafter\def\csname ev@historical:intern/bio/SimNPO-SFT/100M/wmdp\endcsname{75.3}
\expandafter\def\csname ev@historical:intern/bio/SimNPO-SFT/100M/mmlu\endcsname{79.0}
\expandafter\def\csname ev@historical:intern/bio/SimNPO-SFT/250M/wmdp\endcsname{75.7}
\expandafter\def\csname ev@historical:intern/bio/SimNPO-SFT/250M/mmlu\endcsname{79.8}
\expandafter\def\csname ev@historical:intern/bio/TRACE-SFT/10M/wmdp\endcsname{34.2}
\expandafter\def\csname ev@historical:intern/bio/TRACE-SFT/10M/mmlu\endcsname{84.1}
\expandafter\def\csname ev@historical:intern/bio/TRACE-SFT/50M/wmdp\endcsname{69.4}
\expandafter\def\csname ev@historical:intern/bio/TRACE-SFT/50M/mmlu\endcsname{85.0}
\expandafter\def\csname ev@historical:intern/bio/TRACE-SFT/100M/wmdp\endcsname{79.7}
\expandafter\def\csname ev@historical:intern/bio/TRACE-SFT/100M/mmlu\endcsname{85.1}
\expandafter\def\csname ev@historical:intern/bio/TRACE-SFT/250M/wmdp\endcsname{81.1}
\expandafter\def\csname ev@historical:intern/bio/TRACE-SFT/250M/mmlu\endcsname{84.7}
\expandafter\def\csname ev@historical:intern/bio/WGA-SFT/10M/wmdp\endcsname{74.3}
\expandafter\def\csname ev@historical:intern/bio/WGA-SFT/10M/mmlu\endcsname{77.0}
\expandafter\def\csname ev@historical:intern/bio/WGA-SFT/50M/wmdp\endcsname{81.5}
\expandafter\def\csname ev@historical:intern/bio/WGA-SFT/50M/mmlu\endcsname{83.9}
\expandafter\def\csname ev@historical:intern/bio/WGA-SFT/100M/wmdp\endcsname{83.5}
\expandafter\def\csname ev@historical:intern/bio/WGA-SFT/100M/mmlu\endcsname{84.3}
\expandafter\def\csname ev@historical:intern/bio/WGA-SFT/250M/wmdp\endcsname{83.7}
\expandafter\def\csname ev@historical:intern/bio/WGA-SFT/250M/mmlu\endcsname{84.3}
\expandafter\def\csname ev@historical:intern/bio/Routing-F0/0M/routing_z_forget\endcsname{22.9}
\expandafter\def\csname ev@historical:intern/bio/Routing-F0/0M/routing_z_forget_layer_0\endcsname{10.5}
\expandafter\def\csname ev@historical:intern/bio/Routing-F0/0M/routing_z_forget_layer_1\endcsname{12.4}
\expandafter\def\csname ev@historical:intern/bio/Routing-F0/0M/routing_z_forget_layer_2\endcsname{6.1}
\expandafter\def\csname ev@historical:intern/bio/Routing-F0/0M/routing_z_forget_layer_3\endcsname{18.6}
\expandafter\def\csname ev@historical:intern/bio/Routing-F0/0M/routing_z_forget_layer_4\endcsname{25.1}
\expandafter\def\csname ev@historical:intern/bio/Routing-F0/0M/routing_z_forget_layer_5\endcsname{26.1}
\expandafter\def\csname ev@historical:intern/bio/Routing-F0/0M/routing_z_forget_layer_6\endcsname{19.1}
\expandafter\def\csname ev@historical:intern/bio/Routing-F0/0M/routing_z_forget_layer_7\endcsname{24.4}
\expandafter\def\csname ev@historical:intern/bio/Routing-F0/0M/routing_z_forget_layer_8\endcsname{31.1}
\expandafter\def\csname ev@historical:intern/bio/Routing-F0/0M/routing_z_forget_layer_9\endcsname{28.4}
\expandafter\def\csname ev@historical:intern/bio/Routing-F0/0M/routing_z_forget_layer_10\endcsname{16.5}
\expandafter\def\csname ev@historical:intern/bio/Routing-F0/0M/routing_z_forget_layer_11\endcsname{16.7}
\expandafter\def\csname ev@historical:intern/bio/Routing-F0/0M/routing_z_forget_layer_12\endcsname{26.1}
\expandafter\def\csname ev@historical:intern/bio/Routing-F0/0M/routing_z_forget_layer_13\endcsname{25.5}
\expandafter\def\csname ev@historical:intern/bio/Routing-F0/0M/routing_z_forget_layer_14\endcsname{26.5}
\expandafter\def\csname ev@historical:intern/bio/Routing-F0/0M/routing_z_forget_layer_15\endcsname{27.0}
\expandafter\def\csname ev@historical:intern/bio/Routing-F0/0M/routing_z_forget_layer_16\endcsname{17.3}
\expandafter\def\csname ev@historical:intern/bio/Routing-F0/0M/routing_z_forget_layer_17\endcsname{25.5}
\expandafter\def\csname ev@historical:intern/bio/Routing-F0/0M/routing_z_forget_layer_18\endcsname{27.3}
\expandafter\def\csname ev@historical:intern/bio/Routing-F0/0M/routing_z_forget_layer_19\endcsname{23.5}
\expandafter\def\csname ev@historical:intern/bio/Routing-F0/0M/routing_z_forget_layer_20\endcsname{30.7}
\expandafter\def\csname ev@historical:intern/bio/Routing-F0/0M/routing_z_forget_layer_21\endcsname{30.3}
\expandafter\def\csname ev@historical:intern/bio/Routing-F0/0M/routing_z_forget_layer_22\endcsname{18.1}
\expandafter\def\csname ev@historical:intern/bio/Routing-F0/0M/routing_z_forget_layer_23\endcsname{17.0}
\expandafter\def\csname ev@historical:intern/bio/Routing-F0/0M/routing_z_forget_layer_24\endcsname{23.7}
\expandafter\def\csname ev@historical:intern/bio/Routing-F0/0M/routing_z_forget_layer_25\endcsname{23.8}
\expandafter\def\csname ev@historical:intern/bio/Routing-F0/0M/routing_z_forget_layer_26\endcsname{24.4}
\expandafter\def\csname ev@historical:intern/bio/Routing-F0/0M/routing_z_forget_layer_27\endcsname{25.7}
\expandafter\def\csname ev@historical:intern/bio/Routing-F0/0M/routing_z_forget_layer_28\endcsname{16.4}
\expandafter\def\csname ev@historical:intern/bio/Routing-F0/0M/routing_z_forget_layer_29\endcsname{25.5}
\expandafter\def\csname ev@historical:intern/bio/Routing-F0/0M/routing_z_forget_layer_30\endcsname{27.6}
\expandafter\def\csname ev@historical:intern/bio/Routing-F0/0M/routing_z_forget_layer_31\endcsname{22.5}
\expandafter\def\csname ev@historical:intern/bio/Routing-F0/0M/routing_z_forget_layer_32\endcsname{19.7}
\expandafter\def\csname ev@historical:intern/bio/Routing-F0/0M/routing_z_forget_layer_33\endcsname{22.6}
\expandafter\def\csname ev@historical:intern/bio/Routing-F0/0M/routing_z_forget_layer_34\endcsname{18.6}
\expandafter\def\csname ev@historical:intern/bio/Routing-F0/0M/routing_z_forget_layer_35\endcsname{30.3}
\expandafter\def\csname ev@historical:intern/bio/Routing-F0/0M/routing_z_forget_layer_36\endcsname{27.5}
\expandafter\def\csname ev@historical:intern/bio/Routing-F0/0M/routing_z_forget_layer_37\endcsname{26.9}
\expandafter\def\csname ev@historical:intern/bio/Routing-F0/0M/routing_z_forget_layer_38\endcsname{28.1}
\expandafter\def\csname ev@historical:intern/bio/Routing-F0/0M/routing_z_forget_layer_39\endcsname{21.5}
\expandafter\def\csname ev@historical:intern/bio/Routing-F0/0M/routing_z_retain\endcsname{3.7}
\expandafter\def\csname ev@historical:intern/bio/Routing-F0/0M/routing_z_retain_layer_0\endcsname{2.0}
\expandafter\def\csname ev@historical:intern/bio/Routing-F0/0M/routing_z_retain_layer_1\endcsname{1.8}
\expandafter\def\csname ev@historical:intern/bio/Routing-F0/0M/routing_z_retain_layer_2\endcsname{2.1}
\expandafter\def\csname ev@historical:intern/bio/Routing-F0/0M/routing_z_retain_layer_3\endcsname{3.2}
\expandafter\def\csname ev@historical:intern/bio/Routing-F0/0M/routing_z_retain_layer_4\endcsname{3.8}
\expandafter\def\csname ev@historical:intern/bio/Routing-F0/0M/routing_z_retain_layer_5\endcsname{3.8}
\expandafter\def\csname ev@historical:intern/bio/Routing-F0/0M/routing_z_retain_layer_6\endcsname{3.4}
\expandafter\def\csname ev@historical:intern/bio/Routing-F0/0M/routing_z_retain_layer_7\endcsname{3.8}
\expandafter\def\csname ev@historical:intern/bio/Routing-F0/0M/routing_z_retain_layer_8\endcsname{3.1}
\expandafter\def\csname ev@historical:intern/bio/Routing-F0/0M/routing_z_retain_layer_9\endcsname{3.0}
\expandafter\def\csname ev@historical:intern/bio/Routing-F0/0M/routing_z_retain_layer_10\endcsname{3.9}
\expandafter\def\csname ev@historical:intern/bio/Routing-F0/0M/routing_z_retain_layer_11\endcsname{3.2}
\expandafter\def\csname ev@historical:intern/bio/Routing-F0/0M/routing_z_retain_layer_12\endcsname{3.7}
\expandafter\def\csname ev@historical:intern/bio/Routing-F0/0M/routing_z_retain_layer_13\endcsname{3.9}
\expandafter\def\csname ev@historical:intern/bio/Routing-F0/0M/routing_z_retain_layer_14\endcsname{4.6}
\expandafter\def\csname ev@historical:intern/bio/Routing-F0/0M/routing_z_retain_layer_15\endcsname{3.8}
\expandafter\def\csname ev@historical:intern/bio/Routing-F0/0M/routing_z_retain_layer_16\endcsname{2.6}
\expandafter\def\csname ev@historical:intern/bio/Routing-F0/0M/routing_z_retain_layer_17\endcsname{4.1}
\expandafter\def\csname ev@historical:intern/bio/Routing-F0/0M/routing_z_retain_layer_18\endcsname{4.5}
\expandafter\def\csname ev@historical:intern/bio/Routing-F0/0M/routing_z_retain_layer_19\endcsname{5.3}
\expandafter\def\csname ev@historical:intern/bio/Routing-F0/0M/routing_z_retain_layer_20\endcsname{3.9}
\expandafter\def\csname ev@historical:intern/bio/Routing-F0/0M/routing_z_retain_layer_21\endcsname{3.7}
\expandafter\def\csname ev@historical:intern/bio/Routing-F0/0M/routing_z_retain_layer_22\endcsname{3.7}
\expandafter\def\csname ev@historical:intern/bio/Routing-F0/0M/routing_z_retain_layer_23\endcsname{3.0}
\expandafter\def\csname ev@historical:intern/bio/Routing-F0/0M/routing_z_retain_layer_24\endcsname{3.3}
\expandafter\def\csname ev@historical:intern/bio/Routing-F0/0M/routing_z_retain_layer_25\endcsname{4.1}
\expandafter\def\csname ev@historical:intern/bio/Routing-F0/0M/routing_z_retain_layer_26\endcsname{3.9}
\expandafter\def\csname ev@historical:intern/bio/Routing-F0/0M/routing_z_retain_layer_27\endcsname{4.1}
\expandafter\def\csname ev@historical:intern/bio/Routing-F0/0M/routing_z_retain_layer_28\endcsname{3.2}
\expandafter\def\csname ev@historical:intern/bio/Routing-F0/0M/routing_z_retain_layer_29\endcsname{4.1}
\expandafter\def\csname ev@historical:intern/bio/Routing-F0/0M/routing_z_retain_layer_30\endcsname{4.5}
\expandafter\def\csname ev@historical:intern/bio/Routing-F0/0M/routing_z_retain_layer_31\endcsname{5.1}
\expandafter\def\csname ev@historical:intern/bio/Routing-F0/0M/routing_z_retain_layer_32\endcsname{2.2}
\expandafter\def\csname ev@historical:intern/bio/Routing-F0/0M/routing_z_retain_layer_33\endcsname{2.9}
\expandafter\def\csname ev@historical:intern/bio/Routing-F0/0M/routing_z_retain_layer_34\endcsname{3.7}
\expandafter\def\csname ev@historical:intern/bio/Routing-F0/0M/routing_z_retain_layer_35\endcsname{4.8}
\expandafter\def\csname ev@historical:intern/bio/Routing-F0/0M/routing_z_retain_layer_36\endcsname{4.6}
\expandafter\def\csname ev@historical:intern/bio/Routing-F0/0M/routing_z_retain_layer_37\endcsname{5.2}
\expandafter\def\csname ev@historical:intern/bio/Routing-F0/0M/routing_z_retain_layer_38\endcsname{5.6}
\expandafter\def\csname ev@historical:intern/bio/Routing-F0/0M/routing_z_retain_layer_39\endcsname{3.9}
\expandafter\def\csname ev@historical:intern/bio/Routing-F1/0M/routing_z_forget\endcsname{75.9}
\expandafter\def\csname ev@historical:intern/bio/Routing-F1/0M/routing_z_forget_layer_0\endcsname{1.7}
\expandafter\def\csname ev@historical:intern/bio/Routing-F1/0M/routing_z_forget_layer_1\endcsname{71.5}
\expandafter\def\csname ev@historical:intern/bio/Routing-F1/0M/routing_z_forget_layer_2\endcsname{72.9}
\expandafter\def\csname ev@historical:intern/bio/Routing-F1/0M/routing_z_forget_layer_3\endcsname{86.0}
\expandafter\def\csname ev@historical:intern/bio/Routing-F1/0M/routing_z_forget_layer_4\endcsname{86.0}
\expandafter\def\csname ev@historical:intern/bio/Routing-F1/0M/routing_z_forget_layer_5\endcsname{84.2}
\expandafter\def\csname ev@historical:intern/bio/Routing-F1/0M/routing_z_forget_layer_6\endcsname{80.5}
\expandafter\def\csname ev@historical:intern/bio/Routing-F1/0M/routing_z_forget_layer_7\endcsname{83.8}
\expandafter\def\csname ev@historical:intern/bio/Routing-F1/0M/routing_z_forget_layer_8\endcsname{84.8}
\expandafter\def\csname ev@historical:intern/bio/Routing-F1/0M/routing_z_forget_layer_9\endcsname{82.8}
\expandafter\def\csname ev@historical:intern/bio/Routing-F1/0M/routing_z_forget_layer_10\endcsname{76.9}
\expandafter\def\csname ev@historical:intern/bio/Routing-F1/0M/routing_z_forget_layer_11\endcsname{73.7}
\expandafter\def\csname ev@historical:intern/bio/Routing-F1/0M/routing_z_forget_layer_12\endcsname{82.1}
\expandafter\def\csname ev@historical:intern/bio/Routing-F1/0M/routing_z_forget_layer_13\endcsname{81.1}
\expandafter\def\csname ev@historical:intern/bio/Routing-F1/0M/routing_z_forget_layer_14\endcsname{83.7}
\expandafter\def\csname ev@historical:intern/bio/Routing-F1/0M/routing_z_forget_layer_15\endcsname{84.1}
\expandafter\def\csname ev@historical:intern/bio/Routing-F1/0M/routing_z_forget_layer_16\endcsname{79.4}
\expandafter\def\csname ev@historical:intern/bio/Routing-F1/0M/routing_z_forget_layer_17\endcsname{80.2}
\expandafter\def\csname ev@historical:intern/bio/Routing-F1/0M/routing_z_forget_layer_18\endcsname{82.4}
\expandafter\def\csname ev@historical:intern/bio/Routing-F1/0M/routing_z_forget_layer_19\endcsname{80.4}
\expandafter\def\csname ev@historical:intern/bio/Routing-F1/0M/routing_z_forget_layer_20\endcsname{85.4}
\expandafter\def\csname ev@historical:intern/bio/Routing-F1/0M/routing_z_forget_layer_21\endcsname{83.6}
\expandafter\def\csname ev@historical:intern/bio/Routing-F1/0M/routing_z_forget_layer_22\endcsname{77.2}
\expandafter\def\csname ev@historical:intern/bio/Routing-F1/0M/routing_z_forget_layer_23\endcsname{74.1}
\expandafter\def\csname ev@historical:intern/bio/Routing-F1/0M/routing_z_forget_layer_24\endcsname{80.5}
\expandafter\def\csname ev@historical:intern/bio/Routing-F1/0M/routing_z_forget_layer_25\endcsname{79.1}
\expandafter\def\csname ev@historical:intern/bio/Routing-F1/0M/routing_z_forget_layer_26\endcsname{79.0}
\expandafter\def\csname ev@historical:intern/bio/Routing-F1/0M/routing_z_forget_layer_27\endcsname{78.5}
\expandafter\def\csname ev@historical:intern/bio/Routing-F1/0M/routing_z_forget_layer_28\endcsname{72.3}
\expandafter\def\csname ev@historical:intern/bio/Routing-F1/0M/routing_z_forget_layer_29\endcsname{72.4}
\expandafter\def\csname ev@historical:intern/bio/Routing-F1/0M/routing_z_forget_layer_30\endcsname{77.3}
\expandafter\def\csname ev@historical:intern/bio/Routing-F1/0M/routing_z_forget_layer_31\endcsname{73.8}
\expandafter\def\csname ev@historical:intern/bio/Routing-F1/0M/routing_z_forget_layer_32\endcsname{66.3}
\expandafter\def\csname ev@historical:intern/bio/Routing-F1/0M/routing_z_forget_layer_33\endcsname{69.1}
\expandafter\def\csname ev@historical:intern/bio/Routing-F1/0M/routing_z_forget_layer_34\endcsname{67.6}
\expandafter\def\csname ev@historical:intern/bio/Routing-F1/0M/routing_z_forget_layer_35\endcsname{80.7}
\expandafter\def\csname ev@historical:intern/bio/Routing-F1/0M/routing_z_forget_layer_36\endcsname{75.5}
\expandafter\def\csname ev@historical:intern/bio/Routing-F1/0M/routing_z_forget_layer_37\endcsname{76.3}
\expandafter\def\csname ev@historical:intern/bio/Routing-F1/0M/routing_z_forget_layer_38\endcsname{74.3}
\expandafter\def\csname ev@historical:intern/bio/Routing-F1/0M/routing_z_forget_layer_39\endcsname{56.7}
\expandafter\def\csname ev@historical:intern/bio/Routing-F1/0M/routing_z_retain\endcsname{20.2}
\expandafter\def\csname ev@historical:intern/bio/Routing-F1/0M/routing_z_retain_layer_0\endcsname{0.1}
\expandafter\def\csname ev@historical:intern/bio/Routing-F1/0M/routing_z_retain_layer_1\endcsname{10.0}
\expandafter\def\csname ev@historical:intern/bio/Routing-F1/0M/routing_z_retain_layer_2\endcsname{10.3}
\expandafter\def\csname ev@historical:intern/bio/Routing-F1/0M/routing_z_retain_layer_3\endcsname{21.0}
\expandafter\def\csname ev@historical:intern/bio/Routing-F1/0M/routing_z_retain_layer_4\endcsname{25.6}
\expandafter\def\csname ev@historical:intern/bio/Routing-F1/0M/routing_z_retain_layer_5\endcsname{25.2}
\expandafter\def\csname ev@historical:intern/bio/Routing-F1/0M/routing_z_retain_layer_6\endcsname{24.7}
\expandafter\def\csname ev@historical:intern/bio/Routing-F1/0M/routing_z_retain_layer_7\endcsname{25.1}
\expandafter\def\csname ev@historical:intern/bio/Routing-F1/0M/routing_z_retain_layer_8\endcsname{24.3}
\expandafter\def\csname ev@historical:intern/bio/Routing-F1/0M/routing_z_retain_layer_9\endcsname{23.1}
\expandafter\def\csname ev@historical:intern/bio/Routing-F1/0M/routing_z_retain_layer_10\endcsname{22.5}
\expandafter\def\csname ev@historical:intern/bio/Routing-F1/0M/routing_z_retain_layer_11\endcsname{21.7}
\expandafter\def\csname ev@historical:intern/bio/Routing-F1/0M/routing_z_retain_layer_12\endcsname{23.6}
\expandafter\def\csname ev@historical:intern/bio/Routing-F1/0M/routing_z_retain_layer_13\endcsname{23.5}
\expandafter\def\csname ev@historical:intern/bio/Routing-F1/0M/routing_z_retain_layer_14\endcsname{23.7}
\expandafter\def\csname ev@historical:intern/bio/Routing-F1/0M/routing_z_retain_layer_15\endcsname{23.4}
\expandafter\def\csname ev@historical:intern/bio/Routing-F1/0M/routing_z_retain_layer_16\endcsname{22.7}
\expandafter\def\csname ev@historical:intern/bio/Routing-F1/0M/routing_z_retain_layer_17\endcsname{22.7}
\expandafter\def\csname ev@historical:intern/bio/Routing-F1/0M/routing_z_retain_layer_18\endcsname{24.1}
\expandafter\def\csname ev@historical:intern/bio/Routing-F1/0M/routing_z_retain_layer_19\endcsname{23.6}
\expandafter\def\csname ev@historical:intern/bio/Routing-F1/0M/routing_z_retain_layer_20\endcsname{24.1}
\expandafter\def\csname ev@historical:intern/bio/Routing-F1/0M/routing_z_retain_layer_21\endcsname{23.6}
\expandafter\def\csname ev@historical:intern/bio/Routing-F1/0M/routing_z_retain_layer_22\endcsname{22.8}
\expandafter\def\csname ev@historical:intern/bio/Routing-F1/0M/routing_z_retain_layer_23\endcsname{22.5}
\expandafter\def\csname ev@historical:intern/bio/Routing-F1/0M/routing_z_retain_layer_24\endcsname{23.4}
\expandafter\def\csname ev@historical:intern/bio/Routing-F1/0M/routing_z_retain_layer_25\endcsname{23.2}
\expandafter\def\csname ev@historical:intern/bio/Routing-F1/0M/routing_z_retain_layer_26\endcsname{22.4}
\expandafter\def\csname ev@historical:intern/bio/Routing-F1/0M/routing_z_retain_layer_27\endcsname{21.7}
\expandafter\def\csname ev@historical:intern/bio/Routing-F1/0M/routing_z_retain_layer_28\endcsname{21.0}
\expandafter\def\csname ev@historical:intern/bio/Routing-F1/0M/routing_z_retain_layer_29\endcsname{20.2}
\expandafter\def\csname ev@historical:intern/bio/Routing-F1/0M/routing_z_retain_layer_30\endcsname{21.2}
\expandafter\def\csname ev@historical:intern/bio/Routing-F1/0M/routing_z_retain_layer_31\endcsname{20.3}
\expandafter\def\csname ev@historical:intern/bio/Routing-F1/0M/routing_z_retain_layer_32\endcsname{17.4}
\expandafter\def\csname ev@historical:intern/bio/Routing-F1/0M/routing_z_retain_layer_33\endcsname{16.4}
\expandafter\def\csname ev@historical:intern/bio/Routing-F1/0M/routing_z_retain_layer_34\endcsname{16.4}
\expandafter\def\csname ev@historical:intern/bio/Routing-F1/0M/routing_z_retain_layer_35\endcsname{18.0}
\expandafter\def\csname ev@historical:intern/bio/Routing-F1/0M/routing_z_retain_layer_36\endcsname{16.3}
\expandafter\def\csname ev@historical:intern/bio/Routing-F1/0M/routing_z_retain_layer_37\endcsname{15.3}
\expandafter\def\csname ev@historical:intern/bio/Routing-F1/0M/routing_z_retain_layer_38\endcsname{13.4}
\expandafter\def\csname ev@historical:intern/bio/Routing-F1/0M/routing_z_retain_layer_39\endcsname{6.8}
\expandafter\def\csname ev@historical:qwen/bio/B/0M/mmlu\endcsname{82.5}
\expandafter\def\csname ev@historical:qwen/bio/B/0M/wmdp\endcsname{83.8}
\expandafter\def\csname ev@historical:qwen/bio/B/0M/ifeval\endcsname{70.4}
\expandafter\def\csname ev@historical:qwen/bio/S_std(B)/250M/ifeval\endcsname{52.5}
\expandafter\def\csname ev@historical:qwen/bio/F0/0M/mmlu\endcsname{82.3}
\expandafter\def\csname ev@historical:qwen/bio/F0/0M/wmdp\endcsname{82.0}
\expandafter\def\csname ev@historical:qwen/bio/A0/0M/mmlu\endcsname{80.9}
\expandafter\def\csname ev@historical:qwen/bio/A0/0M/wmdp\endcsname{68.7}
\expandafter\def\csname ev@historical:qwen/bio/F1/0M/mmlu\endcsname{80.2}
\expandafter\def\csname ev@historical:qwen/bio/F1/0M/wmdp\endcsname{81.8}
\expandafter\def\csname ev@historical:qwen/bio/A1/0M/mmlu\endcsname{78.6}
\expandafter\def\csname ev@historical:qwen/bio/A1/0M/wmdp\endcsname{63.0}
\expandafter\def\csname ev@historical:qwen/bio/F3/250M/ifeval\endcsname{51.2}
\expandafter\def\csname ev@historical:qwen/bio/A3/250M/ifeval\endcsname{51.4}
\expandafter\def\csname ev@historical:qwen/bio/R/0M/mmlu\endcsname{82.0}
\expandafter\def\csname ev@historical:qwen/bio/R/0M/wmdp\endcsname{41.8}
\expandafter\def\csname ev@historical:qwen/bio/R-SFT/250M/ifeval\endcsname{55.6}
\expandafter\def\csname ev@historical:qwen/bio/A3-direct/250M/ifeval\endcsname{56.6}
\expandafter\def\csname ev@historical:qwen/bio/CEU-SFT/250M/ifeval\endcsname{62.7}
\expandafter\def\csname ev@historical:qwen/bio/GA-SFT/250M/ifeval\endcsname{63.0}
\expandafter\def\csname ev@historical:qwen/bio/GD-SFT/250M/ifeval\endcsname{44.2}
\expandafter\def\csname ev@historical:qwen/bio/PDU-SFT/250M/ifeval\endcsname{61.4}
\expandafter\def\csname ev@historical:qwen/bio/PureNPO-SFT/250M/ifeval\endcsname{42.9}
\expandafter\def\csname ev@historical:qwen/bio/SatImp-SFT/250M/ifeval\endcsname{59.7}
\expandafter\def\csname ev@historical:qwen/bio/SimNPO-SFT/250M/ifeval\endcsname{51.6}
\expandafter\def\csname ev@historical:qwen/bio/TRACE-SFT/250M/ifeval\endcsname{52.5}
\expandafter\def\csname ev@historical:qwen/bio/WGA-SFT/250M/ifeval\endcsname{61.4}
\expandafter\def\csname ev@historical:qwen/bio/S_std(B)/10M/wmdp\endcsname{83.3}
\expandafter\def\csname ev@historical:qwen/bio/S_std(B)/10M/mmlu\endcsname{82.5}
\expandafter\def\csname ev@historical:qwen/bio/S_std(B)/50M/wmdp\endcsname{83.3}
\expandafter\def\csname ev@historical:qwen/bio/S_std(B)/50M/mmlu\endcsname{82.2}
\expandafter\def\csname ev@historical:qwen/bio/S_std(B)/100M/wmdp\endcsname{83.7}
\expandafter\def\csname ev@historical:qwen/bio/S_std(B)/100M/mmlu\endcsname{82.3}
\expandafter\def\csname ev@historical:qwen/bio/S_std(B)/250M/wmdp\endcsname{83.9}
\expandafter\def\csname ev@historical:qwen/bio/S_std(B)/250M/mmlu\endcsname{82.3}
\expandafter\def\csname ev@historical:qwen/bio/A2-masked/0M/wmdp\endcsname{32.9}
\expandafter\def\csname ev@historical:qwen/bio/A2-masked/0M/mmlu\endcsname{77.2}
\expandafter\def\csname ev@historical:qwen/bio/F3/10M/wmdp\endcsname{81.0}
\expandafter\def\csname ev@historical:qwen/bio/F3/10M/mmlu\endcsname{80.0}
\expandafter\def\csname ev@historical:qwen/bio/A3/10M/wmdp\endcsname{34.0}
\expandafter\def\csname ev@historical:qwen/bio/A3/10M/mmlu\endcsname{77.3}
\expandafter\def\csname ev@historical:qwen/bio/F3/50M/wmdp\endcsname{80.4}
\expandafter\def\csname ev@historical:qwen/bio/F3/50M/mmlu\endcsname{80.1}
\expandafter\def\csname ev@historical:qwen/bio/A3/50M/wmdp\endcsname{34.6}
\expandafter\def\csname ev@historical:qwen/bio/A3/50M/mmlu\endcsname{77.7}
\expandafter\def\csname ev@historical:qwen/bio/F3/100M/wmdp\endcsname{81.5}
\expandafter\def\csname ev@historical:qwen/bio/F3/100M/mmlu\endcsname{80.5}
\expandafter\def\csname ev@historical:qwen/bio/A3/100M/wmdp\endcsname{39.7}
\expandafter\def\csname ev@historical:qwen/bio/A3/100M/mmlu\endcsname{78.2}
\expandafter\def\csname ev@historical:qwen/bio/F3/250M/wmdp\endcsname{82.1}
\expandafter\def\csname ev@historical:qwen/bio/F3/250M/mmlu\endcsname{80.8}
\expandafter\def\csname ev@historical:qwen/bio/A3/250M/wmdp\endcsname{43.4}
\expandafter\def\csname ev@historical:qwen/bio/A3/250M/mmlu\endcsname{78.5}
\expandafter\def\csname ev@historical:qwen/bio/A3-direct/10M/wmdp\endcsname{33.9}
\expandafter\def\csname ev@historical:qwen/bio/A3-direct/10M/mmlu\endcsname{77.3}
\expandafter\def\csname ev@historical:qwen/bio/A3-direct/50M/wmdp\endcsname{50.4}
\expandafter\def\csname ev@historical:qwen/bio/A3-direct/50M/mmlu\endcsname{78.3}
\expandafter\def\csname ev@historical:qwen/bio/A3-direct/100M/wmdp\endcsname{58.8}
\expandafter\def\csname ev@historical:qwen/bio/A3-direct/100M/mmlu\endcsname{78.9}
\expandafter\def\csname ev@historical:qwen/bio/A3-direct/250M/wmdp\endcsname{62.6}
\expandafter\def\csname ev@historical:qwen/bio/A3-direct/250M/mmlu\endcsname{79.4}
\expandafter\def\csname ev@historical:qwen/bio/R-SFT/10M/wmdp\endcsname{40.9}
\expandafter\def\csname ev@historical:qwen/bio/R-SFT/10M/mmlu\endcsname{82.1}
\expandafter\def\csname ev@historical:qwen/bio/R-SFT/50M/wmdp\endcsname{60.7}
\expandafter\def\csname ev@historical:qwen/bio/R-SFT/50M/mmlu\endcsname{82.0}
\expandafter\def\csname ev@historical:qwen/bio/R-SFT/100M/wmdp\endcsname{75.3}
\expandafter\def\csname ev@historical:qwen/bio/R-SFT/100M/mmlu\endcsname{82.1}
\expandafter\def\csname ev@historical:qwen/bio/R-SFT/250M/wmdp\endcsname{78.2}
\expandafter\def\csname ev@historical:qwen/bio/R-SFT/250M/mmlu\endcsname{82.3}
\expandafter\def\csname ev@historical:qwen/bio/CEU-SFT/10M/wmdp\endcsname{81.2}
\expandafter\def\csname ev@historical:qwen/bio/CEU-SFT/10M/mmlu\endcsname{81.1}
\expandafter\def\csname ev@historical:qwen/bio/CEU-SFT/50M/wmdp\endcsname{82.4}
\expandafter\def\csname ev@historical:qwen/bio/CEU-SFT/50M/mmlu\endcsname{81.9}
\expandafter\def\csname ev@historical:qwen/bio/CEU-SFT/100M/wmdp\endcsname{83.0}
\expandafter\def\csname ev@historical:qwen/bio/CEU-SFT/100M/mmlu\endcsname{82.0}
\expandafter\def\csname ev@historical:qwen/bio/CEU-SFT/250M/wmdp\endcsname{83.5}
\expandafter\def\csname ev@historical:qwen/bio/CEU-SFT/250M/mmlu\endcsname{81.9}
\expandafter\def\csname ev@historical:qwen/bio/GA-SFT/10M/wmdp\endcsname{24.0}
\expandafter\def\csname ev@historical:qwen/bio/GA-SFT/10M/mmlu\endcsname{28.3}
\expandafter\def\csname ev@historical:qwen/bio/GA-SFT/50M/wmdp\endcsname{57.3}
\expandafter\def\csname ev@historical:qwen/bio/GA-SFT/50M/mmlu\endcsname{64.8}
\expandafter\def\csname ev@historical:qwen/bio/GA-SFT/100M/wmdp\endcsname{80.4}
\expandafter\def\csname ev@historical:qwen/bio/GA-SFT/100M/mmlu\endcsname{79.8}
\expandafter\def\csname ev@historical:qwen/bio/GA-SFT/250M/wmdp\endcsname{81.8}
\expandafter\def\csname ev@historical:qwen/bio/GA-SFT/250M/mmlu\endcsname{80.5}
\expandafter\def\csname ev@historical:qwen/bio/GD-SFT/10M/wmdp\endcsname{81.2}
\expandafter\def\csname ev@historical:qwen/bio/GD-SFT/10M/mmlu\endcsname{81.4}
\expandafter\def\csname ev@historical:qwen/bio/GD-SFT/50M/wmdp\endcsname{81.2}
\expandafter\def\csname ev@historical:qwen/bio/GD-SFT/50M/mmlu\endcsname{81.5}
\expandafter\def\csname ev@historical:qwen/bio/GD-SFT/100M/wmdp\endcsname{81.7}
\expandafter\def\csname ev@historical:qwen/bio/GD-SFT/100M/mmlu\endcsname{81.5}
\expandafter\def\csname ev@historical:qwen/bio/GD-SFT/250M/wmdp\endcsname{81.9}
\expandafter\def\csname ev@historical:qwen/bio/GD-SFT/250M/mmlu\endcsname{81.7}
\expandafter\def\csname ev@historical:qwen/bio/PDU-SFT/10M/wmdp\endcsname{50.6}
\expandafter\def\csname ev@historical:qwen/bio/PDU-SFT/10M/mmlu\endcsname{78.5}
\expandafter\def\csname ev@historical:qwen/bio/PDU-SFT/50M/wmdp\endcsname{78.8}
\expandafter\def\csname ev@historical:qwen/bio/PDU-SFT/50M/mmlu\endcsname{80.8}
\expandafter\def\csname ev@historical:qwen/bio/PDU-SFT/100M/wmdp\endcsname{81.7}
\expandafter\def\csname ev@historical:qwen/bio/PDU-SFT/100M/mmlu\endcsname{81.4}
\expandafter\def\csname ev@historical:qwen/bio/PDU-SFT/250M/wmdp\endcsname{81.5}
\expandafter\def\csname ev@historical:qwen/bio/PDU-SFT/250M/mmlu\endcsname{81.4}
\expandafter\def\csname ev@historical:qwen/bio/PureNPO-SFT/10M/wmdp\endcsname{76.7}
\expandafter\def\csname ev@historical:qwen/bio/PureNPO-SFT/10M/mmlu\endcsname{77.5}
\expandafter\def\csname ev@historical:qwen/bio/PureNPO-SFT/50M/wmdp\endcsname{76.7}
\expandafter\def\csname ev@historical:qwen/bio/PureNPO-SFT/50M/mmlu\endcsname{77.9}
\expandafter\def\csname ev@historical:qwen/bio/PureNPO-SFT/100M/wmdp\endcsname{76.1}
\expandafter\def\csname ev@historical:qwen/bio/PureNPO-SFT/100M/mmlu\endcsname{78.2}
\expandafter\def\csname ev@historical:qwen/bio/PureNPO-SFT/250M/wmdp\endcsname{75.7}
\expandafter\def\csname ev@historical:qwen/bio/PureNPO-SFT/250M/mmlu\endcsname{78.1}
\expandafter\def\csname ev@historical:qwen/bio/ExternalRMU-SFT/10M/wmdp\endcsname{44.1}
\expandafter\def\csname ev@historical:qwen/bio/ExternalRMU-SFT/10M/mmlu\endcsname{82.2}
\expandafter\def\csname ev@historical:qwen/bio/ExternalRMU-SFT/50M/wmdp\endcsname{64.1}
\expandafter\def\csname ev@historical:qwen/bio/ExternalRMU-SFT/50M/mmlu\endcsname{82.2}
\expandafter\def\csname ev@historical:qwen/bio/ExternalRMU-SFT/100M/wmdp\endcsname{76.0}
\expandafter\def\csname ev@historical:qwen/bio/ExternalRMU-SFT/100M/mmlu\endcsname{82.3}
\expandafter\def\csname ev@historical:qwen/bio/ExternalRMU-SFT/250M/wmdp\endcsname{78.4}
\expandafter\def\csname ev@historical:qwen/bio/ExternalRMU-SFT/250M/mmlu\endcsname{82.0}
\expandafter\def\csname ev@historical:qwen/bio/SatImp-SFT/10M/wmdp\endcsname{80.6}
\expandafter\def\csname ev@historical:qwen/bio/SatImp-SFT/10M/mmlu\endcsname{80.7}
\expandafter\def\csname ev@historical:qwen/bio/SatImp-SFT/50M/wmdp\endcsname{81.9}
\expandafter\def\csname ev@historical:qwen/bio/SatImp-SFT/50M/mmlu\endcsname{81.8}
\expandafter\def\csname ev@historical:qwen/bio/SatImp-SFT/100M/wmdp\endcsname{82.9}
\expandafter\def\csname ev@historical:qwen/bio/SatImp-SFT/100M/mmlu\endcsname{82.0}
\expandafter\def\csname ev@historical:qwen/bio/SatImp-SFT/250M/wmdp\endcsname{82.9}
\expandafter\def\csname ev@historical:qwen/bio/SatImp-SFT/250M/mmlu\endcsname{82.0}
\expandafter\def\csname ev@historical:qwen/bio/SimNPO-SFT/10M/wmdp\endcsname{76.8}
\expandafter\def\csname ev@historical:qwen/bio/SimNPO-SFT/10M/mmlu\endcsname{77.3}
\expandafter\def\csname ev@historical:qwen/bio/SimNPO-SFT/50M/wmdp\endcsname{77.1}
\expandafter\def\csname ev@historical:qwen/bio/SimNPO-SFT/50M/mmlu\endcsname{77.7}
\expandafter\def\csname ev@historical:qwen/bio/SimNPO-SFT/100M/wmdp\endcsname{77.3}
\expandafter\def\csname ev@historical:qwen/bio/SimNPO-SFT/100M/mmlu\endcsname{78.3}
\expandafter\def\csname ev@historical:qwen/bio/SimNPO-SFT/250M/wmdp\endcsname{77.0}
\expandafter\def\csname ev@historical:qwen/bio/SimNPO-SFT/250M/mmlu\endcsname{78.5}
\expandafter\def\csname ev@historical:qwen/bio/TRACE-SFT/10M/wmdp\endcsname{63.2}
\expandafter\def\csname ev@historical:qwen/bio/TRACE-SFT/10M/mmlu\endcsname{82.0}
\expandafter\def\csname ev@historical:qwen/bio/TRACE-SFT/50M/wmdp\endcsname{70.9}
\expandafter\def\csname ev@historical:qwen/bio/TRACE-SFT/50M/mmlu\endcsname{81.8}
\expandafter\def\csname ev@historical:qwen/bio/TRACE-SFT/100M/wmdp\endcsname{77.0}
\expandafter\def\csname ev@historical:qwen/bio/TRACE-SFT/100M/mmlu\endcsname{82.2}
\expandafter\def\csname ev@historical:qwen/bio/TRACE-SFT/250M/wmdp\endcsname{79.2}
\expandafter\def\csname ev@historical:qwen/bio/TRACE-SFT/250M/mmlu\endcsname{82.1}
\expandafter\def\csname ev@historical:qwen/bio/WGA-SFT/10M/wmdp\endcsname{81.1}
\expandafter\def\csname ev@historical:qwen/bio/WGA-SFT/10M/mmlu\endcsname{80.9}
\expandafter\def\csname ev@historical:qwen/bio/WGA-SFT/50M/wmdp\endcsname{82.1}
\expandafter\def\csname ev@historical:qwen/bio/WGA-SFT/50M/mmlu\endcsname{81.9}
\expandafter\def\csname ev@historical:qwen/bio/WGA-SFT/100M/wmdp\endcsname{83.0}
\expandafter\def\csname ev@historical:qwen/bio/WGA-SFT/100M/mmlu\endcsname{82.1}
\expandafter\def\csname ev@historical:qwen/bio/WGA-SFT/250M/wmdp\endcsname{82.8}
\expandafter\def\csname ev@historical:qwen/bio/WGA-SFT/250M/mmlu\endcsname{82.0}
\expandafter\def\csname ev@historical:qwen/bio/Routing-F0/0M/routing_z_forget\endcsname{17.1}
\expandafter\def\csname ev@historical:qwen/bio/Routing-F0/0M/routing_z_forget_layer_0\endcsname{8.0}
\expandafter\def\csname ev@historical:qwen/bio/Routing-F0/0M/routing_z_forget_layer_1\endcsname{7.4}
\expandafter\def\csname ev@historical:qwen/bio/Routing-F0/0M/routing_z_forget_layer_2\endcsname{5.6}
\expandafter\def\csname ev@historical:qwen/bio/Routing-F0/0M/routing_z_forget_layer_3\endcsname{12.4}
\expandafter\def\csname ev@historical:qwen/bio/Routing-F0/0M/routing_z_forget_layer_4\endcsname{15.9}
\expandafter\def\csname ev@historical:qwen/bio/Routing-F0/0M/routing_z_forget_layer_5\endcsname{16.0}
\expandafter\def\csname ev@historical:qwen/bio/Routing-F0/0M/routing_z_forget_layer_6\endcsname{13.2}
\expandafter\def\csname ev@historical:qwen/bio/Routing-F0/0M/routing_z_forget_layer_7\endcsname{17.0}
\expandafter\def\csname ev@historical:qwen/bio/Routing-F0/0M/routing_z_forget_layer_8\endcsname{23.0}
\expandafter\def\csname ev@historical:qwen/bio/Routing-F0/0M/routing_z_forget_layer_9\endcsname{21.6}
\expandafter\def\csname ev@historical:qwen/bio/Routing-F0/0M/routing_z_forget_layer_10\endcsname{11.7}
\expandafter\def\csname ev@historical:qwen/bio/Routing-F0/0M/routing_z_forget_layer_11\endcsname{11.6}
\expandafter\def\csname ev@historical:qwen/bio/Routing-F0/0M/routing_z_forget_layer_12\endcsname{18.5}
\expandafter\def\csname ev@historical:qwen/bio/Routing-F0/0M/routing_z_forget_layer_13\endcsname{19.1}
\expandafter\def\csname ev@historical:qwen/bio/Routing-F0/0M/routing_z_forget_layer_14\endcsname{20.9}
\expandafter\def\csname ev@historical:qwen/bio/Routing-F0/0M/routing_z_forget_layer_15\endcsname{20.8}
\expandafter\def\csname ev@historical:qwen/bio/Routing-F0/0M/routing_z_forget_layer_16\endcsname{13.0}
\expandafter\def\csname ev@historical:qwen/bio/Routing-F0/0M/routing_z_forget_layer_17\endcsname{20.2}
\expandafter\def\csname ev@historical:qwen/bio/Routing-F0/0M/routing_z_forget_layer_18\endcsname{20.7}
\expandafter\def\csname ev@historical:qwen/bio/Routing-F0/0M/routing_z_forget_layer_19\endcsname{19.0}
\expandafter\def\csname ev@historical:qwen/bio/Routing-F0/0M/routing_z_forget_layer_20\endcsname{24.0}
\expandafter\def\csname ev@historical:qwen/bio/Routing-F0/0M/routing_z_forget_layer_21\endcsname{23.5}
\expandafter\def\csname ev@historical:qwen/bio/Routing-F0/0M/routing_z_forget_layer_22\endcsname{13.4}
\expandafter\def\csname ev@historical:qwen/bio/Routing-F0/0M/routing_z_forget_layer_23\endcsname{12.1}
\expandafter\def\csname ev@historical:qwen/bio/Routing-F0/0M/routing_z_forget_layer_24\endcsname{18.0}
\expandafter\def\csname ev@historical:qwen/bio/Routing-F0/0M/routing_z_forget_layer_25\endcsname{18.6}
\expandafter\def\csname ev@historical:qwen/bio/Routing-F0/0M/routing_z_forget_layer_26\endcsname{18.8}
\expandafter\def\csname ev@historical:qwen/bio/Routing-F0/0M/routing_z_forget_layer_27\endcsname{20.1}
\expandafter\def\csname ev@historical:qwen/bio/Routing-F0/0M/routing_z_forget_layer_28\endcsname{12.1}
\expandafter\def\csname ev@historical:qwen/bio/Routing-F0/0M/routing_z_forget_layer_29\endcsname{20.1}
\expandafter\def\csname ev@historical:qwen/bio/Routing-F0/0M/routing_z_forget_layer_30\endcsname{21.2}
\expandafter\def\csname ev@historical:qwen/bio/Routing-F0/0M/routing_z_forget_layer_31\endcsname{18.6}
\expandafter\def\csname ev@historical:qwen/bio/Routing-F0/0M/routing_z_forget_layer_32\endcsname{15.7}
\expandafter\def\csname ev@historical:qwen/bio/Routing-F0/0M/routing_z_forget_layer_33\endcsname{19.0}
\expandafter\def\csname ev@historical:qwen/bio/Routing-F0/0M/routing_z_forget_layer_34\endcsname{13.8}
\expandafter\def\csname ev@historical:qwen/bio/Routing-F0/0M/routing_z_forget_layer_35\endcsname{23.3}
\expandafter\def\csname ev@historical:qwen/bio/Routing-F0/0M/routing_z_forget_layer_36\endcsname{20.7}
\expandafter\def\csname ev@historical:qwen/bio/Routing-F0/0M/routing_z_forget_layer_37\endcsname{19.8}
\expandafter\def\csname ev@historical:qwen/bio/Routing-F0/0M/routing_z_forget_layer_38\endcsname{20.9}
\expandafter\def\csname ev@historical:qwen/bio/Routing-F0/0M/routing_z_forget_layer_39\endcsname{16.1}
\expandafter\def\csname ev@historical:qwen/bio/Routing-F0/0M/routing_z_retain\endcsname{3.1}
\expandafter\def\csname ev@historical:qwen/bio/Routing-F0/0M/routing_z_retain_layer_0\endcsname{2.1}
\expandafter\def\csname ev@historical:qwen/bio/Routing-F0/0M/routing_z_retain_layer_1\endcsname{1.6}
\expandafter\def\csname ev@historical:qwen/bio/Routing-F0/0M/routing_z_retain_layer_2\endcsname{2.1}
\expandafter\def\csname ev@historical:qwen/bio/Routing-F0/0M/routing_z_retain_layer_3\endcsname{2.6}
\expandafter\def\csname ev@historical:qwen/bio/Routing-F0/0M/routing_z_retain_layer_4\endcsname{2.8}
\expandafter\def\csname ev@historical:qwen/bio/Routing-F0/0M/routing_z_retain_layer_5\endcsname{2.3}
\expandafter\def\csname ev@historical:qwen/bio/Routing-F0/0M/routing_z_retain_layer_6\endcsname{2.6}
\expandafter\def\csname ev@historical:qwen/bio/Routing-F0/0M/routing_z_retain_layer_7\endcsname{3.3}
\expandafter\def\csname ev@historical:qwen/bio/Routing-F0/0M/routing_z_retain_layer_8\endcsname{2.8}
\expandafter\def\csname ev@historical:qwen/bio/Routing-F0/0M/routing_z_retain_layer_9\endcsname{3.7}
\expandafter\def\csname ev@historical:qwen/bio/Routing-F0/0M/routing_z_retain_layer_10\endcsname{3.0}
\expandafter\def\csname ev@historical:qwen/bio/Routing-F0/0M/routing_z_retain_layer_11\endcsname{2.4}
\expandafter\def\csname ev@historical:qwen/bio/Routing-F0/0M/routing_z_retain_layer_12\endcsname{3.1}
\expandafter\def\csname ev@historical:qwen/bio/Routing-F0/0M/routing_z_retain_layer_13\endcsname{3.1}
\expandafter\def\csname ev@historical:qwen/bio/Routing-F0/0M/routing_z_retain_layer_14\endcsname{3.3}
\expandafter\def\csname ev@historical:qwen/bio/Routing-F0/0M/routing_z_retain_layer_15\endcsname{3.4}
\expandafter\def\csname ev@historical:qwen/bio/Routing-F0/0M/routing_z_retain_layer_16\endcsname{2.2}
\expandafter\def\csname ev@historical:qwen/bio/Routing-F0/0M/routing_z_retain_layer_17\endcsname{3.2}
\expandafter\def\csname ev@historical:qwen/bio/Routing-F0/0M/routing_z_retain_layer_18\endcsname{3.6}
\expandafter\def\csname ev@historical:qwen/bio/Routing-F0/0M/routing_z_retain_layer_19\endcsname{4.5}
\expandafter\def\csname ev@historical:qwen/bio/Routing-F0/0M/routing_z_retain_layer_20\endcsname{3.1}
\expandafter\def\csname ev@historical:qwen/bio/Routing-F0/0M/routing_z_retain_layer_21\endcsname{3.3}
\expandafter\def\csname ev@historical:qwen/bio/Routing-F0/0M/routing_z_retain_layer_22\endcsname{3.2}
\expandafter\def\csname ev@historical:qwen/bio/Routing-F0/0M/routing_z_retain_layer_23\endcsname{2.4}
\expandafter\def\csname ev@historical:qwen/bio/Routing-F0/0M/routing_z_retain_layer_24\endcsname{3.0}
\expandafter\def\csname ev@historical:qwen/bio/Routing-F0/0M/routing_z_retain_layer_25\endcsname{3.8}
\expandafter\def\csname ev@historical:qwen/bio/Routing-F0/0M/routing_z_retain_layer_26\endcsname{3.4}
\expandafter\def\csname ev@historical:qwen/bio/Routing-F0/0M/routing_z_retain_layer_27\endcsname{3.8}
\expandafter\def\csname ev@historical:qwen/bio/Routing-F0/0M/routing_z_retain_layer_28\endcsname{2.2}
\expandafter\def\csname ev@historical:qwen/bio/Routing-F0/0M/routing_z_retain_layer_29\endcsname{3.7}
\expandafter\def\csname ev@historical:qwen/bio/Routing-F0/0M/routing_z_retain_layer_30\endcsname{3.7}
\expandafter\def\csname ev@historical:qwen/bio/Routing-F0/0M/routing_z_retain_layer_31\endcsname{4.2}
\expandafter\def\csname ev@historical:qwen/bio/Routing-F0/0M/routing_z_retain_layer_32\endcsname{2.5}
\expandafter\def\csname ev@historical:qwen/bio/Routing-F0/0M/routing_z_retain_layer_33\endcsname{3.2}
\expandafter\def\csname ev@historical:qwen/bio/Routing-F0/0M/routing_z_retain_layer_34\endcsname{3.1}
\expandafter\def\csname ev@historical:qwen/bio/Routing-F0/0M/routing_z_retain_layer_35\endcsname{3.9}
\expandafter\def\csname ev@historical:qwen/bio/Routing-F0/0M/routing_z_retain_layer_36\endcsname{3.4}
\expandafter\def\csname ev@historical:qwen/bio/Routing-F0/0M/routing_z_retain_layer_37\endcsname{3.0}
\expandafter\def\csname ev@historical:qwen/bio/Routing-F0/0M/routing_z_retain_layer_38\endcsname{4.2}
\expandafter\def\csname ev@historical:qwen/bio/Routing-F0/0M/routing_z_retain_layer_39\endcsname{3.3}
\expandafter\def\csname ev@historical:qwen/bio/Routing-F1/0M/routing_z_forget\endcsname{72.1}
\expandafter\def\csname ev@historical:qwen/bio/Routing-F1/0M/routing_z_forget_layer_0\endcsname{12.2}
\expandafter\def\csname ev@historical:qwen/bio/Routing-F1/0M/routing_z_forget_layer_1\endcsname{27.3}
\expandafter\def\csname ev@historical:qwen/bio/Routing-F1/0M/routing_z_forget_layer_2\endcsname{7.8}
\expandafter\def\csname ev@historical:qwen/bio/Routing-F1/0M/routing_z_forget_layer_3\endcsname{48.5}
\expandafter\def\csname ev@historical:qwen/bio/Routing-F1/0M/routing_z_forget_layer_4\endcsname{51.6}
\expandafter\def\csname ev@historical:qwen/bio/Routing-F1/0M/routing_z_forget_layer_5\endcsname{52.1}
\expandafter\def\csname ev@historical:qwen/bio/Routing-F1/0M/routing_z_forget_layer_6\endcsname{43.9}
\expandafter\def\csname ev@historical:qwen/bio/Routing-F1/0M/routing_z_forget_layer_7\endcsname{62.2}
\expandafter\def\csname ev@historical:qwen/bio/Routing-F1/0M/routing_z_forget_layer_8\endcsname{74.9}
\expandafter\def\csname ev@historical:qwen/bio/Routing-F1/0M/routing_z_forget_layer_9\endcsname{70.9}
\expandafter\def\csname ev@historical:qwen/bio/Routing-F1/0M/routing_z_forget_layer_10\endcsname{64.9}
\expandafter\def\csname ev@historical:qwen/bio/Routing-F1/0M/routing_z_forget_layer_11\endcsname{58.9}
\expandafter\def\csname ev@historical:qwen/bio/Routing-F1/0M/routing_z_forget_layer_12\endcsname{69.3}
\expandafter\def\csname ev@historical:qwen/bio/Routing-F1/0M/routing_z_forget_layer_13\endcsname{77.4}
\expandafter\def\csname ev@historical:qwen/bio/Routing-F1/0M/routing_z_forget_layer_14\endcsname{77.6}
\expandafter\def\csname ev@historical:qwen/bio/Routing-F1/0M/routing_z_forget_layer_15\endcsname{80.8}
\expandafter\def\csname ev@historical:qwen/bio/Routing-F1/0M/routing_z_forget_layer_16\endcsname{77.9}
\expandafter\def\csname ev@historical:qwen/bio/Routing-F1/0M/routing_z_forget_layer_17\endcsname{81.3}
\expandafter\def\csname ev@historical:qwen/bio/Routing-F1/0M/routing_z_forget_layer_18\endcsname{82.9}
\expandafter\def\csname ev@historical:qwen/bio/Routing-F1/0M/routing_z_forget_layer_19\endcsname{81.7}
\expandafter\def\csname ev@historical:qwen/bio/Routing-F1/0M/routing_z_forget_layer_20\endcsname{86.9}
\expandafter\def\csname ev@historical:qwen/bio/Routing-F1/0M/routing_z_forget_layer_21\endcsname{85.0}
\expandafter\def\csname ev@historical:qwen/bio/Routing-F1/0M/routing_z_forget_layer_22\endcsname{81.2}
\expandafter\def\csname ev@historical:qwen/bio/Routing-F1/0M/routing_z_forget_layer_23\endcsname{81.4}
\expandafter\def\csname ev@historical:qwen/bio/Routing-F1/0M/routing_z_forget_layer_24\endcsname{85.3}
\expandafter\def\csname ev@historical:qwen/bio/Routing-F1/0M/routing_z_forget_layer_25\endcsname{85.9}
\expandafter\def\csname ev@historical:qwen/bio/Routing-F1/0M/routing_z_forget_layer_26\endcsname{85.1}
\expandafter\def\csname ev@historical:qwen/bio/Routing-F1/0M/routing_z_forget_layer_27\endcsname{84.4}
\expandafter\def\csname ev@historical:qwen/bio/Routing-F1/0M/routing_z_forget_layer_28\endcsname{84.8}
\expandafter\def\csname ev@historical:qwen/bio/Routing-F1/0M/routing_z_forget_layer_29\endcsname{84.6}
\expandafter\def\csname ev@historical:qwen/bio/Routing-F1/0M/routing_z_forget_layer_30\endcsname{89.0}
\expandafter\def\csname ev@historical:qwen/bio/Routing-F1/0M/routing_z_forget_layer_31\endcsname{87.5}
\expandafter\def\csname ev@historical:qwen/bio/Routing-F1/0M/routing_z_forget_layer_32\endcsname{79.8}
\expandafter\def\csname ev@historical:qwen/bio/Routing-F1/0M/routing_z_forget_layer_33\endcsname{77.2}
\expandafter\def\csname ev@historical:qwen/bio/Routing-F1/0M/routing_z_forget_layer_34\endcsname{79.2}
\expandafter\def\csname ev@historical:qwen/bio/Routing-F1/0M/routing_z_forget_layer_35\endcsname{88.8}
\expandafter\def\csname ev@historical:qwen/bio/Routing-F1/0M/routing_z_forget_layer_36\endcsname{85.7}
\expandafter\def\csname ev@historical:qwen/bio/Routing-F1/0M/routing_z_forget_layer_37\endcsname{86.4}
\expandafter\def\csname ev@historical:qwen/bio/Routing-F1/0M/routing_z_forget_layer_38\endcsname{85.4}
\expandafter\def\csname ev@historical:qwen/bio/Routing-F1/0M/routing_z_forget_layer_39\endcsname{76.6}
\expandafter\def\csname ev@historical:qwen/bio/Routing-F1/0M/routing_z_retain\endcsname{1.8}
\expandafter\def\csname ev@historical:qwen/bio/Routing-F1/0M/routing_z_retain_layer_0\endcsname{0.7}
\expandafter\def\csname ev@historical:qwen/bio/Routing-F1/0M/routing_z_retain_layer_1\endcsname{1.0}
\expandafter\def\csname ev@historical:qwen/bio/Routing-F1/0M/routing_z_retain_layer_2\endcsname{0.7}
\expandafter\def\csname ev@historical:qwen/bio/Routing-F1/0M/routing_z_retain_layer_3\endcsname{2.0}
\expandafter\def\csname ev@historical:qwen/bio/Routing-F1/0M/routing_z_retain_layer_4\endcsname{1.6}
\expandafter\def\csname ev@historical:qwen/bio/Routing-F1/0M/routing_z_retain_layer_5\endcsname{1.5}
\expandafter\def\csname ev@historical:qwen/bio/Routing-F1/0M/routing_z_retain_layer_6\endcsname{1.4}
\expandafter\def\csname ev@historical:qwen/bio/Routing-F1/0M/routing_z_retain_layer_7\endcsname{2.0}
\expandafter\def\csname ev@historical:qwen/bio/Routing-F1/0M/routing_z_retain_layer_8\endcsname{2.4}
\expandafter\def\csname ev@historical:qwen/bio/Routing-F1/0M/routing_z_retain_layer_9\endcsname{1.9}
\expandafter\def\csname ev@historical:qwen/bio/Routing-F1/0M/routing_z_retain_layer_10\endcsname{1.6}
\expandafter\def\csname ev@historical:qwen/bio/Routing-F1/0M/routing_z_retain_layer_11\endcsname{1.3}
\expandafter\def\csname ev@historical:qwen/bio/Routing-F1/0M/routing_z_retain_layer_12\endcsname{1.6}
\expandafter\def\csname ev@historical:qwen/bio/Routing-F1/0M/routing_z_retain_layer_13\endcsname{2.0}
\expandafter\def\csname ev@historical:qwen/bio/Routing-F1/0M/routing_z_retain_layer_14\endcsname{2.1}
\expandafter\def\csname ev@historical:qwen/bio/Routing-F1/0M/routing_z_retain_layer_15\endcsname{2.0}
\expandafter\def\csname ev@historical:qwen/bio/Routing-F1/0M/routing_z_retain_layer_16\endcsname{1.6}
\expandafter\def\csname ev@historical:qwen/bio/Routing-F1/0M/routing_z_retain_layer_17\endcsname{1.9}
\expandafter\def\csname ev@historical:qwen/bio/Routing-F1/0M/routing_z_retain_layer_18\endcsname{2.2}
\expandafter\def\csname ev@historical:qwen/bio/Routing-F1/0M/routing_z_retain_layer_19\endcsname{2.1}
\expandafter\def\csname ev@historical:qwen/bio/Routing-F1/0M/routing_z_retain_layer_20\endcsname{2.4}
\expandafter\def\csname ev@historical:qwen/bio/Routing-F1/0M/routing_z_retain_layer_21\endcsname{2.2}
\expandafter\def\csname ev@historical:qwen/bio/Routing-F1/0M/routing_z_retain_layer_22\endcsname{1.9}
\expandafter\def\csname ev@historical:qwen/bio/Routing-F1/0M/routing_z_retain_layer_23\endcsname{1.8}
\expandafter\def\csname ev@historical:qwen/bio/Routing-F1/0M/routing_z_retain_layer_24\endcsname{2.0}
\expandafter\def\csname ev@historical:qwen/bio/Routing-F1/0M/routing_z_retain_layer_25\endcsname{2.1}
\expandafter\def\csname ev@historical:qwen/bio/Routing-F1/0M/routing_z_retain_layer_26\endcsname{2.1}
\expandafter\def\csname ev@historical:qwen/bio/Routing-F1/0M/routing_z_retain_layer_27\endcsname{2.0}
\expandafter\def\csname ev@historical:qwen/bio/Routing-F1/0M/routing_z_retain_layer_28\endcsname{1.8}
\expandafter\def\csname ev@historical:qwen/bio/Routing-F1/0M/routing_z_retain_layer_29\endcsname{2.0}
\expandafter\def\csname ev@historical:qwen/bio/Routing-F1/0M/routing_z_retain_layer_30\endcsname{2.4}
\expandafter\def\csname ev@historical:qwen/bio/Routing-F1/0M/routing_z_retain_layer_31\endcsname{2.2}
\expandafter\def\csname ev@historical:qwen/bio/Routing-F1/0M/routing_z_retain_layer_32\endcsname{1.9}
\expandafter\def\csname ev@historical:qwen/bio/Routing-F1/0M/routing_z_retain_layer_33\endcsname{1.7}
\expandafter\def\csname ev@historical:qwen/bio/Routing-F1/0M/routing_z_retain_layer_34\endcsname{1.8}
\expandafter\def\csname ev@historical:qwen/bio/Routing-F1/0M/routing_z_retain_layer_35\endcsname{2.2}
\expandafter\def\csname ev@historical:qwen/bio/Routing-F1/0M/routing_z_retain_layer_36\endcsname{1.9}
\expandafter\def\csname ev@historical:qwen/bio/Routing-F1/0M/routing_z_retain_layer_37\endcsname{1.9}
\expandafter\def\csname ev@historical:qwen/bio/Routing-F1/0M/routing_z_retain_layer_38\endcsname{1.9}
\expandafter\def\csname ev@historical:qwen/bio/Routing-F1/0M/routing_z_retain_layer_39\endcsname{1.6}
\expandafter\def\csname ev@historical:gemma/bio/B/0M/mmlu\endcsname{74.5}
\expandafter\def\csname ev@historical:gemma/bio/B/0M/wmdp\endcsname{79.2}
\expandafter\def\csname ev@historical:gemma/bio/B/0M/ifeval\endcsname{27.5}
\expandafter\def\csname ev@historical:gemma/bio/S_std(B)/250M/ifeval\endcsname{27.5}
\expandafter\def\csname ev@historical:gemma/bio/F0/0M/mmlu\endcsname{73.9}
\expandafter\def\csname ev@historical:gemma/bio/F0/0M/wmdp\endcsname{79.2}
\expandafter\def\csname ev@historical:gemma/bio/A0/0M/mmlu\endcsname{72.7}
\expandafter\def\csname ev@historical:gemma/bio/A0/0M/wmdp\endcsname{73.2}
\expandafter\def\csname ev@historical:gemma/bio/CEU-SFT/250M/ifeval\endcsname{25.9}
\expandafter\def\csname ev@historical:gemma/bio/GA-SFT/250M/ifeval\endcsname{16.1}
\expandafter\def\csname ev@historical:gemma/bio/GD-SFT/250M/ifeval\endcsname{28.1}
\expandafter\def\csname ev@historical:gemma/bio/PDU-SFT/250M/ifeval\endcsname{25.9}
\expandafter\def\csname ev@historical:gemma/bio/PureNPO-SFT/250M/ifeval\endcsname{23.8}
\expandafter\def\csname ev@historical:gemma/bio/ExternalRMU-SFT/250M/ifeval\endcsname{27.0}
\expandafter\def\csname ev@historical:gemma/bio/SatImp-SFT/250M/ifeval\endcsname{26.4}
\expandafter\def\csname ev@historical:gemma/bio/SimNPO-SFT/250M/ifeval\endcsname{24.0}
\expandafter\def\csname ev@historical:gemma/bio/WGA-SFT/250M/ifeval\endcsname{26.6}
\expandafter\def\csname ev@historical:gemma/bio/S_std(B)/10M/wmdp\endcsname{79.4}
\expandafter\def\csname ev@historical:gemma/bio/S_std(B)/10M/mmlu\endcsname{74.2}
\expandafter\def\csname ev@historical:gemma/bio/S_std(B)/50M/wmdp\endcsname{79.0}
\expandafter\def\csname ev@historical:gemma/bio/S_std(B)/50M/mmlu\endcsname{74.3}
\expandafter\def\csname ev@historical:gemma/bio/S_std(B)/100M/wmdp\endcsname{78.5}
\expandafter\def\csname ev@historical:gemma/bio/S_std(B)/100M/mmlu\endcsname{74.7}
\expandafter\def\csname ev@historical:gemma/bio/S_std(B)/250M/wmdp\endcsname{79.0}
\expandafter\def\csname ev@historical:gemma/bio/S_std(B)/250M/mmlu\endcsname{74.8}
\expandafter\def\csname ev@historical:gemma/bio/A2-masked/0M/wmdp\endcsname{32.4}
\expandafter\def\csname ev@historical:gemma/bio/A2-masked/0M/mmlu\endcsname{68.8}
\expandafter\def\csname ev@historical:gemma/bio/F3/10M/wmdp\endcsname{70.5}
\expandafter\def\csname ev@historical:gemma/bio/F3/10M/mmlu\endcsname{72.7}
\expandafter\def\csname ev@historical:gemma/bio/A3/10M/wmdp\endcsname{32.2}
\expandafter\def\csname ev@historical:gemma/bio/A3/10M/mmlu\endcsname{68.7}
\expandafter\def\csname ev@historical:gemma/bio/F3/50M/wmdp\endcsname{73.1}
\expandafter\def\csname ev@historical:gemma/bio/F3/50M/mmlu\endcsname{73.1}
\expandafter\def\csname ev@historical:gemma/bio/A3/50M/wmdp\endcsname{35.7}
\expandafter\def\csname ev@historical:gemma/bio/A3/50M/mmlu\endcsname{69.6}
\expandafter\def\csname ev@historical:gemma/bio/F3/100M/wmdp\endcsname{74.6}
\expandafter\def\csname ev@historical:gemma/bio/F3/100M/mmlu\endcsname{73.6}
\expandafter\def\csname ev@historical:gemma/bio/A3/100M/wmdp\endcsname{38.3}
\expandafter\def\csname ev@historical:gemma/bio/A3/100M/mmlu\endcsname{70.2}
\expandafter\def\csname ev@historical:gemma/bio/F3/250M/wmdp\endcsname{76.5}
\expandafter\def\csname ev@historical:gemma/bio/F3/250M/mmlu\endcsname{73.9}
\expandafter\def\csname ev@historical:gemma/bio/A3/250M/wmdp\endcsname{40.4}
\expandafter\def\csname ev@historical:gemma/bio/A3/250M/mmlu\endcsname{70.5}
\expandafter\def\csname ev@historical:gemma/bio/A3-direct/10M/wmdp\endcsname{33.1}
\expandafter\def\csname ev@historical:gemma/bio/A3-direct/10M/mmlu\endcsname{69.0}
\expandafter\def\csname ev@historical:gemma/bio/A3-direct/50M/wmdp\endcsname{40.2}
\expandafter\def\csname ev@historical:gemma/bio/A3-direct/50M/mmlu\endcsname{70.0}
\expandafter\def\csname ev@historical:gemma/bio/A3-direct/100M/wmdp\endcsname{49.2}
\expandafter\def\csname ev@historical:gemma/bio/A3-direct/100M/mmlu\endcsname{70.8}
\expandafter\def\csname ev@historical:gemma/bio/A3-direct/250M/wmdp\endcsname{55.1}
\expandafter\def\csname ev@historical:gemma/bio/A3-direct/250M/mmlu\endcsname{71.5}
\expandafter\def\csname ev@historical:gemma/bio/R-SFT/10M/wmdp\endcsname{49.5}
\expandafter\def\csname ev@historical:gemma/bio/R-SFT/10M/mmlu\endcsname{72.0}
\expandafter\def\csname ev@historical:gemma/bio/R-SFT/50M/wmdp\endcsname{59.1}
\expandafter\def\csname ev@historical:gemma/bio/R-SFT/50M/mmlu\endcsname{73.2}
\expandafter\def\csname ev@historical:gemma/bio/R-SFT/100M/wmdp\endcsname{67.8}
\expandafter\def\csname ev@historical:gemma/bio/R-SFT/100M/mmlu\endcsname{73.9}
\expandafter\def\csname ev@historical:gemma/bio/R-SFT/250M/wmdp\endcsname{71.6}
\expandafter\def\csname ev@historical:gemma/bio/R-SFT/250M/mmlu\endcsname{74.2}
\expandafter\def\csname ev@historical:gemma/bio/CEU-SFT/10M/wmdp\endcsname{75.3}
\expandafter\def\csname ev@historical:gemma/bio/CEU-SFT/10M/mmlu\endcsname{72.0}
\expandafter\def\csname ev@historical:gemma/bio/CEU-SFT/50M/wmdp\endcsname{77.3}
\expandafter\def\csname ev@historical:gemma/bio/CEU-SFT/50M/mmlu\endcsname{73.1}
\expandafter\def\csname ev@historical:gemma/bio/CEU-SFT/100M/wmdp\endcsname{78.1}
\expandafter\def\csname ev@historical:gemma/bio/CEU-SFT/100M/mmlu\endcsname{73.7}
\expandafter\def\csname ev@historical:gemma/bio/CEU-SFT/250M/wmdp\endcsname{78.3}
\expandafter\def\csname ev@historical:gemma/bio/CEU-SFT/250M/mmlu\endcsname{74.4}
\expandafter\def\csname ev@historical:gemma/bio/GA-SFT/10M/wmdp\endcsname{24.7}
\expandafter\def\csname ev@historical:gemma/bio/GA-SFT/10M/mmlu\endcsname{22.9}
\expandafter\def\csname ev@historical:gemma/bio/GA-SFT/50M/wmdp\endcsname{24.7}
\expandafter\def\csname ev@historical:gemma/bio/GA-SFT/50M/mmlu\endcsname{22.9}
\expandafter\def\csname ev@historical:gemma/bio/GA-SFT/100M/wmdp\endcsname{24.4}
\expandafter\def\csname ev@historical:gemma/bio/GA-SFT/100M/mmlu\endcsname{26.7}
\expandafter\def\csname ev@historical:gemma/bio/GA-SFT/250M/wmdp\endcsname{24.0}
\expandafter\def\csname ev@historical:gemma/bio/GA-SFT/250M/mmlu\endcsname{27.3}
\expandafter\def\csname ev@historical:gemma/bio/GD-SFT/10M/wmdp\endcsname{28.6}
\expandafter\def\csname ev@historical:gemma/bio/GD-SFT/10M/mmlu\endcsname{70.8}
\expandafter\def\csname ev@historical:gemma/bio/GD-SFT/50M/wmdp\endcsname{64.4}
\expandafter\def\csname ev@historical:gemma/bio/GD-SFT/50M/mmlu\endcsname{73.0}
\expandafter\def\csname ev@historical:gemma/bio/GD-SFT/100M/wmdp\endcsname{75.3}
\expandafter\def\csname ev@historical:gemma/bio/GD-SFT/100M/mmlu\endcsname{74.2}
\expandafter\def\csname ev@historical:gemma/bio/GD-SFT/250M/wmdp\endcsname{77.8}
\expandafter\def\csname ev@historical:gemma/bio/GD-SFT/250M/mmlu\endcsname{74.8}
\expandafter\def\csname ev@historical:gemma/bio/PDU-SFT/10M/wmdp\endcsname{69.4}
\expandafter\def\csname ev@historical:gemma/bio/PDU-SFT/10M/mmlu\endcsname{73.1}
\expandafter\def\csname ev@historical:gemma/bio/PDU-SFT/50M/wmdp\endcsname{76.1}
\expandafter\def\csname ev@historical:gemma/bio/PDU-SFT/50M/mmlu\endcsname{73.2}
\expandafter\def\csname ev@historical:gemma/bio/PDU-SFT/100M/wmdp\endcsname{77.0}
\expandafter\def\csname ev@historical:gemma/bio/PDU-SFT/100M/mmlu\endcsname{74.0}
\expandafter\def\csname ev@historical:gemma/bio/PDU-SFT/250M/wmdp\endcsname{78.3}
\expandafter\def\csname ev@historical:gemma/bio/PDU-SFT/250M/mmlu\endcsname{74.7}
\expandafter\def\csname ev@historical:gemma/bio/PureNPO-SFT/10M/wmdp\endcsname{65.2}
\expandafter\def\csname ev@historical:gemma/bio/PureNPO-SFT/10M/mmlu\endcsname{69.2}
\expandafter\def\csname ev@historical:gemma/bio/PureNPO-SFT/50M/wmdp\endcsname{65.4}
\expandafter\def\csname ev@historical:gemma/bio/PureNPO-SFT/50M/mmlu\endcsname{69.5}
\expandafter\def\csname ev@historical:gemma/bio/PureNPO-SFT/100M/wmdp\endcsname{66.8}
\expandafter\def\csname ev@historical:gemma/bio/PureNPO-SFT/100M/mmlu\endcsname{69.8}
\expandafter\def\csname ev@historical:gemma/bio/PureNPO-SFT/250M/wmdp\endcsname{67.6}
\expandafter\def\csname ev@historical:gemma/bio/PureNPO-SFT/250M/mmlu\endcsname{70.2}
\expandafter\def\csname ev@historical:gemma/bio/ExternalRMU-SFT/10M/wmdp\endcsname{29.2}
\expandafter\def\csname ev@historical:gemma/bio/ExternalRMU-SFT/10M/mmlu\endcsname{70.1}
\expandafter\def\csname ev@historical:gemma/bio/ExternalRMU-SFT/50M/wmdp\endcsname{39.2}
\expandafter\def\csname ev@historical:gemma/bio/ExternalRMU-SFT/50M/mmlu\endcsname{71.6}
\expandafter\def\csname ev@historical:gemma/bio/ExternalRMU-SFT/100M/wmdp\endcsname{53.0}
\expandafter\def\csname ev@historical:gemma/bio/ExternalRMU-SFT/100M/mmlu\endcsname{72.6}
\expandafter\def\csname ev@historical:gemma/bio/ExternalRMU-SFT/250M/wmdp\endcsname{60.9}
\expandafter\def\csname ev@historical:gemma/bio/ExternalRMU-SFT/250M/mmlu\endcsname{73.4}
\expandafter\def\csname ev@historical:gemma/bio/SatImp-SFT/10M/wmdp\endcsname{63.2}
\expandafter\def\csname ev@historical:gemma/bio/SatImp-SFT/10M/mmlu\endcsname{58.9}
\expandafter\def\csname ev@historical:gemma/bio/SatImp-SFT/50M/wmdp\endcsname{76.0}
\expandafter\def\csname ev@historical:gemma/bio/SatImp-SFT/50M/mmlu\endcsname{73.0}
\expandafter\def\csname ev@historical:gemma/bio/SatImp-SFT/100M/wmdp\endcsname{77.8}
\expandafter\def\csname ev@historical:gemma/bio/SatImp-SFT/100M/mmlu\endcsname{73.7}
\expandafter\def\csname ev@historical:gemma/bio/SatImp-SFT/250M/wmdp\endcsname{78.5}
\expandafter\def\csname ev@historical:gemma/bio/SatImp-SFT/250M/mmlu\endcsname{74.3}
\expandafter\def\csname ev@historical:gemma/bio/SimNPO-SFT/10M/wmdp\endcsname{64.3}
\expandafter\def\csname ev@historical:gemma/bio/SimNPO-SFT/10M/mmlu\endcsname{68.3}
\expandafter\def\csname ev@historical:gemma/bio/SimNPO-SFT/50M/wmdp\endcsname{64.5}
\expandafter\def\csname ev@historical:gemma/bio/SimNPO-SFT/50M/mmlu\endcsname{68.6}
\expandafter\def\csname ev@historical:gemma/bio/SimNPO-SFT/100M/wmdp\endcsname{63.9}
\expandafter\def\csname ev@historical:gemma/bio/SimNPO-SFT/100M/mmlu\endcsname{68.8}
\expandafter\def\csname ev@historical:gemma/bio/SimNPO-SFT/250M/wmdp\endcsname{66.6}
\expandafter\def\csname ev@historical:gemma/bio/SimNPO-SFT/250M/mmlu\endcsname{68.9}
\expandafter\def\csname ev@historical:gemma/bio/TRACE-SFT/10M/wmdp\endcsname{55.0}
\expandafter\def\csname ev@historical:gemma/bio/TRACE-SFT/10M/mmlu\endcsname{73.8}
\expandafter\def\csname ev@historical:gemma/bio/TRACE-SFT/50M/wmdp\endcsname{73.6}
\expandafter\def\csname ev@historical:gemma/bio/TRACE-SFT/50M/mmlu\endcsname{73.7}
\expandafter\def\csname ev@historical:gemma/bio/TRACE-SFT/100M/wmdp\endcsname{75.6}
\expandafter\def\csname ev@historical:gemma/bio/TRACE-SFT/100M/mmlu\endcsname{74.2}
\expandafter\def\csname ev@historical:gemma/bio/TRACE-SFT/250M/wmdp\endcsname{76.2}
\expandafter\def\csname ev@historical:gemma/bio/TRACE-SFT/250M/mmlu\endcsname{74.7}
\expandafter\def\csname ev@historical:gemma/bio/WGA-SFT/10M/wmdp\endcsname{66.1}
\expandafter\def\csname ev@historical:gemma/bio/WGA-SFT/10M/mmlu\endcsname{62.0}
\expandafter\def\csname ev@historical:gemma/bio/WGA-SFT/50M/wmdp\endcsname{76.8}
\expandafter\def\csname ev@historical:gemma/bio/WGA-SFT/50M/mmlu\endcsname{73.0}
\expandafter\def\csname ev@historical:gemma/bio/WGA-SFT/100M/wmdp\endcsname{78.4}
\expandafter\def\csname ev@historical:gemma/bio/WGA-SFT/100M/mmlu\endcsname{73.6}
\expandafter\def\csname ev@historical:gemma/bio/WGA-SFT/250M/wmdp\endcsname{78.5}
\expandafter\def\csname ev@historical:gemma/bio/WGA-SFT/250M/mmlu\endcsname{74.3}
\expandafter\def\csname ev@historical:gemma/bio/Routing-F0/0M/routing_z_forget\endcsname{26.2}
\expandafter\def\csname ev@historical:gemma/bio/Routing-F0/0M/routing_z_forget_layer_0\endcsname{16.1}
\expandafter\def\csname ev@historical:gemma/bio/Routing-F0/0M/routing_z_forget_layer_1\endcsname{15.9}
\expandafter\def\csname ev@historical:gemma/bio/Routing-F0/0M/routing_z_forget_layer_2\endcsname{28.2}
\expandafter\def\csname ev@historical:gemma/bio/Routing-F0/0M/routing_z_forget_layer_3\endcsname{29.5}
\expandafter\def\csname ev@historical:gemma/bio/Routing-F0/0M/routing_z_forget_layer_4\endcsname{19.6}
\expandafter\def\csname ev@historical:gemma/bio/Routing-F0/0M/routing_z_forget_layer_5\endcsname{33.1}
\expandafter\def\csname ev@historical:gemma/bio/Routing-F0/0M/routing_z_forget_layer_6\endcsname{25.7}
\expandafter\def\csname ev@historical:gemma/bio/Routing-F0/0M/routing_z_forget_layer_7\endcsname{20.8}
\expandafter\def\csname ev@historical:gemma/bio/Routing-F0/0M/routing_z_forget_layer_8\endcsname{27.0}
\expandafter\def\csname ev@historical:gemma/bio/Routing-F0/0M/routing_z_forget_layer_9\endcsname{20.1}
\expandafter\def\csname ev@historical:gemma/bio/Routing-F0/0M/routing_z_forget_layer_10\endcsname{30.6}
\expandafter\def\csname ev@historical:gemma/bio/Routing-F0/0M/routing_z_forget_layer_11\endcsname{30.9}
\expandafter\def\csname ev@historical:gemma/bio/Routing-F0/0M/routing_z_forget_layer_12\endcsname{23.4}
\expandafter\def\csname ev@historical:gemma/bio/Routing-F0/0M/routing_z_forget_layer_13\endcsname{26.5}
\expandafter\def\csname ev@historical:gemma/bio/Routing-F0/0M/routing_z_forget_layer_14\endcsname{23.0}
\expandafter\def\csname ev@historical:gemma/bio/Routing-F0/0M/routing_z_forget_layer_15\endcsname{23.1}
\expandafter\def\csname ev@historical:gemma/bio/Routing-F0/0M/routing_z_forget_layer_16\endcsname{21.6}
\expandafter\def\csname ev@historical:gemma/bio/Routing-F0/0M/routing_z_forget_layer_17\endcsname{28.5}
\expandafter\def\csname ev@historical:gemma/bio/Routing-F0/0M/routing_z_forget_layer_18\endcsname{24.1}
\expandafter\def\csname ev@historical:gemma/bio/Routing-F0/0M/routing_z_forget_layer_19\endcsname{23.6}
\expandafter\def\csname ev@historical:gemma/bio/Routing-F0/0M/routing_z_forget_layer_20\endcsname{27.4}
\expandafter\def\csname ev@historical:gemma/bio/Routing-F0/0M/routing_z_forget_layer_21\endcsname{29.7}
\expandafter\def\csname ev@historical:gemma/bio/Routing-F0/0M/routing_z_forget_layer_22\endcsname{31.2}
\expandafter\def\csname ev@historical:gemma/bio/Routing-F0/0M/routing_z_forget_layer_23\endcsname{29.7}
\expandafter\def\csname ev@historical:gemma/bio/Routing-F0/0M/routing_z_forget_layer_24\endcsname{32.1}
\expandafter\def\csname ev@historical:gemma/bio/Routing-F0/0M/routing_z_forget_layer_25\endcsname{32.1}
\expandafter\def\csname ev@historical:gemma/bio/Routing-F0/0M/routing_z_forget_layer_26\endcsname{31.5}
\expandafter\def\csname ev@historical:gemma/bio/Routing-F0/0M/routing_z_forget_layer_27\endcsname{28.3}
\expandafter\def\csname ev@historical:gemma/bio/Routing-F0/0M/routing_z_forget_layer_28\endcsname{30.1}
\expandafter\def\csname ev@historical:gemma/bio/Routing-F0/0M/routing_z_forget_layer_29\endcsname{21.2}
\expandafter\def\csname ev@historical:gemma/bio/Routing-F0/0M/routing_z_retain\endcsname{8.9}
\expandafter\def\csname ev@historical:gemma/bio/Routing-F0/0M/routing_z_retain_layer_0\endcsname{5.2}
\expandafter\def\csname ev@historical:gemma/bio/Routing-F0/0M/routing_z_retain_layer_1\endcsname{5.2}
\expandafter\def\csname ev@historical:gemma/bio/Routing-F0/0M/routing_z_retain_layer_2\endcsname{8.6}
\expandafter\def\csname ev@historical:gemma/bio/Routing-F0/0M/routing_z_retain_layer_3\endcsname{9.7}
\expandafter\def\csname ev@historical:gemma/bio/Routing-F0/0M/routing_z_retain_layer_4\endcsname{4.4}
\expandafter\def\csname ev@historical:gemma/bio/Routing-F0/0M/routing_z_retain_layer_5\endcsname{10.1}
\expandafter\def\csname ev@historical:gemma/bio/Routing-F0/0M/routing_z_retain_layer_6\endcsname{7.4}
\expandafter\def\csname ev@historical:gemma/bio/Routing-F0/0M/routing_z_retain_layer_7\endcsname{8.3}
\expandafter\def\csname ev@historical:gemma/bio/Routing-F0/0M/routing_z_retain_layer_8\endcsname{9.7}
\expandafter\def\csname ev@historical:gemma/bio/Routing-F0/0M/routing_z_retain_layer_9\endcsname{6.1}
\expandafter\def\csname ev@historical:gemma/bio/Routing-F0/0M/routing_z_retain_layer_10\endcsname{11.1}
\expandafter\def\csname ev@historical:gemma/bio/Routing-F0/0M/routing_z_retain_layer_11\endcsname{12.0}
\expandafter\def\csname ev@historical:gemma/bio/Routing-F0/0M/routing_z_retain_layer_12\endcsname{11.0}
\expandafter\def\csname ev@historical:gemma/bio/Routing-F0/0M/routing_z_retain_layer_13\endcsname{8.1}
\expandafter\def\csname ev@historical:gemma/bio/Routing-F0/0M/routing_z_retain_layer_14\endcsname{6.2}
\expandafter\def\csname ev@historical:gemma/bio/Routing-F0/0M/routing_z_retain_layer_15\endcsname{14.0}
\expandafter\def\csname ev@historical:gemma/bio/Routing-F0/0M/routing_z_retain_layer_16\endcsname{8.8}
\expandafter\def\csname ev@historical:gemma/bio/Routing-F0/0M/routing_z_retain_layer_17\endcsname{15.0}
\expandafter\def\csname ev@historical:gemma/bio/Routing-F0/0M/routing_z_retain_layer_18\endcsname{9.2}
\expandafter\def\csname ev@historical:gemma/bio/Routing-F0/0M/routing_z_retain_layer_19\endcsname{9.9}
\expandafter\def\csname ev@historical:gemma/bio/Routing-F0/0M/routing_z_retain_layer_20\endcsname{8.1}
\expandafter\def\csname ev@historical:gemma/bio/Routing-F0/0M/routing_z_retain_layer_21\endcsname{8.9}
\expandafter\def\csname ev@historical:gemma/bio/Routing-F0/0M/routing_z_retain_layer_22\endcsname{10.7}
\expandafter\def\csname ev@historical:gemma/bio/Routing-F0/0M/routing_z_retain_layer_23\endcsname{5.8}
\expandafter\def\csname ev@historical:gemma/bio/Routing-F0/0M/routing_z_retain_layer_24\endcsname{8.2}
\expandafter\def\csname ev@historical:gemma/bio/Routing-F0/0M/routing_z_retain_layer_25\endcsname{9.7}
\expandafter\def\csname ev@historical:gemma/bio/Routing-F0/0M/routing_z_retain_layer_26\endcsname{8.3}
\expandafter\def\csname ev@historical:gemma/bio/Routing-F0/0M/routing_z_retain_layer_27\endcsname{8.9}
\expandafter\def\csname ev@historical:gemma/bio/Routing-F0/0M/routing_z_retain_layer_28\endcsname{9.7}
\expandafter\def\csname ev@historical:gemma/bio/Routing-F0/0M/routing_z_retain_layer_29\endcsname{7.6}
\expandafter\def\csname ev@historical:gemma/bio/Routing-F1/0M/routing_z_forget\endcsname{65.9}
\expandafter\def\csname ev@historical:gemma/bio/Routing-F1/0M/routing_z_forget_layer_0\endcsname{18.0}
\expandafter\def\csname ev@historical:gemma/bio/Routing-F1/0M/routing_z_forget_layer_1\endcsname{18.2}
\expandafter\def\csname ev@historical:gemma/bio/Routing-F1/0M/routing_z_forget_layer_2\endcsname{35.5}
\expandafter\def\csname ev@historical:gemma/bio/Routing-F1/0M/routing_z_forget_layer_3\endcsname{36.3}
\expandafter\def\csname ev@historical:gemma/bio/Routing-F1/0M/routing_z_forget_layer_4\endcsname{29.2}
\expandafter\def\csname ev@historical:gemma/bio/Routing-F1/0M/routing_z_forget_layer_5\endcsname{48.1}
\expandafter\def\csname ev@historical:gemma/bio/Routing-F1/0M/routing_z_forget_layer_6\endcsname{43.0}
\expandafter\def\csname ev@historical:gemma/bio/Routing-F1/0M/routing_z_forget_layer_7\endcsname{42.2}
\expandafter\def\csname ev@historical:gemma/bio/Routing-F1/0M/routing_z_forget_layer_8\endcsname{55.7}
\expandafter\def\csname ev@historical:gemma/bio/Routing-F1/0M/routing_z_forget_layer_9\endcsname{59.0}
\expandafter\def\csname ev@historical:gemma/bio/Routing-F1/0M/routing_z_forget_layer_10\endcsname{72.1}
\expandafter\def\csname ev@historical:gemma/bio/Routing-F1/0M/routing_z_forget_layer_11\endcsname{73.1}
\expandafter\def\csname ev@historical:gemma/bio/Routing-F1/0M/routing_z_forget_layer_12\endcsname{73.6}
\expandafter\def\csname ev@historical:gemma/bio/Routing-F1/0M/routing_z_forget_layer_13\endcsname{75.8}
\expandafter\def\csname ev@historical:gemma/bio/Routing-F1/0M/routing_z_forget_layer_14\endcsname{76.5}
\expandafter\def\csname ev@historical:gemma/bio/Routing-F1/0M/routing_z_forget_layer_15\endcsname{72.1}
\expandafter\def\csname ev@historical:gemma/bio/Routing-F1/0M/routing_z_forget_layer_16\endcsname{84.6}
\expandafter\def\csname ev@historical:gemma/bio/Routing-F1/0M/routing_z_forget_layer_17\endcsname{82.1}
\expandafter\def\csname ev@historical:gemma/bio/Routing-F1/0M/routing_z_forget_layer_18\endcsname{80.7}
\expandafter\def\csname ev@historical:gemma/bio/Routing-F1/0M/routing_z_forget_layer_19\endcsname{81.3}
\expandafter\def\csname ev@historical:gemma/bio/Routing-F1/0M/routing_z_forget_layer_20\endcsname{81.1}
\expandafter\def\csname ev@historical:gemma/bio/Routing-F1/0M/routing_z_forget_layer_21\endcsname{82.3}
\expandafter\def\csname ev@historical:gemma/bio/Routing-F1/0M/routing_z_forget_layer_22\endcsname{85.5}
\expandafter\def\csname ev@historical:gemma/bio/Routing-F1/0M/routing_z_forget_layer_23\endcsname{82.3}
\expandafter\def\csname ev@historical:gemma/bio/Routing-F1/0M/routing_z_forget_layer_24\endcsname{81.9}
\expandafter\def\csname ev@historical:gemma/bio/Routing-F1/0M/routing_z_forget_layer_25\endcsname{83.4}
\expandafter\def\csname ev@historical:gemma/bio/Routing-F1/0M/routing_z_forget_layer_26\endcsname{81.5}
\expandafter\def\csname ev@historical:gemma/bio/Routing-F1/0M/routing_z_forget_layer_27\endcsname{80.9}
\expandafter\def\csname ev@historical:gemma/bio/Routing-F1/0M/routing_z_forget_layer_28\endcsname{85.3}
\expandafter\def\csname ev@historical:gemma/bio/Routing-F1/0M/routing_z_forget_layer_29\endcsname{75.7}
\expandafter\def\csname ev@historical:gemma/bio/Routing-F1/0M/routing_z_retain\endcsname{8.5}
\expandafter\def\csname ev@historical:gemma/bio/Routing-F1/0M/routing_z_retain_layer_0\endcsname{4.4}
\expandafter\def\csname ev@historical:gemma/bio/Routing-F1/0M/routing_z_retain_layer_1\endcsname{4.0}
\expandafter\def\csname ev@historical:gemma/bio/Routing-F1/0M/routing_z_retain_layer_2\endcsname{7.6}
\expandafter\def\csname ev@historical:gemma/bio/Routing-F1/0M/routing_z_retain_layer_3\endcsname{8.4}
\expandafter\def\csname ev@historical:gemma/bio/Routing-F1/0M/routing_z_retain_layer_4\endcsname{5.0}
\expandafter\def\csname ev@historical:gemma/bio/Routing-F1/0M/routing_z_retain_layer_5\endcsname{8.5}
\expandafter\def\csname ev@historical:gemma/bio/Routing-F1/0M/routing_z_retain_layer_6\endcsname{6.7}
\expandafter\def\csname ev@historical:gemma/bio/Routing-F1/0M/routing_z_retain_layer_7\endcsname{6.2}
\expandafter\def\csname ev@historical:gemma/bio/Routing-F1/0M/routing_z_retain_layer_8\endcsname{6.7}
\expandafter\def\csname ev@historical:gemma/bio/Routing-F1/0M/routing_z_retain_layer_9\endcsname{6.5}
\expandafter\def\csname ev@historical:gemma/bio/Routing-F1/0M/routing_z_retain_layer_10\endcsname{10.0}
\expandafter\def\csname ev@historical:gemma/bio/Routing-F1/0M/routing_z_retain_layer_11\endcsname{10.6}
\expandafter\def\csname ev@historical:gemma/bio/Routing-F1/0M/routing_z_retain_layer_12\endcsname{11.5}
\expandafter\def\csname ev@historical:gemma/bio/Routing-F1/0M/routing_z_retain_layer_13\endcsname{8.2}
\expandafter\def\csname ev@historical:gemma/bio/Routing-F1/0M/routing_z_retain_layer_14\endcsname{7.3}
\expandafter\def\csname ev@historical:gemma/bio/Routing-F1/0M/routing_z_retain_layer_15\endcsname{15.3}
\expandafter\def\csname ev@historical:gemma/bio/Routing-F1/0M/routing_z_retain_layer_16\endcsname{11.0}
\expandafter\def\csname ev@historical:gemma/bio/Routing-F1/0M/routing_z_retain_layer_17\endcsname{15.3}
\expandafter\def\csname ev@historical:gemma/bio/Routing-F1/0M/routing_z_retain_layer_18\endcsname{9.9}
\expandafter\def\csname ev@historical:gemma/bio/Routing-F1/0M/routing_z_retain_layer_19\endcsname{10.3}
\expandafter\def\csname ev@historical:gemma/bio/Routing-F1/0M/routing_z_retain_layer_20\endcsname{8.6}
\expandafter\def\csname ev@historical:gemma/bio/Routing-F1/0M/routing_z_retain_layer_21\endcsname{9.4}
\expandafter\def\csname ev@historical:gemma/bio/Routing-F1/0M/routing_z_retain_layer_22\endcsname{9.6}
\expandafter\def\csname ev@historical:gemma/bio/Routing-F1/0M/routing_z_retain_layer_23\endcsname{7.4}
\expandafter\def\csname ev@historical:gemma/bio/Routing-F1/0M/routing_z_retain_layer_24\endcsname{7.9}
\expandafter\def\csname ev@historical:gemma/bio/Routing-F1/0M/routing_z_retain_layer_25\endcsname{8.7}
\expandafter\def\csname ev@historical:gemma/bio/Routing-F1/0M/routing_z_retain_layer_26\endcsname{8.2}
\expandafter\def\csname ev@historical:gemma/bio/Routing-F1/0M/routing_z_retain_layer_27\endcsname{7.0}
\expandafter\def\csname ev@historical:gemma/bio/Routing-F1/0M/routing_z_retain_layer_28\endcsname{7.8}
\expandafter\def\csname ev@historical:gemma/bio/Routing-F1/0M/routing_z_retain_layer_29\endcsname{5.7}
\expandafter\def\csname ev@historical:intern/cyber/B/0M/mmlu\endcsname{85.1}
\expandafter\def\csname ev@historical:intern/cyber/B/0M/wmdp\endcsname{59.0}
\expandafter\def\csname ev@historical:intern/cyber/B/0M/ifeval\endcsname{81.3}
\expandafter\def\csname ev@historical:intern/cyber/S_std(B)/250M/ifeval\endcsname{78.6}
\expandafter\def\csname ev@historical:intern/cyber/F0/0M/mmlu\endcsname{85.1}
\expandafter\def\csname ev@historical:intern/cyber/F0/0M/wmdp\endcsname{57.6}
\expandafter\def\csname ev@historical:intern/cyber/A0/0M/mmlu\endcsname{84.8}
\expandafter\def\csname ev@historical:intern/cyber/A0/0M/wmdp\endcsname{50.2}
\expandafter\def\csname ev@historical:intern/cyber/F1/0M/mmlu\endcsname{84.4}
\expandafter\def\csname ev@historical:intern/cyber/F1/0M/wmdp\endcsname{58.0}
\expandafter\def\csname ev@historical:intern/cyber/A1/0M/mmlu\endcsname{84.3}
\expandafter\def\csname ev@historical:intern/cyber/A1/0M/wmdp\endcsname{45.6}
\expandafter\def\csname ev@historical:intern/cyber/A2/0M/mmlu\endcsname{84.1}
\expandafter\def\csname ev@historical:intern/cyber/A2/0M/wmdp\endcsname{26.5}
\expandafter\def\csname ev@historical:intern/cyber/F2/0M/mmlu\endcsname{84.5}
\expandafter\def\csname ev@historical:intern/cyber/F2/0M/wmdp\endcsname{52.4}
\expandafter\def\csname ev@historical:intern/cyber/F3/250M/ifeval\endcsname{78.6}
\expandafter\def\csname ev@historical:intern/cyber/A3/250M/ifeval\endcsname{75.6}
\expandafter\def\csname ev@historical:intern/cyber/R/0M/mmlu\endcsname{84.9}
\expandafter\def\csname ev@historical:intern/cyber/R/0M/wmdp\endcsname{30.1}
\expandafter\def\csname ev@historical:intern/cyber/R-SFT/250M/ifeval\endcsname{78.4}
\expandafter\def\csname ev@historical:intern/cyber/A3-direct/250M/ifeval\endcsname{78.0}
\expandafter\def\csname ev@historical:intern/cyber/CEU-SFT/250M/ifeval\endcsname{79.7}
\expandafter\def\csname ev@historical:intern/cyber/GA-SFT/250M/ifeval\endcsname{83.0}
\expandafter\def\csname ev@historical:intern/cyber/GD-SFT/250M/ifeval\endcsname{78.9}
\expandafter\def\csname ev@historical:intern/cyber/PDU-SFT/250M/ifeval\endcsname{77.8}
\expandafter\def\csname ev@historical:intern/cyber/PureNPO-SFT/250M/ifeval\endcsname{77.3}
\expandafter\def\csname ev@historical:intern/cyber/SatImp-SFT/250M/ifeval\endcsname{81.2}
\expandafter\def\csname ev@historical:intern/cyber/SimNPO-SFT/250M/ifeval\endcsname{76.9}
\expandafter\def\csname ev@historical:intern/cyber/TRACE-SFT/250M/ifeval\endcsname{80.2}
\expandafter\def\csname ev@historical:intern/cyber/WGA-SFT/250M/ifeval\endcsname{80.0}
\expandafter\def\csname ev@historical:intern/cyber/S_std(B)/10M/wmdp\endcsname{59.5}
\expandafter\def\csname ev@historical:intern/cyber/S_std(B)/10M/mmlu\endcsname{85.1}
\expandafter\def\csname ev@historical:intern/cyber/S_std(B)/50M/wmdp\endcsname{58.6}
\expandafter\def\csname ev@historical:intern/cyber/S_std(B)/50M/mmlu\endcsname{85.1}
\expandafter\def\csname ev@historical:intern/cyber/S_std(B)/100M/wmdp\endcsname{58.9}
\expandafter\def\csname ev@historical:intern/cyber/S_std(B)/100M/mmlu\endcsname{84.9}
\expandafter\def\csname ev@historical:intern/cyber/S_std(B)/250M/wmdp\endcsname{58.7}
\expandafter\def\csname ev@historical:intern/cyber/S_std(B)/250M/mmlu\endcsname{84.8}
\expandafter\def\csname ev@historical:intern/cyber/A2-masked/0M/wmdp\endcsname{26.0}
\expandafter\def\csname ev@historical:intern/cyber/A2-masked/0M/mmlu\endcsname{84.0}
\expandafter\def\csname ev@historical:intern/cyber/F3/10M/wmdp\endcsname{52.7}
\expandafter\def\csname ev@historical:intern/cyber/F3/10M/mmlu\endcsname{84.6}
\expandafter\def\csname ev@historical:intern/cyber/A3/10M/wmdp\endcsname{26.2}
\expandafter\def\csname ev@historical:intern/cyber/A3/10M/mmlu\endcsname{84.0}
\expandafter\def\csname ev@historical:intern/cyber/F3/50M/wmdp\endcsname{53.9}
\expandafter\def\csname ev@historical:intern/cyber/F3/50M/mmlu\endcsname{84.6}
\expandafter\def\csname ev@historical:intern/cyber/A3/50M/wmdp\endcsname{27.5}
\expandafter\def\csname ev@historical:intern/cyber/A3/50M/mmlu\endcsname{84.2}
\expandafter\def\csname ev@historical:intern/cyber/F3/100M/wmdp\endcsname{55.1}
\expandafter\def\csname ev@historical:intern/cyber/F3/100M/mmlu\endcsname{84.4}
\expandafter\def\csname ev@historical:intern/cyber/A3/100M/wmdp\endcsname{29.6}
\expandafter\def\csname ev@historical:intern/cyber/A3/100M/mmlu\endcsname{84.1}
\expandafter\def\csname ev@historical:intern/cyber/F3/250M/wmdp\endcsname{55.5}
\expandafter\def\csname ev@historical:intern/cyber/F3/250M/mmlu\endcsname{84.2}
\expandafter\def\csname ev@historical:intern/cyber/A3/250M/wmdp\endcsname{29.1}
\expandafter\def\csname ev@historical:intern/cyber/A3/250M/mmlu\endcsname{84.0}
\expandafter\def\csname ev@historical:intern/cyber/A3-direct/10M/wmdp\endcsname{26.6}
\expandafter\def\csname ev@historical:intern/cyber/A3-direct/10M/mmlu\endcsname{84.1}
\expandafter\def\csname ev@historical:intern/cyber/A3-direct/50M/wmdp\endcsname{29.1}
\expandafter\def\csname ev@historical:intern/cyber/A3-direct/50M/mmlu\endcsname{84.2}
\expandafter\def\csname ev@historical:intern/cyber/A3-direct/100M/wmdp\endcsname{33.0}
\expandafter\def\csname ev@historical:intern/cyber/A3-direct/100M/mmlu\endcsname{84.4}
\expandafter\def\csname ev@historical:intern/cyber/A3-direct/250M/wmdp\endcsname{36.1}
\expandafter\def\csname ev@historical:intern/cyber/A3-direct/250M/mmlu\endcsname{84.0}
\expandafter\def\csname ev@historical:intern/cyber/R-SFT/10M/wmdp\endcsname{31.1}
\expandafter\def\csname ev@historical:intern/cyber/R-SFT/10M/mmlu\endcsname{85.0}
\expandafter\def\csname ev@historical:intern/cyber/R-SFT/50M/wmdp\endcsname{34.2}
\expandafter\def\csname ev@historical:intern/cyber/R-SFT/50M/mmlu\endcsname{84.9}
\expandafter\def\csname ev@historical:intern/cyber/R-SFT/100M/wmdp\endcsname{41.0}
\expandafter\def\csname ev@historical:intern/cyber/R-SFT/100M/mmlu\endcsname{84.5}
\expandafter\def\csname ev@historical:intern/cyber/R-SFT/250M/wmdp\endcsname{44.0}
\expandafter\def\csname ev@historical:intern/cyber/R-SFT/250M/mmlu\endcsname{84.3}
\expandafter\def\csname ev@historical:intern/cyber/CEU-SFT/10M/wmdp\endcsname{43.9}
\expandafter\def\csname ev@historical:intern/cyber/CEU-SFT/10M/mmlu\endcsname{77.7}
\expandafter\def\csname ev@historical:intern/cyber/CEU-SFT/50M/wmdp\endcsname{46.4}
\expandafter\def\csname ev@historical:intern/cyber/CEU-SFT/50M/mmlu\endcsname{81.2}
\expandafter\def\csname ev@historical:intern/cyber/CEU-SFT/100M/wmdp\endcsname{51.5}
\expandafter\def\csname ev@historical:intern/cyber/CEU-SFT/100M/mmlu\endcsname{83.4}
\expandafter\def\csname ev@historical:intern/cyber/CEU-SFT/250M/wmdp\endcsname{52.1}
\expandafter\def\csname ev@historical:intern/cyber/CEU-SFT/250M/mmlu\endcsname{83.7}
\expandafter\def\csname ev@historical:intern/cyber/GA-SFT/10M/wmdp\endcsname{24.6}
\expandafter\def\csname ev@historical:intern/cyber/GA-SFT/10M/mmlu\endcsname{33.6}
\expandafter\def\csname ev@historical:intern/cyber/GA-SFT/50M/wmdp\endcsname{35.9}
\expandafter\def\csname ev@historical:intern/cyber/GA-SFT/50M/mmlu\endcsname{67.5}
\expandafter\def\csname ev@historical:intern/cyber/GA-SFT/100M/wmdp\endcsname{44.0}
\expandafter\def\csname ev@historical:intern/cyber/GA-SFT/100M/mmlu\endcsname{78.7}
\expandafter\def\csname ev@historical:intern/cyber/GA-SFT/250M/wmdp\endcsname{47.5}
\expandafter\def\csname ev@historical:intern/cyber/GA-SFT/250M/mmlu\endcsname{81.2}
\expandafter\def\csname ev@historical:intern/cyber/GD-SFT/10M/wmdp\endcsname{30.4}
\expandafter\def\csname ev@historical:intern/cyber/GD-SFT/10M/mmlu\endcsname{82.0}
\expandafter\def\csname ev@historical:intern/cyber/GD-SFT/50M/wmdp\endcsname{44.2}
\expandafter\def\csname ev@historical:intern/cyber/GD-SFT/50M/mmlu\endcsname{83.0}
\expandafter\def\csname ev@historical:intern/cyber/GD-SFT/100M/wmdp\endcsname{45.8}
\expandafter\def\csname ev@historical:intern/cyber/GD-SFT/100M/mmlu\endcsname{83.2}
\expandafter\def\csname ev@historical:intern/cyber/GD-SFT/250M/wmdp\endcsname{47.4}
\expandafter\def\csname ev@historical:intern/cyber/GD-SFT/250M/mmlu\endcsname{83.0}
\expandafter\def\csname ev@historical:intern/cyber/PDU-SFT/10M/wmdp\endcsname{27.7}
\expandafter\def\csname ev@historical:intern/cyber/PDU-SFT/10M/mmlu\endcsname{84.7}
\expandafter\def\csname ev@historical:intern/cyber/PDU-SFT/50M/wmdp\endcsname{31.4}
\expandafter\def\csname ev@historical:intern/cyber/PDU-SFT/50M/mmlu\endcsname{84.8}
\expandafter\def\csname ev@historical:intern/cyber/PDU-SFT/100M/wmdp\endcsname{36.7}
\expandafter\def\csname ev@historical:intern/cyber/PDU-SFT/100M/mmlu\endcsname{85.0}
\expandafter\def\csname ev@historical:intern/cyber/PDU-SFT/250M/wmdp\endcsname{39.5}
\expandafter\def\csname ev@historical:intern/cyber/PDU-SFT/250M/mmlu\endcsname{85.0}
\expandafter\def\csname ev@historical:intern/cyber/PureNPO-SFT/10M/wmdp\endcsname{41.4}
\expandafter\def\csname ev@historical:intern/cyber/PureNPO-SFT/10M/mmlu\endcsname{83.2}
\expandafter\def\csname ev@historical:intern/cyber/PureNPO-SFT/50M/wmdp\endcsname{46.4}
\expandafter\def\csname ev@historical:intern/cyber/PureNPO-SFT/50M/mmlu\endcsname{83.4}
\expandafter\def\csname ev@historical:intern/cyber/PureNPO-SFT/100M/wmdp\endcsname{50.7}
\expandafter\def\csname ev@historical:intern/cyber/PureNPO-SFT/100M/mmlu\endcsname{84.2}
\expandafter\def\csname ev@historical:intern/cyber/PureNPO-SFT/250M/wmdp\endcsname{50.9}
\expandafter\def\csname ev@historical:intern/cyber/PureNPO-SFT/250M/mmlu\endcsname{84.4}
\expandafter\def\csname ev@historical:intern/cyber/ExternalRMU-SFT/10M/wmdp\endcsname{35.8}
\expandafter\def\csname ev@historical:intern/cyber/ExternalRMU-SFT/10M/mmlu\endcsname{84.7}
\expandafter\def\csname ev@historical:intern/cyber/ExternalRMU-SFT/50M/wmdp\endcsname{40.5}
\expandafter\def\csname ev@historical:intern/cyber/ExternalRMU-SFT/50M/mmlu\endcsname{85.0}
\expandafter\def\csname ev@historical:intern/cyber/ExternalRMU-SFT/100M/wmdp\endcsname{46.5}
\expandafter\def\csname ev@historical:intern/cyber/ExternalRMU-SFT/100M/mmlu\endcsname{84.8}
\expandafter\def\csname ev@historical:intern/cyber/ExternalRMU-SFT/250M/wmdp\endcsname{50.2}
\expandafter\def\csname ev@historical:intern/cyber/ExternalRMU-SFT/250M/mmlu\endcsname{84.4}
\expandafter\def\csname ev@historical:intern/cyber/SatImp-SFT/10M/wmdp\endcsname{38.4}
\expandafter\def\csname ev@historical:intern/cyber/SatImp-SFT/10M/mmlu\endcsname{79.0}
\expandafter\def\csname ev@historical:intern/cyber/SatImp-SFT/50M/wmdp\endcsname{45.1}
\expandafter\def\csname ev@historical:intern/cyber/SatImp-SFT/50M/mmlu\endcsname{82.8}
\expandafter\def\csname ev@historical:intern/cyber/SatImp-SFT/100M/wmdp\endcsname{50.3}
\expandafter\def\csname ev@historical:intern/cyber/SatImp-SFT/100M/mmlu\endcsname{84.0}
\expandafter\def\csname ev@historical:intern/cyber/SatImp-SFT/250M/wmdp\endcsname{51.8}
\expandafter\def\csname ev@historical:intern/cyber/SatImp-SFT/250M/mmlu\endcsname{84.0}
\expandafter\def\csname ev@historical:intern/cyber/SimNPO-SFT/10M/wmdp\endcsname{47.5}
\expandafter\def\csname ev@historical:intern/cyber/SimNPO-SFT/10M/mmlu\endcsname{81.7}
\expandafter\def\csname ev@historical:intern/cyber/SimNPO-SFT/50M/wmdp\endcsname{49.5}
\expandafter\def\csname ev@historical:intern/cyber/SimNPO-SFT/50M/mmlu\endcsname{82.3}
\expandafter\def\csname ev@historical:intern/cyber/SimNPO-SFT/100M/wmdp\endcsname{51.7}
\expandafter\def\csname ev@historical:intern/cyber/SimNPO-SFT/100M/mmlu\endcsname{82.9}
\expandafter\def\csname ev@historical:intern/cyber/SimNPO-SFT/250M/wmdp\endcsname{52.6}
\expandafter\def\csname ev@historical:intern/cyber/SimNPO-SFT/250M/mmlu\endcsname{82.8}
\expandafter\def\csname ev@historical:intern/cyber/TRACE-SFT/10M/wmdp\endcsname{41.9}
\expandafter\def\csname ev@historical:intern/cyber/TRACE-SFT/10M/mmlu\endcsname{83.9}
\expandafter\def\csname ev@historical:intern/cyber/TRACE-SFT/50M/wmdp\endcsname{43.7}
\expandafter\def\csname ev@historical:intern/cyber/TRACE-SFT/50M/mmlu\endcsname{84.4}
\expandafter\def\csname ev@historical:intern/cyber/TRACE-SFT/100M/wmdp\endcsname{45.5}
\expandafter\def\csname ev@historical:intern/cyber/TRACE-SFT/100M/mmlu\endcsname{84.3}
\expandafter\def\csname ev@historical:intern/cyber/TRACE-SFT/250M/wmdp\endcsname{45.5}
\expandafter\def\csname ev@historical:intern/cyber/TRACE-SFT/250M/mmlu\endcsname{84.2}
\expandafter\def\csname ev@historical:intern/cyber/WGA-SFT/10M/wmdp\endcsname{37.9}
\expandafter\def\csname ev@historical:intern/cyber/WGA-SFT/10M/mmlu\endcsname{79.0}
\expandafter\def\csname ev@historical:intern/cyber/WGA-SFT/50M/wmdp\endcsname{44.9}
\expandafter\def\csname ev@historical:intern/cyber/WGA-SFT/50M/mmlu\endcsname{82.9}
\expandafter\def\csname ev@historical:intern/cyber/WGA-SFT/100M/wmdp\endcsname{49.5}
\expandafter\def\csname ev@historical:intern/cyber/WGA-SFT/100M/mmlu\endcsname{84.0}
\expandafter\def\csname ev@historical:intern/cyber/WGA-SFT/250M/wmdp\endcsname{51.1}
\expandafter\def\csname ev@historical:intern/cyber/WGA-SFT/250M/mmlu\endcsname{83.8}
\expandafter\def\csname ev@historical:intern/cyber/Routing-F0/0M/routing_z_forget\endcsname{19.5}
\expandafter\def\csname ev@historical:intern/cyber/Routing-F0/0M/routing_z_forget_layer_0\endcsname{10.6}
\expandafter\def\csname ev@historical:intern/cyber/Routing-F0/0M/routing_z_forget_layer_1\endcsname{14.0}
\expandafter\def\csname ev@historical:intern/cyber/Routing-F0/0M/routing_z_forget_layer_2\endcsname{7.9}
\expandafter\def\csname ev@historical:intern/cyber/Routing-F0/0M/routing_z_forget_layer_3\endcsname{17.4}
\expandafter\def\csname ev@historical:intern/cyber/Routing-F0/0M/routing_z_forget_layer_4\endcsname{19.2}
\expandafter\def\csname ev@historical:intern/cyber/Routing-F0/0M/routing_z_forget_layer_5\endcsname{23.1}
\expandafter\def\csname ev@historical:intern/cyber/Routing-F0/0M/routing_z_forget_layer_6\endcsname{20.4}
\expandafter\def\csname ev@historical:intern/cyber/Routing-F0/0M/routing_z_forget_layer_7\endcsname{26.6}
\expandafter\def\csname ev@historical:intern/cyber/Routing-F0/0M/routing_z_forget_layer_8\endcsname{27.6}
\expandafter\def\csname ev@historical:intern/cyber/Routing-F0/0M/routing_z_forget_layer_9\endcsname{25.8}
\expandafter\def\csname ev@historical:intern/cyber/Routing-F0/0M/routing_z_forget_layer_10\endcsname{16.1}
\expandafter\def\csname ev@historical:intern/cyber/Routing-F0/0M/routing_z_forget_layer_11\endcsname{16.6}
\expandafter\def\csname ev@historical:intern/cyber/Routing-F0/0M/routing_z_forget_layer_12\endcsname{19.3}
\expandafter\def\csname ev@historical:intern/cyber/Routing-F0/0M/routing_z_forget_layer_13\endcsname{21.5}
\expandafter\def\csname ev@historical:intern/cyber/Routing-F0/0M/routing_z_forget_layer_14\endcsname{23.3}
\expandafter\def\csname ev@historical:intern/cyber/Routing-F0/0M/routing_z_forget_layer_15\endcsname{19.5}
\expandafter\def\csname ev@historical:intern/cyber/Routing-F0/0M/routing_z_forget_layer_16\endcsname{15.7}
\expandafter\def\csname ev@historical:intern/cyber/Routing-F0/0M/routing_z_forget_layer_17\endcsname{18.4}
\expandafter\def\csname ev@historical:intern/cyber/Routing-F0/0M/routing_z_forget_layer_18\endcsname{21.6}
\expandafter\def\csname ev@historical:intern/cyber/Routing-F0/0M/routing_z_forget_layer_19\endcsname{19.9}
\expandafter\def\csname ev@historical:intern/cyber/Routing-F0/0M/routing_z_forget_layer_20\endcsname{26.3}
\expandafter\def\csname ev@historical:intern/cyber/Routing-F0/0M/routing_z_forget_layer_21\endcsname{25.6}
\expandafter\def\csname ev@historical:intern/cyber/Routing-F0/0M/routing_z_forget_layer_22\endcsname{15.9}
\expandafter\def\csname ev@historical:intern/cyber/Routing-F0/0M/routing_z_forget_layer_23\endcsname{18.5}
\expandafter\def\csname ev@historical:intern/cyber/Routing-F0/0M/routing_z_forget_layer_24\endcsname{17.7}
\expandafter\def\csname ev@historical:intern/cyber/Routing-F0/0M/routing_z_forget_layer_25\endcsname{19.7}
\expandafter\def\csname ev@historical:intern/cyber/Routing-F0/0M/routing_z_forget_layer_26\endcsname{21.1}
\expandafter\def\csname ev@historical:intern/cyber/Routing-F0/0M/routing_z_forget_layer_27\endcsname{17.3}
\expandafter\def\csname ev@historical:intern/cyber/Routing-F0/0M/routing_z_forget_layer_28\endcsname{14.7}
\expandafter\def\csname ev@historical:intern/cyber/Routing-F0/0M/routing_z_forget_layer_29\endcsname{16.9}
\expandafter\def\csname ev@historical:intern/cyber/Routing-F0/0M/routing_z_forget_layer_30\endcsname{20.4}
\expandafter\def\csname ev@historical:intern/cyber/Routing-F0/0M/routing_z_forget_layer_31\endcsname{20.2}
\expandafter\def\csname ev@historical:intern/cyber/Routing-F0/0M/routing_z_forget_layer_32\endcsname{15.5}
\expandafter\def\csname ev@historical:intern/cyber/Routing-F0/0M/routing_z_forget_layer_33\endcsname{20.2}
\expandafter\def\csname ev@historical:intern/cyber/Routing-F0/0M/routing_z_forget_layer_34\endcsname{16.1}
\expandafter\def\csname ev@historical:intern/cyber/Routing-F0/0M/routing_z_forget_layer_35\endcsname{23.3}
\expandafter\def\csname ev@historical:intern/cyber/Routing-F0/0M/routing_z_forget_layer_36\endcsname{26.3}
\expandafter\def\csname ev@historical:intern/cyber/Routing-F0/0M/routing_z_forget_layer_37\endcsname{21.3}
\expandafter\def\csname ev@historical:intern/cyber/Routing-F0/0M/routing_z_forget_layer_38\endcsname{22.9}
\expandafter\def\csname ev@historical:intern/cyber/Routing-F0/0M/routing_z_forget_layer_39\endcsname{15.6}
\expandafter\def\csname ev@historical:intern/cyber/Routing-F0/0M/routing_z_retain\endcsname{0.7}
\expandafter\def\csname ev@historical:intern/cyber/Routing-F0/0M/routing_z_retain_layer_0\endcsname{0.7}
\expandafter\def\csname ev@historical:intern/cyber/Routing-F0/0M/routing_z_retain_layer_1\endcsname{0.9}
\expandafter\def\csname ev@historical:intern/cyber/Routing-F0/0M/routing_z_retain_layer_2\endcsname{1.1}
\expandafter\def\csname ev@historical:intern/cyber/Routing-F0/0M/routing_z_retain_layer_3\endcsname{0.7}
\expandafter\def\csname ev@historical:intern/cyber/Routing-F0/0M/routing_z_retain_layer_4\endcsname{1.2}
\expandafter\def\csname ev@historical:intern/cyber/Routing-F0/0M/routing_z_retain_layer_5\endcsname{0.5}
\expandafter\def\csname ev@historical:intern/cyber/Routing-F0/0M/routing_z_retain_layer_6\endcsname{0.6}
\expandafter\def\csname ev@historical:intern/cyber/Routing-F0/0M/routing_z_retain_layer_7\endcsname{0.6}
\expandafter\def\csname ev@historical:intern/cyber/Routing-F0/0M/routing_z_retain_layer_8\endcsname{0.4}
\expandafter\def\csname ev@historical:intern/cyber/Routing-F0/0M/routing_z_retain_layer_9\endcsname{0.6}
\expandafter\def\csname ev@historical:intern/cyber/Routing-F0/0M/routing_z_retain_layer_10\endcsname{0.6}
\expandafter\def\csname ev@historical:intern/cyber/Routing-F0/0M/routing_z_retain_layer_11\endcsname{0.6}
\expandafter\def\csname ev@historical:intern/cyber/Routing-F0/0M/routing_z_retain_layer_12\endcsname{0.9}
\expandafter\def\csname ev@historical:intern/cyber/Routing-F0/0M/routing_z_retain_layer_13\endcsname{0.8}
\expandafter\def\csname ev@historical:intern/cyber/Routing-F0/0M/routing_z_retain_layer_14\endcsname{0.6}
\expandafter\def\csname ev@historical:intern/cyber/Routing-F0/0M/routing_z_retain_layer_15\endcsname{0.5}
\expandafter\def\csname ev@historical:intern/cyber/Routing-F0/0M/routing_z_retain_layer_16\endcsname{1.0}
\expandafter\def\csname ev@historical:intern/cyber/Routing-F0/0M/routing_z_retain_layer_17\endcsname{1.6}
\expandafter\def\csname ev@historical:intern/cyber/Routing-F0/0M/routing_z_retain_layer_18\endcsname{0.6}
\expandafter\def\csname ev@historical:intern/cyber/Routing-F0/0M/routing_z_retain_layer_19\endcsname{0.6}
\expandafter\def\csname ev@historical:intern/cyber/Routing-F0/0M/routing_z_retain_layer_20\endcsname{0.5}
\expandafter\def\csname ev@historical:intern/cyber/Routing-F0/0M/routing_z_retain_layer_21\endcsname{0.7}
\expandafter\def\csname ev@historical:intern/cyber/Routing-F0/0M/routing_z_retain_layer_22\endcsname{0.5}
\expandafter\def\csname ev@historical:intern/cyber/Routing-F0/0M/routing_z_retain_layer_23\endcsname{1.0}
\expandafter\def\csname ev@historical:intern/cyber/Routing-F0/0M/routing_z_retain_layer_24\endcsname{0.9}
\expandafter\def\csname ev@historical:intern/cyber/Routing-F0/0M/routing_z_retain_layer_25\endcsname{0.7}
\expandafter\def\csname ev@historical:intern/cyber/Routing-F0/0M/routing_z_retain_layer_26\endcsname{0.7}
\expandafter\def\csname ev@historical:intern/cyber/Routing-F0/0M/routing_z_retain_layer_27\endcsname{0.6}
\expandafter\def\csname ev@historical:intern/cyber/Routing-F0/0M/routing_z_retain_layer_28\endcsname{0.7}
\expandafter\def\csname ev@historical:intern/cyber/Routing-F0/0M/routing_z_retain_layer_29\endcsname{0.7}
\expandafter\def\csname ev@historical:intern/cyber/Routing-F0/0M/routing_z_retain_layer_30\endcsname{0.6}
\expandafter\def\csname ev@historical:intern/cyber/Routing-F0/0M/routing_z_retain_layer_31\endcsname{0.5}
\expandafter\def\csname ev@historical:intern/cyber/Routing-F0/0M/routing_z_retain_layer_32\endcsname{0.4}
\expandafter\def\csname ev@historical:intern/cyber/Routing-F0/0M/routing_z_retain_layer_33\endcsname{0.5}
\expandafter\def\csname ev@historical:intern/cyber/Routing-F0/0M/routing_z_retain_layer_34\endcsname{0.9}
\expandafter\def\csname ev@historical:intern/cyber/Routing-F0/0M/routing_z_retain_layer_35\endcsname{0.8}
\expandafter\def\csname ev@historical:intern/cyber/Routing-F0/0M/routing_z_retain_layer_36\endcsname{0.7}
\expandafter\def\csname ev@historical:intern/cyber/Routing-F0/0M/routing_z_retain_layer_37\endcsname{0.7}
\expandafter\def\csname ev@historical:intern/cyber/Routing-F0/0M/routing_z_retain_layer_38\endcsname{0.8}
\expandafter\def\csname ev@historical:intern/cyber/Routing-F0/0M/routing_z_retain_layer_39\endcsname{0.8}
\expandafter\def\csname ev@historical:intern/cyber/Routing-F1/0M/routing_z_forget\endcsname{46.8}
\expandafter\def\csname ev@historical:intern/cyber/Routing-F1/0M/routing_z_forget_layer_0\endcsname{10.3}
\expandafter\def\csname ev@historical:intern/cyber/Routing-F1/0M/routing_z_forget_layer_1\endcsname{15.1}
\expandafter\def\csname ev@historical:intern/cyber/Routing-F1/0M/routing_z_forget_layer_2\endcsname{8.1}
\expandafter\def\csname ev@historical:intern/cyber/Routing-F1/0M/routing_z_forget_layer_3\endcsname{20.7}
\expandafter\def\csname ev@historical:intern/cyber/Routing-F1/0M/routing_z_forget_layer_4\endcsname{24.1}
\expandafter\def\csname ev@historical:intern/cyber/Routing-F1/0M/routing_z_forget_layer_5\endcsname{33.9}
\expandafter\def\csname ev@historical:intern/cyber/Routing-F1/0M/routing_z_forget_layer_6\endcsname{31.0}
\expandafter\def\csname ev@historical:intern/cyber/Routing-F1/0M/routing_z_forget_layer_7\endcsname{43.1}
\expandafter\def\csname ev@historical:intern/cyber/Routing-F1/0M/routing_z_forget_layer_8\endcsname{48.7}
\expandafter\def\csname ev@historical:intern/cyber/Routing-F1/0M/routing_z_forget_layer_9\endcsname{50.8}
\expandafter\def\csname ev@historical:intern/cyber/Routing-F1/0M/routing_z_forget_layer_10\endcsname{34.6}
\expandafter\def\csname ev@historical:intern/cyber/Routing-F1/0M/routing_z_forget_layer_11\endcsname{35.8}
\expandafter\def\csname ev@historical:intern/cyber/Routing-F1/0M/routing_z_forget_layer_12\endcsname{37.3}
\expandafter\def\csname ev@historical:intern/cyber/Routing-F1/0M/routing_z_forget_layer_13\endcsname{44.5}
\expandafter\def\csname ev@historical:intern/cyber/Routing-F1/0M/routing_z_forget_layer_14\endcsname{48.3}
\expandafter\def\csname ev@historical:intern/cyber/Routing-F1/0M/routing_z_forget_layer_15\endcsname{49.1}
\expandafter\def\csname ev@historical:intern/cyber/Routing-F1/0M/routing_z_forget_layer_16\endcsname{40.1}
\expandafter\def\csname ev@historical:intern/cyber/Routing-F1/0M/routing_z_forget_layer_17\endcsname{44.8}
\expandafter\def\csname ev@historical:intern/cyber/Routing-F1/0M/routing_z_forget_layer_18\endcsname{50.4}
\expandafter\def\csname ev@historical:intern/cyber/Routing-F1/0M/routing_z_forget_layer_19\endcsname{53.2}
\expandafter\def\csname ev@historical:intern/cyber/Routing-F1/0M/routing_z_forget_layer_20\endcsname{59.5}
\expandafter\def\csname ev@historical:intern/cyber/Routing-F1/0M/routing_z_forget_layer_21\endcsname{57.4}
\expandafter\def\csname ev@historical:intern/cyber/Routing-F1/0M/routing_z_forget_layer_22\endcsname{43.7}
\expandafter\def\csname ev@historical:intern/cyber/Routing-F1/0M/routing_z_forget_layer_23\endcsname{43.9}
\expandafter\def\csname ev@historical:intern/cyber/Routing-F1/0M/routing_z_forget_layer_24\endcsname{42.7}
\expandafter\def\csname ev@historical:intern/cyber/Routing-F1/0M/routing_z_forget_layer_25\endcsname{48.0}
\expandafter\def\csname ev@historical:intern/cyber/Routing-F1/0M/routing_z_forget_layer_26\endcsname{49.3}
\expandafter\def\csname ev@historical:intern/cyber/Routing-F1/0M/routing_z_forget_layer_27\endcsname{49.9}
\expandafter\def\csname ev@historical:intern/cyber/Routing-F1/0M/routing_z_forget_layer_28\endcsname{45.6}
\expandafter\def\csname ev@historical:intern/cyber/Routing-F1/0M/routing_z_forget_layer_29\endcsname{53.1}
\expandafter\def\csname ev@historical:intern/cyber/Routing-F1/0M/routing_z_forget_layer_30\endcsname{57.6}
\expandafter\def\csname ev@historical:intern/cyber/Routing-F1/0M/routing_z_forget_layer_31\endcsname{66.3}
\expandafter\def\csname ev@historical:intern/cyber/Routing-F1/0M/routing_z_forget_layer_32\endcsname{58.8}
\expandafter\def\csname ev@historical:intern/cyber/Routing-F1/0M/routing_z_forget_layer_33\endcsname{59.2}
\expandafter\def\csname ev@historical:intern/cyber/Routing-F1/0M/routing_z_forget_layer_34\endcsname{56.5}
\expandafter\def\csname ev@historical:intern/cyber/Routing-F1/0M/routing_z_forget_layer_35\endcsname{71.0}
\expandafter\def\csname ev@historical:intern/cyber/Routing-F1/0M/routing_z_forget_layer_36\endcsname{74.4}
\expandafter\def\csname ev@historical:intern/cyber/Routing-F1/0M/routing_z_forget_layer_37\endcsname{75.0}
\expandafter\def\csname ev@historical:intern/cyber/Routing-F1/0M/routing_z_forget_layer_38\endcsname{76.1}
\expandafter\def\csname ev@historical:intern/cyber/Routing-F1/0M/routing_z_forget_layer_39\endcsname{61.9}
\expandafter\def\csname ev@historical:intern/cyber/Routing-F1/0M/routing_z_retain\endcsname{0.3}
\expandafter\def\csname ev@historical:intern/cyber/Routing-F1/0M/routing_z_retain_layer_0\endcsname{0.3}
\expandafter\def\csname ev@historical:intern/cyber/Routing-F1/0M/routing_z_retain_layer_1\endcsname{0.3}
\expandafter\def\csname ev@historical:intern/cyber/Routing-F1/0M/routing_z_retain_layer_2\endcsname{0.4}
\expandafter\def\csname ev@historical:intern/cyber/Routing-F1/0M/routing_z_retain_layer_3\endcsname{0.3}
\expandafter\def\csname ev@historical:intern/cyber/Routing-F1/0M/routing_z_retain_layer_4\endcsname{0.3}
\expandafter\def\csname ev@historical:intern/cyber/Routing-F1/0M/routing_z_retain_layer_5\endcsname{0.2}
\expandafter\def\csname ev@historical:intern/cyber/Routing-F1/0M/routing_z_retain_layer_6\endcsname{0.3}
\expandafter\def\csname ev@historical:intern/cyber/Routing-F1/0M/routing_z_retain_layer_7\endcsname{0.3}
\expandafter\def\csname ev@historical:intern/cyber/Routing-F1/0M/routing_z_retain_layer_8\endcsname{0.2}
\expandafter\def\csname ev@historical:intern/cyber/Routing-F1/0M/routing_z_retain_layer_9\endcsname{0.3}
\expandafter\def\csname ev@historical:intern/cyber/Routing-F1/0M/routing_z_retain_layer_10\endcsname{0.3}
\expandafter\def\csname ev@historical:intern/cyber/Routing-F1/0M/routing_z_retain_layer_11\endcsname{0.2}
\expandafter\def\csname ev@historical:intern/cyber/Routing-F1/0M/routing_z_retain_layer_12\endcsname{0.3}
\expandafter\def\csname ev@historical:intern/cyber/Routing-F1/0M/routing_z_retain_layer_13\endcsname{0.3}
\expandafter\def\csname ev@historical:intern/cyber/Routing-F1/0M/routing_z_retain_layer_14\endcsname{0.2}
\expandafter\def\csname ev@historical:intern/cyber/Routing-F1/0M/routing_z_retain_layer_15\endcsname{0.2}
\expandafter\def\csname ev@historical:intern/cyber/Routing-F1/0M/routing_z_retain_layer_16\endcsname{0.3}
\expandafter\def\csname ev@historical:intern/cyber/Routing-F1/0M/routing_z_retain_layer_17\endcsname{0.3}
\expandafter\def\csname ev@historical:intern/cyber/Routing-F1/0M/routing_z_retain_layer_18\endcsname{0.3}
\expandafter\def\csname ev@historical:intern/cyber/Routing-F1/0M/routing_z_retain_layer_19\endcsname{0.2}
\expandafter\def\csname ev@historical:intern/cyber/Routing-F1/0M/routing_z_retain_layer_20\endcsname{0.2}
\expandafter\def\csname ev@historical:intern/cyber/Routing-F1/0M/routing_z_retain_layer_21\endcsname{0.3}
\expandafter\def\csname ev@historical:intern/cyber/Routing-F1/0M/routing_z_retain_layer_22\endcsname{0.2}
\expandafter\def\csname ev@historical:intern/cyber/Routing-F1/0M/routing_z_retain_layer_23\endcsname{0.3}
\expandafter\def\csname ev@historical:intern/cyber/Routing-F1/0M/routing_z_retain_layer_24\endcsname{0.2}
\expandafter\def\csname ev@historical:intern/cyber/Routing-F1/0M/routing_z_retain_layer_25\endcsname{0.2}
\expandafter\def\csname ev@historical:intern/cyber/Routing-F1/0M/routing_z_retain_layer_26\endcsname{0.2}
\expandafter\def\csname ev@historical:intern/cyber/Routing-F1/0M/routing_z_retain_layer_27\endcsname{0.2}
\expandafter\def\csname ev@historical:intern/cyber/Routing-F1/0M/routing_z_retain_layer_28\endcsname{0.2}
\expandafter\def\csname ev@historical:intern/cyber/Routing-F1/0M/routing_z_retain_layer_29\endcsname{0.3}
\expandafter\def\csname ev@historical:intern/cyber/Routing-F1/0M/routing_z_retain_layer_30\endcsname{0.3}
\expandafter\def\csname ev@historical:intern/cyber/Routing-F1/0M/routing_z_retain_layer_31\endcsname{0.3}
\expandafter\def\csname ev@historical:intern/cyber/Routing-F1/0M/routing_z_retain_layer_32\endcsname{0.2}
\expandafter\def\csname ev@historical:intern/cyber/Routing-F1/0M/routing_z_retain_layer_33\endcsname{0.2}
\expandafter\def\csname ev@historical:intern/cyber/Routing-F1/0M/routing_z_retain_layer_34\endcsname{0.3}
\expandafter\def\csname ev@historical:intern/cyber/Routing-F1/0M/routing_z_retain_layer_35\endcsname{0.3}
\expandafter\def\csname ev@historical:intern/cyber/Routing-F1/0M/routing_z_retain_layer_36\endcsname{0.3}
\expandafter\def\csname ev@historical:intern/cyber/Routing-F1/0M/routing_z_retain_layer_37\endcsname{0.3}
\expandafter\def\csname ev@historical:intern/cyber/Routing-F1/0M/routing_z_retain_layer_38\endcsname{0.3}
\expandafter\def\csname ev@historical:intern/cyber/Routing-F1/0M/routing_z_retain_layer_39\endcsname{0.2}
\expandafter\def\csname ev@historical:qwen/cyber/B/0M/mmlu\endcsname{82.5}
\expandafter\def\csname ev@historical:qwen/cyber/B/0M/wmdp\endcsname{57.2}
\expandafter\def\csname ev@historical:qwen/cyber/B/0M/ifeval\endcsname{70.4}
\expandafter\def\csname ev@historical:qwen/cyber/S_std(B)/250M/ifeval\endcsname{52.5}
\expandafter\def\csname ev@historical:qwen/cyber/F0/0M/mmlu\endcsname{82.2}
\expandafter\def\csname ev@historical:qwen/cyber/F0/0M/wmdp\endcsname{57.1}
\expandafter\def\csname ev@historical:qwen/cyber/A0/0M/mmlu\endcsname{81.6}
\expandafter\def\csname ev@historical:qwen/cyber/A0/0M/wmdp\endcsname{48.5}
\expandafter\def\csname ev@historical:qwen/cyber/F1/0M/mmlu\endcsname{81.5}
\expandafter\def\csname ev@historical:qwen/cyber/F1/0M/wmdp\endcsname{57.3}
\expandafter\def\csname ev@historical:qwen/cyber/A1/0M/mmlu\endcsname{81.0}
\expandafter\def\csname ev@historical:qwen/cyber/A1/0M/wmdp\endcsname{43.7}
\expandafter\def\csname ev@historical:qwen/cyber/A2/0M/mmlu\endcsname{80.7}
\expandafter\def\csname ev@historical:qwen/cyber/A2/0M/wmdp\endcsname{27.9}
\expandafter\def\csname ev@historical:qwen/cyber/F2/0M/mmlu\endcsname{81.3}
\expandafter\def\csname ev@historical:qwen/cyber/F2/0M/wmdp\endcsname{46.5}
\expandafter\def\csname ev@historical:qwen/cyber/F3/250M/ifeval\endcsname{47.1}
\expandafter\def\csname ev@historical:qwen/cyber/A3/250M/ifeval\endcsname{44.5}
\expandafter\def\csname ev@historical:qwen/cyber/R/0M/mmlu\endcsname{82.2}
\expandafter\def\csname ev@historical:qwen/cyber/R/0M/wmdp\endcsname{32.0}
\expandafter\def\csname ev@historical:qwen/cyber/R-SFT/250M/ifeval\endcsname{47.5}
\expandafter\def\csname ev@historical:qwen/cyber/A3-direct/250M/ifeval\endcsname{47.1}
\expandafter\def\csname ev@historical:qwen/cyber/CEU-SFT/250M/ifeval\endcsname{65.8}
\expandafter\def\csname ev@historical:qwen/cyber/GA-SFT/250M/ifeval\endcsname{56.0}
\expandafter\def\csname ev@historical:qwen/cyber/GD-SFT/250M/ifeval\endcsname{50.1}
\expandafter\def\csname ev@historical:qwen/cyber/PureNPO-SFT/250M/ifeval\endcsname{45.5}
\expandafter\def\csname ev@historical:qwen/cyber/SatImp-SFT/250M/ifeval\endcsname{67.1}
\expandafter\def\csname ev@historical:qwen/cyber/SimNPO-SFT/250M/ifeval\endcsname{50.1}
\expandafter\def\csname ev@historical:qwen/cyber/TRACE-SFT/250M/ifeval\endcsname{56.2}
\expandafter\def\csname ev@historical:qwen/cyber/WGA-SFT/250M/ifeval\endcsname{64.9}
\expandafter\def\csname ev@historical:qwen/cyber/S_std(B)/10M/wmdp\endcsname{56.7}
\expandafter\def\csname ev@historical:qwen/cyber/S_std(B)/10M/mmlu\endcsname{82.5}
\expandafter\def\csname ev@historical:qwen/cyber/S_std(B)/50M/wmdp\endcsname{56.6}
\expandafter\def\csname ev@historical:qwen/cyber/S_std(B)/50M/mmlu\endcsname{82.2}
\expandafter\def\csname ev@historical:qwen/cyber/S_std(B)/100M/wmdp\endcsname{55.5}
\expandafter\def\csname ev@historical:qwen/cyber/S_std(B)/100M/mmlu\endcsname{82.3}
\expandafter\def\csname ev@historical:qwen/cyber/S_std(B)/250M/wmdp\endcsname{55.3}
\expandafter\def\csname ev@historical:qwen/cyber/S_std(B)/250M/mmlu\endcsname{82.3}
\expandafter\def\csname ev@historical:qwen/cyber/A2-masked/0M/wmdp\endcsname{27.8}
\expandafter\def\csname ev@historical:qwen/cyber/A2-masked/0M/mmlu\endcsname{80.6}
\expandafter\def\csname ev@historical:qwen/cyber/F3/10M/wmdp\endcsname{46.5}
\expandafter\def\csname ev@historical:qwen/cyber/F3/10M/mmlu\endcsname{81.6}
\expandafter\def\csname ev@historical:qwen/cyber/A3/10M/wmdp\endcsname{27.5}
\expandafter\def\csname ev@historical:qwen/cyber/A3/10M/mmlu\endcsname{80.7}
\expandafter\def\csname ev@historical:qwen/cyber/F3/50M/wmdp\endcsname{47.7}
\expandafter\def\csname ev@historical:qwen/cyber/F3/50M/mmlu\endcsname{81.6}
\expandafter\def\csname ev@historical:qwen/cyber/A3/50M/wmdp\endcsname{27.3}
\expandafter\def\csname ev@historical:qwen/cyber/A3/50M/mmlu\endcsname{80.9}
\expandafter\def\csname ev@historical:qwen/cyber/F3/100M/wmdp\endcsname{49.3}
\expandafter\def\csname ev@historical:qwen/cyber/F3/100M/mmlu\endcsname{81.9}
\expandafter\def\csname ev@historical:qwen/cyber/A3/100M/wmdp\endcsname{27.0}
\expandafter\def\csname ev@historical:qwen/cyber/A3/100M/mmlu\endcsname{81.1}
\expandafter\def\csname ev@historical:qwen/cyber/F3/250M/wmdp\endcsname{50.1}
\expandafter\def\csname ev@historical:qwen/cyber/F3/250M/mmlu\endcsname{81.8}
\expandafter\def\csname ev@historical:qwen/cyber/A3/250M/wmdp\endcsname{27.0}
\expandafter\def\csname ev@historical:qwen/cyber/A3/250M/mmlu\endcsname{81.3}
\expandafter\def\csname ev@historical:qwen/cyber/A3-direct/10M/wmdp\endcsname{27.3}
\expandafter\def\csname ev@historical:qwen/cyber/A3-direct/10M/mmlu\endcsname{80.8}
\expandafter\def\csname ev@historical:qwen/cyber/A3-direct/50M/wmdp\endcsname{27.7}
\expandafter\def\csname ev@historical:qwen/cyber/A3-direct/50M/mmlu\endcsname{81.0}
\expandafter\def\csname ev@historical:qwen/cyber/A3-direct/100M/wmdp\endcsname{32.9}
\expandafter\def\csname ev@historical:qwen/cyber/A3-direct/100M/mmlu\endcsname{81.4}
\expandafter\def\csname ev@historical:qwen/cyber/A3-direct/250M/wmdp\endcsname{35.3}
\expandafter\def\csname ev@historical:qwen/cyber/A3-direct/250M/mmlu\endcsname{81.4}
\expandafter\def\csname ev@historical:qwen/cyber/R-SFT/10M/wmdp\endcsname{32.5}
\expandafter\def\csname ev@historical:qwen/cyber/R-SFT/10M/mmlu\endcsname{82.1}
\expandafter\def\csname ev@historical:qwen/cyber/R-SFT/50M/wmdp\endcsname{35.1}
\expandafter\def\csname ev@historical:qwen/cyber/R-SFT/50M/mmlu\endcsname{82.2}
\expandafter\def\csname ev@historical:qwen/cyber/R-SFT/100M/wmdp\endcsname{43.0}
\expandafter\def\csname ev@historical:qwen/cyber/R-SFT/100M/mmlu\endcsname{82.1}
\expandafter\def\csname ev@historical:qwen/cyber/R-SFT/250M/wmdp\endcsname{44.2}
\expandafter\def\csname ev@historical:qwen/cyber/R-SFT/250M/mmlu\endcsname{82.1}
\expandafter\def\csname ev@historical:qwen/cyber/CEU-SFT/10M/wmdp\endcsname{35.8}
\expandafter\def\csname ev@historical:qwen/cyber/CEU-SFT/10M/mmlu\endcsname{75.3}
\expandafter\def\csname ev@historical:qwen/cyber/CEU-SFT/50M/wmdp\endcsname{43.1}
\expandafter\def\csname ev@historical:qwen/cyber/CEU-SFT/50M/mmlu\endcsname{80.1}
\expandafter\def\csname ev@historical:qwen/cyber/CEU-SFT/100M/wmdp\endcsname{47.4}
\expandafter\def\csname ev@historical:qwen/cyber/CEU-SFT/100M/mmlu\endcsname{81.3}
\expandafter\def\csname ev@historical:qwen/cyber/CEU-SFT/250M/wmdp\endcsname{48.7}
\expandafter\def\csname ev@historical:qwen/cyber/CEU-SFT/250M/mmlu\endcsname{81.3}
\expandafter\def\csname ev@historical:qwen/cyber/GA-SFT/10M/wmdp\endcsname{36.7}
\expandafter\def\csname ev@historical:qwen/cyber/GA-SFT/10M/mmlu\endcsname{74.2}
\expandafter\def\csname ev@historical:qwen/cyber/GA-SFT/50M/wmdp\endcsname{45.2}
\expandafter\def\csname ev@historical:qwen/cyber/GA-SFT/50M/mmlu\endcsname{80.5}
\expandafter\def\csname ev@historical:qwen/cyber/GA-SFT/100M/wmdp\endcsname{48.0}
\expandafter\def\csname ev@historical:qwen/cyber/GA-SFT/100M/mmlu\endcsname{81.2}
\expandafter\def\csname ev@historical:qwen/cyber/GA-SFT/250M/wmdp\endcsname{46.2}
\expandafter\def\csname ev@historical:qwen/cyber/GA-SFT/250M/mmlu\endcsname{81.0}
\expandafter\def\csname ev@historical:qwen/cyber/GD-SFT/10M/wmdp\endcsname{38.8}
\expandafter\def\csname ev@historical:qwen/cyber/GD-SFT/10M/mmlu\endcsname{79.0}
\expandafter\def\csname ev@historical:qwen/cyber/GD-SFT/50M/wmdp\endcsname{47.5}
\expandafter\def\csname ev@historical:qwen/cyber/GD-SFT/50M/mmlu\endcsname{81.7}
\expandafter\def\csname ev@historical:qwen/cyber/GD-SFT/100M/wmdp\endcsname{48.2}
\expandafter\def\csname ev@historical:qwen/cyber/GD-SFT/100M/mmlu\endcsname{81.6}
\expandafter\def\csname ev@historical:qwen/cyber/GD-SFT/250M/wmdp\endcsname{47.8}
\expandafter\def\csname ev@historical:qwen/cyber/GD-SFT/250M/mmlu\endcsname{81.4}
\expandafter\def\csname ev@historical:qwen/cyber/PDU-SFT/10M/wmdp\endcsname{42.6}
\expandafter\def\csname ev@historical:qwen/cyber/PDU-SFT/10M/mmlu\endcsname{82.0}
\expandafter\def\csname ev@historical:qwen/cyber/PDU-SFT/50M/wmdp\endcsname{46.8}
\expandafter\def\csname ev@historical:qwen/cyber/PDU-SFT/50M/mmlu\endcsname{81.8}
\expandafter\def\csname ev@historical:qwen/cyber/PDU-SFT/100M/wmdp\endcsname{46.0}
\expandafter\def\csname ev@historical:qwen/cyber/PDU-SFT/100M/mmlu\endcsname{82.1}
\expandafter\def\csname ev@historical:qwen/cyber/PDU-SFT/250M/wmdp\endcsname{45.8}
\expandafter\def\csname ev@historical:qwen/cyber/PDU-SFT/250M/mmlu\endcsname{82.2}
\expandafter\def\csname ev@historical:qwen/cyber/PureNPO-SFT/10M/wmdp\endcsname{42.2}
\expandafter\def\csname ev@historical:qwen/cyber/PureNPO-SFT/10M/mmlu\endcsname{81.4}
\expandafter\def\csname ev@historical:qwen/cyber/PureNPO-SFT/50M/wmdp\endcsname{45.8}
\expandafter\def\csname ev@historical:qwen/cyber/PureNPO-SFT/50M/mmlu\endcsname{81.4}
\expandafter\def\csname ev@historical:qwen/cyber/PureNPO-SFT/100M/wmdp\endcsname{48.7}
\expandafter\def\csname ev@historical:qwen/cyber/PureNPO-SFT/100M/mmlu\endcsname{81.5}
\expandafter\def\csname ev@historical:qwen/cyber/PureNPO-SFT/250M/wmdp\endcsname{48.5}
\expandafter\def\csname ev@historical:qwen/cyber/PureNPO-SFT/250M/mmlu\endcsname{81.8}
\expandafter\def\csname ev@historical:qwen/cyber/ExternalRMU-SFT/10M/wmdp\endcsname{28.0}
\expandafter\def\csname ev@historical:qwen/cyber/ExternalRMU-SFT/10M/mmlu\endcsname{81.8}
\expandafter\def\csname ev@historical:qwen/cyber/ExternalRMU-SFT/50M/wmdp\endcsname{31.9}
\expandafter\def\csname ev@historical:qwen/cyber/ExternalRMU-SFT/50M/mmlu\endcsname{81.7}
\expandafter\def\csname ev@historical:qwen/cyber/ExternalRMU-SFT/100M/wmdp\endcsname{36.1}
\expandafter\def\csname ev@historical:qwen/cyber/ExternalRMU-SFT/100M/mmlu\endcsname{82.2}
\expandafter\def\csname ev@historical:qwen/cyber/ExternalRMU-SFT/250M/wmdp\endcsname{40.6}
\expandafter\def\csname ev@historical:qwen/cyber/ExternalRMU-SFT/250M/mmlu\endcsname{82.2}
\expandafter\def\csname ev@historical:qwen/cyber/SatImp-SFT/10M/wmdp\endcsname{33.3}
\expandafter\def\csname ev@historical:qwen/cyber/SatImp-SFT/10M/mmlu\endcsname{74.9}
\expandafter\def\csname ev@historical:qwen/cyber/SatImp-SFT/50M/wmdp\endcsname{38.5}
\expandafter\def\csname ev@historical:qwen/cyber/SatImp-SFT/50M/mmlu\endcsname{79.7}
\expandafter\def\csname ev@historical:qwen/cyber/SatImp-SFT/100M/wmdp\endcsname{43.5}
\expandafter\def\csname ev@historical:qwen/cyber/SatImp-SFT/100M/mmlu\endcsname{80.8}
\expandafter\def\csname ev@historical:qwen/cyber/SatImp-SFT/250M/wmdp\endcsname{44.4}
\expandafter\def\csname ev@historical:qwen/cyber/SatImp-SFT/250M/mmlu\endcsname{81.2}
\expandafter\def\csname ev@historical:qwen/cyber/SimNPO-SFT/10M/wmdp\endcsname{42.7}
\expandafter\def\csname ev@historical:qwen/cyber/SimNPO-SFT/10M/mmlu\endcsname{81.2}
\expandafter\def\csname ev@historical:qwen/cyber/SimNPO-SFT/50M/wmdp\endcsname{46.2}
\expandafter\def\csname ev@historical:qwen/cyber/SimNPO-SFT/50M/mmlu\endcsname{81.4}
\expandafter\def\csname ev@historical:qwen/cyber/SimNPO-SFT/100M/wmdp\endcsname{51.5}
\expandafter\def\csname ev@historical:qwen/cyber/SimNPO-SFT/100M/mmlu\endcsname{81.8}
\expandafter\def\csname ev@historical:qwen/cyber/SimNPO-SFT/250M/wmdp\endcsname{52.0}
\expandafter\def\csname ev@historical:qwen/cyber/SimNPO-SFT/250M/mmlu\endcsname{81.8}
\expandafter\def\csname ev@historical:qwen/cyber/TRACE-SFT/10M/wmdp\endcsname{46.0}
\expandafter\def\csname ev@historical:qwen/cyber/TRACE-SFT/10M/mmlu\endcsname{81.7}
\expandafter\def\csname ev@historical:qwen/cyber/TRACE-SFT/50M/wmdp\endcsname{48.0}
\expandafter\def\csname ev@historical:qwen/cyber/TRACE-SFT/50M/mmlu\endcsname{82.1}
\expandafter\def\csname ev@historical:qwen/cyber/TRACE-SFT/100M/wmdp\endcsname{47.9}
\expandafter\def\csname ev@historical:qwen/cyber/TRACE-SFT/100M/mmlu\endcsname{82.1}
\expandafter\def\csname ev@historical:qwen/cyber/TRACE-SFT/250M/wmdp\endcsname{47.4}
\expandafter\def\csname ev@historical:qwen/cyber/TRACE-SFT/250M/mmlu\endcsname{82.0}
\expandafter\def\csname ev@historical:qwen/cyber/WGA-SFT/10M/wmdp\endcsname{32.1}
\expandafter\def\csname ev@historical:qwen/cyber/WGA-SFT/10M/mmlu\endcsname{73.3}
\expandafter\def\csname ev@historical:qwen/cyber/WGA-SFT/50M/wmdp\endcsname{38.2}
\expandafter\def\csname ev@historical:qwen/cyber/WGA-SFT/50M/mmlu\endcsname{79.0}
\expandafter\def\csname ev@historical:qwen/cyber/WGA-SFT/100M/wmdp\endcsname{42.6}
\expandafter\def\csname ev@historical:qwen/cyber/WGA-SFT/100M/mmlu\endcsname{80.5}
\expandafter\def\csname ev@historical:qwen/cyber/WGA-SFT/250M/wmdp\endcsname{44.5}
\expandafter\def\csname ev@historical:qwen/cyber/WGA-SFT/250M/mmlu\endcsname{81.1}
\expandafter\def\csname ev@historical:qwen/cyber/Routing-F0/0M/routing_z_forget\endcsname{16.0}
\expandafter\def\csname ev@historical:qwen/cyber/Routing-F0/0M/routing_z_forget_layer_0\endcsname{8.5}
\expandafter\def\csname ev@historical:qwen/cyber/Routing-F0/0M/routing_z_forget_layer_1\endcsname{10.6}
\expandafter\def\csname ev@historical:qwen/cyber/Routing-F0/0M/routing_z_forget_layer_2\endcsname{7.3}
\expandafter\def\csname ev@historical:qwen/cyber/Routing-F0/0M/routing_z_forget_layer_3\endcsname{13.6}
\expandafter\def\csname ev@historical:qwen/cyber/Routing-F0/0M/routing_z_forget_layer_4\endcsname{15.6}
\expandafter\def\csname ev@historical:qwen/cyber/Routing-F0/0M/routing_z_forget_layer_5\endcsname{17.4}
\expandafter\def\csname ev@historical:qwen/cyber/Routing-F0/0M/routing_z_forget_layer_6\endcsname{15.9}
\expandafter\def\csname ev@historical:qwen/cyber/Routing-F0/0M/routing_z_forget_layer_7\endcsname{21.4}
\expandafter\def\csname ev@historical:qwen/cyber/Routing-F0/0M/routing_z_forget_layer_8\endcsname{22.7}
\expandafter\def\csname ev@historical:qwen/cyber/Routing-F0/0M/routing_z_forget_layer_9\endcsname{21.7}
\expandafter\def\csname ev@historical:qwen/cyber/Routing-F0/0M/routing_z_forget_layer_10\endcsname{13.2}
\expandafter\def\csname ev@historical:qwen/cyber/Routing-F0/0M/routing_z_forget_layer_11\endcsname{13.4}
\expandafter\def\csname ev@historical:qwen/cyber/Routing-F0/0M/routing_z_forget_layer_12\endcsname{15.3}
\expandafter\def\csname ev@historical:qwen/cyber/Routing-F0/0M/routing_z_forget_layer_13\endcsname{17.1}
\expandafter\def\csname ev@historical:qwen/cyber/Routing-F0/0M/routing_z_forget_layer_14\endcsname{20.3}
\expandafter\def\csname ev@historical:qwen/cyber/Routing-F0/0M/routing_z_forget_layer_15\endcsname{16.5}
\expandafter\def\csname ev@historical:qwen/cyber/Routing-F0/0M/routing_z_forget_layer_16\endcsname{14.1}
\expandafter\def\csname ev@historical:qwen/cyber/Routing-F0/0M/routing_z_forget_layer_17\endcsname{15.5}
\expandafter\def\csname ev@historical:qwen/cyber/Routing-F0/0M/routing_z_forget_layer_18\endcsname{17.7}
\expandafter\def\csname ev@historical:qwen/cyber/Routing-F0/0M/routing_z_forget_layer_19\endcsname{15.7}
\expandafter\def\csname ev@historical:qwen/cyber/Routing-F0/0M/routing_z_forget_layer_20\endcsname{22.1}
\expandafter\def\csname ev@historical:qwen/cyber/Routing-F0/0M/routing_z_forget_layer_21\endcsname{22.3}
\expandafter\def\csname ev@historical:qwen/cyber/Routing-F0/0M/routing_z_forget_layer_22\endcsname{13.3}
\expandafter\def\csname ev@historical:qwen/cyber/Routing-F0/0M/routing_z_forget_layer_23\endcsname{14.8}
\expandafter\def\csname ev@historical:qwen/cyber/Routing-F0/0M/routing_z_forget_layer_24\endcsname{13.4}
\expandafter\def\csname ev@historical:qwen/cyber/Routing-F0/0M/routing_z_forget_layer_25\endcsname{15.8}
\expandafter\def\csname ev@historical:qwen/cyber/Routing-F0/0M/routing_z_forget_layer_26\endcsname{18.6}
\expandafter\def\csname ev@historical:qwen/cyber/Routing-F0/0M/routing_z_forget_layer_27\endcsname{14.8}
\expandafter\def\csname ev@historical:qwen/cyber/Routing-F0/0M/routing_z_forget_layer_28\endcsname{12.9}
\expandafter\def\csname ev@historical:qwen/cyber/Routing-F0/0M/routing_z_forget_layer_29\endcsname{13.9}
\expandafter\def\csname ev@historical:qwen/cyber/Routing-F0/0M/routing_z_forget_layer_30\endcsname{16.7}
\expandafter\def\csname ev@historical:qwen/cyber/Routing-F0/0M/routing_z_forget_layer_31\endcsname{15.0}
\expandafter\def\csname ev@historical:qwen/cyber/Routing-F0/0M/routing_z_forget_layer_32\endcsname{13.4}
\expandafter\def\csname ev@historical:qwen/cyber/Routing-F0/0M/routing_z_forget_layer_33\endcsname{16.0}
\expandafter\def\csname ev@historical:qwen/cyber/Routing-F0/0M/routing_z_forget_layer_34\endcsname{13.4}
\expandafter\def\csname ev@historical:qwen/cyber/Routing-F0/0M/routing_z_forget_layer_35\endcsname{18.8}
\expandafter\def\csname ev@historical:qwen/cyber/Routing-F0/0M/routing_z_forget_layer_36\endcsname{22.2}
\expandafter\def\csname ev@historical:qwen/cyber/Routing-F0/0M/routing_z_forget_layer_37\endcsname{17.7}
\expandafter\def\csname ev@historical:qwen/cyber/Routing-F0/0M/routing_z_forget_layer_38\endcsname{16.8}
\expandafter\def\csname ev@historical:qwen/cyber/Routing-F0/0M/routing_z_forget_layer_39\endcsname{12.9}
\expandafter\def\csname ev@historical:qwen/cyber/Routing-F0/0M/routing_z_retain\endcsname{1.1}
\expandafter\def\csname ev@historical:qwen/cyber/Routing-F0/0M/routing_z_retain_layer_0\endcsname{1.6}
\expandafter\def\csname ev@historical:qwen/cyber/Routing-F0/0M/routing_z_retain_layer_1\endcsname{1.3}
\expandafter\def\csname ev@historical:qwen/cyber/Routing-F0/0M/routing_z_retain_layer_2\endcsname{1.7}
\expandafter\def\csname ev@historical:qwen/cyber/Routing-F0/0M/routing_z_retain_layer_3\endcsname{1.0}
\expandafter\def\csname ev@historical:qwen/cyber/Routing-F0/0M/routing_z_retain_layer_4\endcsname{1.6}
\expandafter\def\csname ev@historical:qwen/cyber/Routing-F0/0M/routing_z_retain_layer_5\endcsname{0.8}
\expandafter\def\csname ev@historical:qwen/cyber/Routing-F0/0M/routing_z_retain_layer_6\endcsname{0.9}
\expandafter\def\csname ev@historical:qwen/cyber/Routing-F0/0M/routing_z_retain_layer_7\endcsname{0.9}
\expandafter\def\csname ev@historical:qwen/cyber/Routing-F0/0M/routing_z_retain_layer_8\endcsname{1.0}
\expandafter\def\csname ev@historical:qwen/cyber/Routing-F0/0M/routing_z_retain_layer_9\endcsname{1.0}
\expandafter\def\csname ev@historical:qwen/cyber/Routing-F0/0M/routing_z_retain_layer_10\endcsname{1.0}
\expandafter\def\csname ev@historical:qwen/cyber/Routing-F0/0M/routing_z_retain_layer_11\endcsname{1.2}
\expandafter\def\csname ev@historical:qwen/cyber/Routing-F0/0M/routing_z_retain_layer_12\endcsname{1.1}
\expandafter\def\csname ev@historical:qwen/cyber/Routing-F0/0M/routing_z_retain_layer_13\endcsname{0.9}
\expandafter\def\csname ev@historical:qwen/cyber/Routing-F0/0M/routing_z_retain_layer_14\endcsname{0.9}
\expandafter\def\csname ev@historical:qwen/cyber/Routing-F0/0M/routing_z_retain_layer_15\endcsname{1.1}
\expandafter\def\csname ev@historical:qwen/cyber/Routing-F0/0M/routing_z_retain_layer_16\endcsname{1.3}
\expandafter\def\csname ev@historical:qwen/cyber/Routing-F0/0M/routing_z_retain_layer_17\endcsname{1.5}
\expandafter\def\csname ev@historical:qwen/cyber/Routing-F0/0M/routing_z_retain_layer_18\endcsname{1.0}
\expandafter\def\csname ev@historical:qwen/cyber/Routing-F0/0M/routing_z_retain_layer_19\endcsname{0.8}
\expandafter\def\csname ev@historical:qwen/cyber/Routing-F0/0M/routing_z_retain_layer_20\endcsname{0.9}
\expandafter\def\csname ev@historical:qwen/cyber/Routing-F0/0M/routing_z_retain_layer_21\endcsname{1.0}
\expandafter\def\csname ev@historical:qwen/cyber/Routing-F0/0M/routing_z_retain_layer_22\endcsname{1.2}
\expandafter\def\csname ev@historical:qwen/cyber/Routing-F0/0M/routing_z_retain_layer_23\endcsname{1.3}
\expandafter\def\csname ev@historical:qwen/cyber/Routing-F0/0M/routing_z_retain_layer_24\endcsname{1.1}
\expandafter\def\csname ev@historical:qwen/cyber/Routing-F0/0M/routing_z_retain_layer_25\endcsname{1.0}
\expandafter\def\csname ev@historical:qwen/cyber/Routing-F0/0M/routing_z_retain_layer_26\endcsname{1.0}
\expandafter\def\csname ev@historical:qwen/cyber/Routing-F0/0M/routing_z_retain_layer_27\endcsname{1.1}
\expandafter\def\csname ev@historical:qwen/cyber/Routing-F0/0M/routing_z_retain_layer_28\endcsname{1.2}
\expandafter\def\csname ev@historical:qwen/cyber/Routing-F0/0M/routing_z_retain_layer_29\endcsname{0.9}
\expandafter\def\csname ev@historical:qwen/cyber/Routing-F0/0M/routing_z_retain_layer_30\endcsname{1.1}
\expandafter\def\csname ev@historical:qwen/cyber/Routing-F0/0M/routing_z_retain_layer_31\endcsname{0.7}
\expandafter\def\csname ev@historical:qwen/cyber/Routing-F0/0M/routing_z_retain_layer_32\endcsname{0.8}
\expandafter\def\csname ev@historical:qwen/cyber/Routing-F0/0M/routing_z_retain_layer_33\endcsname{0.6}
\expandafter\def\csname ev@historical:qwen/cyber/Routing-F0/0M/routing_z_retain_layer_34\endcsname{1.2}
\expandafter\def\csname ev@historical:qwen/cyber/Routing-F0/0M/routing_z_retain_layer_35\endcsname{1.0}
\expandafter\def\csname ev@historical:qwen/cyber/Routing-F0/0M/routing_z_retain_layer_36\endcsname{1.0}
\expandafter\def\csname ev@historical:qwen/cyber/Routing-F0/0M/routing_z_retain_layer_37\endcsname{1.1}
\expandafter\def\csname ev@historical:qwen/cyber/Routing-F0/0M/routing_z_retain_layer_38\endcsname{1.1}
\expandafter\def\csname ev@historical:qwen/cyber/Routing-F0/0M/routing_z_retain_layer_39\endcsname{1.3}
\expandafter\def\csname ev@historical:qwen/cyber/Routing-F1/0M/routing_z_forget\endcsname{35.2}
\expandafter\def\csname ev@historical:qwen/cyber/Routing-F1/0M/routing_z_forget_layer_0\endcsname{7.4}
\expandafter\def\csname ev@historical:qwen/cyber/Routing-F1/0M/routing_z_forget_layer_1\endcsname{8.6}
\expandafter\def\csname ev@historical:qwen/cyber/Routing-F1/0M/routing_z_forget_layer_2\endcsname{5.5}
\expandafter\def\csname ev@historical:qwen/cyber/Routing-F1/0M/routing_z_forget_layer_3\endcsname{11.0}
\expandafter\def\csname ev@historical:qwen/cyber/Routing-F1/0M/routing_z_forget_layer_4\endcsname{11.5}
\expandafter\def\csname ev@historical:qwen/cyber/Routing-F1/0M/routing_z_forget_layer_5\endcsname{18.3}
\expandafter\def\csname ev@historical:qwen/cyber/Routing-F1/0M/routing_z_forget_layer_6\endcsname{17.7}
\expandafter\def\csname ev@historical:qwen/cyber/Routing-F1/0M/routing_z_forget_layer_7\endcsname{27.8}
\expandafter\def\csname ev@historical:qwen/cyber/Routing-F1/0M/routing_z_forget_layer_8\endcsname{33.2}
\expandafter\def\csname ev@historical:qwen/cyber/Routing-F1/0M/routing_z_forget_layer_9\endcsname{34.8}
\expandafter\def\csname ev@historical:qwen/cyber/Routing-F1/0M/routing_z_forget_layer_10\endcsname{22.7}
\expandafter\def\csname ev@historical:qwen/cyber/Routing-F1/0M/routing_z_forget_layer_11\endcsname{21.8}
\expandafter\def\csname ev@historical:qwen/cyber/Routing-F1/0M/routing_z_forget_layer_12\endcsname{23.9}
\expandafter\def\csname ev@historical:qwen/cyber/Routing-F1/0M/routing_z_forget_layer_13\endcsname{30.4}
\expandafter\def\csname ev@historical:qwen/cyber/Routing-F1/0M/routing_z_forget_layer_14\endcsname{35.4}
\expandafter\def\csname ev@historical:qwen/cyber/Routing-F1/0M/routing_z_forget_layer_15\endcsname{34.8}
\expandafter\def\csname ev@historical:qwen/cyber/Routing-F1/0M/routing_z_forget_layer_16\endcsname{30.3}
\expandafter\def\csname ev@historical:qwen/cyber/Routing-F1/0M/routing_z_forget_layer_17\endcsname{34.0}
\expandafter\def\csname ev@historical:qwen/cyber/Routing-F1/0M/routing_z_forget_layer_18\endcsname{36.8}
\expandafter\def\csname ev@historical:qwen/cyber/Routing-F1/0M/routing_z_forget_layer_19\endcsname{38.8}
\expandafter\def\csname ev@historical:qwen/cyber/Routing-F1/0M/routing_z_forget_layer_20\endcsname{45.0}
\expandafter\def\csname ev@historical:qwen/cyber/Routing-F1/0M/routing_z_forget_layer_21\endcsname{44.5}
\expandafter\def\csname ev@historical:qwen/cyber/Routing-F1/0M/routing_z_forget_layer_22\endcsname{29.2}
\expandafter\def\csname ev@historical:qwen/cyber/Routing-F1/0M/routing_z_forget_layer_23\endcsname{30.2}
\expandafter\def\csname ev@historical:qwen/cyber/Routing-F1/0M/routing_z_forget_layer_24\endcsname{29.2}
\expandafter\def\csname ev@historical:qwen/cyber/Routing-F1/0M/routing_z_forget_layer_25\endcsname{33.9}
\expandafter\def\csname ev@historical:qwen/cyber/Routing-F1/0M/routing_z_forget_layer_26\endcsname{37.1}
\expandafter\def\csname ev@historical:qwen/cyber/Routing-F1/0M/routing_z_forget_layer_27\endcsname{37.9}
\expandafter\def\csname ev@historical:qwen/cyber/Routing-F1/0M/routing_z_forget_layer_28\endcsname{33.4}
\expandafter\def\csname ev@historical:qwen/cyber/Routing-F1/0M/routing_z_forget_layer_29\endcsname{39.0}
\expandafter\def\csname ev@historical:qwen/cyber/Routing-F1/0M/routing_z_forget_layer_30\endcsname{44.9}
\expandafter\def\csname ev@historical:qwen/cyber/Routing-F1/0M/routing_z_forget_layer_31\endcsname{50.6}
\expandafter\def\csname ev@historical:qwen/cyber/Routing-F1/0M/routing_z_forget_layer_32\endcsname{47.8}
\expandafter\def\csname ev@historical:qwen/cyber/Routing-F1/0M/routing_z_forget_layer_33\endcsname{47.9}
\expandafter\def\csname ev@historical:qwen/cyber/Routing-F1/0M/routing_z_forget_layer_34\endcsname{43.2}
\expandafter\def\csname ev@historical:qwen/cyber/Routing-F1/0M/routing_z_forget_layer_35\endcsname{62.8}
\expandafter\def\csname ev@historical:qwen/cyber/Routing-F1/0M/routing_z_forget_layer_36\endcsname{69.2}
\expandafter\def\csname ev@historical:qwen/cyber/Routing-F1/0M/routing_z_forget_layer_37\endcsname{68.2}
\expandafter\def\csname ev@historical:qwen/cyber/Routing-F1/0M/routing_z_forget_layer_38\endcsname{73.5}
\expandafter\def\csname ev@historical:qwen/cyber/Routing-F1/0M/routing_z_forget_layer_39\endcsname{55.9}
\expandafter\def\csname ev@historical:qwen/cyber/Routing-F1/0M/routing_z_retain\endcsname{0.1}
\expandafter\def\csname ev@historical:qwen/cyber/Routing-F1/0M/routing_z_retain_layer_0\endcsname{0.6}
\expandafter\def\csname ev@historical:qwen/cyber/Routing-F1/0M/routing_z_retain_layer_1\endcsname{0.1}
\expandafter\def\csname ev@historical:qwen/cyber/Routing-F1/0M/routing_z_retain_layer_2\endcsname{0.2}
\expandafter\def\csname ev@historical:qwen/cyber/Routing-F1/0M/routing_z_retain_layer_3\endcsname{0.1}
\expandafter\def\csname ev@historical:qwen/cyber/Routing-F1/0M/routing_z_retain_layer_4\endcsname{0.1}
\expandafter\def\csname ev@historical:qwen/cyber/Routing-F1/0M/routing_z_retain_layer_5\endcsname{0.1}
\expandafter\def\csname ev@historical:qwen/cyber/Routing-F1/0M/routing_z_retain_layer_6\endcsname{0.1}
\expandafter\def\csname ev@historical:qwen/cyber/Routing-F1/0M/routing_z_retain_layer_7\endcsname{0.1}
\expandafter\def\csname ev@historical:qwen/cyber/Routing-F1/0M/routing_z_retain_layer_8\endcsname{0.0}
\expandafter\def\csname ev@historical:qwen/cyber/Routing-F1/0M/routing_z_retain_layer_9\endcsname{0.1}
\expandafter\def\csname ev@historical:qwen/cyber/Routing-F1/0M/routing_z_retain_layer_10\endcsname{0.1}
\expandafter\def\csname ev@historical:qwen/cyber/Routing-F1/0M/routing_z_retain_layer_11\endcsname{0.1}
\expandafter\def\csname ev@historical:qwen/cyber/Routing-F1/0M/routing_z_retain_layer_12\endcsname{0.1}
\expandafter\def\csname ev@historical:qwen/cyber/Routing-F1/0M/routing_z_retain_layer_13\endcsname{0.1}
\expandafter\def\csname ev@historical:qwen/cyber/Routing-F1/0M/routing_z_retain_layer_14\endcsname{0.1}
\expandafter\def\csname ev@historical:qwen/cyber/Routing-F1/0M/routing_z_retain_layer_15\endcsname{0.1}
\expandafter\def\csname ev@historical:qwen/cyber/Routing-F1/0M/routing_z_retain_layer_16\endcsname{0.1}
\expandafter\def\csname ev@historical:qwen/cyber/Routing-F1/0M/routing_z_retain_layer_17\endcsname{0.1}
\expandafter\def\csname ev@historical:qwen/cyber/Routing-F1/0M/routing_z_retain_layer_18\endcsname{0.1}
\expandafter\def\csname ev@historical:qwen/cyber/Routing-F1/0M/routing_z_retain_layer_19\endcsname{0.1}
\expandafter\def\csname ev@historical:qwen/cyber/Routing-F1/0M/routing_z_retain_layer_20\endcsname{0.1}
\expandafter\def\csname ev@historical:qwen/cyber/Routing-F1/0M/routing_z_retain_layer_21\endcsname{0.1}
\expandafter\def\csname ev@historical:qwen/cyber/Routing-F1/0M/routing_z_retain_layer_22\endcsname{0.1}
\expandafter\def\csname ev@historical:qwen/cyber/Routing-F1/0M/routing_z_retain_layer_23\endcsname{0.1}
\expandafter\def\csname ev@historical:qwen/cyber/Routing-F1/0M/routing_z_retain_layer_24\endcsname{0.0}
\expandafter\def\csname ev@historical:qwen/cyber/Routing-F1/0M/routing_z_retain_layer_25\endcsname{0.1}
\expandafter\def\csname ev@historical:qwen/cyber/Routing-F1/0M/routing_z_retain_layer_26\endcsname{0.0}
\expandafter\def\csname ev@historical:qwen/cyber/Routing-F1/0M/routing_z_retain_layer_27\endcsname{0.1}
\expandafter\def\csname ev@historical:qwen/cyber/Routing-F1/0M/routing_z_retain_layer_28\endcsname{0.1}
\expandafter\def\csname ev@historical:qwen/cyber/Routing-F1/0M/routing_z_retain_layer_29\endcsname{0.1}
\expandafter\def\csname ev@historical:qwen/cyber/Routing-F1/0M/routing_z_retain_layer_30\endcsname{0.1}
\expandafter\def\csname ev@historical:qwen/cyber/Routing-F1/0M/routing_z_retain_layer_31\endcsname{0.1}
\expandafter\def\csname ev@historical:qwen/cyber/Routing-F1/0M/routing_z_retain_layer_32\endcsname{0.1}
\expandafter\def\csname ev@historical:qwen/cyber/Routing-F1/0M/routing_z_retain_layer_33\endcsname{0.1}
\expandafter\def\csname ev@historical:qwen/cyber/Routing-F1/0M/routing_z_retain_layer_34\endcsname{0.2}
\expandafter\def\csname ev@historical:qwen/cyber/Routing-F1/0M/routing_z_retain_layer_35\endcsname{0.2}
\expandafter\def\csname ev@historical:qwen/cyber/Routing-F1/0M/routing_z_retain_layer_36\endcsname{0.2}
\expandafter\def\csname ev@historical:qwen/cyber/Routing-F1/0M/routing_z_retain_layer_37\endcsname{0.2}
\expandafter\def\csname ev@historical:qwen/cyber/Routing-F1/0M/routing_z_retain_layer_38\endcsname{0.2}
\expandafter\def\csname ev@historical:qwen/cyber/Routing-F1/0M/routing_z_retain_layer_39\endcsname{0.1}
\expandafter\def\csname ev@historical:gemma/cyber/B/0M/mmlu\endcsname{74.5}
\expandafter\def\csname ev@historical:gemma/cyber/B/0M/wmdp\endcsname{49.7}
\expandafter\def\csname ev@historical:gemma/cyber/B/0M/ifeval\endcsname{27.5}
\expandafter\def\csname ev@historical:gemma/cyber/S_std(B)/250M/ifeval\endcsname{27.5}
\expandafter\def\csname ev@historical:gemma/cyber/F0/0M/mmlu\endcsname{74.0}
\expandafter\def\csname ev@historical:gemma/cyber/F0/0M/wmdp\endcsname{49.9}
\expandafter\def\csname ev@historical:gemma/cyber/A0/0M/mmlu\endcsname{73.8}
\expandafter\def\csname ev@historical:gemma/cyber/A0/0M/wmdp\endcsname{47.6}
\expandafter\def\csname ev@historical:gemma/cyber/CEU-SFT/250M/ifeval\endcsname{11.1}
\expandafter\def\csname ev@historical:gemma/cyber/GA-SFT/250M/ifeval\endcsname{17.2}
\expandafter\def\csname ev@historical:gemma/cyber/GD-SFT/250M/ifeval\endcsname{24.6}
\expandafter\def\csname ev@historical:gemma/cyber/PDU-SFT/250M/ifeval\endcsname{25.3}
\expandafter\def\csname ev@historical:gemma/cyber/PureNPO-SFT/250M/ifeval\endcsname{23.1}
\expandafter\def\csname ev@historical:gemma/cyber/ExternalRMU-SFT/250M/ifeval\endcsname{27.0}
\expandafter\def\csname ev@historical:gemma/cyber/SatImp-SFT/250M/ifeval\endcsname{14.6}
\expandafter\def\csname ev@historical:gemma/cyber/SimNPO-SFT/250M/ifeval\endcsname{20.1}
\expandafter\def\csname ev@historical:gemma/cyber/TRACE-SFT/250M/ifeval\endcsname{27.0}
\expandafter\def\csname ev@historical:gemma/cyber/WGA-SFT/250M/ifeval\endcsname{13.3}
\expandafter\def\csname ev@historical:gemma/cyber/S_std(B)/10M/wmdp\endcsname{50.3}
\expandafter\def\csname ev@historical:gemma/cyber/S_std(B)/10M/mmlu\endcsname{74.2}
\expandafter\def\csname ev@historical:gemma/cyber/S_std(B)/50M/wmdp\endcsname{50.3}
\expandafter\def\csname ev@historical:gemma/cyber/S_std(B)/50M/mmlu\endcsname{74.3}
\expandafter\def\csname ev@historical:gemma/cyber/S_std(B)/100M/wmdp\endcsname{50.5}
\expandafter\def\csname ev@historical:gemma/cyber/S_std(B)/100M/mmlu\endcsname{74.7}
\expandafter\def\csname ev@historical:gemma/cyber/S_std(B)/250M/wmdp\endcsname{50.9}
\expandafter\def\csname ev@historical:gemma/cyber/S_std(B)/250M/mmlu\endcsname{74.8}
\expandafter\def\csname ev@historical:gemma/cyber/A2-masked/0M/wmdp\endcsname{30.3}
\expandafter\def\csname ev@historical:gemma/cyber/A2-masked/0M/mmlu\endcsname{72.1}
\expandafter\def\csname ev@historical:gemma/cyber/F3/10M/wmdp\endcsname{39.4}
\expandafter\def\csname ev@historical:gemma/cyber/F3/10M/mmlu\endcsname{72.5}
\expandafter\def\csname ev@historical:gemma/cyber/A3/10M/wmdp\endcsname{29.1}
\expandafter\def\csname ev@historical:gemma/cyber/A3/10M/mmlu\endcsname{72.0}
\expandafter\def\csname ev@historical:gemma/cyber/F3/50M/wmdp\endcsname{41.0}
\expandafter\def\csname ev@historical:gemma/cyber/F3/50M/mmlu\endcsname{73.0}
\expandafter\def\csname ev@historical:gemma/cyber/A3/50M/wmdp\endcsname{31.1}
\expandafter\def\csname ev@historical:gemma/cyber/A3/50M/mmlu\endcsname{72.4}
\expandafter\def\csname ev@historical:gemma/cyber/F3/100M/wmdp\endcsname{43.9}
\expandafter\def\csname ev@historical:gemma/cyber/F3/100M/mmlu\endcsname{73.6}
\expandafter\def\csname ev@historical:gemma/cyber/A3/100M/wmdp\endcsname{33.2}
\expandafter\def\csname ev@historical:gemma/cyber/A3/100M/mmlu\endcsname{72.8}
\expandafter\def\csname ev@historical:gemma/cyber/F3/250M/wmdp\endcsname{44.7}
\expandafter\def\csname ev@historical:gemma/cyber/F3/250M/mmlu\endcsname{74.0}
\expandafter\def\csname ev@historical:gemma/cyber/A3/250M/wmdp\endcsname{33.5}
\expandafter\def\csname ev@historical:gemma/cyber/A3/250M/mmlu\endcsname{73.1}
\expandafter\def\csname ev@historical:gemma/cyber/A3-direct/10M/wmdp\endcsname{29.7}
\expandafter\def\csname ev@historical:gemma/cyber/A3-direct/10M/mmlu\endcsname{71.8}
\expandafter\def\csname ev@historical:gemma/cyber/A3-direct/50M/wmdp\endcsname{32.2}
\expandafter\def\csname ev@historical:gemma/cyber/A3-direct/50M/mmlu\endcsname{72.3}
\expandafter\def\csname ev@historical:gemma/cyber/A3-direct/100M/wmdp\endcsname{34.8}
\expandafter\def\csname ev@historical:gemma/cyber/A3-direct/100M/mmlu\endcsname{72.7}
\expandafter\def\csname ev@historical:gemma/cyber/A3-direct/250M/wmdp\endcsname{36.2}
\expandafter\def\csname ev@historical:gemma/cyber/A3-direct/250M/mmlu\endcsname{73.2}
\expandafter\def\csname ev@historical:gemma/cyber/R-SFT/10M/wmdp\endcsname{30.7}
\expandafter\def\csname ev@historical:gemma/cyber/R-SFT/10M/mmlu\endcsname{72.1}
\expandafter\def\csname ev@historical:gemma/cyber/R-SFT/50M/wmdp\endcsname{33.1}
\expandafter\def\csname ev@historical:gemma/cyber/R-SFT/50M/mmlu\endcsname{72.7}
\expandafter\def\csname ev@historical:gemma/cyber/R-SFT/100M/wmdp\endcsname{36.4}
\expandafter\def\csname ev@historical:gemma/cyber/R-SFT/100M/mmlu\endcsname{73.1}
\expandafter\def\csname ev@historical:gemma/cyber/R-SFT/250M/wmdp\endcsname{37.4}
\expandafter\def\csname ev@historical:gemma/cyber/R-SFT/250M/mmlu\endcsname{74.0}
\expandafter\def\csname ev@historical:gemma/cyber/CEU-SFT/10M/wmdp\endcsname{26.7}
\expandafter\def\csname ev@historical:gemma/cyber/CEU-SFT/10M/mmlu\endcsname{32.8}
\expandafter\def\csname ev@historical:gemma/cyber/CEU-SFT/50M/wmdp\endcsname{27.1}
\expandafter\def\csname ev@historical:gemma/cyber/CEU-SFT/50M/mmlu\endcsname{46.8}
\expandafter\def\csname ev@historical:gemma/cyber/CEU-SFT/100M/wmdp\endcsname{28.0}
\expandafter\def\csname ev@historical:gemma/cyber/CEU-SFT/100M/mmlu\endcsname{48.3}
\expandafter\def\csname ev@historical:gemma/cyber/CEU-SFT/250M/wmdp\endcsname{31.9}
\expandafter\def\csname ev@historical:gemma/cyber/CEU-SFT/250M/mmlu\endcsname{66.2}
\expandafter\def\csname ev@historical:gemma/cyber/GA-SFT/10M/wmdp\endcsname{44.4}
\expandafter\def\csname ev@historical:gemma/cyber/GA-SFT/10M/mmlu\endcsname{72.5}
\expandafter\def\csname ev@historical:gemma/cyber/GA-SFT/50M/wmdp\endcsname{47.5}
\expandafter\def\csname ev@historical:gemma/cyber/GA-SFT/50M/mmlu\endcsname{73.6}
\expandafter\def\csname ev@historical:gemma/cyber/GA-SFT/100M/wmdp\endcsname{49.6}
\expandafter\def\csname ev@historical:gemma/cyber/GA-SFT/100M/mmlu\endcsname{73.8}
\expandafter\def\csname ev@historical:gemma/cyber/GA-SFT/250M/wmdp\endcsname{49.6}
\expandafter\def\csname ev@historical:gemma/cyber/GA-SFT/250M/mmlu\endcsname{73.7}
\expandafter\def\csname ev@historical:gemma/cyber/GD-SFT/10M/wmdp\endcsname{25.9}
\expandafter\def\csname ev@historical:gemma/cyber/GD-SFT/10M/mmlu\endcsname{66.5}
\expandafter\def\csname ev@historical:gemma/cyber/GD-SFT/50M/wmdp\endcsname{30.2}
\expandafter\def\csname ev@historical:gemma/cyber/GD-SFT/50M/mmlu\endcsname{61.1}
\expandafter\def\csname ev@historical:gemma/cyber/GD-SFT/100M/wmdp\endcsname{37.4}
\expandafter\def\csname ev@historical:gemma/cyber/GD-SFT/100M/mmlu\endcsname{69.4}
\expandafter\def\csname ev@historical:gemma/cyber/GD-SFT/250M/wmdp\endcsname{44.1}
\expandafter\def\csname ev@historical:gemma/cyber/GD-SFT/250M/mmlu\endcsname{72.9}
\expandafter\def\csname ev@historical:gemma/cyber/PDU-SFT/10M/wmdp\endcsname{24.8}
\expandafter\def\csname ev@historical:gemma/cyber/PDU-SFT/10M/mmlu\endcsname{73.7}
\expandafter\def\csname ev@historical:gemma/cyber/PDU-SFT/50M/wmdp\endcsname{28.5}
\expandafter\def\csname ev@historical:gemma/cyber/PDU-SFT/50M/mmlu\endcsname{73.9}
\expandafter\def\csname ev@historical:gemma/cyber/PDU-SFT/100M/wmdp\endcsname{32.3}
\expandafter\def\csname ev@historical:gemma/cyber/PDU-SFT/100M/mmlu\endcsname{74.0}
\expandafter\def\csname ev@historical:gemma/cyber/PDU-SFT/250M/wmdp\endcsname{34.8}
\expandafter\def\csname ev@historical:gemma/cyber/PDU-SFT/250M/mmlu\endcsname{74.4}
\expandafter\def\csname ev@historical:gemma/cyber/PureNPO-SFT/10M/wmdp\endcsname{44.4}
\expandafter\def\csname ev@historical:gemma/cyber/PureNPO-SFT/10M/mmlu\endcsname{70.8}
\expandafter\def\csname ev@historical:gemma/cyber/PureNPO-SFT/50M/wmdp\endcsname{44.7}
\expandafter\def\csname ev@historical:gemma/cyber/PureNPO-SFT/50M/mmlu\endcsname{71.5}
\expandafter\def\csname ev@historical:gemma/cyber/PureNPO-SFT/100M/wmdp\endcsname{45.9}
\expandafter\def\csname ev@historical:gemma/cyber/PureNPO-SFT/100M/mmlu\endcsname{72.1}
\expandafter\def\csname ev@historical:gemma/cyber/PureNPO-SFT/250M/wmdp\endcsname{46.8}
\expandafter\def\csname ev@historical:gemma/cyber/PureNPO-SFT/250M/mmlu\endcsname{72.6}
\expandafter\def\csname ev@historical:gemma/cyber/ExternalRMU-SFT/10M/wmdp\endcsname{27.0}
\expandafter\def\csname ev@historical:gemma/cyber/ExternalRMU-SFT/10M/mmlu\endcsname{72.6}
\expandafter\def\csname ev@historical:gemma/cyber/ExternalRMU-SFT/50M/wmdp\endcsname{29.4}
\expandafter\def\csname ev@historical:gemma/cyber/ExternalRMU-SFT/50M/mmlu\endcsname{73.0}
\expandafter\def\csname ev@historical:gemma/cyber/ExternalRMU-SFT/100M/wmdp\endcsname{31.4}
\expandafter\def\csname ev@historical:gemma/cyber/ExternalRMU-SFT/100M/mmlu\endcsname{73.2}
\expandafter\def\csname ev@historical:gemma/cyber/ExternalRMU-SFT/250M/wmdp\endcsname{35.8}
\expandafter\def\csname ev@historical:gemma/cyber/ExternalRMU-SFT/250M/mmlu\endcsname{73.7}
\expandafter\def\csname ev@historical:gemma/cyber/SatImp-SFT/10M/wmdp\endcsname{28.2}
\expandafter\def\csname ev@historical:gemma/cyber/SatImp-SFT/10M/mmlu\endcsname{40.5}
\expandafter\def\csname ev@historical:gemma/cyber/SatImp-SFT/50M/wmdp\endcsname{32.7}
\expandafter\def\csname ev@historical:gemma/cyber/SatImp-SFT/50M/mmlu\endcsname{54.3}
\expandafter\def\csname ev@historical:gemma/cyber/SatImp-SFT/100M/wmdp\endcsname{37.0}
\expandafter\def\csname ev@historical:gemma/cyber/SatImp-SFT/100M/mmlu\endcsname{67.4}
\expandafter\def\csname ev@historical:gemma/cyber/SatImp-SFT/250M/wmdp\endcsname{43.4}
\expandafter\def\csname ev@historical:gemma/cyber/SatImp-SFT/250M/mmlu\endcsname{71.1}
\expandafter\def\csname ev@historical:gemma/cyber/SimNPO-SFT/10M/wmdp\endcsname{27.4}
\expandafter\def\csname ev@historical:gemma/cyber/SimNPO-SFT/10M/mmlu\endcsname{67.8}
\expandafter\def\csname ev@historical:gemma/cyber/SimNPO-SFT/50M/wmdp\endcsname{29.4}
\expandafter\def\csname ev@historical:gemma/cyber/SimNPO-SFT/50M/mmlu\endcsname{67.4}
\expandafter\def\csname ev@historical:gemma/cyber/SimNPO-SFT/100M/wmdp\endcsname{33.9}
\expandafter\def\csname ev@historical:gemma/cyber/SimNPO-SFT/100M/mmlu\endcsname{69.2}
\expandafter\def\csname ev@historical:gemma/cyber/SimNPO-SFT/250M/wmdp\endcsname{37.8}
\expandafter\def\csname ev@historical:gemma/cyber/SimNPO-SFT/250M/mmlu\endcsname{71.3}
\expandafter\def\csname ev@historical:gemma/cyber/TRACE-SFT/10M/wmdp\endcsname{50.0}
\expandafter\def\csname ev@historical:gemma/cyber/TRACE-SFT/10M/mmlu\endcsname{74.5}
\expandafter\def\csname ev@historical:gemma/cyber/TRACE-SFT/50M/wmdp\endcsname{49.7}
\expandafter\def\csname ev@historical:gemma/cyber/TRACE-SFT/50M/mmlu\endcsname{74.3}
\expandafter\def\csname ev@historical:gemma/cyber/TRACE-SFT/100M/wmdp\endcsname{50.5}
\expandafter\def\csname ev@historical:gemma/cyber/TRACE-SFT/100M/mmlu\endcsname{74.5}
\expandafter\def\csname ev@historical:gemma/cyber/TRACE-SFT/250M/wmdp\endcsname{50.1}
\expandafter\def\csname ev@historical:gemma/cyber/TRACE-SFT/250M/mmlu\endcsname{74.8}
\expandafter\def\csname ev@historical:gemma/cyber/WGA-SFT/10M/wmdp\endcsname{26.6}
\expandafter\def\csname ev@historical:gemma/cyber/WGA-SFT/10M/mmlu\endcsname{34.4}
\expandafter\def\csname ev@historical:gemma/cyber/WGA-SFT/50M/wmdp\endcsname{31.8}
\expandafter\def\csname ev@historical:gemma/cyber/WGA-SFT/50M/mmlu\endcsname{45.7}
\expandafter\def\csname ev@historical:gemma/cyber/WGA-SFT/100M/wmdp\endcsname{36.1}
\expandafter\def\csname ev@historical:gemma/cyber/WGA-SFT/100M/mmlu\endcsname{56.1}
\expandafter\def\csname ev@historical:gemma/cyber/WGA-SFT/250M/wmdp\endcsname{41.6}
\expandafter\def\csname ev@historical:gemma/cyber/WGA-SFT/250M/mmlu\endcsname{66.8}
\expandafter\def\csname ev@historical:gemma/cyber/Routing-F0/0M/routing_z_forget\endcsname{21.1}
\expandafter\def\csname ev@historical:gemma/cyber/Routing-F0/0M/routing_z_forget_layer_0\endcsname{10.8}
\expandafter\def\csname ev@historical:gemma/cyber/Routing-F0/0M/routing_z_forget_layer_1\endcsname{8.8}
\expandafter\def\csname ev@historical:gemma/cyber/Routing-F0/0M/routing_z_forget_layer_2\endcsname{19.3}
\expandafter\def\csname ev@historical:gemma/cyber/Routing-F0/0M/routing_z_forget_layer_3\endcsname{21.1}
\expandafter\def\csname ev@historical:gemma/cyber/Routing-F0/0M/routing_z_forget_layer_4\endcsname{16.8}
\expandafter\def\csname ev@historical:gemma/cyber/Routing-F0/0M/routing_z_forget_layer_5\endcsname{24.6}
\expandafter\def\csname ev@historical:gemma/cyber/Routing-F0/0M/routing_z_forget_layer_6\endcsname{22.7}
\expandafter\def\csname ev@historical:gemma/cyber/Routing-F0/0M/routing_z_forget_layer_7\endcsname{19.2}
\expandafter\def\csname ev@historical:gemma/cyber/Routing-F0/0M/routing_z_forget_layer_8\endcsname{20.5}
\expandafter\def\csname ev@historical:gemma/cyber/Routing-F0/0M/routing_z_forget_layer_9\endcsname{18.5}
\expandafter\def\csname ev@historical:gemma/cyber/Routing-F0/0M/routing_z_forget_layer_10\endcsname{22.1}
\expandafter\def\csname ev@historical:gemma/cyber/Routing-F0/0M/routing_z_forget_layer_11\endcsname{23.0}
\expandafter\def\csname ev@historical:gemma/cyber/Routing-F0/0M/routing_z_forget_layer_12\endcsname{18.7}
\expandafter\def\csname ev@historical:gemma/cyber/Routing-F0/0M/routing_z_forget_layer_13\endcsname{18.9}
\expandafter\def\csname ev@historical:gemma/cyber/Routing-F0/0M/routing_z_forget_layer_14\endcsname{20.4}
\expandafter\def\csname ev@historical:gemma/cyber/Routing-F0/0M/routing_z_forget_layer_15\endcsname{17.2}
\expandafter\def\csname ev@historical:gemma/cyber/Routing-F0/0M/routing_z_forget_layer_16\endcsname{18.2}
\expandafter\def\csname ev@historical:gemma/cyber/Routing-F0/0M/routing_z_forget_layer_17\endcsname{18.9}
\expandafter\def\csname ev@historical:gemma/cyber/Routing-F0/0M/routing_z_forget_layer_18\endcsname{21.8}
\expandafter\def\csname ev@historical:gemma/cyber/Routing-F0/0M/routing_z_forget_layer_19\endcsname{19.3}
\expandafter\def\csname ev@historical:gemma/cyber/Routing-F0/0M/routing_z_forget_layer_20\endcsname{25.4}
\expandafter\def\csname ev@historical:gemma/cyber/Routing-F0/0M/routing_z_forget_layer_21\endcsname{22.2}
\expandafter\def\csname ev@historical:gemma/cyber/Routing-F0/0M/routing_z_forget_layer_22\endcsname{29.5}
\expandafter\def\csname ev@historical:gemma/cyber/Routing-F0/0M/routing_z_forget_layer_23\endcsname{25.1}
\expandafter\def\csname ev@historical:gemma/cyber/Routing-F0/0M/routing_z_forget_layer_24\endcsname{28.5}
\expandafter\def\csname ev@historical:gemma/cyber/Routing-F0/0M/routing_z_forget_layer_25\endcsname{27.6}
\expandafter\def\csname ev@historical:gemma/cyber/Routing-F0/0M/routing_z_forget_layer_26\endcsname{23.0}
\expandafter\def\csname ev@historical:gemma/cyber/Routing-F0/0M/routing_z_forget_layer_27\endcsname{22.5}
\expandafter\def\csname ev@historical:gemma/cyber/Routing-F0/0M/routing_z_forget_layer_28\endcsname{28.6}
\expandafter\def\csname ev@historical:gemma/cyber/Routing-F0/0M/routing_z_forget_layer_29\endcsname{20.5}
\expandafter\def\csname ev@historical:gemma/cyber/Routing-F0/0M/routing_z_retain\endcsname{2.2}
\expandafter\def\csname ev@historical:gemma/cyber/Routing-F0/0M/routing_z_retain_layer_0\endcsname{2.0}
\expandafter\def\csname ev@historical:gemma/cyber/Routing-F0/0M/routing_z_retain_layer_1\endcsname{1.7}
\expandafter\def\csname ev@historical:gemma/cyber/Routing-F0/0M/routing_z_retain_layer_2\endcsname{0.9}
\expandafter\def\csname ev@historical:gemma/cyber/Routing-F0/0M/routing_z_retain_layer_3\endcsname{2.4}
\expandafter\def\csname ev@historical:gemma/cyber/Routing-F0/0M/routing_z_retain_layer_4\endcsname{1.3}
\expandafter\def\csname ev@historical:gemma/cyber/Routing-F0/0M/routing_z_retain_layer_5\endcsname{0.9}
\expandafter\def\csname ev@historical:gemma/cyber/Routing-F0/0M/routing_z_retain_layer_6\endcsname{1.1}
\expandafter\def\csname ev@historical:gemma/cyber/Routing-F0/0M/routing_z_retain_layer_7\endcsname{1.3}
\expandafter\def\csname ev@historical:gemma/cyber/Routing-F0/0M/routing_z_retain_layer_8\endcsname{1.3}
\expandafter\def\csname ev@historical:gemma/cyber/Routing-F0/0M/routing_z_retain_layer_9\endcsname{2.2}
\expandafter\def\csname ev@historical:gemma/cyber/Routing-F0/0M/routing_z_retain_layer_10\endcsname{0.6}
\expandafter\def\csname ev@historical:gemma/cyber/Routing-F0/0M/routing_z_retain_layer_11\endcsname{2.4}
\expandafter\def\csname ev@historical:gemma/cyber/Routing-F0/0M/routing_z_retain_layer_12\endcsname{5.0}
\expandafter\def\csname ev@historical:gemma/cyber/Routing-F0/0M/routing_z_retain_layer_13\endcsname{1.4}
\expandafter\def\csname ev@historical:gemma/cyber/Routing-F0/0M/routing_z_retain_layer_14\endcsname{7.0}
\expandafter\def\csname ev@historical:gemma/cyber/Routing-F0/0M/routing_z_retain_layer_15\endcsname{2.3}
\expandafter\def\csname ev@historical:gemma/cyber/Routing-F0/0M/routing_z_retain_layer_16\endcsname{2.9}
\expandafter\def\csname ev@historical:gemma/cyber/Routing-F0/0M/routing_z_retain_layer_17\endcsname{2.3}
\expandafter\def\csname ev@historical:gemma/cyber/Routing-F0/0M/routing_z_retain_layer_18\endcsname{4.2}
\expandafter\def\csname ev@historical:gemma/cyber/Routing-F0/0M/routing_z_retain_layer_19\endcsname{0.8}
\expandafter\def\csname ev@historical:gemma/cyber/Routing-F0/0M/routing_z_retain_layer_20\endcsname{1.4}
\expandafter\def\csname ev@historical:gemma/cyber/Routing-F0/0M/routing_z_retain_layer_21\endcsname{1.2}
\expandafter\def\csname ev@historical:gemma/cyber/Routing-F0/0M/routing_z_retain_layer_22\endcsname{1.5}
\expandafter\def\csname ev@historical:gemma/cyber/Routing-F0/0M/routing_z_retain_layer_23\endcsname{1.0}
\expandafter\def\csname ev@historical:gemma/cyber/Routing-F0/0M/routing_z_retain_layer_24\endcsname{3.3}
\expandafter\def\csname ev@historical:gemma/cyber/Routing-F0/0M/routing_z_retain_layer_25\endcsname{1.7}
\expandafter\def\csname ev@historical:gemma/cyber/Routing-F0/0M/routing_z_retain_layer_26\endcsname{2.0}
\expandafter\def\csname ev@historical:gemma/cyber/Routing-F0/0M/routing_z_retain_layer_27\endcsname{3.4}
\expandafter\def\csname ev@historical:gemma/cyber/Routing-F0/0M/routing_z_retain_layer_28\endcsname{2.8}
\expandafter\def\csname ev@historical:gemma/cyber/Routing-F0/0M/routing_z_retain_layer_29\endcsname{4.0}
\expandafter\def\csname ev@historical:gemma/cyber/Routing-F1/0M/routing_z_forget\endcsname{35.7}
\expandafter\def\csname ev@historical:gemma/cyber/Routing-F1/0M/routing_z_forget_layer_0\endcsname{12.9}
\expandafter\def\csname ev@historical:gemma/cyber/Routing-F1/0M/routing_z_forget_layer_1\endcsname{10.3}
\expandafter\def\csname ev@historical:gemma/cyber/Routing-F1/0M/routing_z_forget_layer_2\endcsname{22.6}
\expandafter\def\csname ev@historical:gemma/cyber/Routing-F1/0M/routing_z_forget_layer_3\endcsname{24.7}
\expandafter\def\csname ev@historical:gemma/cyber/Routing-F1/0M/routing_z_forget_layer_4\endcsname{22.6}
\expandafter\def\csname ev@historical:gemma/cyber/Routing-F1/0M/routing_z_forget_layer_5\endcsname{32.9}
\expandafter\def\csname ev@historical:gemma/cyber/Routing-F1/0M/routing_z_forget_layer_6\endcsname{33.8}
\expandafter\def\csname ev@historical:gemma/cyber/Routing-F1/0M/routing_z_forget_layer_7\endcsname{29.3}
\expandafter\def\csname ev@historical:gemma/cyber/Routing-F1/0M/routing_z_forget_layer_8\endcsname{31.9}
\expandafter\def\csname ev@historical:gemma/cyber/Routing-F1/0M/routing_z_forget_layer_9\endcsname{33.1}
\expandafter\def\csname ev@historical:gemma/cyber/Routing-F1/0M/routing_z_forget_layer_10\endcsname{39.5}
\expandafter\def\csname ev@historical:gemma/cyber/Routing-F1/0M/routing_z_forget_layer_11\endcsname{42.1}
\expandafter\def\csname ev@historical:gemma/cyber/Routing-F1/0M/routing_z_forget_layer_12\endcsname{30.5}
\expandafter\def\csname ev@historical:gemma/cyber/Routing-F1/0M/routing_z_forget_layer_13\endcsname{31.7}
\expandafter\def\csname ev@historical:gemma/cyber/Routing-F1/0M/routing_z_forget_layer_14\endcsname{33.0}
\expandafter\def\csname ev@historical:gemma/cyber/Routing-F1/0M/routing_z_forget_layer_15\endcsname{27.5}
\expandafter\def\csname ev@historical:gemma/cyber/Routing-F1/0M/routing_z_forget_layer_16\endcsname{30.9}
\expandafter\def\csname ev@historical:gemma/cyber/Routing-F1/0M/routing_z_forget_layer_17\endcsname{34.8}
\expandafter\def\csname ev@historical:gemma/cyber/Routing-F1/0M/routing_z_forget_layer_18\endcsname{34.8}
\expandafter\def\csname ev@historical:gemma/cyber/Routing-F1/0M/routing_z_forget_layer_19\endcsname{39.4}
\expandafter\def\csname ev@historical:gemma/cyber/Routing-F1/0M/routing_z_forget_layer_20\endcsname{43.0}
\expandafter\def\csname ev@historical:gemma/cyber/Routing-F1/0M/routing_z_forget_layer_21\endcsname{39.9}
\expandafter\def\csname ev@historical:gemma/cyber/Routing-F1/0M/routing_z_forget_layer_22\endcsname{48.6}
\expandafter\def\csname ev@historical:gemma/cyber/Routing-F1/0M/routing_z_forget_layer_23\endcsname{48.2}
\expandafter\def\csname ev@historical:gemma/cyber/Routing-F1/0M/routing_z_forget_layer_24\endcsname{48.8}
\expandafter\def\csname ev@historical:gemma/cyber/Routing-F1/0M/routing_z_forget_layer_25\endcsname{50.0}
\expandafter\def\csname ev@historical:gemma/cyber/Routing-F1/0M/routing_z_forget_layer_26\endcsname{46.6}
\expandafter\def\csname ev@historical:gemma/cyber/Routing-F1/0M/routing_z_forget_layer_27\endcsname{48.1}
\expandafter\def\csname ev@historical:gemma/cyber/Routing-F1/0M/routing_z_forget_layer_28\endcsname{54.8}
\expandafter\def\csname ev@historical:gemma/cyber/Routing-F1/0M/routing_z_forget_layer_29\endcsname{46.1}
\expandafter\def\csname ev@historical:gemma/cyber/Routing-F1/0M/routing_z_retain\endcsname{2.5}
\expandafter\def\csname ev@historical:gemma/cyber/Routing-F1/0M/routing_z_retain_layer_0\endcsname{2.0}
\expandafter\def\csname ev@historical:gemma/cyber/Routing-F1/0M/routing_z_retain_layer_1\endcsname{1.6}
\expandafter\def\csname ev@historical:gemma/cyber/Routing-F1/0M/routing_z_retain_layer_2\endcsname{0.9}
\expandafter\def\csname ev@historical:gemma/cyber/Routing-F1/0M/routing_z_retain_layer_3\endcsname{2.4}
\expandafter\def\csname ev@historical:gemma/cyber/Routing-F1/0M/routing_z_retain_layer_4\endcsname{1.4}
\expandafter\def\csname ev@historical:gemma/cyber/Routing-F1/0M/routing_z_retain_layer_5\endcsname{1.2}
\expandafter\def\csname ev@historical:gemma/cyber/Routing-F1/0M/routing_z_retain_layer_6\endcsname{1.4}
\expandafter\def\csname ev@historical:gemma/cyber/Routing-F1/0M/routing_z_retain_layer_7\endcsname{1.6}
\expandafter\def\csname ev@historical:gemma/cyber/Routing-F1/0M/routing_z_retain_layer_8\endcsname{1.5}
\expandafter\def\csname ev@historical:gemma/cyber/Routing-F1/0M/routing_z_retain_layer_9\endcsname{2.4}
\expandafter\def\csname ev@historical:gemma/cyber/Routing-F1/0M/routing_z_retain_layer_10\endcsname{1.0}
\expandafter\def\csname ev@historical:gemma/cyber/Routing-F1/0M/routing_z_retain_layer_11\endcsname{3.1}
\expandafter\def\csname ev@historical:gemma/cyber/Routing-F1/0M/routing_z_retain_layer_12\endcsname{4.5}
\expandafter\def\csname ev@historical:gemma/cyber/Routing-F1/0M/routing_z_retain_layer_13\endcsname{1.8}
\expandafter\def\csname ev@historical:gemma/cyber/Routing-F1/0M/routing_z_retain_layer_14\endcsname{7.4}
\expandafter\def\csname ev@historical:gemma/cyber/Routing-F1/0M/routing_z_retain_layer_15\endcsname{2.8}
\expandafter\def\csname ev@historical:gemma/cyber/Routing-F1/0M/routing_z_retain_layer_16\endcsname{3.3}
\expandafter\def\csname ev@historical:gemma/cyber/Routing-F1/0M/routing_z_retain_layer_17\endcsname{2.8}
\expandafter\def\csname ev@historical:gemma/cyber/Routing-F1/0M/routing_z_retain_layer_18\endcsname{4.3}
\expandafter\def\csname ev@historical:gemma/cyber/Routing-F1/0M/routing_z_retain_layer_19\endcsname{1.2}
\expandafter\def\csname ev@historical:gemma/cyber/Routing-F1/0M/routing_z_retain_layer_20\endcsname{1.8}
\expandafter\def\csname ev@historical:gemma/cyber/Routing-F1/0M/routing_z_retain_layer_21\endcsname{1.5}
\expandafter\def\csname ev@historical:gemma/cyber/Routing-F1/0M/routing_z_retain_layer_22\endcsname{2.0}
\expandafter\def\csname ev@historical:gemma/cyber/Routing-F1/0M/routing_z_retain_layer_23\endcsname{1.7}
\expandafter\def\csname ev@historical:gemma/cyber/Routing-F1/0M/routing_z_retain_layer_24\endcsname{3.2}
\expandafter\def\csname ev@historical:gemma/cyber/Routing-F1/0M/routing_z_retain_layer_25\endcsname{2.3}
\expandafter\def\csname ev@historical:gemma/cyber/Routing-F1/0M/routing_z_retain_layer_26\endcsname{2.5}
\expandafter\def\csname ev@historical:gemma/cyber/Routing-F1/0M/routing_z_retain_layer_27\endcsname{3.5}
\expandafter\def\csname ev@historical:gemma/cyber/Routing-F1/0M/routing_z_retain_layer_28\endcsname{3.4}
\expandafter\def\csname ev@historical:gemma/cyber/Routing-F1/0M/routing_z_retain_layer_29\endcsname{3.7}
\expandafter\def\csname ev@intern/bio/B/0M/arc_challenge\endcsname{74.3}
\expandafter\def\csname ev@intern/bio/B/0M/gsm8k\endcsname{87.1}
\expandafter\def\csname ev@intern/bio/B/0M/hellaswag\endcsname{85.8}
\expandafter\def\csname ev@intern/bio/B/0M/mmlu\endcsname{85.0}
\expandafter\def\csname ev@intern/bio/B/0M/mmlu_adj\endcsname{88.7}
\expandafter\def\csname ev@intern/bio/B/0M/mmlu_rest\endcsname{84.4}
\expandafter\def\csname ev@intern/bio/B/0M/wmdp\endcsname{83.4}
\expandafter\def\csname ev@intern/bio/S_std(B)/250M/arc_challenge\endcsname{74.4}
\expandafter\def\csname ev@intern/bio/S_std(B)/250M/gsm8k\endcsname{86.9}
\expandafter\def\csname ev@intern/bio/S_std(B)/250M/hellaswag\endcsname{85.8}
\expandafter\def\csname ev@intern/bio/S_std(B)/250M/mmlu\endcsname{84.8}
\expandafter\def\csname ev@intern/bio/S_std(B)/250M/mmlu_adj\endcsname{89.2}
\expandafter\def\csname ev@intern/bio/S_std(B)/250M/mmlu_rest\endcsname{84.0}
\expandafter\def\csname ev@intern/bio/S_std(B)/250M/wmdp\endcsname{84.5}
\expandafter\def\csname ev@intern/bio/S_std(B)/250M/ifeval\endcsname{78.9}
\expandafter\def\csname ev@intern/bio/F0/0M/arc_challenge\endcsname{73.8}
\expandafter\def\csname ev@intern/bio/F0/0M/gsm8k\endcsname{87.1}
\expandafter\def\csname ev@intern/bio/F0/0M/hellaswag\endcsname{85.7}
\expandafter\def\csname ev@intern/bio/F0/0M/mmlu\endcsname{85.0}
\expandafter\def\csname ev@intern/bio/F0/0M/mmlu_adj\endcsname{88.8}
\expandafter\def\csname ev@intern/bio/F0/0M/mmlu_rest\endcsname{84.3}
\expandafter\def\csname ev@intern/bio/F0/0M/wmdp\endcsname{83.0}
\expandafter\def\csname ev@intern/bio/A0/0M/arc_challenge\endcsname{71.5}
\expandafter\def\csname ev@intern/bio/A0/0M/gsm8k\endcsname{87.3}
\expandafter\def\csname ev@intern/bio/A0/0M/hellaswag\endcsname{85.2}
\expandafter\def\csname ev@intern/bio/A0/0M/mmlu\endcsname{83.2}
\expandafter\def\csname ev@intern/bio/A0/0M/mmlu_adj\endcsname{79.9}
\expandafter\def\csname ev@intern/bio/A0/0M/mmlu_rest\endcsname{83.8}
\expandafter\def\csname ev@intern/bio/A0/0M/wmdp\endcsname{69.2}
\expandafter\def\csname ev@intern/bio/F1/0M/arc_challenge\endcsname{71.2}
\expandafter\def\csname ev@intern/bio/F1/0M/gsm8k\endcsname{84.3}
\expandafter\def\csname ev@intern/bio/F1/0M/hellaswag\endcsname{84.7}
\expandafter\def\csname ev@intern/bio/F1/0M/mmlu\endcsname{83.2}
\expandafter\def\csname ev@intern/bio/F1/0M/mmlu_adj\endcsname{82.5}
\expandafter\def\csname ev@intern/bio/F1/0M/mmlu_rest\endcsname{83.4}
\expandafter\def\csname ev@intern/bio/F1/0M/wmdp\endcsname{79.4}
\expandafter\def\csname ev@intern/bio/F1/0M/ifeval\endcsname{78.4}
\expandafter\def\csname ev@intern/bio/A1/0M/arc_challenge\endcsname{70.7}
\expandafter\def\csname ev@intern/bio/A1/0M/gsm8k\endcsname{84.6}
\expandafter\def\csname ev@intern/bio/A1/0M/hellaswag\endcsname{84.7}
\expandafter\def\csname ev@intern/bio/A1/0M/mmlu\endcsname{82.3}
\expandafter\def\csname ev@intern/bio/A1/0M/mmlu_adj\endcsname{75.8}
\expandafter\def\csname ev@intern/bio/A1/0M/mmlu_rest\endcsname{83.4}
\expandafter\def\csname ev@intern/bio/A1/0M/wmdp\endcsname{64.3}
\expandafter\def\csname ev@intern/bio/A1/0M/ifeval\endcsname{79.3}
\expandafter\def\csname ev@intern/bio/A2/0M/arc_challenge\endcsname{70.7}
\expandafter\def\csname ev@intern/bio/A2/0M/gsm8k\endcsname{83.8}
\expandafter\def\csname ev@intern/bio/A2/0M/hellaswag\endcsname{84.9}
\expandafter\def\csname ev@intern/bio/A2/0M/mmlu\endcsname{81.0}
\expandafter\def\csname ev@intern/bio/A2/0M/mmlu_adj\endcsname{68.2}
\expandafter\def\csname ev@intern/bio/A2/0M/mmlu_rest\endcsname{83.2}
\expandafter\def\csname ev@intern/bio/A2/0M/wmdp\endcsname{29.6}
\expandafter\def\csname ev@intern/bio/A2/0M/ifeval\endcsname{80.8}
\expandafter\def\csname ev@intern/bio/F2/0M/arc_challenge\endcsname{70.7}
\expandafter\def\csname ev@intern/bio/F2/0M/gsm8k\endcsname{84.8}
\expandafter\def\csname ev@intern/bio/F2/0M/hellaswag\endcsname{84.8}
\expandafter\def\csname ev@intern/bio/F2/0M/mmlu\endcsname{82.4}
\expandafter\def\csname ev@intern/bio/F2/0M/mmlu_adj\endcsname{76.5}
\expandafter\def\csname ev@intern/bio/F2/0M/mmlu_rest\endcsname{83.4}
\expandafter\def\csname ev@intern/bio/F2/0M/wmdp\endcsname{61.9}
\expandafter\def\csname ev@intern/bio/F2/0M/ifeval\endcsname{80.8}
\expandafter\def\csname ev@intern/bio/F3/250M/arc_challenge\endcsname{72.0}
\expandafter\def\csname ev@intern/bio/F3/250M/gsm8k\endcsname{85.2}
\expandafter\def\csname ev@intern/bio/F3/250M/hellaswag\endcsname{85.3}
\expandafter\def\csname ev@intern/bio/F3/250M/mmlu\endcsname{84.1}
\expandafter\def\csname ev@intern/bio/F3/250M/mmlu_adj\endcsname{84.7}
\expandafter\def\csname ev@intern/bio/F3/250M/mmlu_rest\endcsname{84.0}
\expandafter\def\csname ev@intern/bio/F3/250M/wmdp\endcsname{78.6}
\expandafter\def\csname ev@intern/bio/F3/250M/ifeval\endcsname{78.6}
\expandafter\def\csname ev@intern/bio/A3/250M/arc_challenge\endcsname{71.3}
\expandafter\def\csname ev@intern/bio/A3/250M/gsm8k\endcsname{86.2}
\expandafter\def\csname ev@intern/bio/A3/250M/hellaswag\endcsname{85.1}
\expandafter\def\csname ev@intern/bio/A3/250M/mmlu\endcsname{82.5}
\expandafter\def\csname ev@intern/bio/A3/250M/mmlu_adj\endcsname{76.8}
\expandafter\def\csname ev@intern/bio/A3/250M/mmlu_rest\endcsname{83.5}
\expandafter\def\csname ev@intern/bio/A3/250M/wmdp\endcsname{55.0}
\expandafter\def\csname ev@intern/bio/A3/250M/ifeval\endcsname{78.6}
\expandafter\def\csname ev@intern/bio/R/0M/arc_challenge\endcsname{74.2}
\expandafter\def\csname ev@intern/bio/R/0M/gsm8k\endcsname{86.8}
\expandafter\def\csname ev@intern/bio/R/0M/hellaswag\endcsname{85.6}
\expandafter\def\csname ev@intern/bio/R/0M/mmlu\endcsname{84.5}
\expandafter\def\csname ev@intern/bio/R/0M/mmlu_adj\endcsname{85.8}
\expandafter\def\csname ev@intern/bio/R/0M/mmlu_rest\endcsname{84.2}
\expandafter\def\csname ev@intern/bio/R/0M/wmdp\endcsname{33.3}
\expandafter\def\csname ev@intern/bio/R-SFT/250M/arc_challenge\endcsname{74.2}
\expandafter\def\csname ev@intern/bio/R-SFT/250M/gsm8k\endcsname{86.5}
\expandafter\def\csname ev@intern/bio/R-SFT/250M/hellaswag\endcsname{85.8}
\expandafter\def\csname ev@intern/bio/R-SFT/250M/mmlu\endcsname{84.6}
\expandafter\def\csname ev@intern/bio/R-SFT/250M/mmlu_adj\endcsname{89.2}
\expandafter\def\csname ev@intern/bio/R-SFT/250M/mmlu_rest\endcsname{83.8}
\expandafter\def\csname ev@intern/bio/R-SFT/250M/wmdp\endcsname{81.0}
\expandafter\def\csname ev@intern/bio/R-SFT/250M/ifeval\endcsname{78.6}
\expandafter\def\csname ev@intern/bio/A3-direct/250M/arc_challenge\endcsname{71.2}
\expandafter\def\csname ev@intern/bio/A3-direct/250M/gsm8k\endcsname{86.4}
\expandafter\def\csname ev@intern/bio/A3-direct/250M/hellaswag\endcsname{85.4}
\expandafter\def\csname ev@intern/bio/A3-direct/250M/mmlu\endcsname{83.0}
\expandafter\def\csname ev@intern/bio/A3-direct/250M/mmlu_adj\endcsname{79.2}
\expandafter\def\csname ev@intern/bio/A3-direct/250M/mmlu_rest\endcsname{83.7}
\expandafter\def\csname ev@intern/bio/A3-direct/250M/wmdp\endcsname{64.6}
\expandafter\def\csname ev@intern/bio/A3-direct/250M/ifeval\endcsname{79.3}
\expandafter\def\csname ev@intern/bio/CEU/0M/arc_challenge\endcsname{29.0}
\expandafter\def\csname ev@intern/bio/CEU/0M/gsm8k\endcsname{0.0}
\expandafter\def\csname ev@intern/bio/CEU/0M/hellaswag\endcsname{30.6}
\expandafter\def\csname ev@intern/bio/CEU/0M/mmlu\endcsname{75.7}
\expandafter\def\csname ev@intern/bio/CEU/0M/mmlu_adj\endcsname{79.8}
\expandafter\def\csname ev@intern/bio/CEU/0M/mmlu_rest\endcsname{75.0}
\expandafter\def\csname ev@intern/bio/CEU/0M/wmdp\endcsname{70.2}
\expandafter\def\csname ev@intern/bio/CEU-SFT/250M/arc_challenge\endcsname{71.5}
\expandafter\def\csname ev@intern/bio/CEU-SFT/250M/gsm8k\endcsname{82.7}
\expandafter\def\csname ev@intern/bio/CEU-SFT/250M/hellaswag\endcsname{82.9}
\expandafter\def\csname ev@intern/bio/CEU-SFT/250M/mmlu\endcsname{84.7}
\expandafter\def\csname ev@intern/bio/CEU-SFT/250M/mmlu_adj\endcsname{88.9}
\expandafter\def\csname ev@intern/bio/CEU-SFT/250M/mmlu_rest\endcsname{83.9}
\expandafter\def\csname ev@intern/bio/CEU-SFT/250M/wmdp\endcsname{84.4}
\expandafter\def\csname ev@intern/bio/CEU-SFT/250M/ifeval\endcsname{79.7}
\expandafter\def\csname ev@intern/bio/GA/0M/arc_challenge\endcsname{25.6}
\expandafter\def\csname ev@intern/bio/GA/0M/gsm8k\endcsname{0.0}
\expandafter\def\csname ev@intern/bio/GA/0M/hellaswag\endcsname{25.5}
\expandafter\def\csname ev@intern/bio/GA/0M/mmlu\endcsname{23.0}
\expandafter\def\csname ev@intern/bio/GA/0M/mmlu_adj\endcsname{22.9}
\expandafter\def\csname ev@intern/bio/GA/0M/mmlu_rest\endcsname{23.0}
\expandafter\def\csname ev@intern/bio/GA/0M/wmdp\endcsname{24.7}
\expandafter\def\csname ev@intern/bio/PDU/0M/arc_challenge\endcsname{75.3}
\expandafter\def\csname ev@intern/bio/PDU/0M/gsm8k\endcsname{87.1}
\expandafter\def\csname ev@intern/bio/PDU/0M/hellaswag\endcsname{86.2}
\expandafter\def\csname ev@intern/bio/PDU/0M/mmlu\endcsname{83.8}
\expandafter\def\csname ev@intern/bio/PDU/0M/mmlu_adj\endcsname{78.8}
\expandafter\def\csname ev@intern/bio/PDU/0M/mmlu_rest\endcsname{84.6}
\expandafter\def\csname ev@intern/bio/PDU/0M/wmdp\endcsname{54.8}
\expandafter\def\csname ev@intern/bio/PDU/0M/ifeval\endcsname{81.9}
\expandafter\def\csname ev@intern/bio/GA-SFT/250M/arc_challenge\endcsname{50.8}
\expandafter\def\csname ev@intern/bio/GA-SFT/250M/gsm8k\endcsname{74.5}
\expandafter\def\csname ev@intern/bio/GA-SFT/250M/hellaswag\endcsname{23.9}
\expandafter\def\csname ev@intern/bio/GA-SFT/250M/mmlu\endcsname{46.9}
\expandafter\def\csname ev@intern/bio/GA-SFT/250M/mmlu_adj\endcsname{23.5}
\expandafter\def\csname ev@intern/bio/GA-SFT/250M/mmlu_rest\endcsname{51.0}
\expandafter\def\csname ev@intern/bio/GA-SFT/250M/wmdp\endcsname{42.1}
\expandafter\def\csname ev@intern/bio/GA-SFT/250M/ifeval\endcsname{79.3}
\expandafter\def\csname ev@intern/bio/GD/0M/arc_challenge\endcsname{73.5}
\expandafter\def\csname ev@intern/bio/GD/0M/gsm8k\endcsname{85.1}
\expandafter\def\csname ev@intern/bio/GD/0M/hellaswag\endcsname{85.0}
\expandafter\def\csname ev@intern/bio/GD/0M/mmlu\endcsname{82.5}
\expandafter\def\csname ev@intern/bio/GD/0M/mmlu_adj\endcsname{81.5}
\expandafter\def\csname ev@intern/bio/GD/0M/mmlu_rest\endcsname{82.6}
\expandafter\def\csname ev@intern/bio/GD/0M/wmdp\endcsname{66.9}
\expandafter\def\csname ev@intern/bio/GD-SFT/250M/arc_challenge\endcsname{74.7}
\expandafter\def\csname ev@intern/bio/GD-SFT/250M/gsm8k\endcsname{87.3}
\expandafter\def\csname ev@intern/bio/GD-SFT/250M/hellaswag\endcsname{85.5}
\expandafter\def\csname ev@intern/bio/GD-SFT/250M/mmlu\endcsname{84.4}
\expandafter\def\csname ev@intern/bio/GD-SFT/250M/mmlu_adj\endcsname{88.1}
\expandafter\def\csname ev@intern/bio/GD-SFT/250M/mmlu_rest\endcsname{83.8}
\expandafter\def\csname ev@intern/bio/GD-SFT/250M/wmdp\endcsname{83.0}
\expandafter\def\csname ev@intern/bio/GD-SFT/250M/ifeval\endcsname{78.7}
\expandafter\def\csname ev@intern/bio/PureNPO/0M/arc_challenge\endcsname{67.4}
\expandafter\def\csname ev@intern/bio/PureNPO/0M/gsm8k\endcsname{82.2}
\expandafter\def\csname ev@intern/bio/PureNPO/0M/hellaswag\endcsname{83.2}
\expandafter\def\csname ev@intern/bio/PureNPO/0M/mmlu\endcsname{78.7}
\expandafter\def\csname ev@intern/bio/PureNPO/0M/mmlu_adj\endcsname{84.2}
\expandafter\def\csname ev@intern/bio/PureNPO/0M/mmlu_rest\endcsname{77.7}
\expandafter\def\csname ev@intern/bio/PureNPO/0M/wmdp\endcsname{76.9}
\expandafter\def\csname ev@intern/bio/PureNPO-SFT/250M/arc_challenge\endcsname{69.3}
\expandafter\def\csname ev@intern/bio/PureNPO-SFT/250M/gsm8k\endcsname{84.8}
\expandafter\def\csname ev@intern/bio/PureNPO-SFT/250M/hellaswag\endcsname{81.8}
\expandafter\def\csname ev@intern/bio/PureNPO-SFT/250M/mmlu\endcsname{80.7}
\expandafter\def\csname ev@intern/bio/PureNPO-SFT/250M/mmlu_adj\endcsname{85.3}
\expandafter\def\csname ev@intern/bio/PureNPO-SFT/250M/mmlu_rest\endcsname{79.9}
\expandafter\def\csname ev@intern/bio/PureNPO-SFT/250M/wmdp\endcsname{77.6}
\expandafter\def\csname ev@intern/bio/PureNPO-SFT/250M/ifeval\endcsname{70.4}
\expandafter\def\csname ev@intern/bio/PDU-SFT/250M/arc_challenge\endcsname{74.9}
\expandafter\def\csname ev@intern/bio/PDU-SFT/250M/gsm8k\endcsname{87.8}
\expandafter\def\csname ev@intern/bio/PDU-SFT/250M/hellaswag\endcsname{86.0}
\expandafter\def\csname ev@intern/bio/PDU-SFT/250M/mmlu\endcsname{84.8}
\expandafter\def\csname ev@intern/bio/PDU-SFT/250M/mmlu_adj\endcsname{87.2}
\expandafter\def\csname ev@intern/bio/PDU-SFT/250M/mmlu_rest\endcsname{84.4}
\expandafter\def\csname ev@intern/bio/PDU-SFT/250M/wmdp\endcsname{82.7}
\expandafter\def\csname ev@intern/bio/PDU-SFT/250M/ifeval\endcsname{78.4}
\expandafter\def\csname ev@intern/bio/ExternalRMU/0M/arc_challenge\endcsname{74.6}
\expandafter\def\csname ev@intern/bio/ExternalRMU/0M/gsm8k\endcsname{86.1}
\expandafter\def\csname ev@intern/bio/ExternalRMU/0M/hellaswag\endcsname{85.4}
\expandafter\def\csname ev@intern/bio/ExternalRMU/0M/mmlu\endcsname{84.3}
\expandafter\def\csname ev@intern/bio/ExternalRMU/0M/mmlu_adj\endcsname{86.2}
\expandafter\def\csname ev@intern/bio/ExternalRMU/0M/mmlu_rest\endcsname{84.0}
\expandafter\def\csname ev@intern/bio/ExternalRMU/0M/wmdp\endcsname{37.4}
\expandafter\def\csname ev@intern/bio/ExternalRMU-SFT/250M/arc_challenge\endcsname{74.3}
\expandafter\def\csname ev@intern/bio/ExternalRMU-SFT/250M/gsm8k\endcsname{86.3}
\expandafter\def\csname ev@intern/bio/ExternalRMU-SFT/250M/hellaswag\endcsname{85.5}
\expandafter\def\csname ev@intern/bio/ExternalRMU-SFT/250M/mmlu\endcsname{84.6}
\expandafter\def\csname ev@intern/bio/ExternalRMU-SFT/250M/mmlu_adj\endcsname{88.8}
\expandafter\def\csname ev@intern/bio/ExternalRMU-SFT/250M/mmlu_rest\endcsname{83.9}
\expandafter\def\csname ev@intern/bio/ExternalRMU-SFT/250M/wmdp\endcsname{82.0}
\expandafter\def\csname ev@intern/bio/ExternalRMU-SFT/250M/ifeval\endcsname{79.3}
\expandafter\def\csname ev@intern/bio/SatImp/0M/arc_challenge\endcsname{35.1}
\expandafter\def\csname ev@intern/bio/SatImp/0M/gsm8k\endcsname{0.3}
\expandafter\def\csname ev@intern/bio/SatImp/0M/hellaswag\endcsname{38.0}
\expandafter\def\csname ev@intern/bio/SatImp/0M/mmlu\endcsname{77.6}
\expandafter\def\csname ev@intern/bio/SatImp/0M/mmlu_adj\endcsname{81.4}
\expandafter\def\csname ev@intern/bio/SatImp/0M/mmlu_rest\endcsname{76.9}
\expandafter\def\csname ev@intern/bio/SatImp/0M/wmdp\endcsname{73.8}
\expandafter\def\csname ev@intern/bio/SatImp-SFT/250M/arc_challenge\endcsname{73.7}
\expandafter\def\csname ev@intern/bio/SatImp-SFT/250M/gsm8k\endcsname{85.1}
\expandafter\def\csname ev@intern/bio/SatImp-SFT/250M/hellaswag\endcsname{84.5}
\expandafter\def\csname ev@intern/bio/SatImp-SFT/250M/mmlu\endcsname{84.4}
\expandafter\def\csname ev@intern/bio/SatImp-SFT/250M/mmlu_adj\endcsname{88.7}
\expandafter\def\csname ev@intern/bio/SatImp-SFT/250M/mmlu_rest\endcsname{83.7}
\expandafter\def\csname ev@intern/bio/SatImp-SFT/250M/wmdp\endcsname{83.1}
\expandafter\def\csname ev@intern/bio/SatImp-SFT/250M/ifeval\endcsname{79.5}
\expandafter\def\csname ev@intern/bio/SimNPO/0M/arc_challenge\endcsname{69.4}
\expandafter\def\csname ev@intern/bio/SimNPO/0M/gsm8k\endcsname{81.5}
\expandafter\def\csname ev@intern/bio/SimNPO/0M/hellaswag\endcsname{83.5}
\expandafter\def\csname ev@intern/bio/SimNPO/0M/mmlu\endcsname{77.8}
\expandafter\def\csname ev@intern/bio/SimNPO/0M/mmlu_adj\endcsname{83.5}
\expandafter\def\csname ev@intern/bio/SimNPO/0M/mmlu_rest\endcsname{76.8}
\expandafter\def\csname ev@intern/bio/SimNPO/0M/wmdp\endcsname{74.6}
\expandafter\def\csname ev@intern/bio/SimNPO-SFT/250M/arc_challenge\endcsname{70.2}
\expandafter\def\csname ev@intern/bio/SimNPO-SFT/250M/gsm8k\endcsname{85.6}
\expandafter\def\csname ev@intern/bio/SimNPO-SFT/250M/hellaswag\endcsname{82.5}
\expandafter\def\csname ev@intern/bio/SimNPO-SFT/250M/mmlu\endcsname{79.9}
\expandafter\def\csname ev@intern/bio/SimNPO-SFT/250M/mmlu_adj\endcsname{84.0}
\expandafter\def\csname ev@intern/bio/SimNPO-SFT/250M/mmlu_rest\endcsname{79.2}
\expandafter\def\csname ev@intern/bio/SimNPO-SFT/250M/wmdp\endcsname{75.6}
\expandafter\def\csname ev@intern/bio/SimNPO-SFT/250M/ifeval\endcsname{65.8}
\expandafter\def\csname ev@intern/bio/TRACE/0M/arc_challenge\endcsname{72.6}
\expandafter\def\csname ev@intern/bio/TRACE/0M/gsm8k\endcsname{87.3}
\expandafter\def\csname ev@intern/bio/TRACE/0M/hellaswag\endcsname{85.2}
\expandafter\def\csname ev@intern/bio/TRACE/0M/mmlu\endcsname{84.1}
\expandafter\def\csname ev@intern/bio/TRACE/0M/mmlu_adj\endcsname{81.8}
\expandafter\def\csname ev@intern/bio/TRACE/0M/mmlu_rest\endcsname{84.5}
\expandafter\def\csname ev@intern/bio/TRACE/0M/wmdp\endcsname{33.9}
\expandafter\def\csname ev@intern/bio/TRACE/0M/ifeval\endcsname{80.6}
\expandafter\def\csname ev@intern/bio/TRACE-SFT/250M/arc_challenge\endcsname{74.3}
\expandafter\def\csname ev@intern/bio/TRACE-SFT/250M/gsm8k\endcsname{86.7}
\expandafter\def\csname ev@intern/bio/TRACE-SFT/250M/hellaswag\endcsname{85.4}
\expandafter\def\csname ev@intern/bio/TRACE-SFT/250M/mmlu\endcsname{84.8}
\expandafter\def\csname ev@intern/bio/TRACE-SFT/250M/mmlu_adj\endcsname{88.4}
\expandafter\def\csname ev@intern/bio/TRACE-SFT/250M/mmlu_rest\endcsname{84.1}
\expandafter\def\csname ev@intern/bio/TRACE-SFT/250M/wmdp\endcsname{81.2}
\expandafter\def\csname ev@intern/bio/TRACE-SFT/250M/ifeval\endcsname{77.8}
\expandafter\def\csname ev@intern/bio/WGA/0M/arc_challenge\endcsname{35.0}
\expandafter\def\csname ev@intern/bio/WGA/0M/gsm8k\endcsname{0.5}
\expandafter\def\csname ev@intern/bio/WGA/0M/hellaswag\endcsname{37.0}
\expandafter\def\csname ev@intern/bio/WGA/0M/mmlu\endcsname{76.9}
\expandafter\def\csname ev@intern/bio/WGA/0M/mmlu_adj\endcsname{81.1}
\expandafter\def\csname ev@intern/bio/WGA/0M/mmlu_rest\endcsname{76.2}
\expandafter\def\csname ev@intern/bio/WGA/0M/wmdp\endcsname{74.0}
\expandafter\def\csname ev@intern/bio/WGA-SFT/250M/arc_challenge\endcsname{73.5}
\expandafter\def\csname ev@intern/bio/WGA-SFT/250M/gsm8k\endcsname{85.4}
\expandafter\def\csname ev@intern/bio/WGA-SFT/250M/hellaswag\endcsname{84.5}
\expandafter\def\csname ev@intern/bio/WGA-SFT/250M/mmlu\endcsname{84.6}
\expandafter\def\csname ev@intern/bio/WGA-SFT/250M/mmlu_adj\endcsname{88.5}
\expandafter\def\csname ev@intern/bio/WGA-SFT/250M/mmlu_rest\endcsname{83.9}
\expandafter\def\csname ev@intern/bio/WGA-SFT/250M/wmdp\endcsname{83.8}
\expandafter\def\csname ev@intern/bio/WGA-SFT/250M/ifeval\endcsname{78.9}
\expandafter\def\csname ev@variant:intern/bio/A2-h0s1/0M/arc_challenge\endcsname{65.8}
\expandafter\def\csname ev@variant:intern/bio/A2-h0s1/0M/gsm8k\endcsname{80.6}
\expandafter\def\csname ev@variant:intern/bio/A2-h0s1/0M/hellaswag\endcsname{83.0}
\expandafter\def\csname ev@variant:intern/bio/A2-h0s1/0M/mmlu\endcsname{76.2}
\expandafter\def\csname ev@variant:intern/bio/A2-h0s1/0M/mmlu_adj\endcsname{57.5}
\expandafter\def\csname ev@variant:intern/bio/A2-h0s1/0M/mmlu_rest\endcsname{79.5}
\expandafter\def\csname ev@variant:intern/bio/A2-h0s1/0M/wmdp\endcsname{29.0}
\expandafter\def\csname ev@variant:intern/bio/A1-h0s1/0M/arc_challenge\endcsname{64.8}
\expandafter\def\csname ev@variant:intern/bio/A1-h0s1/0M/gsm8k\endcsname{80.7}
\expandafter\def\csname ev@variant:intern/bio/A1-h0s1/0M/hellaswag\endcsname{82.7}
\expandafter\def\csname ev@variant:intern/bio/A1-h0s1/0M/mmlu\endcsname{78.0}
\expandafter\def\csname ev@variant:intern/bio/A1-h0s1/0M/mmlu_adj\endcsname{65.7}
\expandafter\def\csname ev@variant:intern/bio/A1-h0s1/0M/mmlu_rest\endcsname{80.2}
\expandafter\def\csname ev@variant:intern/bio/A1-h0s1/0M/wmdp\endcsname{57.1}
\expandafter\def\csname ev@variant:intern/bio/F1-h0s1/0M/arc_challenge\endcsname{68.4}
\expandafter\def\csname ev@variant:intern/bio/F1-h0s1/0M/gsm8k\endcsname{80.1}
\expandafter\def\csname ev@variant:intern/bio/F1-h0s1/0M/hellaswag\endcsname{83.0}
\expandafter\def\csname ev@variant:intern/bio/F1-h0s1/0M/mmlu\endcsname{81.0}
\expandafter\def\csname ev@variant:intern/bio/F1-h0s1/0M/mmlu_adj\endcsname{84.2}
\expandafter\def\csname ev@variant:intern/bio/F1-h0s1/0M/mmlu_rest\endcsname{80.5}
\expandafter\def\csname ev@variant:intern/bio/F1-h0s1/0M/wmdp\endcsname{81.4}
\expandafter\def\csname ev@variant:intern/bio/F2-h0s1/0M/arc_challenge\endcsname{68.3}
\expandafter\def\csname ev@variant:intern/bio/F2-h0s1/0M/gsm8k\endcsname{79.6}
\expandafter\def\csname ev@variant:intern/bio/F2-h0s1/0M/hellaswag\endcsname{83.2}
\expandafter\def\csname ev@variant:intern/bio/F2-h0s1/0M/mmlu\endcsname{80.4}
\expandafter\def\csname ev@variant:intern/bio/F2-h0s1/0M/mmlu_adj\endcsname{82.4}
\expandafter\def\csname ev@variant:intern/bio/F2-h0s1/0M/mmlu_rest\endcsname{80.1}
\expandafter\def\csname ev@variant:intern/bio/F2-h0s1/0M/wmdp\endcsname{79.1}
\expandafter\def\csname ev@variant:intern/bio/F3-h0s1/250M/arc_challenge\endcsname{71.0}
\expandafter\def\csname ev@variant:intern/bio/F3-h0s1/250M/gsm8k\endcsname{83.1}
\expandafter\def\csname ev@variant:intern/bio/F3-h0s1/250M/hellaswag\endcsname{84.5}
\expandafter\def\csname ev@variant:intern/bio/F3-h0s1/250M/mmlu\endcsname{82.1}
\expandafter\def\csname ev@variant:intern/bio/F3-h0s1/250M/mmlu_adj\endcsname{85.8}
\expandafter\def\csname ev@variant:intern/bio/F3-h0s1/250M/mmlu_rest\endcsname{81.5}
\expandafter\def\csname ev@variant:intern/bio/F3-h0s1/250M/wmdp\endcsname{81.7}
\expandafter\def\csname ev@variant:intern/bio/F3-h0s1/250M/ifeval\endcsname{78.9}
\expandafter\def\csname ev@variant:intern/bio/A3-h0s1/250M/arc_challenge\endcsname{68.5}
\expandafter\def\csname ev@variant:intern/bio/A3-h0s1/250M/gsm8k\endcsname{84.5}
\expandafter\def\csname ev@variant:intern/bio/A3-h0s1/250M/hellaswag\endcsname{84.4}
\expandafter\def\csname ev@variant:intern/bio/A3-h0s1/250M/mmlu\endcsname{79.3}
\expandafter\def\csname ev@variant:intern/bio/A3-h0s1/250M/mmlu_adj\endcsname{69.0}
\expandafter\def\csname ev@variant:intern/bio/A3-h0s1/250M/mmlu_rest\endcsname{81.1}
\expandafter\def\csname ev@variant:intern/bio/A3-h0s1/250M/wmdp\endcsname{43.0}
\expandafter\def\csname ev@variant:intern/bio/A3-h0s1/250M/ifeval\endcsname{77.6}
\expandafter\def\csname ev@variant:intern/bio/A3-direct-h0s1/250M/arc_challenge\endcsname{69.2}
\expandafter\def\csname ev@variant:intern/bio/A3-direct-h0s1/250M/gsm8k\endcsname{84.8}
\expandafter\def\csname ev@variant:intern/bio/A3-direct-h0s1/250M/hellaswag\endcsname{84.9}
\expandafter\def\csname ev@variant:intern/bio/A3-direct-h0s1/250M/mmlu\endcsname{80.3}
\expandafter\def\csname ev@variant:intern/bio/A3-direct-h0s1/250M/mmlu_adj\endcsname{74.6}
\expandafter\def\csname ev@variant:intern/bio/A3-direct-h0s1/250M/mmlu_rest\endcsname{81.2}
\expandafter\def\csname ev@variant:intern/bio/A3-direct-h0s1/250M/wmdp\endcsname{59.9}
\expandafter\def\csname ev@variant:intern/bio/A3-direct-h0s1/250M/ifeval\endcsname{78.6}
\expandafter\def\csname ev@variant:intern/bio/Random-excision-0/0M/arc_challenge\endcsname{71.6}
\expandafter\def\csname ev@variant:intern/bio/Random-excision-0/0M/gsm8k\endcsname{85.6}
\expandafter\def\csname ev@variant:intern/bio/Random-excision-0/0M/hellaswag\endcsname{84.0}
\expandafter\def\csname ev@variant:intern/bio/Random-excision-0/0M/mmlu\endcsname{82.6}
\expandafter\def\csname ev@variant:intern/bio/Random-excision-0/0M/mmlu_adj\endcsname{81.9}
\expandafter\def\csname ev@variant:intern/bio/Random-excision-0/0M/mmlu_rest\endcsname{82.7}
\expandafter\def\csname ev@variant:intern/bio/Random-excision-0/0M/wmdp\endcsname{80.1}
\expandafter\def\csname ev@variant:intern/bio/Random-excision-1/0M/arc_challenge\endcsname{71.0}
\expandafter\def\csname ev@variant:intern/bio/Random-excision-1/0M/gsm8k\endcsname{85.1}
\expandafter\def\csname ev@variant:intern/bio/Random-excision-1/0M/hellaswag\endcsname{84.3}
\expandafter\def\csname ev@variant:intern/bio/Random-excision-1/0M/mmlu\endcsname{82.7}
\expandafter\def\csname ev@variant:intern/bio/Random-excision-1/0M/mmlu_adj\endcsname{82.0}
\expandafter\def\csname ev@variant:intern/bio/Random-excision-1/0M/mmlu_rest\endcsname{82.8}
\expandafter\def\csname ev@variant:intern/bio/Random-excision-1/0M/wmdp\endcsname{78.2}
\expandafter\def\csname ev@variant:intern/bio/Random-excision-2/0M/arc_challenge\endcsname{70.4}
\expandafter\def\csname ev@variant:intern/bio/Random-excision-2/0M/gsm8k\endcsname{84.0}
\expandafter\def\csname ev@variant:intern/bio/Random-excision-2/0M/hellaswag\endcsname{84.0}
\expandafter\def\csname ev@variant:intern/bio/Random-excision-2/0M/mmlu\endcsname{82.2}
\expandafter\def\csname ev@variant:intern/bio/Random-excision-2/0M/mmlu_adj\endcsname{81.1}
\expandafter\def\csname ev@variant:intern/bio/Random-excision-2/0M/mmlu_rest\endcsname{82.4}
\expandafter\def\csname ev@variant:intern/bio/Random-excision-2/0M/wmdp\endcsname{79.3}
\expandafter\def\csname ev@variant:intern/bio/Random-excision-3/0M/arc_challenge\endcsname{71.2}
\expandafter\def\csname ev@variant:intern/bio/Random-excision-3/0M/gsm8k\endcsname{84.1}
\expandafter\def\csname ev@variant:intern/bio/Random-excision-3/0M/hellaswag\endcsname{84.1}
\expandafter\def\csname ev@variant:intern/bio/Random-excision-3/0M/mmlu\endcsname{82.4}
\expandafter\def\csname ev@variant:intern/bio/Random-excision-3/0M/mmlu_adj\endcsname{81.0}
\expandafter\def\csname ev@variant:intern/bio/Random-excision-3/0M/mmlu_rest\endcsname{82.6}
\expandafter\def\csname ev@variant:intern/bio/Random-excision-3/0M/wmdp\endcsname{77.9}
\expandafter\def\csname ev@variant:intern/bio/Random-excision-4/0M/arc_challenge\endcsname{70.7}
\expandafter\def\csname ev@variant:intern/bio/Random-excision-4/0M/gsm8k\endcsname{84.9}
\expandafter\def\csname ev@variant:intern/bio/Random-excision-4/0M/hellaswag\endcsname{84.3}
\expandafter\def\csname ev@variant:intern/bio/Random-excision-4/0M/mmlu\endcsname{82.3}
\expandafter\def\csname ev@variant:intern/bio/Random-excision-4/0M/mmlu_adj\endcsname{82.2}
\expandafter\def\csname ev@variant:intern/bio/Random-excision-4/0M/mmlu_rest\endcsname{82.3}
\expandafter\def\csname ev@variant:intern/bio/Random-excision-4/0M/wmdp\endcsname{79.2}
\expandafter\def\csname ev@variant:intern/bio/F1-gradient_mask_only-h0s1/0M/arc_challenge\endcsname{74.3}
\expandafter\def\csname ev@variant:intern/bio/F1-gradient_mask_only-h0s1/0M/gsm8k\endcsname{87.6}
\expandafter\def\csname ev@variant:intern/bio/F1-gradient_mask_only-h0s1/0M/hellaswag\endcsname{85.6}
\expandafter\def\csname ev@variant:intern/bio/F1-gradient_mask_only-h0s1/0M/mmlu\endcsname{85.2}
\expandafter\def\csname ev@variant:intern/bio/F1-gradient_mask_only-h0s1/0M/mmlu_adj\endcsname{89.2}
\expandafter\def\csname ev@variant:intern/bio/F1-gradient_mask_only-h0s1/0M/mmlu_rest\endcsname{84.5}
\expandafter\def\csname ev@variant:intern/bio/F1-gradient_mask_only-h0s1/0M/wmdp\endcsname{83.3}
\expandafter\def\csname ev@variant:intern/bio/F1-gradient_mask_only/0M/arc_challenge\endcsname{74.1}
\expandafter\def\csname ev@variant:intern/bio/F1-gradient_mask_only/0M/gsm8k\endcsname{86.7}
\expandafter\def\csname ev@variant:intern/bio/F1-gradient_mask_only/0M/hellaswag\endcsname{85.7}
\expandafter\def\csname ev@variant:intern/bio/F1-gradient_mask_only/0M/mmlu\endcsname{85.0}
\expandafter\def\csname ev@variant:intern/bio/F1-gradient_mask_only/0M/mmlu_adj\endcsname{88.8}
\expandafter\def\csname ev@variant:intern/bio/F1-gradient_mask_only/0M/mmlu_rest\endcsname{84.3}
\expandafter\def\csname ev@variant:intern/bio/F1-gradient_mask_only/0M/wmdp\endcsname{84.2}
\expandafter\def\csname ev@variant:intern/bio/A1-gradient_mask_only-h0s1/0M/arc_challenge\endcsname{71.1}
\expandafter\def\csname ev@variant:intern/bio/A1-gradient_mask_only-h0s1/0M/gsm8k\endcsname{86.5}
\expandafter\def\csname ev@variant:intern/bio/A1-gradient_mask_only-h0s1/0M/hellaswag\endcsname{85.1}
\expandafter\def\csname ev@variant:intern/bio/A1-gradient_mask_only-h0s1/0M/mmlu\endcsname{83.4}
\expandafter\def\csname ev@variant:intern/bio/A1-gradient_mask_only-h0s1/0M/mmlu_adj\endcsname{79.2}
\expandafter\def\csname ev@variant:intern/bio/A1-gradient_mask_only-h0s1/0M/mmlu_rest\endcsname{84.2}
\expandafter\def\csname ev@variant:intern/bio/A1-gradient_mask_only-h0s1/0M/wmdp\endcsname{68.5}
\expandafter\def\csname ev@variant:intern/bio/A1-gradient_mask_only/0M/arc_challenge\endcsname{71.5}
\expandafter\def\csname ev@variant:intern/bio/A1-gradient_mask_only/0M/gsm8k\endcsname{86.8}
\expandafter\def\csname ev@variant:intern/bio/A1-gradient_mask_only/0M/hellaswag\endcsname{85.0}
\expandafter\def\csname ev@variant:intern/bio/A1-gradient_mask_only/0M/mmlu\endcsname{83.3}
\expandafter\def\csname ev@variant:intern/bio/A1-gradient_mask_only/0M/mmlu_adj\endcsname{79.3}
\expandafter\def\csname ev@variant:intern/bio/A1-gradient_mask_only/0M/mmlu_rest\endcsname{84.1}
\expandafter\def\csname ev@variant:intern/bio/A1-gradient_mask_only/0M/wmdp\endcsname{68.4}
\expandafter\def\csname ev@variant:intern/bio/F1-without_std_activation_mask-h0s1/0M/arc_challenge\endcsname{69.4}
\expandafter\def\csname ev@variant:intern/bio/F1-without_std_activation_mask-h0s1/0M/gsm8k\endcsname{79.2}
\expandafter\def\csname ev@variant:intern/bio/F1-without_std_activation_mask-h0s1/0M/hellaswag\endcsname{83.6}
\expandafter\def\csname ev@variant:intern/bio/F1-without_std_activation_mask-h0s1/0M/mmlu\endcsname{81.5}
\expandafter\def\csname ev@variant:intern/bio/F1-without_std_activation_mask-h0s1/0M/mmlu_adj\endcsname{85.8}
\expandafter\def\csname ev@variant:intern/bio/F1-without_std_activation_mask-h0s1/0M/mmlu_rest\endcsname{80.7}
\expandafter\def\csname ev@variant:intern/bio/F1-without_std_activation_mask-h0s1/0M/wmdp\endcsname{81.8}
\expandafter\def\csname ev@variant:intern/bio/F1-without_std_activation_mask/0M/arc_challenge\endcsname{72.1}
\expandafter\def\csname ev@variant:intern/bio/F1-without_std_activation_mask/0M/gsm8k\endcsname{79.8}
\expandafter\def\csname ev@variant:intern/bio/F1-without_std_activation_mask/0M/hellaswag\endcsname{84.5}
\expandafter\def\csname ev@variant:intern/bio/F1-without_std_activation_mask/0M/mmlu\endcsname{83.3}
\expandafter\def\csname ev@variant:intern/bio/F1-without_std_activation_mask/0M/mmlu_adj\endcsname{84.5}
\expandafter\def\csname ev@variant:intern/bio/F1-without_std_activation_mask/0M/mmlu_rest\endcsname{83.1}
\expandafter\def\csname ev@variant:intern/bio/F1-without_std_activation_mask/0M/wmdp\endcsname{80.8}
\expandafter\def\csname ev@variant:intern/bio/A1-without_std_activation_mask-h0s1/0M/arc_challenge\endcsname{68.2}
\expandafter\def\csname ev@variant:intern/bio/A1-without_std_activation_mask-h0s1/0M/gsm8k\endcsname{79.2}
\expandafter\def\csname ev@variant:intern/bio/A1-without_std_activation_mask-h0s1/0M/hellaswag\endcsname{83.5}
\expandafter\def\csname ev@variant:intern/bio/A1-without_std_activation_mask-h0s1/0M/mmlu\endcsname{79.9}
\expandafter\def\csname ev@variant:intern/bio/A1-without_std_activation_mask-h0s1/0M/mmlu_adj\endcsname{75.0}
\expandafter\def\csname ev@variant:intern/bio/A1-without_std_activation_mask-h0s1/0M/mmlu_rest\endcsname{80.7}
\expandafter\def\csname ev@variant:intern/bio/A1-without_std_activation_mask-h0s1/0M/wmdp\endcsname{62.0}
\expandafter\def\csname ev@variant:intern/bio/A1-without_std_activation_mask/0M/arc_challenge\endcsname{70.5}
\expandafter\def\csname ev@variant:intern/bio/A1-without_std_activation_mask/0M/gsm8k\endcsname{80.5}
\expandafter\def\csname ev@variant:intern/bio/A1-without_std_activation_mask/0M/hellaswag\endcsname{84.6}
\expandafter\def\csname ev@variant:intern/bio/A1-without_std_activation_mask/0M/mmlu\endcsname{82.2}
\expandafter\def\csname ev@variant:intern/bio/A1-without_std_activation_mask/0M/mmlu_adj\endcsname{77.5}
\expandafter\def\csname ev@variant:intern/bio/A1-without_std_activation_mask/0M/mmlu_rest\endcsname{83.0}
\expandafter\def\csname ev@variant:intern/bio/A1-without_std_activation_mask/0M/wmdp\endcsname{65.9}
\expandafter\def\csname ev@variant:intern/bio/F1-without_routing_loss-h0s1/0M/arc_challenge\endcsname{73.7}
\expandafter\def\csname ev@variant:intern/bio/F1-without_routing_loss-h0s1/0M/gsm8k\endcsname{87.3}
\expandafter\def\csname ev@variant:intern/bio/F1-without_routing_loss-h0s1/0M/hellaswag\endcsname{85.5}
\expandafter\def\csname ev@variant:intern/bio/F1-without_routing_loss-h0s1/0M/mmlu\endcsname{85.1}
\expandafter\def\csname ev@variant:intern/bio/F1-without_routing_loss-h0s1/0M/mmlu_adj\endcsname{88.8}
\expandafter\def\csname ev@variant:intern/bio/F1-without_routing_loss-h0s1/0M/mmlu_rest\endcsname{84.4}
\expandafter\def\csname ev@variant:intern/bio/F1-without_routing_loss-h0s1/0M/wmdp\endcsname{83.6}
\expandafter\def\csname ev@variant:intern/bio/F1-without_routing_loss/0M/arc_challenge\endcsname{73.4}
\expandafter\def\csname ev@variant:intern/bio/F1-without_routing_loss/0M/gsm8k\endcsname{86.4}
\expandafter\def\csname ev@variant:intern/bio/F1-without_routing_loss/0M/hellaswag\endcsname{85.6}
\expandafter\def\csname ev@variant:intern/bio/F1-without_routing_loss/0M/mmlu\endcsname{85.1}
\expandafter\def\csname ev@variant:intern/bio/F1-without_routing_loss/0M/mmlu_adj\endcsname{89.0}
\expandafter\def\csname ev@variant:intern/bio/F1-without_routing_loss/0M/mmlu_rest\endcsname{84.4}
\expandafter\def\csname ev@variant:intern/bio/F1-without_routing_loss/0M/wmdp\endcsname{84.0}
\expandafter\def\csname ev@variant:intern/bio/A1-without_routing_loss-h0s1/0M/arc_challenge\endcsname{71.4}
\expandafter\def\csname ev@variant:intern/bio/A1-without_routing_loss-h0s1/0M/gsm8k\endcsname{87.1}
\expandafter\def\csname ev@variant:intern/bio/A1-without_routing_loss-h0s1/0M/hellaswag\endcsname{85.1}
\expandafter\def\csname ev@variant:intern/bio/A1-without_routing_loss-h0s1/0M/mmlu\endcsname{83.7}
\expandafter\def\csname ev@variant:intern/bio/A1-without_routing_loss-h0s1/0M/mmlu_adj\endcsname{81.1}
\expandafter\def\csname ev@variant:intern/bio/A1-without_routing_loss-h0s1/0M/mmlu_rest\endcsname{84.2}
\expandafter\def\csname ev@variant:intern/bio/A1-without_routing_loss-h0s1/0M/wmdp\endcsname{70.3}
\expandafter\def\csname ev@variant:intern/bio/A1-without_routing_loss/0M/arc_challenge\endcsname{71.0}
\expandafter\def\csname ev@variant:intern/bio/A1-without_routing_loss/0M/gsm8k\endcsname{86.9}
\expandafter\def\csname ev@variant:intern/bio/A1-without_routing_loss/0M/hellaswag\endcsname{85.0}
\expandafter\def\csname ev@variant:intern/bio/A1-without_routing_loss/0M/mmlu\endcsname{83.6}
\expandafter\def\csname ev@variant:intern/bio/A1-without_routing_loss/0M/mmlu_adj\endcsname{81.1}
\expandafter\def\csname ev@variant:intern/bio/A1-without_routing_loss/0M/mmlu_rest\endcsname{84.0}
\expandafter\def\csname ev@variant:intern/bio/A1-without_routing_loss/0M/wmdp\endcsname{70.2}
\expandafter\def\csname ev@qwen/bio/B/0M/arc_challenge\endcsname{73.5}
\expandafter\def\csname ev@qwen/bio/B/0M/gsm8k\endcsname{85.8}
\expandafter\def\csname ev@qwen/bio/B/0M/hellaswag\endcsname{85.6}
\expandafter\def\csname ev@qwen/bio/B/0M/mmlu\endcsname{82.4}
\expandafter\def\csname ev@qwen/bio/B/0M/mmlu_adj\endcsname{86.9}
\expandafter\def\csname ev@qwen/bio/B/0M/mmlu_rest\endcsname{81.7}
\expandafter\def\csname ev@qwen/bio/B/0M/wmdp\endcsname{83.2}
\expandafter\def\csname ev@qwen/bio/S_std(B)/250M/arc_challenge\endcsname{73.6}
\expandafter\def\csname ev@qwen/bio/S_std(B)/250M/gsm8k\endcsname{85.1}
\expandafter\def\csname ev@qwen/bio/S_std(B)/250M/hellaswag\endcsname{85.4}
\expandafter\def\csname ev@qwen/bio/S_std(B)/250M/mmlu\endcsname{82.2}
\expandafter\def\csname ev@qwen/bio/S_std(B)/250M/mmlu_adj\endcsname{86.8}
\expandafter\def\csname ev@qwen/bio/S_std(B)/250M/mmlu_rest\endcsname{81.4}
\expandafter\def\csname ev@qwen/bio/S_std(B)/250M/wmdp\endcsname{84.0}
\expandafter\def\csname ev@qwen/bio/S_std(B)/250M/ifeval\endcsname{43.1}
\expandafter\def\csname ev@qwen/bio/F0/0M/arc_challenge\endcsname{73.5}
\expandafter\def\csname ev@qwen/bio/F0/0M/gsm8k\endcsname{86.9}
\expandafter\def\csname ev@qwen/bio/F0/0M/hellaswag\endcsname{85.5}
\expandafter\def\csname ev@qwen/bio/F0/0M/mmlu\endcsname{82.5}
\expandafter\def\csname ev@qwen/bio/F0/0M/mmlu_adj\endcsname{86.3}
\expandafter\def\csname ev@qwen/bio/F0/0M/mmlu_rest\endcsname{81.8}
\expandafter\def\csname ev@qwen/bio/F0/0M/wmdp\endcsname{82.3}
\expandafter\def\csname ev@qwen/bio/A0/0M/arc_challenge\endcsname{71.3}
\expandafter\def\csname ev@qwen/bio/A0/0M/gsm8k\endcsname{87.0}
\expandafter\def\csname ev@qwen/bio/A0/0M/hellaswag\endcsname{85.0}
\expandafter\def\csname ev@qwen/bio/A0/0M/mmlu\endcsname{81.1}
\expandafter\def\csname ev@qwen/bio/A0/0M/mmlu_adj\endcsname{79.9}
\expandafter\def\csname ev@qwen/bio/A0/0M/mmlu_rest\endcsname{81.3}
\expandafter\def\csname ev@qwen/bio/A0/0M/wmdp\endcsname{69.2}
\expandafter\def\csname ev@qwen/bio/F1/0M/arc_challenge\endcsname{70.1}
\expandafter\def\csname ev@qwen/bio/F1/0M/gsm8k\endcsname{83.1}
\expandafter\def\csname ev@qwen/bio/F1/0M/hellaswag\endcsname{83.9}
\expandafter\def\csname ev@qwen/bio/F1/0M/mmlu\endcsname{80.4}
\expandafter\def\csname ev@qwen/bio/F1/0M/mmlu_adj\endcsname{84.1}
\expandafter\def\csname ev@qwen/bio/F1/0M/mmlu_rest\endcsname{79.7}
\expandafter\def\csname ev@qwen/bio/F1/0M/wmdp\endcsname{81.8}
\expandafter\def\csname ev@qwen/bio/A1/0M/arc_challenge\endcsname{68.4}
\expandafter\def\csname ev@qwen/bio/A1/0M/gsm8k\endcsname{83.5}
\expandafter\def\csname ev@qwen/bio/A1/0M/hellaswag\endcsname{83.8}
\expandafter\def\csname ev@qwen/bio/A1/0M/mmlu\endcsname{78.6}
\expandafter\def\csname ev@qwen/bio/A1/0M/mmlu_adj\endcsname{74.3}
\expandafter\def\csname ev@qwen/bio/A1/0M/mmlu_rest\endcsname{79.3}
\expandafter\def\csname ev@qwen/bio/A1/0M/wmdp\endcsname{62.9}
\expandafter\def\csname ev@qwen/bio/A2/0M/arc_challenge\endcsname{68.8}
\expandafter\def\csname ev@qwen/bio/A2/0M/gsm8k\endcsname{83.3}
\expandafter\def\csname ev@qwen/bio/A2/0M/hellaswag\endcsname{83.6}
\expandafter\def\csname ev@qwen/bio/A2/0M/mmlu\endcsname{77.2}
\expandafter\def\csname ev@qwen/bio/A2/0M/mmlu_adj\endcsname{66.9}
\expandafter\def\csname ev@qwen/bio/A2/0M/mmlu_rest\endcsname{79.0}
\expandafter\def\csname ev@qwen/bio/A2/0M/wmdp\endcsname{32.4}
\expandafter\def\csname ev@qwen/bio/F2/0M/arc_challenge\endcsname{70.1}
\expandafter\def\csname ev@qwen/bio/F2/0M/gsm8k\endcsname{83.4}
\expandafter\def\csname ev@qwen/bio/F2/0M/hellaswag\endcsname{83.7}
\expandafter\def\csname ev@qwen/bio/F2/0M/mmlu\endcsname{80.1}
\expandafter\def\csname ev@qwen/bio/F2/0M/mmlu_adj\endcsname{84.3}
\expandafter\def\csname ev@qwen/bio/F2/0M/mmlu_rest\endcsname{79.3}
\expandafter\def\csname ev@qwen/bio/F2/0M/wmdp\endcsname{80.8}
\expandafter\def\csname ev@qwen/bio/F3/250M/arc_challenge\endcsname{70.7}
\expandafter\def\csname ev@qwen/bio/F3/250M/gsm8k\endcsname{85.9}
\expandafter\def\csname ev@qwen/bio/F3/250M/hellaswag\endcsname{84.2}
\expandafter\def\csname ev@qwen/bio/F3/250M/mmlu\endcsname{80.7}
\expandafter\def\csname ev@qwen/bio/F3/250M/mmlu_adj\endcsname{84.9}
\expandafter\def\csname ev@qwen/bio/F3/250M/mmlu_rest\endcsname{80.0}
\expandafter\def\csname ev@qwen/bio/F3/250M/wmdp\endcsname{81.7}
\expandafter\def\csname ev@qwen/bio/F3/250M/ifeval\endcsname{43.2}
\expandafter\def\csname ev@qwen/bio/A3/250M/arc_challenge\endcsname{70.3}
\expandafter\def\csname ev@qwen/bio/A3/250M/gsm8k\endcsname{85.7}
\expandafter\def\csname ev@qwen/bio/A3/250M/hellaswag\endcsname{84.1}
\expandafter\def\csname ev@qwen/bio/A3/250M/mmlu\endcsname{78.7}
\expandafter\def\csname ev@qwen/bio/A3/250M/mmlu_adj\endcsname{72.8}
\expandafter\def\csname ev@qwen/bio/A3/250M/mmlu_rest\endcsname{79.7}
\expandafter\def\csname ev@qwen/bio/A3/250M/wmdp\endcsname{42.8}
\expandafter\def\csname ev@qwen/bio/A3/250M/ifeval\endcsname{44.7}
\expandafter\def\csname ev@qwen/bio/R/0M/arc_challenge\endcsname{73.6}
\expandafter\def\csname ev@qwen/bio/R/0M/gsm8k\endcsname{85.4}
\expandafter\def\csname ev@qwen/bio/R/0M/hellaswag\endcsname{85.4}
\expandafter\def\csname ev@qwen/bio/R/0M/mmlu\endcsname{81.8}
\expandafter\def\csname ev@qwen/bio/R/0M/mmlu_adj\endcsname{84.8}
\expandafter\def\csname ev@qwen/bio/R/0M/mmlu_rest\endcsname{81.3}
\expandafter\def\csname ev@qwen/bio/R/0M/wmdp\endcsname{41.6}
\expandafter\def\csname ev@qwen/bio/R-SFT/250M/arc_challenge\endcsname{73.7}
\expandafter\def\csname ev@qwen/bio/R-SFT/250M/gsm8k\endcsname{86.3}
\expandafter\def\csname ev@qwen/bio/R-SFT/250M/hellaswag\endcsname{85.4}
\expandafter\def\csname ev@qwen/bio/R-SFT/250M/mmlu\endcsname{82.0}
\expandafter\def\csname ev@qwen/bio/R-SFT/250M/mmlu_adj\endcsname{86.4}
\expandafter\def\csname ev@qwen/bio/R-SFT/250M/mmlu_rest\endcsname{81.3}
\expandafter\def\csname ev@qwen/bio/R-SFT/250M/wmdp\endcsname{78.8}
\expandafter\def\csname ev@qwen/bio/R-SFT/250M/ifeval\endcsname{45.7}
\expandafter\def\csname ev@qwen/bio/A3-direct/250M/arc_challenge\endcsname{69.8}
\expandafter\def\csname ev@qwen/bio/A3-direct/250M/gsm8k\endcsname{86.1}
\expandafter\def\csname ev@qwen/bio/A3-direct/250M/hellaswag\endcsname{84.6}
\expandafter\def\csname ev@qwen/bio/A3-direct/250M/mmlu\endcsname{79.4}
\expandafter\def\csname ev@qwen/bio/A3-direct/250M/mmlu_adj\endcsname{77.3}
\expandafter\def\csname ev@qwen/bio/A3-direct/250M/mmlu_rest\endcsname{79.8}
\expandafter\def\csname ev@qwen/bio/A3-direct/250M/wmdp\endcsname{63.0}
\expandafter\def\csname ev@qwen/bio/A3-direct/250M/ifeval\endcsname{45.3}
\expandafter\def\csname ev@qwen/bio/CEU/0M/arc_challenge\endcsname{40.1}
\expandafter\def\csname ev@qwen/bio/CEU/0M/gsm8k\endcsname{3.8}
\expandafter\def\csname ev@qwen/bio/CEU/0M/hellaswag\endcsname{39.3}
\expandafter\def\csname ev@qwen/bio/CEU/0M/mmlu\endcsname{81.0}
\expandafter\def\csname ev@qwen/bio/CEU/0M/mmlu_adj\endcsname{85.3}
\expandafter\def\csname ev@qwen/bio/CEU/0M/mmlu_rest\endcsname{80.3}
\expandafter\def\csname ev@qwen/bio/CEU/0M/wmdp\endcsname{81.2}
\expandafter\def\csname ev@qwen/bio/CEU-SFT/250M/arc_challenge\endcsname{72.7}
\expandafter\def\csname ev@qwen/bio/CEU-SFT/250M/gsm8k\endcsname{80.6}
\expandafter\def\csname ev@qwen/bio/CEU-SFT/250M/hellaswag\endcsname{84.2}
\expandafter\def\csname ev@qwen/bio/CEU-SFT/250M/mmlu\endcsname{82.1}
\expandafter\def\csname ev@qwen/bio/CEU-SFT/250M/mmlu_adj\endcsname{86.7}
\expandafter\def\csname ev@qwen/bio/CEU-SFT/250M/mmlu_rest\endcsname{81.3}
\expandafter\def\csname ev@qwen/bio/CEU-SFT/250M/wmdp\endcsname{83.4}
\expandafter\def\csname ev@qwen/bio/CEU-SFT/250M/ifeval\endcsname{50.1}
\expandafter\def\csname ev@qwen/bio/GA/0M/arc_challenge\endcsname{26.0}
\expandafter\def\csname ev@qwen/bio/GA/0M/gsm8k\endcsname{0.0}
\expandafter\def\csname ev@qwen/bio/GA/0M/hellaswag\endcsname{26.1}
\expandafter\def\csname ev@qwen/bio/GA/0M/mmlu\endcsname{28.1}
\expandafter\def\csname ev@qwen/bio/GA/0M/mmlu_adj\endcsname{28.0}
\expandafter\def\csname ev@qwen/bio/GA/0M/mmlu_rest\endcsname{28.1}
\expandafter\def\csname ev@qwen/bio/GA/0M/wmdp\endcsname{25.0}
\expandafter\def\csname ev@qwen/bio/GA-SFT/250M/arc_challenge\endcsname{69.6}
\expandafter\def\csname ev@qwen/bio/GA-SFT/250M/gsm8k\endcsname{32.7}
\expandafter\def\csname ev@qwen/bio/GA-SFT/250M/hellaswag\endcsname{52.7}
\expandafter\def\csname ev@qwen/bio/GA-SFT/250M/mmlu\endcsname{80.5}
\expandafter\def\csname ev@qwen/bio/GA-SFT/250M/mmlu_adj\endcsname{85.3}
\expandafter\def\csname ev@qwen/bio/GA-SFT/250M/mmlu_rest\endcsname{79.7}
\expandafter\def\csname ev@qwen/bio/GA-SFT/250M/wmdp\endcsname{82.5}
\expandafter\def\csname ev@qwen/bio/GA-SFT/250M/ifeval\endcsname{51.9}
\expandafter\def\csname ev@qwen/bio/GD/0M/arc_challenge\endcsname{72.0}
\expandafter\def\csname ev@qwen/bio/GD/0M/gsm8k\endcsname{74.7}
\expandafter\def\csname ev@qwen/bio/GD/0M/hellaswag\endcsname{85.2}
\expandafter\def\csname ev@qwen/bio/GD/0M/mmlu\endcsname{81.2}
\expandafter\def\csname ev@qwen/bio/GD/0M/mmlu_adj\endcsname{85.8}
\expandafter\def\csname ev@qwen/bio/GD/0M/mmlu_rest\endcsname{80.4}
\expandafter\def\csname ev@qwen/bio/GD/0M/wmdp\endcsname{81.0}
\expandafter\def\csname ev@qwen/bio/GD-SFT/250M/arc_challenge\endcsname{73.5}
\expandafter\def\csname ev@qwen/bio/GD-SFT/250M/gsm8k\endcsname{67.1}
\expandafter\def\csname ev@qwen/bio/GD-SFT/250M/hellaswag\endcsname{85.3}
\expandafter\def\csname ev@qwen/bio/GD-SFT/250M/mmlu\endcsname{81.5}
\expandafter\def\csname ev@qwen/bio/GD-SFT/250M/mmlu_adj\endcsname{86.2}
\expandafter\def\csname ev@qwen/bio/GD-SFT/250M/mmlu_rest\endcsname{80.7}
\expandafter\def\csname ev@qwen/bio/GD-SFT/250M/wmdp\endcsname{82.2}
\expandafter\def\csname ev@qwen/bio/GD-SFT/250M/ifeval\endcsname{38.8}
\expandafter\def\csname ev@qwen/bio/PDU/0M/arc_challenge\endcsname{72.5}
\expandafter\def\csname ev@qwen/bio/PDU/0M/gsm8k\endcsname{67.7}
\expandafter\def\csname ev@qwen/bio/PDU/0M/hellaswag\endcsname{84.7}
\expandafter\def\csname ev@qwen/bio/PDU/0M/mmlu\endcsname{78.2}
\expandafter\def\csname ev@qwen/bio/PDU/0M/mmlu_adj\endcsname{77.0}
\expandafter\def\csname ev@qwen/bio/PDU/0M/mmlu_rest\endcsname{78.4}
\expandafter\def\csname ev@qwen/bio/PDU/0M/wmdp\endcsname{47.8}
\expandafter\def\csname ev@qwen/bio/PDU-SFT/250M/arc_challenge\endcsname{73.5}
\expandafter\def\csname ev@qwen/bio/PDU-SFT/250M/gsm8k\endcsname{77.4}
\expandafter\def\csname ev@qwen/bio/PDU-SFT/250M/hellaswag\endcsname{85.2}
\expandafter\def\csname ev@qwen/bio/PDU-SFT/250M/mmlu\endcsname{81.4}
\expandafter\def\csname ev@qwen/bio/PDU-SFT/250M/mmlu_adj\endcsname{86.2}
\expandafter\def\csname ev@qwen/bio/PDU-SFT/250M/mmlu_rest\endcsname{80.5}
\expandafter\def\csname ev@qwen/bio/PDU-SFT/250M/wmdp\endcsname{81.8}
\expandafter\def\csname ev@qwen/bio/PDU-SFT/250M/ifeval\endcsname{50.3}
\expandafter\def\csname ev@qwen/bio/PureNPO/0M/arc_challenge\endcsname{71.5}
\expandafter\def\csname ev@qwen/bio/PureNPO/0M/gsm8k\endcsname{83.6}
\expandafter\def\csname ev@qwen/bio/PureNPO/0M/hellaswag\endcsname{83.4}
\expandafter\def\csname ev@qwen/bio/PureNPO/0M/mmlu\endcsname{77.4}
\expandafter\def\csname ev@qwen/bio/PureNPO/0M/mmlu_adj\endcsname{83.1}
\expandafter\def\csname ev@qwen/bio/PureNPO/0M/mmlu_rest\endcsname{76.4}
\expandafter\def\csname ev@qwen/bio/PureNPO/0M/wmdp\endcsname{76.4}
\expandafter\def\csname ev@qwen/bio/PureNPO-SFT/250M/arc_challenge\endcsname{72.0}
\expandafter\def\csname ev@qwen/bio/PureNPO-SFT/250M/gsm8k\endcsname{84.6}
\expandafter\def\csname ev@qwen/bio/PureNPO-SFT/250M/hellaswag\endcsname{81.9}
\expandafter\def\csname ev@qwen/bio/PureNPO-SFT/250M/mmlu\endcsname{78.3}
\expandafter\def\csname ev@qwen/bio/PureNPO-SFT/250M/mmlu_adj\endcsname{84.3}
\expandafter\def\csname ev@qwen/bio/PureNPO-SFT/250M/mmlu_rest\endcsname{77.2}
\expandafter\def\csname ev@qwen/bio/PureNPO-SFT/250M/wmdp\endcsname{76.0}
\expandafter\def\csname ev@qwen/bio/PureNPO-SFT/250M/ifeval\endcsname{37.9}
\expandafter\def\csname ev@qwen/bio/ExternalRMU/0M/arc_challenge\endcsname{73.9}
\expandafter\def\csname ev@qwen/bio/ExternalRMU/0M/gsm8k\endcsname{84.7}
\expandafter\def\csname ev@qwen/bio/ExternalRMU/0M/hellaswag\endcsname{85.4}
\expandafter\def\csname ev@qwen/bio/ExternalRMU/0M/mmlu\endcsname{82.4}
\expandafter\def\csname ev@qwen/bio/ExternalRMU/0M/mmlu_adj\endcsname{85.9}
\expandafter\def\csname ev@qwen/bio/ExternalRMU/0M/mmlu_rest\endcsname{81.8}
\expandafter\def\csname ev@qwen/bio/ExternalRMU/0M/wmdp\endcsname{44.1}
\expandafter\def\csname ev@qwen/bio/ExternalRMU/0M/ifeval\endcsname{56.0}
\expandafter\def\csname ev@qwen/bio/ExternalRMU-SFT/250M/arc_challenge\endcsname{73.7}
\expandafter\def\csname ev@qwen/bio/ExternalRMU-SFT/250M/gsm8k\endcsname{85.0}
\expandafter\def\csname ev@qwen/bio/ExternalRMU-SFT/250M/hellaswag\endcsname{85.5}
\expandafter\def\csname ev@qwen/bio/ExternalRMU-SFT/250M/mmlu\endcsname{82.0}
\expandafter\def\csname ev@qwen/bio/ExternalRMU-SFT/250M/mmlu_adj\endcsname{86.7}
\expandafter\def\csname ev@qwen/bio/ExternalRMU-SFT/250M/mmlu_rest\endcsname{81.2}
\expandafter\def\csname ev@qwen/bio/ExternalRMU-SFT/250M/wmdp\endcsname{78.3}
\expandafter\def\csname ev@qwen/bio/ExternalRMU-SFT/250M/ifeval\endcsname{40.7}
\expandafter\def\csname ev@qwen/bio/SatImp/0M/arc_challenge\endcsname{43.3}
\expandafter\def\csname ev@qwen/bio/SatImp/0M/gsm8k\endcsname{70.3}
\expandafter\def\csname ev@qwen/bio/SatImp/0M/hellaswag\endcsname{44.1}
\expandafter\def\csname ev@qwen/bio/SatImp/0M/mmlu\endcsname{80.6}
\expandafter\def\csname ev@qwen/bio/SatImp/0M/mmlu_adj\endcsname{84.4}
\expandafter\def\csname ev@qwen/bio/SatImp/0M/mmlu_rest\endcsname{80.0}
\expandafter\def\csname ev@qwen/bio/SatImp/0M/wmdp\endcsname{80.4}
\expandafter\def\csname ev@qwen/bio/SatImp-SFT/250M/arc_challenge\endcsname{72.9}
\expandafter\def\csname ev@qwen/bio/SatImp-SFT/250M/gsm8k\endcsname{80.4}
\expandafter\def\csname ev@qwen/bio/SatImp-SFT/250M/hellaswag\endcsname{84.5}
\expandafter\def\csname ev@qwen/bio/SatImp-SFT/250M/mmlu\endcsname{82.1}
\expandafter\def\csname ev@qwen/bio/SatImp-SFT/250M/mmlu_adj\endcsname{86.3}
\expandafter\def\csname ev@qwen/bio/SatImp-SFT/250M/mmlu_rest\endcsname{81.3}
\expandafter\def\csname ev@qwen/bio/SatImp-SFT/250M/wmdp\endcsname{82.7}
\expandafter\def\csname ev@qwen/bio/SatImp-SFT/250M/ifeval\endcsname{48.2}
\expandafter\def\csname ev@qwen/bio/SimNPO/0M/arc_challenge\endcsname{71.0}
\expandafter\def\csname ev@qwen/bio/SimNPO/0M/gsm8k\endcsname{82.6}
\expandafter\def\csname ev@qwen/bio/SimNPO/0M/hellaswag\endcsname{83.7}
\expandafter\def\csname ev@qwen/bio/SimNPO/0M/mmlu\endcsname{77.2}
\expandafter\def\csname ev@qwen/bio/SimNPO/0M/mmlu_adj\endcsname{83.0}
\expandafter\def\csname ev@qwen/bio/SimNPO/0M/mmlu_rest\endcsname{76.2}
\expandafter\def\csname ev@qwen/bio/SimNPO/0M/wmdp\endcsname{76.4}
\expandafter\def\csname ev@qwen/bio/SimNPO-SFT/250M/arc_challenge\endcsname{70.8}
\expandafter\def\csname ev@qwen/bio/SimNPO-SFT/250M/gsm8k\endcsname{86.3}
\expandafter\def\csname ev@qwen/bio/SimNPO-SFT/250M/hellaswag\endcsname{82.3}
\expandafter\def\csname ev@qwen/bio/SimNPO-SFT/250M/mmlu\endcsname{78.3}
\expandafter\def\csname ev@qwen/bio/SimNPO-SFT/250M/mmlu_adj\endcsname{84.3}
\expandafter\def\csname ev@qwen/bio/SimNPO-SFT/250M/mmlu_rest\endcsname{77.2}
\expandafter\def\csname ev@qwen/bio/SimNPO-SFT/250M/wmdp\endcsname{77.5}
\expandafter\def\csname ev@qwen/bio/SimNPO-SFT/250M/ifeval\endcsname{42.5}
\expandafter\def\csname ev@qwen/bio/TRACE/0M/arc_challenge\endcsname{72.3}
\expandafter\def\csname ev@qwen/bio/TRACE/0M/gsm8k\endcsname{85.7}
\expandafter\def\csname ev@qwen/bio/TRACE/0M/hellaswag\endcsname{84.5}
\expandafter\def\csname ev@qwen/bio/TRACE/0M/mmlu\endcsname{81.9}
\expandafter\def\csname ev@qwen/bio/TRACE/0M/mmlu_adj\endcsname{83.4}
\expandafter\def\csname ev@qwen/bio/TRACE/0M/mmlu_rest\endcsname{81.6}
\expandafter\def\csname ev@qwen/bio/TRACE/0M/wmdp\endcsname{61.4}
\expandafter\def\csname ev@qwen/bio/TRACE-SFT/250M/arc_challenge\endcsname{73.5}
\expandafter\def\csname ev@qwen/bio/TRACE-SFT/250M/gsm8k\endcsname{85.1}
\expandafter\def\csname ev@qwen/bio/TRACE-SFT/250M/hellaswag\endcsname{85.1}
\expandafter\def\csname ev@qwen/bio/TRACE-SFT/250M/mmlu\endcsname{82.2}
\expandafter\def\csname ev@qwen/bio/TRACE-SFT/250M/mmlu_adj\endcsname{86.2}
\expandafter\def\csname ev@qwen/bio/TRACE-SFT/250M/mmlu_rest\endcsname{81.4}
\expandafter\def\csname ev@qwen/bio/TRACE-SFT/250M/wmdp\endcsname{78.9}
\expandafter\def\csname ev@qwen/bio/TRACE-SFT/250M/ifeval\endcsname{46.8}
\expandafter\def\csname ev@qwen/bio/WGA/0M/arc_challenge\endcsname{43.5}
\expandafter\def\csname ev@qwen/bio/WGA/0M/gsm8k\endcsname{71.2}
\expandafter\def\csname ev@qwen/bio/WGA/0M/hellaswag\endcsname{44.2}
\expandafter\def\csname ev@qwen/bio/WGA/0M/mmlu\endcsname{80.7}
\expandafter\def\csname ev@qwen/bio/WGA/0M/mmlu_adj\endcsname{84.6}
\expandafter\def\csname ev@qwen/bio/WGA/0M/mmlu_rest\endcsname{80.0}
\expandafter\def\csname ev@qwen/bio/WGA/0M/wmdp\endcsname{80.9}
\expandafter\def\csname ev@qwen/bio/WGA-SFT/250M/arc_challenge\endcsname{72.8}
\expandafter\def\csname ev@qwen/bio/WGA-SFT/250M/gsm8k\endcsname{81.0}
\expandafter\def\csname ev@qwen/bio/WGA-SFT/250M/hellaswag\endcsname{84.6}
\expandafter\def\csname ev@qwen/bio/WGA-SFT/250M/mmlu\endcsname{82.0}
\expandafter\def\csname ev@qwen/bio/WGA-SFT/250M/mmlu_adj\endcsname{86.5}
\expandafter\def\csname ev@qwen/bio/WGA-SFT/250M/mmlu_rest\endcsname{81.2}
\expandafter\def\csname ev@qwen/bio/WGA-SFT/250M/wmdp\endcsname{82.7}
\expandafter\def\csname ev@qwen/bio/WGA-SFT/250M/ifeval\endcsname{49.2}
\expandafter\def\csname ev@gemma/bio/B/0M/arc_challenge\endcsname{73.2}
\expandafter\def\csname ev@gemma/bio/B/0M/gsm8k\endcsname{71.4}
\expandafter\def\csname ev@gemma/bio/B/0M/hellaswag\endcsname{85.6}
\expandafter\def\csname ev@gemma/bio/B/0M/mmlu\endcsname{74.3}
\expandafter\def\csname ev@gemma/bio/B/0M/mmlu_adj\endcsname{81.5}
\expandafter\def\csname ev@gemma/bio/B/0M/mmlu_rest\endcsname{73.1}
\expandafter\def\csname ev@gemma/bio/B/0M/wmdp\endcsname{78.9}
\expandafter\def\csname ev@gemma/bio/S_std(B)/250M/arc_challenge\endcsname{73.7}
\expandafter\def\csname ev@gemma/bio/S_std(B)/250M/gsm8k\endcsname{72.2}
\expandafter\def\csname ev@gemma/bio/S_std(B)/250M/hellaswag\endcsname{85.5}
\expandafter\def\csname ev@gemma/bio/S_std(B)/250M/mmlu\endcsname{74.9}
\expandafter\def\csname ev@gemma/bio/S_std(B)/250M/mmlu_adj\endcsname{82.4}
\expandafter\def\csname ev@gemma/bio/S_std(B)/250M/mmlu_rest\endcsname{73.6}
\expandafter\def\csname ev@gemma/bio/S_std(B)/250M/wmdp\endcsname{78.8}
\expandafter\def\csname ev@gemma/bio/S_std(B)/250M/ifeval\endcsname{26.8}
\expandafter\def\csname ev@gemma/bio/F0/0M/arc_challenge\endcsname{72.7}
\expandafter\def\csname ev@gemma/bio/F0/0M/gsm8k\endcsname{70.4}
\expandafter\def\csname ev@gemma/bio/F0/0M/hellaswag\endcsname{85.5}
\expandafter\def\csname ev@gemma/bio/F0/0M/mmlu\endcsname{74.0}
\expandafter\def\csname ev@gemma/bio/F0/0M/mmlu_adj\endcsname{80.4}
\expandafter\def\csname ev@gemma/bio/F0/0M/mmlu_rest\endcsname{72.9}
\expandafter\def\csname ev@gemma/bio/F0/0M/wmdp\endcsname{78.9}
\expandafter\def\csname ev@gemma/bio/A0/0M/arc_challenge\endcsname{69.4}
\expandafter\def\csname ev@gemma/bio/A0/0M/gsm8k\endcsname{69.9}
\expandafter\def\csname ev@gemma/bio/A0/0M/hellaswag\endcsname{85.3}
\expandafter\def\csname ev@gemma/bio/A0/0M/mmlu\endcsname{72.6}
\expandafter\def\csname ev@gemma/bio/A0/0M/mmlu_adj\endcsname{77.5}
\expandafter\def\csname ev@gemma/bio/A0/0M/mmlu_rest\endcsname{71.8}
\expandafter\def\csname ev@gemma/bio/A0/0M/wmdp\endcsname{73.2}
\expandafter\def\csname ev@gemma/bio/F1/0M/arc_challenge\endcsname{71.6}
\expandafter\def\csname ev@gemma/bio/F1/0M/gsm8k\endcsname{70.7}
\expandafter\def\csname ev@gemma/bio/F1/0M/hellaswag\endcsname{85.4}
\expandafter\def\csname ev@gemma/bio/F1/0M/mmlu\endcsname{74.0}
\expandafter\def\csname ev@gemma/bio/F1/0M/mmlu_adj\endcsname{80.6}
\expandafter\def\csname ev@gemma/bio/F1/0M/mmlu_rest\endcsname{72.8}
\expandafter\def\csname ev@gemma/bio/F1/0M/wmdp\endcsname{76.2}
\expandafter\def\csname ev@gemma/bio/A1/0M/arc_challenge\endcsname{68.1}
\expandafter\def\csname ev@gemma/bio/A1/0M/gsm8k\endcsname{69.2}
\expandafter\def\csname ev@gemma/bio/A1/0M/hellaswag\endcsname{85.1}
\expandafter\def\csname ev@gemma/bio/A1/0M/mmlu\endcsname{72.1}
\expandafter\def\csname ev@gemma/bio/A1/0M/mmlu_adj\endcsname{74.9}
\expandafter\def\csname ev@gemma/bio/A1/0M/mmlu_rest\endcsname{71.6}
\expandafter\def\csname ev@gemma/bio/A1/0M/wmdp\endcsname{70.2}
\expandafter\def\csname ev@gemma/bio/A2/0M/arc_challenge\endcsname{66.7}
\expandafter\def\csname ev@gemma/bio/A2/0M/gsm8k\endcsname{66.0}
\expandafter\def\csname ev@gemma/bio/A2/0M/hellaswag\endcsname{84.6}
\expandafter\def\csname ev@gemma/bio/A2/0M/mmlu\endcsname{68.7}
\expandafter\def\csname ev@gemma/bio/A2/0M/mmlu_adj\endcsname{59.8}
\expandafter\def\csname ev@gemma/bio/A2/0M/mmlu_rest\endcsname{70.3}
\expandafter\def\csname ev@gemma/bio/A2/0M/wmdp\endcsname{33.6}
\expandafter\def\csname ev@gemma/bio/F2/0M/arc_challenge\endcsname{71.0}
\expandafter\def\csname ev@gemma/bio/F2/0M/gsm8k\endcsname{69.1}
\expandafter\def\csname ev@gemma/bio/F2/0M/hellaswag\endcsname{85.0}
\expandafter\def\csname ev@gemma/bio/F2/0M/mmlu\endcsname{72.6}
\expandafter\def\csname ev@gemma/bio/F2/0M/mmlu_adj\endcsname{78.8}
\expandafter\def\csname ev@gemma/bio/F2/0M/mmlu_rest\endcsname{71.5}
\expandafter\def\csname ev@gemma/bio/F2/0M/wmdp\endcsname{71.0}
\expandafter\def\csname ev@gemma/bio/F3/250M/arc_challenge\endcsname{71.8}
\expandafter\def\csname ev@gemma/bio/F3/250M/gsm8k\endcsname{69.2}
\expandafter\def\csname ev@gemma/bio/F3/250M/hellaswag\endcsname{85.2}
\expandafter\def\csname ev@gemma/bio/F3/250M/mmlu\endcsname{73.9}
\expandafter\def\csname ev@gemma/bio/F3/250M/mmlu_adj\endcsname{80.3}
\expandafter\def\csname ev@gemma/bio/F3/250M/mmlu_rest\endcsname{72.8}
\expandafter\def\csname ev@gemma/bio/F3/250M/wmdp\endcsname{76.4}
\expandafter\def\csname ev@gemma/bio/F3/250M/ifeval\endcsname{26.1}
\expandafter\def\csname ev@gemma/bio/A3/250M/arc_challenge\endcsname{67.9}
\expandafter\def\csname ev@gemma/bio/A3/250M/gsm8k\endcsname{68.2}
\expandafter\def\csname ev@gemma/bio/A3/250M/hellaswag\endcsname{84.8}
\expandafter\def\csname ev@gemma/bio/A3/250M/mmlu\endcsname{70.5}
\expandafter\def\csname ev@gemma/bio/A3/250M/mmlu_adj\endcsname{66.5}
\expandafter\def\csname ev@gemma/bio/A3/250M/mmlu_rest\endcsname{71.2}
\expandafter\def\csname ev@gemma/bio/A3/250M/wmdp\endcsname{41.3}
\expandafter\def\csname ev@gemma/bio/A3/250M/ifeval\endcsname{28.1}
\expandafter\def\csname ev@gemma/bio/A3-direct/250M/arc_challenge\endcsname{69.0}
\expandafter\def\csname ev@gemma/bio/A3-direct/250M/gsm8k\endcsname{69.7}
\expandafter\def\csname ev@gemma/bio/A3-direct/250M/hellaswag\endcsname{85.5}
\expandafter\def\csname ev@gemma/bio/A3-direct/250M/mmlu\endcsname{71.7}
\expandafter\def\csname ev@gemma/bio/A3-direct/250M/mmlu_adj\endcsname{73.4}
\expandafter\def\csname ev@gemma/bio/A3-direct/250M/mmlu_rest\endcsname{71.4}
\expandafter\def\csname ev@gemma/bio/A3-direct/250M/wmdp\endcsname{55.7}
\expandafter\def\csname ev@gemma/bio/A3-direct/250M/ifeval\endcsname{27.7}
\expandafter\def\csname ev@gemma/bio/R/0M/arc_challenge\endcsname{72.7}
\expandafter\def\csname ev@gemma/bio/R/0M/gsm8k\endcsname{68.5}
\expandafter\def\csname ev@gemma/bio/R/0M/hellaswag\endcsname{85.2}
\expandafter\def\csname ev@gemma/bio/R/0M/mmlu\endcsname{72.2}
\expandafter\def\csname ev@gemma/bio/R/0M/mmlu_adj\endcsname{74.7}
\expandafter\def\csname ev@gemma/bio/R/0M/mmlu_rest\endcsname{71.8}
\expandafter\def\csname ev@gemma/bio/R/0M/wmdp\endcsname{48.5}
\expandafter\def\csname ev@gemma/bio/R/0M/ifeval\endcsname{27.5}
\expandafter\def\csname ev@gemma/bio/R-SFT/250M/arc_challenge\endcsname{73.6}
\expandafter\def\csname ev@gemma/bio/R-SFT/250M/gsm8k\endcsname{72.4}
\expandafter\def\csname ev@gemma/bio/R-SFT/250M/hellaswag\endcsname{85.2}
\expandafter\def\csname ev@gemma/bio/R-SFT/250M/mmlu\endcsname{74.1}
\expandafter\def\csname ev@gemma/bio/R-SFT/250M/mmlu_adj\endcsname{80.7}
\expandafter\def\csname ev@gemma/bio/R-SFT/250M/mmlu_rest\endcsname{73.0}
\expandafter\def\csname ev@gemma/bio/R-SFT/250M/wmdp\endcsname{71.5}
\expandafter\def\csname ev@gemma/bio/R-SFT/250M/ifeval\endcsname{28.8}
\expandafter\def\csname ev@gemma/bio/CEU/0M/arc_challenge\endcsname{34.5}
\expandafter\def\csname ev@gemma/bio/CEU/0M/gsm8k\endcsname{0.0}
\expandafter\def\csname ev@gemma/bio/CEU/0M/hellaswag\endcsname{33.5}
\expandafter\def\csname ev@gemma/bio/CEU/0M/mmlu\endcsname{72.0}
\expandafter\def\csname ev@gemma/bio/CEU/0M/mmlu_adj\endcsname{79.6}
\expandafter\def\csname ev@gemma/bio/CEU/0M/mmlu_rest\endcsname{70.7}
\expandafter\def\csname ev@gemma/bio/CEU/0M/wmdp\endcsname{75.1}
\expandafter\def\csname ev@gemma/bio/CEU-SFT/250M/arc_challenge\endcsname{71.8}
\expandafter\def\csname ev@gemma/bio/CEU-SFT/250M/gsm8k\endcsname{70.9}
\expandafter\def\csname ev@gemma/bio/CEU-SFT/250M/hellaswag\endcsname{83.5}
\expandafter\def\csname ev@gemma/bio/CEU-SFT/250M/mmlu\endcsname{74.5}
\expandafter\def\csname ev@gemma/bio/CEU-SFT/250M/mmlu_adj\endcsname{82.3}
\expandafter\def\csname ev@gemma/bio/CEU-SFT/250M/mmlu_rest\endcsname{73.1}
\expandafter\def\csname ev@gemma/bio/CEU-SFT/250M/wmdp\endcsname{78.3}
\expandafter\def\csname ev@gemma/bio/CEU-SFT/250M/ifeval\endcsname{26.2}
\expandafter\def\csname ev@gemma/bio/GA/0M/arc_challenge\endcsname{27.2}
\expandafter\def\csname ev@gemma/bio/GA/0M/gsm8k\endcsname{0.0}
\expandafter\def\csname ev@gemma/bio/GA/0M/hellaswag\endcsname{26.6}
\expandafter\def\csname ev@gemma/bio/GA/0M/mmlu\endcsname{22.9}
\expandafter\def\csname ev@gemma/bio/GA/0M/mmlu_adj\endcsname{22.6}
\expandafter\def\csname ev@gemma/bio/GA/0M/mmlu_rest\endcsname{22.9}
\expandafter\def\csname ev@gemma/bio/GA/0M/wmdp\endcsname{24.7}
\expandafter\def\csname ev@gemma/bio/GA-SFT/250M/arc_challenge\endcsname{26.4}
\expandafter\def\csname ev@gemma/bio/GA-SFT/250M/gsm8k\endcsname{0.0}
\expandafter\def\csname ev@gemma/bio/GA-SFT/250M/hellaswag\endcsname{26.3}
\expandafter\def\csname ev@gemma/bio/GA-SFT/250M/mmlu\endcsname{27.3}
\expandafter\def\csname ev@gemma/bio/GA-SFT/250M/mmlu_adj\endcsname{28.5}
\expandafter\def\csname ev@gemma/bio/GA-SFT/250M/mmlu_rest\endcsname{27.1}
\expandafter\def\csname ev@gemma/bio/GA-SFT/250M/wmdp\endcsname{24.0}
\expandafter\def\csname ev@gemma/bio/GA-SFT/250M/ifeval\endcsname{16.1}
\expandafter\def\csname ev@gemma/bio/GD/0M/arc_challenge\endcsname{65.5}
\expandafter\def\csname ev@gemma/bio/GD/0M/gsm8k\endcsname{40.5}
\expandafter\def\csname ev@gemma/bio/GD/0M/hellaswag\endcsname{79.2}
\expandafter\def\csname ev@gemma/bio/GD/0M/mmlu\endcsname{70.7}
\expandafter\def\csname ev@gemma/bio/GD/0M/mmlu_adj\endcsname{71.4}
\expandafter\def\csname ev@gemma/bio/GD/0M/mmlu_rest\endcsname{70.6}
\expandafter\def\csname ev@gemma/bio/GD/0M/wmdp\endcsname{28.6}
\expandafter\def\csname ev@gemma/bio/GD-SFT/250M/arc_challenge\endcsname{70.6}
\expandafter\def\csname ev@gemma/bio/GD-SFT/250M/gsm8k\endcsname{67.9}
\expandafter\def\csname ev@gemma/bio/GD-SFT/250M/hellaswag\endcsname{83.9}
\expandafter\def\csname ev@gemma/bio/GD-SFT/250M/mmlu\endcsname{74.8}
\expandafter\def\csname ev@gemma/bio/GD-SFT/250M/mmlu_adj\endcsname{81.3}
\expandafter\def\csname ev@gemma/bio/GD-SFT/250M/mmlu_rest\endcsname{73.6}
\expandafter\def\csname ev@gemma/bio/GD-SFT/250M/wmdp\endcsname{77.4}
\expandafter\def\csname ev@gemma/bio/GD-SFT/250M/ifeval\endcsname{27.5}
\expandafter\def\csname ev@gemma/bio/PDU/0M/arc_challenge\endcsname{71.9}
\expandafter\def\csname ev@gemma/bio/PDU/0M/gsm8k\endcsname{68.5}
\expandafter\def\csname ev@gemma/bio/PDU/0M/hellaswag\endcsname{84.4}
\expandafter\def\csname ev@gemma/bio/PDU/0M/mmlu\endcsname{73.3}
\expandafter\def\csname ev@gemma/bio/PDU/0M/mmlu_adj\endcsname{79.5}
\expandafter\def\csname ev@gemma/bio/PDU/0M/mmlu_rest\endcsname{72.2}
\expandafter\def\csname ev@gemma/bio/PDU/0M/wmdp\endcsname{69.4}
\expandafter\def\csname ev@gemma/bio/PDU-SFT/250M/arc_challenge\endcsname{72.0}
\expandafter\def\csname ev@gemma/bio/PDU-SFT/250M/gsm8k\endcsname{71.4}
\expandafter\def\csname ev@gemma/bio/PDU-SFT/250M/hellaswag\endcsname{84.8}
\expandafter\def\csname ev@gemma/bio/PDU-SFT/250M/mmlu\endcsname{74.7}
\expandafter\def\csname ev@gemma/bio/PDU-SFT/250M/mmlu_adj\endcsname{81.0}
\expandafter\def\csname ev@gemma/bio/PDU-SFT/250M/mmlu_rest\endcsname{73.6}
\expandafter\def\csname ev@gemma/bio/PDU-SFT/250M/wmdp\endcsname{78.5}
\expandafter\def\csname ev@gemma/bio/PDU-SFT/250M/ifeval\endcsname{26.2}
\expandafter\def\csname ev@gemma/bio/PureNPO/0M/arc_challenge\endcsname{68.4}
\expandafter\def\csname ev@gemma/bio/PureNPO/0M/gsm8k\endcsname{57.1}
\expandafter\def\csname ev@gemma/bio/PureNPO/0M/hellaswag\endcsname{84.1}
\expandafter\def\csname ev@gemma/bio/PureNPO/0M/mmlu\endcsname{69.2}
\expandafter\def\csname ev@gemma/bio/PureNPO/0M/mmlu_adj\endcsname{76.0}
\expandafter\def\csname ev@gemma/bio/PureNPO/0M/mmlu_rest\endcsname{68.0}
\expandafter\def\csname ev@gemma/bio/PureNPO/0M/wmdp\endcsname{65.0}
\expandafter\def\csname ev@gemma/bio/PureNPO-SFT/250M/arc_challenge\endcsname{69.6}
\expandafter\def\csname ev@gemma/bio/PureNPO-SFT/250M/gsm8k\endcsname{60.8}
\expandafter\def\csname ev@gemma/bio/PureNPO-SFT/250M/hellaswag\endcsname{83.0}
\expandafter\def\csname ev@gemma/bio/PureNPO-SFT/250M/mmlu\endcsname{70.3}
\expandafter\def\csname ev@gemma/bio/PureNPO-SFT/250M/mmlu_adj\endcsname{77.5}
\expandafter\def\csname ev@gemma/bio/PureNPO-SFT/250M/mmlu_rest\endcsname{69.1}
\expandafter\def\csname ev@gemma/bio/PureNPO-SFT/250M/wmdp\endcsname{67.8}
\expandafter\def\csname ev@gemma/bio/PureNPO-SFT/250M/ifeval\endcsname{22.4}
\expandafter\def\csname ev@gemma/bio/ExternalRMU/0M/arc_challenge\endcsname{68.0}
\expandafter\def\csname ev@gemma/bio/ExternalRMU/0M/gsm8k\endcsname{68.4}
\expandafter\def\csname ev@gemma/bio/ExternalRMU/0M/hellaswag\endcsname{78.6}
\expandafter\def\csname ev@gemma/bio/ExternalRMU/0M/mmlu\endcsname{70.0}
\expandafter\def\csname ev@gemma/bio/ExternalRMU/0M/mmlu_adj\endcsname{60.4}
\expandafter\def\csname ev@gemma/bio/ExternalRMU/0M/mmlu_rest\endcsname{71.7}
\expandafter\def\csname ev@gemma/bio/ExternalRMU/0M/wmdp\endcsname{29.2}
\expandafter\def\csname ev@gemma/bio/ExternalRMU-SFT/250M/arc_challenge\endcsname{69.9}
\expandafter\def\csname ev@gemma/bio/ExternalRMU-SFT/250M/gsm8k\endcsname{72.6}
\expandafter\def\csname ev@gemma/bio/ExternalRMU-SFT/250M/hellaswag\endcsname{81.7}
\expandafter\def\csname ev@gemma/bio/ExternalRMU-SFT/250M/mmlu\endcsname{73.2}
\expandafter\def\csname ev@gemma/bio/ExternalRMU-SFT/250M/mmlu_adj\endcsname{76.7}
\expandafter\def\csname ev@gemma/bio/ExternalRMU-SFT/250M/mmlu_rest\endcsname{72.6}
\expandafter\def\csname ev@gemma/bio/ExternalRMU-SFT/250M/wmdp\endcsname{60.8}
\expandafter\def\csname ev@gemma/bio/ExternalRMU-SFT/250M/ifeval\endcsname{27.7}
\expandafter\def\csname ev@gemma/bio/SatImp/0M/arc_challenge\endcsname{35.1}
\expandafter\def\csname ev@gemma/bio/SatImp/0M/gsm8k\endcsname{0.0}
\expandafter\def\csname ev@gemma/bio/SatImp/0M/hellaswag\endcsname{33.4}
\expandafter\def\csname ev@gemma/bio/SatImp/0M/mmlu\endcsname{57.1}
\expandafter\def\csname ev@gemma/bio/SatImp/0M/mmlu_adj\endcsname{60.4}
\expandafter\def\csname ev@gemma/bio/SatImp/0M/mmlu_rest\endcsname{56.5}
\expandafter\def\csname ev@gemma/bio/SatImp/0M/wmdp\endcsname{61.6}
\expandafter\def\csname ev@gemma/bio/SatImp-SFT/250M/arc_challenge\endcsname{72.4}
\expandafter\def\csname ev@gemma/bio/SatImp-SFT/250M/gsm8k\endcsname{71.3}
\expandafter\def\csname ev@gemma/bio/SatImp-SFT/250M/hellaswag\endcsname{84.1}
\expandafter\def\csname ev@gemma/bio/SatImp-SFT/250M/mmlu\endcsname{74.2}
\expandafter\def\csname ev@gemma/bio/SatImp-SFT/250M/mmlu_adj\endcsname{81.0}
\expandafter\def\csname ev@gemma/bio/SatImp-SFT/250M/mmlu_rest\endcsname{73.0}
\expandafter\def\csname ev@gemma/bio/SatImp-SFT/250M/wmdp\endcsname{78.6}
\expandafter\def\csname ev@gemma/bio/SatImp-SFT/250M/ifeval\endcsname{25.7}
\expandafter\def\csname ev@gemma/bio/SimNPO/0M/arc_challenge\endcsname{69.2}
\expandafter\def\csname ev@gemma/bio/SimNPO/0M/gsm8k\endcsname{58.5}
\expandafter\def\csname ev@gemma/bio/SimNPO/0M/hellaswag\endcsname{84.1}
\expandafter\def\csname ev@gemma/bio/SimNPO/0M/mmlu\endcsname{68.4}
\expandafter\def\csname ev@gemma/bio/SimNPO/0M/mmlu_adj\endcsname{74.1}
\expandafter\def\csname ev@gemma/bio/SimNPO/0M/mmlu_rest\endcsname{67.4}
\expandafter\def\csname ev@gemma/bio/SimNPO/0M/wmdp\endcsname{64.3}
\expandafter\def\csname ev@gemma/bio/SimNPO-SFT/250M/arc_challenge\endcsname{68.8}
\expandafter\def\csname ev@gemma/bio/SimNPO-SFT/250M/gsm8k\endcsname{61.9}
\expandafter\def\csname ev@gemma/bio/SimNPO-SFT/250M/hellaswag\endcsname{83.0}
\expandafter\def\csname ev@gemma/bio/SimNPO-SFT/250M/mmlu\endcsname{69.0}
\expandafter\def\csname ev@gemma/bio/SimNPO-SFT/250M/mmlu_adj\endcsname{75.4}
\expandafter\def\csname ev@gemma/bio/SimNPO-SFT/250M/mmlu_rest\endcsname{67.8}
\expandafter\def\csname ev@gemma/bio/SimNPO-SFT/250M/wmdp\endcsname{66.9}
\expandafter\def\csname ev@gemma/bio/SimNPO-SFT/250M/ifeval\endcsname{24.4}
\expandafter\def\csname ev@gemma/bio/TRACE/0M/arc_challenge\endcsname{57.0}
\expandafter\def\csname ev@gemma/bio/TRACE/0M/gsm8k\endcsname{69.1}
\expandafter\def\csname ev@gemma/bio/TRACE/0M/hellaswag\endcsname{71.8}
\expandafter\def\csname ev@gemma/bio/TRACE/0M/mmlu\endcsname{74.0}
\expandafter\def\csname ev@gemma/bio/TRACE/0M/mmlu_adj\endcsname{79.6}
\expandafter\def\csname ev@gemma/bio/TRACE/0M/mmlu_rest\endcsname{73.0}
\expandafter\def\csname ev@gemma/bio/TRACE/0M/wmdp\endcsname{52.7}
\expandafter\def\csname ev@gemma/bio/TRACE/0M/ifeval\endcsname{26.6}
\expandafter\def\csname ev@gemma/bio/TRACE-SFT/250M/arc_challenge\endcsname{69.4}
\expandafter\def\csname ev@gemma/bio/TRACE-SFT/250M/gsm8k\endcsname{72.8}
\expandafter\def\csname ev@gemma/bio/TRACE-SFT/250M/hellaswag\endcsname{80.4}
\expandafter\def\csname ev@gemma/bio/TRACE-SFT/250M/mmlu\endcsname{74.6}
\expandafter\def\csname ev@gemma/bio/TRACE-SFT/250M/mmlu_adj\endcsname{80.5}
\expandafter\def\csname ev@gemma/bio/TRACE-SFT/250M/mmlu_rest\endcsname{73.6}
\expandafter\def\csname ev@gemma/bio/TRACE-SFT/250M/wmdp\endcsname{76.0}
\expandafter\def\csname ev@gemma/bio/TRACE-SFT/250M/ifeval\endcsname{27.0}
\expandafter\def\csname ev@gemma/bio/WGA/0M/arc_challenge\endcsname{35.5}
\expandafter\def\csname ev@gemma/bio/WGA/0M/gsm8k\endcsname{0.0}
\expandafter\def\csname ev@gemma/bio/WGA/0M/hellaswag\endcsname{34.3}
\expandafter\def\csname ev@gemma/bio/WGA/0M/mmlu\endcsname{61.0}
\expandafter\def\csname ev@gemma/bio/WGA/0M/mmlu_adj\endcsname{66.0}
\expandafter\def\csname ev@gemma/bio/WGA/0M/mmlu_rest\endcsname{60.1}
\expandafter\def\csname ev@gemma/bio/WGA/0M/wmdp\endcsname{66.3}
\expandafter\def\csname ev@gemma/bio/WGA-SFT/250M/arc_challenge\endcsname{72.7}
\expandafter\def\csname ev@gemma/bio/WGA-SFT/250M/gsm8k\endcsname{72.3}
\expandafter\def\csname ev@gemma/bio/WGA-SFT/250M/hellaswag\endcsname{84.2}
\expandafter\def\csname ev@gemma/bio/WGA-SFT/250M/mmlu\endcsname{74.3}
\expandafter\def\csname ev@gemma/bio/WGA-SFT/250M/mmlu_adj\endcsname{81.7}
\expandafter\def\csname ev@gemma/bio/WGA-SFT/250M/mmlu_rest\endcsname{73.0}
\expandafter\def\csname ev@gemma/bio/WGA-SFT/250M/wmdp\endcsname{78.8}
\expandafter\def\csname ev@gemma/bio/WGA-SFT/250M/ifeval\endcsname{26.4}
\expandafter\def\csname ev@intern/cyber/B/0M/arc_challenge\endcsname{74.3}
\expandafter\def\csname ev@intern/cyber/B/0M/gsm8k\endcsname{87.1}
\expandafter\def\csname ev@intern/cyber/B/0M/hellaswag\endcsname{85.8}
\expandafter\def\csname ev@intern/cyber/B/0M/mmlu\endcsname{85.0}
\expandafter\def\csname ev@intern/cyber/B/0M/mmlu_adj\endcsname{87.4}
\expandafter\def\csname ev@intern/cyber/B/0M/mmlu_rest\endcsname{85.0}
\expandafter\def\csname ev@intern/cyber/B/0M/wmdp\endcsname{58.9}
\expandafter\def\csname ev@intern/cyber/S_std(B)/250M/arc_challenge\endcsname{74.4}
\expandafter\def\csname ev@intern/cyber/S_std(B)/250M/gsm8k\endcsname{86.9}
\expandafter\def\csname ev@intern/cyber/S_std(B)/250M/hellaswag\endcsname{85.8}
\expandafter\def\csname ev@intern/cyber/S_std(B)/250M/mmlu\endcsname{84.8}
\expandafter\def\csname ev@intern/cyber/S_std(B)/250M/mmlu_adj\endcsname{87.6}
\expandafter\def\csname ev@intern/cyber/S_std(B)/250M/mmlu_rest\endcsname{84.7}
\expandafter\def\csname ev@intern/cyber/S_std(B)/250M/wmdp\endcsname{59.1}
\expandafter\def\csname ev@intern/cyber/S_std(B)/250M/ifeval\endcsname{78.9}
\expandafter\def\csname ev@intern/cyber/F0/0M/arc_challenge\endcsname{74.5}
\expandafter\def\csname ev@intern/cyber/F0/0M/gsm8k\endcsname{88.0}
\expandafter\def\csname ev@intern/cyber/F0/0M/hellaswag\endcsname{85.7}
\expandafter\def\csname ev@intern/cyber/F0/0M/mmlu\endcsname{85.0}
\expandafter\def\csname ev@intern/cyber/F0/0M/mmlu_adj\endcsname{86.4}
\expandafter\def\csname ev@intern/cyber/F0/0M/mmlu_rest\endcsname{85.0}
\expandafter\def\csname ev@intern/cyber/F0/0M/wmdp\endcsname{57.5}
\expandafter\def\csname ev@intern/cyber/A0/0M/arc_challenge\endcsname{74.3}
\expandafter\def\csname ev@intern/cyber/A0/0M/gsm8k\endcsname{87.4}
\expandafter\def\csname ev@intern/cyber/A0/0M/hellaswag\endcsname{85.4}
\expandafter\def\csname ev@intern/cyber/A0/0M/mmlu\endcsname{84.6}
\expandafter\def\csname ev@intern/cyber/A0/0M/mmlu_adj\endcsname{81.3}
\expandafter\def\csname ev@intern/cyber/A0/0M/mmlu_rest\endcsname{84.7}
\expandafter\def\csname ev@intern/cyber/A0/0M/wmdp\endcsname{49.8}
\expandafter\def\csname ev@intern/cyber/F1/0M/arc_challenge\endcsname{73.7}
\expandafter\def\csname ev@intern/cyber/F1/0M/gsm8k\endcsname{84.6}
\expandafter\def\csname ev@intern/cyber/F1/0M/hellaswag\endcsname{83.7}
\expandafter\def\csname ev@intern/cyber/F1/0M/mmlu\endcsname{84.5}
\expandafter\def\csname ev@intern/cyber/F1/0M/mmlu_adj\endcsname{85.4}
\expandafter\def\csname ev@intern/cyber/F1/0M/mmlu_rest\endcsname{84.4}
\expandafter\def\csname ev@intern/cyber/F1/0M/wmdp\endcsname{57.9}
\expandafter\def\csname ev@intern/cyber/A1/0M/arc_challenge\endcsname{73.5}
\expandafter\def\csname ev@intern/cyber/A1/0M/gsm8k\endcsname{84.5}
\expandafter\def\csname ev@intern/cyber/A1/0M/hellaswag\endcsname{83.3}
\expandafter\def\csname ev@intern/cyber/A1/0M/mmlu\endcsname{84.2}
\expandafter\def\csname ev@intern/cyber/A1/0M/mmlu_adj\endcsname{78.4}
\expandafter\def\csname ev@intern/cyber/A1/0M/mmlu_rest\endcsname{84.4}
\expandafter\def\csname ev@intern/cyber/A1/0M/wmdp\endcsname{45.1}
\expandafter\def\csname ev@intern/cyber/A2/0M/arc_challenge\endcsname{73.6}
\expandafter\def\csname ev@intern/cyber/A2/0M/gsm8k\endcsname{83.9}
\expandafter\def\csname ev@intern/cyber/A2/0M/hellaswag\endcsname{81.7}
\expandafter\def\csname ev@intern/cyber/A2/0M/mmlu\endcsname{84.1}
\expandafter\def\csname ev@intern/cyber/A2/0M/mmlu_adj\endcsname{72.1}
\expandafter\def\csname ev@intern/cyber/A2/0M/mmlu_rest\endcsname{84.4}
\expandafter\def\csname ev@intern/cyber/A2/0M/wmdp\endcsname{26.3}
\expandafter\def\csname ev@intern/cyber/F2/0M/arc_challenge\endcsname{73.3}
\expandafter\def\csname ev@intern/cyber/F2/0M/gsm8k\endcsname{84.0}
\expandafter\def\csname ev@intern/cyber/F2/0M/hellaswag\endcsname{82.3}
\expandafter\def\csname ev@intern/cyber/F2/0M/mmlu\endcsname{84.4}
\expandafter\def\csname ev@intern/cyber/F2/0M/mmlu_adj\endcsname{85.7}
\expandafter\def\csname ev@intern/cyber/F2/0M/mmlu_rest\endcsname{84.4}
\expandafter\def\csname ev@intern/cyber/F2/0M/wmdp\endcsname{52.8}
\expandafter\def\csname ev@intern/cyber/F3/250M/arc_challenge\endcsname{73.7}
\expandafter\def\csname ev@intern/cyber/F3/250M/gsm8k\endcsname{86.5}
\expandafter\def\csname ev@intern/cyber/F3/250M/hellaswag\endcsname{84.7}
\expandafter\def\csname ev@intern/cyber/F3/250M/mmlu\endcsname{84.0}
\expandafter\def\csname ev@intern/cyber/F3/250M/mmlu_adj\endcsname{85.2}
\expandafter\def\csname ev@intern/cyber/F3/250M/mmlu_rest\endcsname{84.0}
\expandafter\def\csname ev@intern/cyber/F3/250M/wmdp\endcsname{55.8}
\expandafter\def\csname ev@intern/cyber/F3/250M/ifeval\endcsname{79.1}
\expandafter\def\csname ev@intern/cyber/A3/250M/arc_challenge\endcsname{73.7}
\expandafter\def\csname ev@intern/cyber/A3/250M/gsm8k\endcsname{85.6}
\expandafter\def\csname ev@intern/cyber/A3/250M/hellaswag\endcsname{84.3}
\expandafter\def\csname ev@intern/cyber/A3/250M/mmlu\endcsname{84.0}
\expandafter\def\csname ev@intern/cyber/A3/250M/mmlu_adj\endcsname{77.4}
\expandafter\def\csname ev@intern/cyber/A3/250M/mmlu_rest\endcsname{84.2}
\expandafter\def\csname ev@intern/cyber/A3/250M/wmdp\endcsname{29.8}
\expandafter\def\csname ev@intern/cyber/A3/250M/ifeval\endcsname{76.2}
\expandafter\def\csname ev@intern/cyber/R/0M/arc_challenge\endcsname{74.1}
\expandafter\def\csname ev@intern/cyber/R/0M/gsm8k\endcsname{86.0}
\expandafter\def\csname ev@intern/cyber/R/0M/hellaswag\endcsname{85.5}
\expandafter\def\csname ev@intern/cyber/R/0M/mmlu\endcsname{84.9}
\expandafter\def\csname ev@intern/cyber/R/0M/mmlu_adj\endcsname{83.2}
\expandafter\def\csname ev@intern/cyber/R/0M/mmlu_rest\endcsname{84.9}
\expandafter\def\csname ev@intern/cyber/R/0M/wmdp\endcsname{30.9}
\expandafter\def\csname ev@intern/cyber/R-SFT/250M/arc_challenge\endcsname{74.2}
\expandafter\def\csname ev@intern/cyber/R-SFT/250M/gsm8k\endcsname{86.0}
\expandafter\def\csname ev@intern/cyber/R-SFT/250M/hellaswag\endcsname{85.7}
\expandafter\def\csname ev@intern/cyber/R-SFT/250M/mmlu\endcsname{84.2}
\expandafter\def\csname ev@intern/cyber/R-SFT/250M/mmlu_adj\endcsname{85.9}
\expandafter\def\csname ev@intern/cyber/R-SFT/250M/mmlu_rest\endcsname{84.1}
\expandafter\def\csname ev@intern/cyber/R-SFT/250M/wmdp\endcsname{44.7}
\expandafter\def\csname ev@intern/cyber/R-SFT/250M/ifeval\endcsname{78.0}
\expandafter\def\csname ev@intern/cyber/A3-direct/250M/arc_challenge\endcsname{73.9}
\expandafter\def\csname ev@intern/cyber/A3-direct/250M/gsm8k\endcsname{86.0}
\expandafter\def\csname ev@intern/cyber/A3-direct/250M/hellaswag\endcsname{84.6}
\expandafter\def\csname ev@intern/cyber/A3-direct/250M/mmlu\endcsname{84.2}
\expandafter\def\csname ev@intern/cyber/A3-direct/250M/mmlu_adj\endcsname{79.8}
\expandafter\def\csname ev@intern/cyber/A3-direct/250M/mmlu_rest\endcsname{84.3}
\expandafter\def\csname ev@intern/cyber/A3-direct/250M/wmdp\endcsname{35.9}
\expandafter\def\csname ev@intern/cyber/A3-direct/250M/ifeval\endcsname{78.0}
\expandafter\def\csname ev@intern/cyber/CEU/0M/arc_challenge\endcsname{34.1}
\expandafter\def\csname ev@intern/cyber/CEU/0M/gsm8k\endcsname{0.0}
\expandafter\def\csname ev@intern/cyber/CEU/0M/hellaswag\endcsname{36.0}
\expandafter\def\csname ev@intern/cyber/CEU/0M/mmlu\endcsname{77.8}
\expandafter\def\csname ev@intern/cyber/CEU/0M/mmlu_adj\endcsname{79.8}
\expandafter\def\csname ev@intern/cyber/CEU/0M/mmlu_rest\endcsname{77.7}
\expandafter\def\csname ev@intern/cyber/CEU/0M/wmdp\endcsname{44.8}
\expandafter\def\csname ev@intern/cyber/CEU-SFT/250M/arc_challenge\endcsname{42.4}
\expandafter\def\csname ev@intern/cyber/CEU-SFT/250M/gsm8k\endcsname{0.0}
\expandafter\def\csname ev@intern/cyber/CEU-SFT/250M/hellaswag\endcsname{52.6}
\expandafter\def\csname ev@intern/cyber/CEU-SFT/250M/mmlu\endcsname{83.4}
\expandafter\def\csname ev@intern/cyber/CEU-SFT/250M/mmlu_adj\endcsname{85.4}
\expandafter\def\csname ev@intern/cyber/CEU-SFT/250M/mmlu_rest\endcsname{83.4}
\expandafter\def\csname ev@intern/cyber/CEU-SFT/250M/wmdp\endcsname{52.0}
\expandafter\def\csname ev@intern/cyber/CEU-SFT/250M/ifeval\endcsname{78.9}
\expandafter\def\csname ev@intern/cyber/GA/0M/arc_challenge\endcsname{26.3}
\expandafter\def\csname ev@intern/cyber/GA/0M/gsm8k\endcsname{0.0}
\expandafter\def\csname ev@intern/cyber/GA/0M/hellaswag\endcsname{26.7}
\expandafter\def\csname ev@intern/cyber/GA/0M/mmlu\endcsname{35.5}
\expandafter\def\csname ev@intern/cyber/GA/0M/mmlu_adj\endcsname{28.6}
\expandafter\def\csname ev@intern/cyber/GA/0M/mmlu_rest\endcsname{35.7}
\expandafter\def\csname ev@intern/cyber/GA/0M/wmdp\endcsname{24.6}
\expandafter\def\csname ev@intern/cyber/GA-SFT/250M/arc_challenge\endcsname{61.0}
\expandafter\def\csname ev@intern/cyber/GA-SFT/250M/gsm8k\endcsname{79.4}
\expandafter\def\csname ev@intern/cyber/GA-SFT/250M/hellaswag\endcsname{66.0}
\expandafter\def\csname ev@intern/cyber/GA-SFT/250M/mmlu\endcsname{81.1}
\expandafter\def\csname ev@intern/cyber/GA-SFT/250M/mmlu_adj\endcsname{82.3}
\expandafter\def\csname ev@intern/cyber/GA-SFT/250M/mmlu_rest\endcsname{81.0}
\expandafter\def\csname ev@intern/cyber/GA-SFT/250M/wmdp\endcsname{47.6}
\expandafter\def\csname ev@intern/cyber/GA-SFT/250M/ifeval\endcsname{81.9}
\expandafter\def\csname ev@intern/cyber/GD/0M/arc_challenge\endcsname{74.3}
\expandafter\def\csname ev@intern/cyber/GD/0M/gsm8k\endcsname{0.4}
\expandafter\def\csname ev@intern/cyber/GD/0M/hellaswag\endcsname{84.2}
\expandafter\def\csname ev@intern/cyber/GD/0M/mmlu\endcsname{81.8}
\expandafter\def\csname ev@intern/cyber/GD/0M/mmlu_adj\endcsname{73.5}
\expandafter\def\csname ev@intern/cyber/GD/0M/mmlu_rest\endcsname{82.1}
\expandafter\def\csname ev@intern/cyber/GD/0M/wmdp\endcsname{31.0}
\expandafter\def\csname ev@intern/cyber/GD-SFT/250M/arc_challenge\endcsname{74.1}
\expandafter\def\csname ev@intern/cyber/GD-SFT/250M/gsm8k\endcsname{86.2}
\expandafter\def\csname ev@intern/cyber/GD-SFT/250M/hellaswag\endcsname{84.6}
\expandafter\def\csname ev@intern/cyber/GD-SFT/250M/mmlu\endcsname{83.2}
\expandafter\def\csname ev@intern/cyber/GD-SFT/250M/mmlu_adj\endcsname{85.7}
\expandafter\def\csname ev@intern/cyber/GD-SFT/250M/mmlu_rest\endcsname{83.2}
\expandafter\def\csname ev@intern/cyber/GD-SFT/250M/wmdp\endcsname{47.4}
\expandafter\def\csname ev@intern/cyber/GD-SFT/250M/ifeval\endcsname{78.4}
\expandafter\def\csname ev@intern/cyber/PDU/0M/arc_challenge\endcsname{74.5}
\expandafter\def\csname ev@intern/cyber/PDU/0M/gsm8k\endcsname{85.8}
\expandafter\def\csname ev@intern/cyber/PDU/0M/hellaswag\endcsname{86.4}
\expandafter\def\csname ev@intern/cyber/PDU/0M/mmlu\endcsname{84.7}
\expandafter\def\csname ev@intern/cyber/PDU/0M/mmlu_adj\endcsname{75.0}
\expandafter\def\csname ev@intern/cyber/PDU/0M/mmlu_rest\endcsname{85.0}
\expandafter\def\csname ev@intern/cyber/PDU/0M/wmdp\endcsname{27.8}
\expandafter\def\csname ev@intern/cyber/PDU-SFT/250M/arc_challenge\endcsname{74.6}
\expandafter\def\csname ev@intern/cyber/PDU-SFT/250M/gsm8k\endcsname{87.3}
\expandafter\def\csname ev@intern/cyber/PDU-SFT/250M/hellaswag\endcsname{86.1}
\expandafter\def\csname ev@intern/cyber/PDU-SFT/250M/mmlu\endcsname{84.8}
\expandafter\def\csname ev@intern/cyber/PDU-SFT/250M/mmlu_adj\endcsname{84.2}
\expandafter\def\csname ev@intern/cyber/PDU-SFT/250M/mmlu_rest\endcsname{84.9}
\expandafter\def\csname ev@intern/cyber/PDU-SFT/250M/wmdp\endcsname{40.1}
\expandafter\def\csname ev@intern/cyber/PDU-SFT/250M/ifeval\endcsname{77.8}
\expandafter\def\csname ev@intern/cyber/PureNPO/0M/arc_challenge\endcsname{72.3}
\expandafter\def\csname ev@intern/cyber/PureNPO/0M/gsm8k\endcsname{87.6}
\expandafter\def\csname ev@intern/cyber/PureNPO/0M/hellaswag\endcsname{86.5}
\expandafter\def\csname ev@intern/cyber/PureNPO/0M/mmlu\endcsname{83.2}
\expandafter\def\csname ev@intern/cyber/PureNPO/0M/mmlu_adj\endcsname{82.5}
\expandafter\def\csname ev@intern/cyber/PureNPO/0M/mmlu_rest\endcsname{83.2}
\expandafter\def\csname ev@intern/cyber/PureNPO/0M/wmdp\endcsname{41.4}
\expandafter\def\csname ev@intern/cyber/PureNPO-SFT/250M/arc_challenge\endcsname{73.0}
\expandafter\def\csname ev@intern/cyber/PureNPO-SFT/250M/gsm8k\endcsname{89.5}
\expandafter\def\csname ev@intern/cyber/PureNPO-SFT/250M/hellaswag\endcsname{85.8}
\expandafter\def\csname ev@intern/cyber/PureNPO-SFT/250M/mmlu\endcsname{84.5}
\expandafter\def\csname ev@intern/cyber/PureNPO-SFT/250M/mmlu_adj\endcsname{86.9}
\expandafter\def\csname ev@intern/cyber/PureNPO-SFT/250M/mmlu_rest\endcsname{84.4}
\expandafter\def\csname ev@intern/cyber/PureNPO-SFT/250M/wmdp\endcsname{50.6}
\expandafter\def\csname ev@intern/cyber/PureNPO-SFT/250M/ifeval\endcsname{75.8}
\expandafter\def\csname ev@intern/cyber/ExternalRMU/0M/arc_challenge\endcsname{74.3}
\expandafter\def\csname ev@intern/cyber/ExternalRMU/0M/gsm8k\endcsname{85.0}
\expandafter\def\csname ev@intern/cyber/ExternalRMU/0M/hellaswag\endcsname{83.4}
\expandafter\def\csname ev@intern/cyber/ExternalRMU/0M/mmlu\endcsname{84.8}
\expandafter\def\csname ev@intern/cyber/ExternalRMU/0M/mmlu_adj\endcsname{84.0}
\expandafter\def\csname ev@intern/cyber/ExternalRMU/0M/mmlu_rest\endcsname{84.8}
\expandafter\def\csname ev@intern/cyber/ExternalRMU/0M/wmdp\endcsname{35.7}
\expandafter\def\csname ev@intern/cyber/ExternalRMU/0M/ifeval\endcsname{81.0}
\expandafter\def\csname ev@intern/cyber/ExternalRMU-SFT/250M/arc_challenge\endcsname{74.3}
\expandafter\def\csname ev@intern/cyber/ExternalRMU-SFT/250M/gsm8k\endcsname{86.4}
\expandafter\def\csname ev@intern/cyber/ExternalRMU-SFT/250M/hellaswag\endcsname{84.9}
\expandafter\def\csname ev@intern/cyber/ExternalRMU-SFT/250M/mmlu\endcsname{84.3}
\expandafter\def\csname ev@intern/cyber/ExternalRMU-SFT/250M/mmlu_adj\endcsname{86.4}
\expandafter\def\csname ev@intern/cyber/ExternalRMU-SFT/250M/mmlu_rest\endcsname{84.2}
\expandafter\def\csname ev@intern/cyber/ExternalRMU-SFT/250M/wmdp\endcsname{51.2}
\expandafter\def\csname ev@intern/cyber/ExternalRMU-SFT/250M/ifeval\endcsname{77.5}
\expandafter\def\csname ev@intern/cyber/SatImp/0M/arc_challenge\endcsname{48.5}
\expandafter\def\csname ev@intern/cyber/SatImp/0M/gsm8k\endcsname{0.0}
\expandafter\def\csname ev@intern/cyber/SatImp/0M/hellaswag\endcsname{54.2}
\expandafter\def\csname ev@intern/cyber/SatImp/0M/mmlu\endcsname{78.8}
\expandafter\def\csname ev@intern/cyber/SatImp/0M/mmlu_adj\endcsname{73.8}
\expandafter\def\csname ev@intern/cyber/SatImp/0M/mmlu_rest\endcsname{78.9}
\expandafter\def\csname ev@intern/cyber/SatImp/0M/wmdp\endcsname{37.6}
\expandafter\def\csname ev@intern/cyber/SatImp-SFT/250M/arc_challenge\endcsname{72.3}
\expandafter\def\csname ev@intern/cyber/SatImp-SFT/250M/gsm8k\endcsname{86.1}
\expandafter\def\csname ev@intern/cyber/SatImp-SFT/250M/hellaswag\endcsname{83.0}
\expandafter\def\csname ev@intern/cyber/SatImp-SFT/250M/mmlu\endcsname{83.9}
\expandafter\def\csname ev@intern/cyber/SatImp-SFT/250M/mmlu_adj\endcsname{86.4}
\expandafter\def\csname ev@intern/cyber/SatImp-SFT/250M/mmlu_rest\endcsname{83.8}
\expandafter\def\csname ev@intern/cyber/SatImp-SFT/250M/wmdp\endcsname{50.9}
\expandafter\def\csname ev@intern/cyber/SatImp-SFT/250M/ifeval\endcsname{80.8}
\expandafter\def\csname ev@intern/cyber/SimNPO/0M/arc_challenge\endcsname{72.7}
\expandafter\def\csname ev@intern/cyber/SimNPO/0M/gsm8k\endcsname{86.0}
\expandafter\def\csname ev@intern/cyber/SimNPO/0M/hellaswag\endcsname{82.4}
\expandafter\def\csname ev@intern/cyber/SimNPO/0M/mmlu\endcsname{81.6}
\expandafter\def\csname ev@intern/cyber/SimNPO/0M/mmlu_adj\endcsname{81.5}
\expandafter\def\csname ev@intern/cyber/SimNPO/0M/mmlu_rest\endcsname{81.6}
\expandafter\def\csname ev@intern/cyber/SimNPO/0M/wmdp\endcsname{47.5}
\expandafter\def\csname ev@intern/cyber/SimNPO-SFT/250M/arc_challenge\endcsname{73.2}
\expandafter\def\csname ev@intern/cyber/SimNPO-SFT/250M/gsm8k\endcsname{87.0}
\expandafter\def\csname ev@intern/cyber/SimNPO-SFT/250M/hellaswag\endcsname{82.9}
\expandafter\def\csname ev@intern/cyber/SimNPO-SFT/250M/mmlu\endcsname{82.8}
\expandafter\def\csname ev@intern/cyber/SimNPO-SFT/250M/mmlu_adj\endcsname{83.7}
\expandafter\def\csname ev@intern/cyber/SimNPO-SFT/250M/mmlu_rest\endcsname{82.8}
\expandafter\def\csname ev@intern/cyber/SimNPO-SFT/250M/wmdp\endcsname{52.1}
\expandafter\def\csname ev@intern/cyber/SimNPO-SFT/250M/ifeval\endcsname{75.8}
\expandafter\def\csname ev@intern/cyber/TRACE/0M/arc_challenge\endcsname{70.0}
\expandafter\def\csname ev@intern/cyber/TRACE/0M/gsm8k\endcsname{85.8}
\expandafter\def\csname ev@intern/cyber/TRACE/0M/hellaswag\endcsname{82.8}
\expandafter\def\csname ev@intern/cyber/TRACE/0M/mmlu\endcsname{83.6}
\expandafter\def\csname ev@intern/cyber/TRACE/0M/mmlu_adj\endcsname{77.4}
\expandafter\def\csname ev@intern/cyber/TRACE/0M/mmlu_rest\endcsname{83.8}
\expandafter\def\csname ev@intern/cyber/TRACE/0M/wmdp\endcsname{42.4}
\expandafter\def\csname ev@intern/cyber/TRACE-SFT/250M/arc_challenge\endcsname{74.0}
\expandafter\def\csname ev@intern/cyber/TRACE-SFT/250M/gsm8k\endcsname{86.4}
\expandafter\def\csname ev@intern/cyber/TRACE-SFT/250M/hellaswag\endcsname{85.7}
\expandafter\def\csname ev@intern/cyber/TRACE-SFT/250M/mmlu\endcsname{84.1}
\expandafter\def\csname ev@intern/cyber/TRACE-SFT/250M/mmlu_adj\endcsname{83.7}
\expandafter\def\csname ev@intern/cyber/TRACE-SFT/250M/mmlu_rest\endcsname{84.1}
\expandafter\def\csname ev@intern/cyber/TRACE-SFT/250M/wmdp\endcsname{45.1}
\expandafter\def\csname ev@intern/cyber/TRACE-SFT/250M/ifeval\endcsname{77.5}
\expandafter\def\csname ev@intern/cyber/WGA/0M/arc_challenge\endcsname{48.4}
\expandafter\def\csname ev@intern/cyber/WGA/0M/gsm8k\endcsname{0.0}
\expandafter\def\csname ev@intern/cyber/WGA/0M/hellaswag\endcsname{53.4}
\expandafter\def\csname ev@intern/cyber/WGA/0M/mmlu\endcsname{78.7}
\expandafter\def\csname ev@intern/cyber/WGA/0M/mmlu_adj\endcsname{74.3}
\expandafter\def\csname ev@intern/cyber/WGA/0M/mmlu_rest\endcsname{78.8}
\expandafter\def\csname ev@intern/cyber/WGA/0M/wmdp\endcsname{37.6}
\expandafter\def\csname ev@intern/cyber/WGA-SFT/250M/arc_challenge\endcsname{72.1}
\expandafter\def\csname ev@intern/cyber/WGA-SFT/250M/gsm8k\endcsname{86.2}
\expandafter\def\csname ev@intern/cyber/WGA-SFT/250M/hellaswag\endcsname{82.7}
\expandafter\def\csname ev@intern/cyber/WGA-SFT/250M/mmlu\endcsname{83.7}
\expandafter\def\csname ev@intern/cyber/WGA-SFT/250M/mmlu_adj\endcsname{85.2}
\expandafter\def\csname ev@intern/cyber/WGA-SFT/250M/mmlu_rest\endcsname{83.7}
\expandafter\def\csname ev@intern/cyber/WGA-SFT/250M/wmdp\endcsname{51.3}
\expandafter\def\csname ev@intern/cyber/WGA-SFT/250M/ifeval\endcsname{82.6}
\expandafter\def\csname ev@qwen/cyber/B/0M/arc_challenge\endcsname{73.5}
\expandafter\def\csname ev@qwen/cyber/B/0M/gsm8k\endcsname{85.8}
\expandafter\def\csname ev@qwen/cyber/B/0M/hellaswag\endcsname{85.6}
\expandafter\def\csname ev@qwen/cyber/B/0M/mmlu\endcsname{82.4}
\expandafter\def\csname ev@qwen/cyber/B/0M/mmlu_adj\endcsname{81.8}
\expandafter\def\csname ev@qwen/cyber/B/0M/mmlu_rest\endcsname{82.5}
\expandafter\def\csname ev@qwen/cyber/B/0M/wmdp\endcsname{57.6}
\expandafter\def\csname ev@qwen/cyber/S_std(B)/250M/arc_challenge\endcsname{73.6}
\expandafter\def\csname ev@qwen/cyber/S_std(B)/250M/gsm8k\endcsname{85.1}
\expandafter\def\csname ev@qwen/cyber/S_std(B)/250M/hellaswag\endcsname{85.4}
\expandafter\def\csname ev@qwen/cyber/S_std(B)/250M/mmlu\endcsname{82.2}
\expandafter\def\csname ev@qwen/cyber/S_std(B)/250M/mmlu_adj\endcsname{80.1}
\expandafter\def\csname ev@qwen/cyber/S_std(B)/250M/mmlu_rest\endcsname{82.3}
\expandafter\def\csname ev@qwen/cyber/S_std(B)/250M/wmdp\endcsname{55.3}
\expandafter\def\csname ev@qwen/cyber/S_std(B)/250M/ifeval\endcsname{43.1}
\expandafter\def\csname ev@qwen/cyber/F0/0M/arc_challenge\endcsname{73.3}
\expandafter\def\csname ev@qwen/cyber/F0/0M/gsm8k\endcsname{85.4}
\expandafter\def\csname ev@qwen/cyber/F0/0M/hellaswag\endcsname{85.5}
\expandafter\def\csname ev@qwen/cyber/F0/0M/mmlu\endcsname{82.3}
\expandafter\def\csname ev@qwen/cyber/F0/0M/mmlu_adj\endcsname{80.3}
\expandafter\def\csname ev@qwen/cyber/F0/0M/mmlu_rest\endcsname{82.4}
\expandafter\def\csname ev@qwen/cyber/F0/0M/wmdp\endcsname{56.9}
\expandafter\def\csname ev@qwen/cyber/A0/0M/arc_challenge\endcsname{73.7}
\expandafter\def\csname ev@qwen/cyber/A0/0M/gsm8k\endcsname{85.2}
\expandafter\def\csname ev@qwen/cyber/A0/0M/hellaswag\endcsname{85.3}
\expandafter\def\csname ev@qwen/cyber/A0/0M/mmlu\endcsname{81.8}
\expandafter\def\csname ev@qwen/cyber/A0/0M/mmlu_adj\endcsname{76.0}
\expandafter\def\csname ev@qwen/cyber/A0/0M/mmlu_rest\endcsname{81.9}
\expandafter\def\csname ev@qwen/cyber/A0/0M/wmdp\endcsname{48.9}
\expandafter\def\csname ev@qwen/cyber/F1/0M/arc_challenge\endcsname{71.8}
\expandafter\def\csname ev@qwen/cyber/F1/0M/gsm8k\endcsname{85.5}
\expandafter\def\csname ev@qwen/cyber/F1/0M/hellaswag\endcsname{83.6}
\expandafter\def\csname ev@qwen/cyber/F1/0M/mmlu\endcsname{81.4}
\expandafter\def\csname ev@qwen/cyber/F1/0M/mmlu_adj\endcsname{77.2}
\expandafter\def\csname ev@qwen/cyber/F1/0M/mmlu_rest\endcsname{81.5}
\expandafter\def\csname ev@qwen/cyber/F1/0M/wmdp\endcsname{57.6}
\expandafter\def\csname ev@qwen/cyber/A1/0M/arc_challenge\endcsname{72.8}
\expandafter\def\csname ev@qwen/cyber/A1/0M/gsm8k\endcsname{85.0}
\expandafter\def\csname ev@qwen/cyber/A1/0M/hellaswag\endcsname{83.3}
\expandafter\def\csname ev@qwen/cyber/A1/0M/mmlu\endcsname{81.0}
\expandafter\def\csname ev@qwen/cyber/A1/0M/mmlu_adj\endcsname{68.9}
\expandafter\def\csname ev@qwen/cyber/A1/0M/mmlu_rest\endcsname{81.4}
\expandafter\def\csname ev@qwen/cyber/A1/0M/wmdp\endcsname{43.2}
\expandafter\def\csname ev@qwen/cyber/A2/0M/arc_challenge\endcsname{72.3}
\expandafter\def\csname ev@qwen/cyber/A2/0M/gsm8k\endcsname{83.2}
\expandafter\def\csname ev@qwen/cyber/A2/0M/hellaswag\endcsname{82.4}
\expandafter\def\csname ev@qwen/cyber/A2/0M/mmlu\endcsname{80.7}
\expandafter\def\csname ev@qwen/cyber/A2/0M/mmlu_adj\endcsname{58.5}
\expandafter\def\csname ev@qwen/cyber/A2/0M/mmlu_rest\endcsname{81.4}
\expandafter\def\csname ev@qwen/cyber/A2/0M/wmdp\endcsname{28.2}
\expandafter\def\csname ev@qwen/cyber/F2/0M/arc_challenge\endcsname{72.9}
\expandafter\def\csname ev@qwen/cyber/F2/0M/gsm8k\endcsname{84.4}
\expandafter\def\csname ev@qwen/cyber/F2/0M/hellaswag\endcsname{82.9}
\expandafter\def\csname ev@qwen/cyber/F2/0M/mmlu\endcsname{81.4}
\expandafter\def\csname ev@qwen/cyber/F2/0M/mmlu_adj\endcsname{76.9}
\expandafter\def\csname ev@qwen/cyber/F2/0M/mmlu_rest\endcsname{81.5}
\expandafter\def\csname ev@qwen/cyber/F2/0M/wmdp\endcsname{46.5}
\expandafter\def\csname ev@qwen/cyber/F3/250M/arc_challenge\endcsname{72.5}
\expandafter\def\csname ev@qwen/cyber/F3/250M/gsm8k\endcsname{85.6}
\expandafter\def\csname ev@qwen/cyber/F3/250M/hellaswag\endcsname{84.3}
\expandafter\def\csname ev@qwen/cyber/F3/250M/mmlu\endcsname{82.0}
\expandafter\def\csname ev@qwen/cyber/F3/250M/mmlu_adj\endcsname{81.5}
\expandafter\def\csname ev@qwen/cyber/F3/250M/mmlu_rest\endcsname{82.0}
\expandafter\def\csname ev@qwen/cyber/F3/250M/wmdp\endcsname{50.3}
\expandafter\def\csname ev@qwen/cyber/F3/250M/ifeval\endcsname{39.2}
\expandafter\def\csname ev@qwen/cyber/A3/250M/arc_challenge\endcsname{73.0}
\expandafter\def\csname ev@qwen/cyber/A3/250M/gsm8k\endcsname{85.4}
\expandafter\def\csname ev@qwen/cyber/A3/250M/hellaswag\endcsname{83.9}
\expandafter\def\csname ev@qwen/cyber/A3/250M/mmlu\endcsname{81.3}
\expandafter\def\csname ev@qwen/cyber/A3/250M/mmlu_adj\endcsname{65.3}
\expandafter\def\csname ev@qwen/cyber/A3/250M/mmlu_rest\endcsname{81.8}
\expandafter\def\csname ev@qwen/cyber/A3/250M/wmdp\endcsname{27.3}
\expandafter\def\csname ev@qwen/cyber/A3/250M/ifeval\endcsname{38.1}
\expandafter\def\csname ev@qwen/cyber/R/0M/arc_challenge\endcsname{73.3}
\expandafter\def\csname ev@qwen/cyber/R/0M/gsm8k\endcsname{86.0}
\expandafter\def\csname ev@qwen/cyber/R/0M/hellaswag\endcsname{85.3}
\expandafter\def\csname ev@qwen/cyber/R/0M/mmlu\endcsname{82.3}
\expandafter\def\csname ev@qwen/cyber/R/0M/mmlu_adj\endcsname{81.1}
\expandafter\def\csname ev@qwen/cyber/R/0M/mmlu_rest\endcsname{82.3}
\expandafter\def\csname ev@qwen/cyber/R/0M/wmdp\endcsname{31.8}
\expandafter\def\csname ev@qwen/cyber/R-SFT/250M/arc_challenge\endcsname{73.3}
\expandafter\def\csname ev@qwen/cyber/R-SFT/250M/gsm8k\endcsname{85.1}
\expandafter\def\csname ev@qwen/cyber/R-SFT/250M/hellaswag\endcsname{85.4}
\expandafter\def\csname ev@qwen/cyber/R-SFT/250M/mmlu\endcsname{82.3}
\expandafter\def\csname ev@qwen/cyber/R-SFT/250M/mmlu_adj\endcsname{79.6}
\expandafter\def\csname ev@qwen/cyber/R-SFT/250M/mmlu_rest\endcsname{82.4}
\expandafter\def\csname ev@qwen/cyber/R-SFT/250M/wmdp\endcsname{43.5}
\expandafter\def\csname ev@qwen/cyber/R-SFT/250M/ifeval\endcsname{39.6}
\expandafter\def\csname ev@qwen/cyber/A3-direct/250M/arc_challenge\endcsname{72.7}
\expandafter\def\csname ev@qwen/cyber/A3-direct/250M/gsm8k\endcsname{84.9}
\expandafter\def\csname ev@qwen/cyber/A3-direct/250M/hellaswag\endcsname{84.2}
\expandafter\def\csname ev@qwen/cyber/A3-direct/250M/mmlu\endcsname{81.6}
\expandafter\def\csname ev@qwen/cyber/A3-direct/250M/mmlu_adj\endcsname{74.5}
\expandafter\def\csname ev@qwen/cyber/A3-direct/250M/mmlu_rest\endcsname{81.8}
\expandafter\def\csname ev@qwen/cyber/A3-direct/250M/wmdp\endcsname{35.9}
\expandafter\def\csname ev@qwen/cyber/A3-direct/250M/ifeval\endcsname{41.0}
\expandafter\def\csname ev@qwen/cyber/CEU/0M/arc_challenge\endcsname{35.5}
\expandafter\def\csname ev@qwen/cyber/CEU/0M/gsm8k\endcsname{0.0}
\expandafter\def\csname ev@qwen/cyber/CEU/0M/hellaswag\endcsname{33.4}
\expandafter\def\csname ev@qwen/cyber/CEU/0M/mmlu\endcsname{75.4}
\expandafter\def\csname ev@qwen/cyber/CEU/0M/mmlu_adj\endcsname{69.9}
\expandafter\def\csname ev@qwen/cyber/CEU/0M/mmlu_rest\endcsname{75.6}
\expandafter\def\csname ev@qwen/cyber/CEU/0M/wmdp\endcsname{37.8}
\expandafter\def\csname ev@qwen/cyber/CEU-SFT/250M/arc_challenge\endcsname{69.3}
\expandafter\def\csname ev@qwen/cyber/CEU-SFT/250M/gsm8k\endcsname{62.9}
\expandafter\def\csname ev@qwen/cyber/CEU-SFT/250M/hellaswag\endcsname{69.5}
\expandafter\def\csname ev@qwen/cyber/CEU-SFT/250M/mmlu\endcsname{81.5}
\expandafter\def\csname ev@qwen/cyber/CEU-SFT/250M/mmlu_adj\endcsname{80.3}
\expandafter\def\csname ev@qwen/cyber/CEU-SFT/250M/mmlu_rest\endcsname{81.6}
\expandafter\def\csname ev@qwen/cyber/CEU-SFT/250M/wmdp\endcsname{48.1}
\expandafter\def\csname ev@qwen/cyber/CEU-SFT/250M/ifeval\endcsname{54.2}
\expandafter\def\csname ev@qwen/cyber/GA/0M/arc_challenge\endcsname{30.4}
\expandafter\def\csname ev@qwen/cyber/GA/0M/gsm8k\endcsname{0.0}
\expandafter\def\csname ev@qwen/cyber/GA/0M/hellaswag\endcsname{26.9}
\expandafter\def\csname ev@qwen/cyber/GA/0M/mmlu\endcsname{74.4}
\expandafter\def\csname ev@qwen/cyber/GA/0M/mmlu_adj\endcsname{61.4}
\expandafter\def\csname ev@qwen/cyber/GA/0M/mmlu_rest\endcsname{74.8}
\expandafter\def\csname ev@qwen/cyber/GA/0M/wmdp\endcsname{34.7}
\expandafter\def\csname ev@qwen/cyber/GA-SFT/250M/arc_challenge\endcsname{73.1}
\expandafter\def\csname ev@qwen/cyber/GA-SFT/250M/gsm8k\endcsname{6.3}
\expandafter\def\csname ev@qwen/cyber/GA-SFT/250M/hellaswag\endcsname{73.7}
\expandafter\def\csname ev@qwen/cyber/GA-SFT/250M/mmlu\endcsname{80.8}
\expandafter\def\csname ev@qwen/cyber/GA-SFT/250M/mmlu_adj\endcsname{76.0}
\expandafter\def\csname ev@qwen/cyber/GA-SFT/250M/mmlu_rest\endcsname{81.0}
\expandafter\def\csname ev@qwen/cyber/GA-SFT/250M/wmdp\endcsname{46.0}
\expandafter\def\csname ev@qwen/cyber/GA-SFT/250M/ifeval\endcsname{46.6}
\expandafter\def\csname ev@qwen/cyber/GD/0M/arc_challenge\endcsname{74.7}
\expandafter\def\csname ev@qwen/cyber/GD/0M/gsm8k\endcsname{0.0}
\expandafter\def\csname ev@qwen/cyber/GD/0M/hellaswag\endcsname{84.3}
\expandafter\def\csname ev@qwen/cyber/GD/0M/mmlu\endcsname{78.6}
\expandafter\def\csname ev@qwen/cyber/GD/0M/mmlu_adj\endcsname{63.1}
\expandafter\def\csname ev@qwen/cyber/GD/0M/mmlu_rest\endcsname{79.1}
\expandafter\def\csname ev@qwen/cyber/GD/0M/wmdp\endcsname{36.9}
\expandafter\def\csname ev@qwen/cyber/GD-SFT/250M/arc_challenge\endcsname{73.7}
\expandafter\def\csname ev@qwen/cyber/GD-SFT/250M/gsm8k\endcsname{8.0}
\expandafter\def\csname ev@qwen/cyber/GD-SFT/250M/hellaswag\endcsname{84.9}
\expandafter\def\csname ev@qwen/cyber/GD-SFT/250M/mmlu\endcsname{81.3}
\expandafter\def\csname ev@qwen/cyber/GD-SFT/250M/mmlu_adj\endcsname{77.4}
\expandafter\def\csname ev@qwen/cyber/GD-SFT/250M/mmlu_rest\endcsname{81.4}
\expandafter\def\csname ev@qwen/cyber/GD-SFT/250M/wmdp\endcsname{46.8}
\expandafter\def\csname ev@qwen/cyber/GD-SFT/250M/ifeval\endcsname{43.8}
\expandafter\def\csname ev@qwen/cyber/PDU/0M/arc_challenge\endcsname{73.2}
\expandafter\def\csname ev@qwen/cyber/PDU/0M/gsm8k\endcsname{76.0}
\expandafter\def\csname ev@qwen/cyber/PDU/0M/hellaswag\endcsname{85.4}
\expandafter\def\csname ev@qwen/cyber/PDU/0M/mmlu\endcsname{81.9}
\expandafter\def\csname ev@qwen/cyber/PDU/0M/mmlu_adj\endcsname{80.8}
\expandafter\def\csname ev@qwen/cyber/PDU/0M/mmlu_rest\endcsname{81.9}
\expandafter\def\csname ev@qwen/cyber/PDU/0M/wmdp\endcsname{42.4}
\expandafter\def\csname ev@qwen/cyber/PDU/0M/ifeval\endcsname{50.1}
\expandafter\def\csname ev@qwen/cyber/PDU-SFT/250M/arc_challenge\endcsname{73.9}
\expandafter\def\csname ev@qwen/cyber/PDU-SFT/250M/gsm8k\endcsname{81.3}
\expandafter\def\csname ev@qwen/cyber/PDU-SFT/250M/hellaswag\endcsname{85.5}
\expandafter\def\csname ev@qwen/cyber/PDU-SFT/250M/mmlu\endcsname{82.3}
\expandafter\def\csname ev@qwen/cyber/PDU-SFT/250M/mmlu_adj\endcsname{80.6}
\expandafter\def\csname ev@qwen/cyber/PDU-SFT/250M/mmlu_rest\endcsname{82.4}
\expandafter\def\csname ev@qwen/cyber/PDU-SFT/250M/wmdp\endcsname{45.5}
\expandafter\def\csname ev@qwen/cyber/PDU-SFT/250M/ifeval\endcsname{48.2}
\expandafter\def\csname ev@qwen/cyber/PureNPO/0M/arc_challenge\endcsname{74.2}
\expandafter\def\csname ev@qwen/cyber/PureNPO/0M/gsm8k\endcsname{85.3}
\expandafter\def\csname ev@qwen/cyber/PureNPO/0M/hellaswag\endcsname{87.6}
\expandafter\def\csname ev@qwen/cyber/PureNPO/0M/mmlu\endcsname{81.3}
\expandafter\def\csname ev@qwen/cyber/PureNPO/0M/mmlu_adj\endcsname{77.2}
\expandafter\def\csname ev@qwen/cyber/PureNPO/0M/mmlu_rest\endcsname{81.4}
\expandafter\def\csname ev@qwen/cyber/PureNPO/0M/wmdp\endcsname{41.9}
\expandafter\def\csname ev@qwen/cyber/PureNPO-SFT/250M/arc_challenge\endcsname{76.0}
\expandafter\def\csname ev@qwen/cyber/PureNPO-SFT/250M/gsm8k\endcsname{88.2}
\expandafter\def\csname ev@qwen/cyber/PureNPO-SFT/250M/hellaswag\endcsname{87.3}
\expandafter\def\csname ev@qwen/cyber/PureNPO-SFT/250M/mmlu\endcsname{81.9}
\expandafter\def\csname ev@qwen/cyber/PureNPO-SFT/250M/mmlu_adj\endcsname{79.8}
\expandafter\def\csname ev@qwen/cyber/PureNPO-SFT/250M/mmlu_rest\endcsname{82.0}
\expandafter\def\csname ev@qwen/cyber/PureNPO-SFT/250M/wmdp\endcsname{49.4}
\expandafter\def\csname ev@qwen/cyber/PureNPO-SFT/250M/ifeval\endcsname{37.7}
\expandafter\def\csname ev@qwen/cyber/ExternalRMU/0M/arc_challenge\endcsname{72.6}
\expandafter\def\csname ev@qwen/cyber/ExternalRMU/0M/gsm8k\endcsname{76.9}
\expandafter\def\csname ev@qwen/cyber/ExternalRMU/0M/hellaswag\endcsname{84.9}
\expandafter\def\csname ev@qwen/cyber/ExternalRMU/0M/mmlu\endcsname{81.8}
\expandafter\def\csname ev@qwen/cyber/ExternalRMU/0M/mmlu_adj\endcsname{76.5}
\expandafter\def\csname ev@qwen/cyber/ExternalRMU/0M/mmlu_rest\endcsname{82.0}
\expandafter\def\csname ev@qwen/cyber/ExternalRMU/0M/wmdp\endcsname{27.5}
\expandafter\def\csname ev@qwen/cyber/ExternalRMU/0M/ifeval\endcsname{55.8}
\expandafter\def\csname ev@qwen/cyber/ExternalRMU-SFT/250M/arc_challenge\endcsname{73.3}
\expandafter\def\csname ev@qwen/cyber/ExternalRMU-SFT/250M/gsm8k\endcsname{78.5}
\expandafter\def\csname ev@qwen/cyber/ExternalRMU-SFT/250M/hellaswag\endcsname{85.2}
\expandafter\def\csname ev@qwen/cyber/ExternalRMU-SFT/250M/mmlu\endcsname{82.2}
\expandafter\def\csname ev@qwen/cyber/ExternalRMU-SFT/250M/mmlu_adj\endcsname{81.1}
\expandafter\def\csname ev@qwen/cyber/ExternalRMU-SFT/250M/mmlu_rest\endcsname{82.2}
\expandafter\def\csname ev@qwen/cyber/ExternalRMU-SFT/250M/wmdp\endcsname{40.6}
\expandafter\def\csname ev@qwen/cyber/ExternalRMU-SFT/250M/ifeval\endcsname{38.5}
\expandafter\def\csname ev@qwen/cyber/SatImp/0M/arc_challenge\endcsname{39.1}
\expandafter\def\csname ev@qwen/cyber/SatImp/0M/gsm8k\endcsname{0.0}
\expandafter\def\csname ev@qwen/cyber/SatImp/0M/hellaswag\endcsname{37.2}
\expandafter\def\csname ev@qwen/cyber/SatImp/0M/mmlu\endcsname{74.4}
\expandafter\def\csname ev@qwen/cyber/SatImp/0M/mmlu_adj\endcsname{66.3}
\expandafter\def\csname ev@qwen/cyber/SatImp/0M/mmlu_rest\endcsname{74.7}
\expandafter\def\csname ev@qwen/cyber/SatImp/0M/wmdp\endcsname{32.8}
\expandafter\def\csname ev@qwen/cyber/SatImp-SFT/250M/arc_challenge\endcsname{70.5}
\expandafter\def\csname ev@qwen/cyber/SatImp-SFT/250M/gsm8k\endcsname{75.4}
\expandafter\def\csname ev@qwen/cyber/SatImp-SFT/250M/hellaswag\endcsname{79.1}
\expandafter\def\csname ev@qwen/cyber/SatImp-SFT/250M/mmlu\endcsname{81.1}
\expandafter\def\csname ev@qwen/cyber/SatImp-SFT/250M/mmlu_adj\endcsname{79.6}
\expandafter\def\csname ev@qwen/cyber/SatImp-SFT/250M/mmlu_rest\endcsname{81.1}
\expandafter\def\csname ev@qwen/cyber/SatImp-SFT/250M/wmdp\endcsname{44.5}
\expandafter\def\csname ev@qwen/cyber/SatImp-SFT/250M/ifeval\endcsname{50.5}
\expandafter\def\csname ev@qwen/cyber/SimNPO/0M/arc_challenge\endcsname{74.9}
\expandafter\def\csname ev@qwen/cyber/SimNPO/0M/gsm8k\endcsname{74.3}
\expandafter\def\csname ev@qwen/cyber/SimNPO/0M/hellaswag\endcsname{88.5}
\expandafter\def\csname ev@qwen/cyber/SimNPO/0M/mmlu\endcsname{81.1}
\expandafter\def\csname ev@qwen/cyber/SimNPO/0M/mmlu_adj\endcsname{76.0}
\expandafter\def\csname ev@qwen/cyber/SimNPO/0M/mmlu_rest\endcsname{81.2}
\expandafter\def\csname ev@qwen/cyber/SimNPO/0M/wmdp\endcsname{42.5}
\expandafter\def\csname ev@qwen/cyber/SimNPO-SFT/250M/arc_challenge\endcsname{74.5}
\expandafter\def\csname ev@qwen/cyber/SimNPO-SFT/250M/gsm8k\endcsname{87.9}
\expandafter\def\csname ev@qwen/cyber/SimNPO-SFT/250M/hellaswag\endcsname{87.8}
\expandafter\def\csname ev@qwen/cyber/SimNPO-SFT/250M/mmlu\endcsname{81.9}
\expandafter\def\csname ev@qwen/cyber/SimNPO-SFT/250M/mmlu_adj\endcsname{79.4}
\expandafter\def\csname ev@qwen/cyber/SimNPO-SFT/250M/mmlu_rest\endcsname{82.0}
\expandafter\def\csname ev@qwen/cyber/SimNPO-SFT/250M/wmdp\endcsname{52.5}
\expandafter\def\csname ev@qwen/cyber/SimNPO-SFT/250M/ifeval\endcsname{40.7}
\expandafter\def\csname ev@qwen/cyber/TRACE/0M/arc_challenge\endcsname{73.2}
\expandafter\def\csname ev@qwen/cyber/TRACE/0M/gsm8k\endcsname{73.3}
\expandafter\def\csname ev@qwen/cyber/TRACE/0M/hellaswag\endcsname{84.0}
\expandafter\def\csname ev@qwen/cyber/TRACE/0M/mmlu\endcsname{81.6}
\expandafter\def\csname ev@qwen/cyber/TRACE/0M/mmlu_adj\endcsname{79.8}
\expandafter\def\csname ev@qwen/cyber/TRACE/0M/mmlu_rest\endcsname{81.6}
\expandafter\def\csname ev@qwen/cyber/TRACE/0M/wmdp\endcsname{46.0}
\expandafter\def\csname ev@qwen/cyber/TRACE-SFT/250M/arc_challenge\endcsname{73.9}
\expandafter\def\csname ev@qwen/cyber/TRACE-SFT/250M/gsm8k\endcsname{61.1}
\expandafter\def\csname ev@qwen/cyber/TRACE-SFT/250M/hellaswag\endcsname{85.4}
\expandafter\def\csname ev@qwen/cyber/TRACE-SFT/250M/mmlu\endcsname{82.1}
\expandafter\def\csname ev@qwen/cyber/TRACE-SFT/250M/mmlu_adj\endcsname{80.6}
\expandafter\def\csname ev@qwen/cyber/TRACE-SFT/250M/mmlu_rest\endcsname{82.1}
\expandafter\def\csname ev@qwen/cyber/TRACE-SFT/250M/wmdp\endcsname{47.1}
\expandafter\def\csname ev@qwen/cyber/TRACE-SFT/250M/ifeval\endcsname{48.1}
\expandafter\def\csname ev@qwen/cyber/WGA/0M/arc_challenge\endcsname{37.5}
\expandafter\def\csname ev@qwen/cyber/WGA/0M/gsm8k\endcsname{0.0}
\expandafter\def\csname ev@qwen/cyber/WGA/0M/hellaswag\endcsname{35.7}
\expandafter\def\csname ev@qwen/cyber/WGA/0M/mmlu\endcsname{72.9}
\expandafter\def\csname ev@qwen/cyber/WGA/0M/mmlu_adj\endcsname{62.9}
\expandafter\def\csname ev@qwen/cyber/WGA/0M/mmlu_rest\endcsname{73.2}
\expandafter\def\csname ev@qwen/cyber/WGA/0M/wmdp\endcsname{32.2}
\expandafter\def\csname ev@qwen/cyber/WGA-SFT/250M/arc_challenge\endcsname{70.0}
\expandafter\def\csname ev@qwen/cyber/WGA-SFT/250M/gsm8k\endcsname{72.2}
\expandafter\def\csname ev@qwen/cyber/WGA-SFT/250M/hellaswag\endcsname{77.9}
\expandafter\def\csname ev@qwen/cyber/WGA-SFT/250M/mmlu\endcsname{81.0}
\expandafter\def\csname ev@qwen/cyber/WGA-SFT/250M/mmlu_adj\endcsname{79.8}
\expandafter\def\csname ev@qwen/cyber/WGA-SFT/250M/mmlu_rest\endcsname{81.0}
\expandafter\def\csname ev@qwen/cyber/WGA-SFT/250M/wmdp\endcsname{45.1}
\expandafter\def\csname ev@qwen/cyber/WGA-SFT/250M/ifeval\endcsname{51.8}
\expandafter\def\csname ev@gemma/cyber/B/0M/arc_challenge\endcsname{73.2}
\expandafter\def\csname ev@gemma/cyber/B/0M/gsm8k\endcsname{71.4}
\expandafter\def\csname ev@gemma/cyber/B/0M/hellaswag\endcsname{85.6}
\expandafter\def\csname ev@gemma/cyber/B/0M/mmlu\endcsname{74.3}
\expandafter\def\csname ev@gemma/cyber/B/0M/mmlu_adj\endcsname{69.7}
\expandafter\def\csname ev@gemma/cyber/B/0M/mmlu_rest\endcsname{74.5}
\expandafter\def\csname ev@gemma/cyber/B/0M/wmdp\endcsname{50.8}
\expandafter\def\csname ev@gemma/cyber/S_std(B)/250M/arc_challenge\endcsname{73.7}
\expandafter\def\csname ev@gemma/cyber/S_std(B)/250M/gsm8k\endcsname{72.2}
\expandafter\def\csname ev@gemma/cyber/S_std(B)/250M/hellaswag\endcsname{85.5}
\expandafter\def\csname ev@gemma/cyber/S_std(B)/250M/mmlu\endcsname{74.9}
\expandafter\def\csname ev@gemma/cyber/S_std(B)/250M/mmlu_adj\endcsname{71.8}
\expandafter\def\csname ev@gemma/cyber/S_std(B)/250M/mmlu_rest\endcsname{75.0}
\expandafter\def\csname ev@gemma/cyber/S_std(B)/250M/wmdp\endcsname{51.1}
\expandafter\def\csname ev@gemma/cyber/S_std(B)/250M/ifeval\endcsname{26.8}
\expandafter\def\csname ev@gemma/cyber/F0/0M/arc_challenge\endcsname{73.5}
\expandafter\def\csname ev@gemma/cyber/F0/0M/gsm8k\endcsname{71.2}
\expandafter\def\csname ev@gemma/cyber/F0/0M/hellaswag\endcsname{85.6}
\expandafter\def\csname ev@gemma/cyber/F0/0M/mmlu\endcsname{74.1}
\expandafter\def\csname ev@gemma/cyber/F0/0M/mmlu_adj\endcsname{69.4}
\expandafter\def\csname ev@gemma/cyber/F0/0M/mmlu_rest\endcsname{74.2}
\expandafter\def\csname ev@gemma/cyber/F0/0M/wmdp\endcsname{50.0}
\expandafter\def\csname ev@gemma/cyber/A0/0M/arc_challenge\endcsname{72.9}
\expandafter\def\csname ev@gemma/cyber/A0/0M/gsm8k\endcsname{68.0}
\expandafter\def\csname ev@gemma/cyber/A0/0M/hellaswag\endcsname{85.4}
\expandafter\def\csname ev@gemma/cyber/A0/0M/mmlu\endcsname{73.8}
\expandafter\def\csname ev@gemma/cyber/A0/0M/mmlu_adj\endcsname{67.7}
\expandafter\def\csname ev@gemma/cyber/A0/0M/mmlu_rest\endcsname{74.0}
\expandafter\def\csname ev@gemma/cyber/A0/0M/wmdp\endcsname{48.5}
\expandafter\def\csname ev@gemma/cyber/F1/0M/arc_challenge\endcsname{73.0}
\expandafter\def\csname ev@gemma/cyber/F1/0M/gsm8k\endcsname{71.1}
\expandafter\def\csname ev@gemma/cyber/F1/0M/hellaswag\endcsname{85.6}
\expandafter\def\csname ev@gemma/cyber/F1/0M/mmlu\endcsname{74.0}
\expandafter\def\csname ev@gemma/cyber/F1/0M/mmlu_adj\endcsname{70.6}
\expandafter\def\csname ev@gemma/cyber/F1/0M/mmlu_rest\endcsname{74.1}
\expandafter\def\csname ev@gemma/cyber/F1/0M/wmdp\endcsname{49.6}
\expandafter\def\csname ev@gemma/cyber/A1/0M/arc_challenge\endcsname{73.0}
\expandafter\def\csname ev@gemma/cyber/A1/0M/gsm8k\endcsname{67.8}
\expandafter\def\csname ev@gemma/cyber/A1/0M/hellaswag\endcsname{85.6}
\expandafter\def\csname ev@gemma/cyber/A1/0M/mmlu\endcsname{73.6}
\expandafter\def\csname ev@gemma/cyber/A1/0M/mmlu_adj\endcsname{67.2}
\expandafter\def\csname ev@gemma/cyber/A1/0M/mmlu_rest\endcsname{73.8}
\expandafter\def\csname ev@gemma/cyber/A1/0M/wmdp\endcsname{48.1}
\expandafter\def\csname ev@gemma/cyber/A2/0M/arc_challenge\endcsname{72.0}
\expandafter\def\csname ev@gemma/cyber/A2/0M/gsm8k\endcsname{53.6}
\expandafter\def\csname ev@gemma/cyber/A2/0M/hellaswag\endcsname{83.9}
\expandafter\def\csname ev@gemma/cyber/A2/0M/mmlu\endcsname{71.9}
\expandafter\def\csname ev@gemma/cyber/A2/0M/mmlu_adj\endcsname{47.3}
\expandafter\def\csname ev@gemma/cyber/A2/0M/mmlu_rest\endcsname{72.7}
\expandafter\def\csname ev@gemma/cyber/A2/0M/wmdp\endcsname{29.4}
\expandafter\def\csname ev@gemma/cyber/F2/0M/arc_challenge\endcsname{72.6}
\expandafter\def\csname ev@gemma/cyber/F2/0M/gsm8k\endcsname{63.1}
\expandafter\def\csname ev@gemma/cyber/F2/0M/hellaswag\endcsname{84.6}
\expandafter\def\csname ev@gemma/cyber/F2/0M/mmlu\endcsname{72.5}
\expandafter\def\csname ev@gemma/cyber/F2/0M/mmlu_adj\endcsname{61.9}
\expandafter\def\csname ev@gemma/cyber/F2/0M/mmlu_rest\endcsname{72.8}
\expandafter\def\csname ev@gemma/cyber/F2/0M/wmdp\endcsname{38.9}
\expandafter\def\csname ev@gemma/cyber/F3/250M/arc_challenge\endcsname{72.8}
\expandafter\def\csname ev@gemma/cyber/F3/250M/gsm8k\endcsname{72.1}
\expandafter\def\csname ev@gemma/cyber/F3/250M/hellaswag\endcsname{85.4}
\expandafter\def\csname ev@gemma/cyber/F3/250M/mmlu\endcsname{74.0}
\expandafter\def\csname ev@gemma/cyber/F3/250M/mmlu_adj\endcsname{67.2}
\expandafter\def\csname ev@gemma/cyber/F3/250M/mmlu_rest\endcsname{74.2}
\expandafter\def\csname ev@gemma/cyber/F3/250M/wmdp\endcsname{45.1}
\expandafter\def\csname ev@gemma/cyber/F3/250M/ifeval\endcsname{27.5}
\expandafter\def\csname ev@gemma/cyber/A3/250M/arc_challenge\endcsname{72.7}
\expandafter\def\csname ev@gemma/cyber/A3/250M/gsm8k\endcsname{69.9}
\expandafter\def\csname ev@gemma/cyber/A3/250M/hellaswag\endcsname{85.2}
\expandafter\def\csname ev@gemma/cyber/A3/250M/mmlu\endcsname{73.3}
\expandafter\def\csname ev@gemma/cyber/A3/250M/mmlu_adj\endcsname{59.2}
\expandafter\def\csname ev@gemma/cyber/A3/250M/mmlu_rest\endcsname{73.8}
\expandafter\def\csname ev@gemma/cyber/A3/250M/wmdp\endcsname{34.0}
\expandafter\def\csname ev@gemma/cyber/A3/250M/ifeval\endcsname{26.2}
\expandafter\def\csname ev@gemma/cyber/A3-direct/250M/arc_challenge\endcsname{72.4}
\expandafter\def\csname ev@gemma/cyber/A3-direct/250M/gsm8k\endcsname{70.0}
\expandafter\def\csname ev@gemma/cyber/A3-direct/250M/hellaswag\endcsname{85.6}
\expandafter\def\csname ev@gemma/cyber/A3-direct/250M/mmlu\endcsname{73.3}
\expandafter\def\csname ev@gemma/cyber/A3-direct/250M/mmlu_adj\endcsname{60.0}
\expandafter\def\csname ev@gemma/cyber/A3-direct/250M/mmlu_rest\endcsname{73.7}
\expandafter\def\csname ev@gemma/cyber/A3-direct/250M/wmdp\endcsname{35.7}
\expandafter\def\csname ev@gemma/cyber/A3-direct/250M/ifeval\endcsname{26.8}
\expandafter\def\csname ev@gemma/cyber/R/0M/arc_challenge\endcsname{71.8}
\expandafter\def\csname ev@gemma/cyber/R/0M/gsm8k\endcsname{56.0}
\expandafter\def\csname ev@gemma/cyber/R/0M/hellaswag\endcsname{84.3}
\expandafter\def\csname ev@gemma/cyber/R/0M/mmlu\endcsname{72.1}
\expandafter\def\csname ev@gemma/cyber/R/0M/mmlu_adj\endcsname{50.7}
\expandafter\def\csname ev@gemma/cyber/R/0M/mmlu_rest\endcsname{72.7}
\expandafter\def\csname ev@gemma/cyber/R/0M/wmdp\endcsname{30.9}
\expandafter\def\csname ev@gemma/cyber/R-SFT/250M/arc_challenge\endcsname{73.4}
\expandafter\def\csname ev@gemma/cyber/R-SFT/250M/gsm8k\endcsname{71.1}
\expandafter\def\csname ev@gemma/cyber/R-SFT/250M/hellaswag\endcsname{85.5}
\expandafter\def\csname ev@gemma/cyber/R-SFT/250M/mmlu\endcsname{73.7}
\expandafter\def\csname ev@gemma/cyber/R-SFT/250M/mmlu_adj\endcsname{64.6}
\expandafter\def\csname ev@gemma/cyber/R-SFT/250M/mmlu_rest\endcsname{74.0}
\expandafter\def\csname ev@gemma/cyber/R-SFT/250M/wmdp\endcsname{37.6}
\expandafter\def\csname ev@gemma/cyber/R-SFT/250M/ifeval\endcsname{27.0}
\expandafter\def\csname ev@gemma/cyber/CEU/0M/arc_challenge\endcsname{30.4}
\expandafter\def\csname ev@gemma/cyber/CEU/0M/gsm8k\endcsname{0.0}
\expandafter\def\csname ev@gemma/cyber/CEU/0M/hellaswag\endcsname{30.1}
\expandafter\def\csname ev@gemma/cyber/CEU/0M/mmlu\endcsname{32.3}
\expandafter\def\csname ev@gemma/cyber/CEU/0M/mmlu_adj\endcsname{25.5}
\expandafter\def\csname ev@gemma/cyber/CEU/0M/mmlu_rest\endcsname{32.5}
\expandafter\def\csname ev@gemma/cyber/CEU/0M/wmdp\endcsname{26.6}
\expandafter\def\csname ev@gemma/cyber/CEU-SFT/250M/arc_challenge\endcsname{54.1}
\expandafter\def\csname ev@gemma/cyber/CEU-SFT/250M/gsm8k\endcsname{0.0}
\expandafter\def\csname ev@gemma/cyber/CEU-SFT/250M/hellaswag\endcsname{65.5}
\expandafter\def\csname ev@gemma/cyber/CEU-SFT/250M/mmlu\endcsname{66.3}
\expandafter\def\csname ev@gemma/cyber/CEU-SFT/250M/mmlu_adj\endcsname{51.5}
\expandafter\def\csname ev@gemma/cyber/CEU-SFT/250M/mmlu_rest\endcsname{66.8}
\expandafter\def\csname ev@gemma/cyber/CEU-SFT/250M/wmdp\endcsname{31.7}
\expandafter\def\csname ev@gemma/cyber/CEU-SFT/250M/ifeval\endcsname{10.5}
\expandafter\def\csname ev@gemma/cyber/GA/0M/arc_challenge\endcsname{28.2}
\expandafter\def\csname ev@gemma/cyber/GA/0M/gsm8k\endcsname{0.0}
\expandafter\def\csname ev@gemma/cyber/GA/0M/hellaswag\endcsname{29.0}
\expandafter\def\csname ev@gemma/cyber/GA/0M/mmlu\endcsname{72.4}
\expandafter\def\csname ev@gemma/cyber/GA/0M/mmlu_adj\endcsname{64.6}
\expandafter\def\csname ev@gemma/cyber/GA/0M/mmlu_rest\endcsname{72.6}
\expandafter\def\csname ev@gemma/cyber/GA/0M/wmdp\endcsname{43.6}
\expandafter\def\csname ev@gemma/cyber/GA-SFT/250M/arc_challenge\endcsname{64.7}
\expandafter\def\csname ev@gemma/cyber/GA-SFT/250M/gsm8k\endcsname{71.0}
\expandafter\def\csname ev@gemma/cyber/GA-SFT/250M/hellaswag\endcsname{71.8}
\expandafter\def\csname ev@gemma/cyber/GA-SFT/250M/mmlu\endcsname{73.7}
\expandafter\def\csname ev@gemma/cyber/GA-SFT/250M/mmlu_adj\endcsname{68.7}
\expandafter\def\csname ev@gemma/cyber/GA-SFT/250M/mmlu_rest\endcsname{73.9}
\expandafter\def\csname ev@gemma/cyber/GA-SFT/250M/wmdp\endcsname{49.2}
\expandafter\def\csname ev@gemma/cyber/GA-SFT/250M/ifeval\endcsname{17.2}
\expandafter\def\csname ev@gemma/cyber/GD/0M/arc_challenge\endcsname{60.2}
\expandafter\def\csname ev@gemma/cyber/GD/0M/gsm8k\endcsname{0.0}
\expandafter\def\csname ev@gemma/cyber/GD/0M/hellaswag\endcsname{73.9}
\expandafter\def\csname ev@gemma/cyber/GD/0M/mmlu\endcsname{66.6}
\expandafter\def\csname ev@gemma/cyber/GD/0M/mmlu_adj\endcsname{56.1}
\expandafter\def\csname ev@gemma/cyber/GD/0M/mmlu_rest\endcsname{66.9}
\expandafter\def\csname ev@gemma/cyber/GD/0M/wmdp\endcsname{27.3}
\expandafter\def\csname ev@gemma/cyber/GD-SFT/250M/arc_challenge\endcsname{67.6}
\expandafter\def\csname ev@gemma/cyber/GD-SFT/250M/gsm8k\endcsname{65.7}
\expandafter\def\csname ev@gemma/cyber/GD-SFT/250M/hellaswag\endcsname{80.1}
\expandafter\def\csname ev@gemma/cyber/GD-SFT/250M/mmlu\endcsname{73.1}
\expandafter\def\csname ev@gemma/cyber/GD-SFT/250M/mmlu_adj\endcsname{69.7}
\expandafter\def\csname ev@gemma/cyber/GD-SFT/250M/mmlu_rest\endcsname{73.2}
\expandafter\def\csname ev@gemma/cyber/GD-SFT/250M/wmdp\endcsname{45.6}
\expandafter\def\csname ev@gemma/cyber/GD-SFT/250M/ifeval\endcsname{26.6}
\expandafter\def\csname ev@gemma/cyber/PDU/0M/arc_challenge\endcsname{71.3}
\expandafter\def\csname ev@gemma/cyber/PDU/0M/gsm8k\endcsname{69.8}
\expandafter\def\csname ev@gemma/cyber/PDU/0M/hellaswag\endcsname{84.5}
\expandafter\def\csname ev@gemma/cyber/PDU/0M/mmlu\endcsname{73.7}
\expandafter\def\csname ev@gemma/cyber/PDU/0M/mmlu_adj\endcsname{61.9}
\expandafter\def\csname ev@gemma/cyber/PDU/0M/mmlu_rest\endcsname{74.1}
\expandafter\def\csname ev@gemma/cyber/PDU/0M/wmdp\endcsname{25.1}
\expandafter\def\csname ev@gemma/cyber/PDU-SFT/250M/arc_challenge\endcsname{71.8}
\expandafter\def\csname ev@gemma/cyber/PDU-SFT/250M/gsm8k\endcsname{72.9}
\expandafter\def\csname ev@gemma/cyber/PDU-SFT/250M/hellaswag\endcsname{84.6}
\expandafter\def\csname ev@gemma/cyber/PDU-SFT/250M/mmlu\endcsname{74.5}
\expandafter\def\csname ev@gemma/cyber/PDU-SFT/250M/mmlu_adj\endcsname{67.7}
\expandafter\def\csname ev@gemma/cyber/PDU-SFT/250M/mmlu_rest\endcsname{74.7}
\expandafter\def\csname ev@gemma/cyber/PDU-SFT/250M/wmdp\endcsname{35.5}
\expandafter\def\csname ev@gemma/cyber/PDU-SFT/250M/ifeval\endcsname{25.7}
\expandafter\def\csname ev@gemma/cyber/PureNPO/0M/arc_challenge\endcsname{67.9}
\expandafter\def\csname ev@gemma/cyber/PureNPO/0M/gsm8k\endcsname{56.2}
\expandafter\def\csname ev@gemma/cyber/PureNPO/0M/hellaswag\endcsname{84.1}
\expandafter\def\csname ev@gemma/cyber/PureNPO/0M/mmlu\endcsname{70.9}
\expandafter\def\csname ev@gemma/cyber/PureNPO/0M/mmlu_adj\endcsname{61.9}
\expandafter\def\csname ev@gemma/cyber/PureNPO/0M/mmlu_rest\endcsname{71.2}
\expandafter\def\csname ev@gemma/cyber/PureNPO/0M/wmdp\endcsname{44.1}
\expandafter\def\csname ev@gemma/cyber/PureNPO-SFT/250M/arc_challenge\endcsname{70.1}
\expandafter\def\csname ev@gemma/cyber/PureNPO-SFT/250M/gsm8k\endcsname{65.9}
\expandafter\def\csname ev@gemma/cyber/PureNPO-SFT/250M/hellaswag\endcsname{84.9}
\expandafter\def\csname ev@gemma/cyber/PureNPO-SFT/250M/mmlu\endcsname{72.6}
\expandafter\def\csname ev@gemma/cyber/PureNPO-SFT/250M/mmlu_adj\endcsname{63.6}
\expandafter\def\csname ev@gemma/cyber/PureNPO-SFT/250M/mmlu_rest\endcsname{72.8}
\expandafter\def\csname ev@gemma/cyber/PureNPO-SFT/250M/wmdp\endcsname{46.5}
\expandafter\def\csname ev@gemma/cyber/PureNPO-SFT/250M/ifeval\endcsname{23.8}
\expandafter\def\csname ev@gemma/cyber/ExternalRMU/0M/arc_challenge\endcsname{67.9}
\expandafter\def\csname ev@gemma/cyber/ExternalRMU/0M/gsm8k\endcsname{65.4}
\expandafter\def\csname ev@gemma/cyber/ExternalRMU/0M/hellaswag\endcsname{78.7}
\expandafter\def\csname ev@gemma/cyber/ExternalRMU/0M/mmlu\endcsname{72.6}
\expandafter\def\csname ev@gemma/cyber/ExternalRMU/0M/mmlu_adj\endcsname{55.1}
\expandafter\def\csname ev@gemma/cyber/ExternalRMU/0M/mmlu_rest\endcsname{73.1}
\expandafter\def\csname ev@gemma/cyber/ExternalRMU/0M/wmdp\endcsname{26.8}
\expandafter\def\csname ev@gemma/cyber/ExternalRMU-SFT/250M/arc_challenge\endcsname{70.6}
\expandafter\def\csname ev@gemma/cyber/ExternalRMU-SFT/250M/gsm8k\endcsname{71.0}
\expandafter\def\csname ev@gemma/cyber/ExternalRMU-SFT/250M/hellaswag\endcsname{82.0}
\expandafter\def\csname ev@gemma/cyber/ExternalRMU-SFT/250M/mmlu\endcsname{73.8}
\expandafter\def\csname ev@gemma/cyber/ExternalRMU-SFT/250M/mmlu_adj\endcsname{63.4}
\expandafter\def\csname ev@gemma/cyber/ExternalRMU-SFT/250M/mmlu_rest\endcsname{74.2}
\expandafter\def\csname ev@gemma/cyber/ExternalRMU-SFT/250M/wmdp\endcsname{35.7}
\expandafter\def\csname ev@gemma/cyber/ExternalRMU-SFT/250M/ifeval\endcsname{27.4}
\expandafter\def\csname ev@gemma/cyber/SatImp/0M/arc_challenge\endcsname{30.5}
\expandafter\def\csname ev@gemma/cyber/SatImp/0M/gsm8k\endcsname{0.0}
\expandafter\def\csname ev@gemma/cyber/SatImp/0M/hellaswag\endcsname{34.1}
\expandafter\def\csname ev@gemma/cyber/SatImp/0M/mmlu\endcsname{40.4}
\expandafter\def\csname ev@gemma/cyber/SatImp/0M/mmlu_adj\endcsname{33.7}
\expandafter\def\csname ev@gemma/cyber/SatImp/0M/mmlu_rest\endcsname{40.6}
\expandafter\def\csname ev@gemma/cyber/SatImp/0M/wmdp\endcsname{28.2}
\expandafter\def\csname ev@gemma/cyber/SatImp-SFT/250M/arc_challenge\endcsname{66.5}
\expandafter\def\csname ev@gemma/cyber/SatImp-SFT/250M/gsm8k\endcsname{25.4}
\expandafter\def\csname ev@gemma/cyber/SatImp-SFT/250M/hellaswag\endcsname{80.3}
\expandafter\def\csname ev@gemma/cyber/SatImp-SFT/250M/mmlu\endcsname{71.0}
\expandafter\def\csname ev@gemma/cyber/SatImp-SFT/250M/mmlu_adj\endcsname{64.6}
\expandafter\def\csname ev@gemma/cyber/SatImp-SFT/250M/mmlu_rest\endcsname{71.2}
\expandafter\def\csname ev@gemma/cyber/SatImp-SFT/250M/wmdp\endcsname{43.5}
\expandafter\def\csname ev@gemma/cyber/SatImp-SFT/250M/ifeval\endcsname{15.5}
\expandafter\def\csname ev@gemma/cyber/SimNPO/0M/arc_challenge\endcsname{66.1}
\expandafter\def\csname ev@gemma/cyber/SimNPO/0M/gsm8k\endcsname{48.7}
\expandafter\def\csname ev@gemma/cyber/SimNPO/0M/hellaswag\endcsname{82.2}
\expandafter\def\csname ev@gemma/cyber/SimNPO/0M/mmlu\endcsname{67.9}
\expandafter\def\csname ev@gemma/cyber/SimNPO/0M/mmlu_adj\endcsname{61.4}
\expandafter\def\csname ev@gemma/cyber/SimNPO/0M/mmlu_rest\endcsname{68.1}
\expandafter\def\csname ev@gemma/cyber/SimNPO/0M/wmdp\endcsname{26.8}
\expandafter\def\csname ev@gemma/cyber/SimNPO-SFT/250M/arc_challenge\endcsname{69.1}
\expandafter\def\csname ev@gemma/cyber/SimNPO-SFT/250M/gsm8k\endcsname{65.3}
\expandafter\def\csname ev@gemma/cyber/SimNPO-SFT/250M/hellaswag\endcsname{83.5}
\expandafter\def\csname ev@gemma/cyber/SimNPO-SFT/250M/mmlu\endcsname{71.4}
\expandafter\def\csname ev@gemma/cyber/SimNPO-SFT/250M/mmlu_adj\endcsname{62.6}
\expandafter\def\csname ev@gemma/cyber/SimNPO-SFT/250M/mmlu_rest\endcsname{71.6}
\expandafter\def\csname ev@gemma/cyber/SimNPO-SFT/250M/wmdp\endcsname{38.0}
\expandafter\def\csname ev@gemma/cyber/SimNPO-SFT/250M/ifeval\endcsname{20.1}
\expandafter\def\csname ev@gemma/cyber/TRACE/0M/arc_challenge\endcsname{68.7}
\expandafter\def\csname ev@gemma/cyber/TRACE/0M/gsm8k\endcsname{71.4}
\expandafter\def\csname ev@gemma/cyber/TRACE/0M/hellaswag\endcsname{78.5}
\expandafter\def\csname ev@gemma/cyber/TRACE/0M/mmlu\endcsname{74.6}
\expandafter\def\csname ev@gemma/cyber/TRACE/0M/mmlu_adj\endcsname{71.1}
\expandafter\def\csname ev@gemma/cyber/TRACE/0M/mmlu_rest\endcsname{74.7}
\expandafter\def\csname ev@gemma/cyber/TRACE/0M/wmdp\endcsname{49.7}
\expandafter\def\csname ev@gemma/cyber/TRACE-SFT/250M/arc_challenge\endcsname{70.8}
\expandafter\def\csname ev@gemma/cyber/TRACE-SFT/250M/gsm8k\endcsname{71.3}
\expandafter\def\csname ev@gemma/cyber/TRACE-SFT/250M/hellaswag\endcsname{81.5}
\expandafter\def\csname ev@gemma/cyber/TRACE-SFT/250M/mmlu\endcsname{74.8}
\expandafter\def\csname ev@gemma/cyber/TRACE-SFT/250M/mmlu_adj\endcsname{69.2}
\expandafter\def\csname ev@gemma/cyber/TRACE-SFT/250M/mmlu_rest\endcsname{75.0}
\expandafter\def\csname ev@gemma/cyber/TRACE-SFT/250M/wmdp\endcsname{50.4}
\expandafter\def\csname ev@gemma/cyber/TRACE-SFT/250M/ifeval\endcsname{27.9}
\expandafter\def\csname ev@gemma/cyber/WGA/0M/arc_challenge\endcsname{27.5}
\expandafter\def\csname ev@gemma/cyber/WGA/0M/gsm8k\endcsname{0.0}
\expandafter\def\csname ev@gemma/cyber/WGA/0M/hellaswag\endcsname{30.6}
\expandafter\def\csname ev@gemma/cyber/WGA/0M/mmlu\endcsname{34.1}
\expandafter\def\csname ev@gemma/cyber/WGA/0M/mmlu_adj\endcsname{26.9}
\expandafter\def\csname ev@gemma/cyber/WGA/0M/mmlu_rest\endcsname{34.3}
\expandafter\def\csname ev@gemma/cyber/WGA/0M/wmdp\endcsname{26.3}
\expandafter\def\csname ev@gemma/cyber/WGA-SFT/250M/arc_challenge\endcsname{62.1}
\expandafter\def\csname ev@gemma/cyber/WGA-SFT/250M/gsm8k\endcsname{23.8}
\expandafter\def\csname ev@gemma/cyber/WGA-SFT/250M/hellaswag\endcsname{76.4}
\expandafter\def\csname ev@gemma/cyber/WGA-SFT/250M/mmlu\endcsname{66.9}
\expandafter\def\csname ev@gemma/cyber/WGA-SFT/250M/mmlu_adj\endcsname{61.2}
\expandafter\def\csname ev@gemma/cyber/WGA-SFT/250M/mmlu_rest\endcsname{67.0}
\expandafter\def\csname ev@gemma/cyber/WGA-SFT/250M/wmdp\endcsname{42.1}
\expandafter\def\csname ev@gemma/cyber/WGA-SFT/250M/ifeval\endcsname{13.3}
\expandafter\def\csname ev@intern/bio/Routing-F0/0M/routing_z_forget\endcsname{22.9}
\expandafter\def\csname ev@intern/bio/Routing-F0/0M/routing_z_forget_layer_0\endcsname{10.5}
\expandafter\def\csname ev@intern/bio/Routing-F0/0M/routing_z_forget_layer_1\endcsname{12.4}
\expandafter\def\csname ev@intern/bio/Routing-F0/0M/routing_z_forget_layer_2\endcsname{6.1}
\expandafter\def\csname ev@intern/bio/Routing-F0/0M/routing_z_forget_layer_3\endcsname{18.6}
\expandafter\def\csname ev@intern/bio/Routing-F0/0M/routing_z_forget_layer_4\endcsname{25.1}
\expandafter\def\csname ev@intern/bio/Routing-F0/0M/routing_z_forget_layer_5\endcsname{26.1}
\expandafter\def\csname ev@intern/bio/Routing-F0/0M/routing_z_forget_layer_6\endcsname{19.1}
\expandafter\def\csname ev@intern/bio/Routing-F0/0M/routing_z_forget_layer_7\endcsname{24.4}
\expandafter\def\csname ev@intern/bio/Routing-F0/0M/routing_z_forget_layer_8\endcsname{31.1}
\expandafter\def\csname ev@intern/bio/Routing-F0/0M/routing_z_forget_layer_9\endcsname{28.4}
\expandafter\def\csname ev@intern/bio/Routing-F0/0M/routing_z_forget_layer_10\endcsname{16.5}
\expandafter\def\csname ev@intern/bio/Routing-F0/0M/routing_z_forget_layer_11\endcsname{16.7}
\expandafter\def\csname ev@intern/bio/Routing-F0/0M/routing_z_forget_layer_12\endcsname{26.1}
\expandafter\def\csname ev@intern/bio/Routing-F0/0M/routing_z_forget_layer_13\endcsname{25.5}
\expandafter\def\csname ev@intern/bio/Routing-F0/0M/routing_z_forget_layer_14\endcsname{26.5}
\expandafter\def\csname ev@intern/bio/Routing-F0/0M/routing_z_forget_layer_15\endcsname{27.0}
\expandafter\def\csname ev@intern/bio/Routing-F0/0M/routing_z_forget_layer_16\endcsname{17.3}
\expandafter\def\csname ev@intern/bio/Routing-F0/0M/routing_z_forget_layer_17\endcsname{25.5}
\expandafter\def\csname ev@intern/bio/Routing-F0/0M/routing_z_forget_layer_18\endcsname{27.3}
\expandafter\def\csname ev@intern/bio/Routing-F0/0M/routing_z_forget_layer_19\endcsname{23.5}
\expandafter\def\csname ev@intern/bio/Routing-F0/0M/routing_z_forget_layer_20\endcsname{30.7}
\expandafter\def\csname ev@intern/bio/Routing-F0/0M/routing_z_forget_layer_21\endcsname{30.3}
\expandafter\def\csname ev@intern/bio/Routing-F0/0M/routing_z_forget_layer_22\endcsname{18.1}
\expandafter\def\csname ev@intern/bio/Routing-F0/0M/routing_z_forget_layer_23\endcsname{17.0}
\expandafter\def\csname ev@intern/bio/Routing-F0/0M/routing_z_forget_layer_24\endcsname{23.7}
\expandafter\def\csname ev@intern/bio/Routing-F0/0M/routing_z_forget_layer_25\endcsname{23.8}
\expandafter\def\csname ev@intern/bio/Routing-F0/0M/routing_z_forget_layer_26\endcsname{24.4}
\expandafter\def\csname ev@intern/bio/Routing-F0/0M/routing_z_forget_layer_27\endcsname{25.7}
\expandafter\def\csname ev@intern/bio/Routing-F0/0M/routing_z_forget_layer_28\endcsname{16.4}
\expandafter\def\csname ev@intern/bio/Routing-F0/0M/routing_z_forget_layer_29\endcsname{25.5}
\expandafter\def\csname ev@intern/bio/Routing-F0/0M/routing_z_forget_layer_30\endcsname{27.6}
\expandafter\def\csname ev@intern/bio/Routing-F0/0M/routing_z_forget_layer_31\endcsname{22.5}
\expandafter\def\csname ev@intern/bio/Routing-F0/0M/routing_z_forget_layer_32\endcsname{19.7}
\expandafter\def\csname ev@intern/bio/Routing-F0/0M/routing_z_forget_layer_33\endcsname{22.6}
\expandafter\def\csname ev@intern/bio/Routing-F0/0M/routing_z_forget_layer_34\endcsname{18.6}
\expandafter\def\csname ev@intern/bio/Routing-F0/0M/routing_z_forget_layer_35\endcsname{30.3}
\expandafter\def\csname ev@intern/bio/Routing-F0/0M/routing_z_forget_layer_36\endcsname{27.5}
\expandafter\def\csname ev@intern/bio/Routing-F0/0M/routing_z_forget_layer_37\endcsname{26.9}
\expandafter\def\csname ev@intern/bio/Routing-F0/0M/routing_z_forget_layer_38\endcsname{28.1}
\expandafter\def\csname ev@intern/bio/Routing-F0/0M/routing_z_forget_layer_39\endcsname{21.5}
\expandafter\def\csname ev@intern/bio/Routing-F0/0M/routing_z_retain\endcsname{3.7}
\expandafter\def\csname ev@intern/bio/Routing-F0/0M/routing_z_retain_layer_0\endcsname{2.0}
\expandafter\def\csname ev@intern/bio/Routing-F0/0M/routing_z_retain_layer_1\endcsname{1.8}
\expandafter\def\csname ev@intern/bio/Routing-F0/0M/routing_z_retain_layer_2\endcsname{2.1}
\expandafter\def\csname ev@intern/bio/Routing-F0/0M/routing_z_retain_layer_3\endcsname{3.2}
\expandafter\def\csname ev@intern/bio/Routing-F0/0M/routing_z_retain_layer_4\endcsname{3.8}
\expandafter\def\csname ev@intern/bio/Routing-F0/0M/routing_z_retain_layer_5\endcsname{3.8}
\expandafter\def\csname ev@intern/bio/Routing-F0/0M/routing_z_retain_layer_6\endcsname{3.4}
\expandafter\def\csname ev@intern/bio/Routing-F0/0M/routing_z_retain_layer_7\endcsname{3.8}
\expandafter\def\csname ev@intern/bio/Routing-F0/0M/routing_z_retain_layer_8\endcsname{3.1}
\expandafter\def\csname ev@intern/bio/Routing-F0/0M/routing_z_retain_layer_9\endcsname{3.0}
\expandafter\def\csname ev@intern/bio/Routing-F0/0M/routing_z_retain_layer_10\endcsname{3.9}
\expandafter\def\csname ev@intern/bio/Routing-F0/0M/routing_z_retain_layer_11\endcsname{3.2}
\expandafter\def\csname ev@intern/bio/Routing-F0/0M/routing_z_retain_layer_12\endcsname{3.7}
\expandafter\def\csname ev@intern/bio/Routing-F0/0M/routing_z_retain_layer_13\endcsname{3.9}
\expandafter\def\csname ev@intern/bio/Routing-F0/0M/routing_z_retain_layer_14\endcsname{4.6}
\expandafter\def\csname ev@intern/bio/Routing-F0/0M/routing_z_retain_layer_15\endcsname{3.8}
\expandafter\def\csname ev@intern/bio/Routing-F0/0M/routing_z_retain_layer_16\endcsname{2.6}
\expandafter\def\csname ev@intern/bio/Routing-F0/0M/routing_z_retain_layer_17\endcsname{4.1}
\expandafter\def\csname ev@intern/bio/Routing-F0/0M/routing_z_retain_layer_18\endcsname{4.5}
\expandafter\def\csname ev@intern/bio/Routing-F0/0M/routing_z_retain_layer_19\endcsname{5.3}
\expandafter\def\csname ev@intern/bio/Routing-F0/0M/routing_z_retain_layer_20\endcsname{3.9}
\expandafter\def\csname ev@intern/bio/Routing-F0/0M/routing_z_retain_layer_21\endcsname{3.7}
\expandafter\def\csname ev@intern/bio/Routing-F0/0M/routing_z_retain_layer_22\endcsname{3.7}
\expandafter\def\csname ev@intern/bio/Routing-F0/0M/routing_z_retain_layer_23\endcsname{3.0}
\expandafter\def\csname ev@intern/bio/Routing-F0/0M/routing_z_retain_layer_24\endcsname{3.3}
\expandafter\def\csname ev@intern/bio/Routing-F0/0M/routing_z_retain_layer_25\endcsname{4.1}
\expandafter\def\csname ev@intern/bio/Routing-F0/0M/routing_z_retain_layer_26\endcsname{3.9}
\expandafter\def\csname ev@intern/bio/Routing-F0/0M/routing_z_retain_layer_27\endcsname{4.1}
\expandafter\def\csname ev@intern/bio/Routing-F0/0M/routing_z_retain_layer_28\endcsname{3.2}
\expandafter\def\csname ev@intern/bio/Routing-F0/0M/routing_z_retain_layer_29\endcsname{4.1}
\expandafter\def\csname ev@intern/bio/Routing-F0/0M/routing_z_retain_layer_30\endcsname{4.5}
\expandafter\def\csname ev@intern/bio/Routing-F0/0M/routing_z_retain_layer_31\endcsname{5.1}
\expandafter\def\csname ev@intern/bio/Routing-F0/0M/routing_z_retain_layer_32\endcsname{2.2}
\expandafter\def\csname ev@intern/bio/Routing-F0/0M/routing_z_retain_layer_33\endcsname{2.9}
\expandafter\def\csname ev@intern/bio/Routing-F0/0M/routing_z_retain_layer_34\endcsname{3.7}
\expandafter\def\csname ev@intern/bio/Routing-F0/0M/routing_z_retain_layer_35\endcsname{4.8}
\expandafter\def\csname ev@intern/bio/Routing-F0/0M/routing_z_retain_layer_36\endcsname{4.6}
\expandafter\def\csname ev@intern/bio/Routing-F0/0M/routing_z_retain_layer_37\endcsname{5.2}
\expandafter\def\csname ev@intern/bio/Routing-F0/0M/routing_z_retain_layer_38\endcsname{5.6}
\expandafter\def\csname ev@intern/bio/Routing-F0/0M/routing_z_retain_layer_39\endcsname{3.9}
\expandafter\def\csname ev@intern/bio/Routing-F1/0M/routing_z_forget\endcsname{75.9}
\expandafter\def\csname ev@intern/bio/Routing-F1/0M/routing_z_forget_layer_0\endcsname{1.7}
\expandafter\def\csname ev@intern/bio/Routing-F1/0M/routing_z_forget_layer_1\endcsname{71.5}
\expandafter\def\csname ev@intern/bio/Routing-F1/0M/routing_z_forget_layer_2\endcsname{72.9}
\expandafter\def\csname ev@intern/bio/Routing-F1/0M/routing_z_forget_layer_3\endcsname{86.0}
\expandafter\def\csname ev@intern/bio/Routing-F1/0M/routing_z_forget_layer_4\endcsname{86.0}
\expandafter\def\csname ev@intern/bio/Routing-F1/0M/routing_z_forget_layer_5\endcsname{84.2}
\expandafter\def\csname ev@intern/bio/Routing-F1/0M/routing_z_forget_layer_6\endcsname{80.5}
\expandafter\def\csname ev@intern/bio/Routing-F1/0M/routing_z_forget_layer_7\endcsname{83.8}
\expandafter\def\csname ev@intern/bio/Routing-F1/0M/routing_z_forget_layer_8\endcsname{84.8}
\expandafter\def\csname ev@intern/bio/Routing-F1/0M/routing_z_forget_layer_9\endcsname{82.8}
\expandafter\def\csname ev@intern/bio/Routing-F1/0M/routing_z_forget_layer_10\endcsname{76.9}
\expandafter\def\csname ev@intern/bio/Routing-F1/0M/routing_z_forget_layer_11\endcsname{73.7}
\expandafter\def\csname ev@intern/bio/Routing-F1/0M/routing_z_forget_layer_12\endcsname{82.1}
\expandafter\def\csname ev@intern/bio/Routing-F1/0M/routing_z_forget_layer_13\endcsname{81.1}
\expandafter\def\csname ev@intern/bio/Routing-F1/0M/routing_z_forget_layer_14\endcsname{83.7}
\expandafter\def\csname ev@intern/bio/Routing-F1/0M/routing_z_forget_layer_15\endcsname{84.1}
\expandafter\def\csname ev@intern/bio/Routing-F1/0M/routing_z_forget_layer_16\endcsname{79.4}
\expandafter\def\csname ev@intern/bio/Routing-F1/0M/routing_z_forget_layer_17\endcsname{80.2}
\expandafter\def\csname ev@intern/bio/Routing-F1/0M/routing_z_forget_layer_18\endcsname{82.4}
\expandafter\def\csname ev@intern/bio/Routing-F1/0M/routing_z_forget_layer_19\endcsname{80.4}
\expandafter\def\csname ev@intern/bio/Routing-F1/0M/routing_z_forget_layer_20\endcsname{85.4}
\expandafter\def\csname ev@intern/bio/Routing-F1/0M/routing_z_forget_layer_21\endcsname{83.6}
\expandafter\def\csname ev@intern/bio/Routing-F1/0M/routing_z_forget_layer_22\endcsname{77.2}
\expandafter\def\csname ev@intern/bio/Routing-F1/0M/routing_z_forget_layer_23\endcsname{74.1}
\expandafter\def\csname ev@intern/bio/Routing-F1/0M/routing_z_forget_layer_24\endcsname{80.5}
\expandafter\def\csname ev@intern/bio/Routing-F1/0M/routing_z_forget_layer_25\endcsname{79.1}
\expandafter\def\csname ev@intern/bio/Routing-F1/0M/routing_z_forget_layer_26\endcsname{79.0}
\expandafter\def\csname ev@intern/bio/Routing-F1/0M/routing_z_forget_layer_27\endcsname{78.5}
\expandafter\def\csname ev@intern/bio/Routing-F1/0M/routing_z_forget_layer_28\endcsname{72.3}
\expandafter\def\csname ev@intern/bio/Routing-F1/0M/routing_z_forget_layer_29\endcsname{72.4}
\expandafter\def\csname ev@intern/bio/Routing-F1/0M/routing_z_forget_layer_30\endcsname{77.3}
\expandafter\def\csname ev@intern/bio/Routing-F1/0M/routing_z_forget_layer_31\endcsname{73.8}
\expandafter\def\csname ev@intern/bio/Routing-F1/0M/routing_z_forget_layer_32\endcsname{66.3}
\expandafter\def\csname ev@intern/bio/Routing-F1/0M/routing_z_forget_layer_33\endcsname{69.1}
\expandafter\def\csname ev@intern/bio/Routing-F1/0M/routing_z_forget_layer_34\endcsname{67.6}
\expandafter\def\csname ev@intern/bio/Routing-F1/0M/routing_z_forget_layer_35\endcsname{80.7}
\expandafter\def\csname ev@intern/bio/Routing-F1/0M/routing_z_forget_layer_36\endcsname{75.5}
\expandafter\def\csname ev@intern/bio/Routing-F1/0M/routing_z_forget_layer_37\endcsname{76.3}
\expandafter\def\csname ev@intern/bio/Routing-F1/0M/routing_z_forget_layer_38\endcsname{74.3}
\expandafter\def\csname ev@intern/bio/Routing-F1/0M/routing_z_forget_layer_39\endcsname{56.7}
\expandafter\def\csname ev@intern/bio/Routing-F1/0M/routing_z_retain\endcsname{20.2}
\expandafter\def\csname ev@intern/bio/Routing-F1/0M/routing_z_retain_layer_0\endcsname{0.1}
\expandafter\def\csname ev@intern/bio/Routing-F1/0M/routing_z_retain_layer_1\endcsname{10.0}
\expandafter\def\csname ev@intern/bio/Routing-F1/0M/routing_z_retain_layer_2\endcsname{10.3}
\expandafter\def\csname ev@intern/bio/Routing-F1/0M/routing_z_retain_layer_3\endcsname{21.0}
\expandafter\def\csname ev@intern/bio/Routing-F1/0M/routing_z_retain_layer_4\endcsname{25.6}
\expandafter\def\csname ev@intern/bio/Routing-F1/0M/routing_z_retain_layer_5\endcsname{25.2}
\expandafter\def\csname ev@intern/bio/Routing-F1/0M/routing_z_retain_layer_6\endcsname{24.7}
\expandafter\def\csname ev@intern/bio/Routing-F1/0M/routing_z_retain_layer_7\endcsname{25.1}
\expandafter\def\csname ev@intern/bio/Routing-F1/0M/routing_z_retain_layer_8\endcsname{24.3}
\expandafter\def\csname ev@intern/bio/Routing-F1/0M/routing_z_retain_layer_9\endcsname{23.1}
\expandafter\def\csname ev@intern/bio/Routing-F1/0M/routing_z_retain_layer_10\endcsname{22.5}
\expandafter\def\csname ev@intern/bio/Routing-F1/0M/routing_z_retain_layer_11\endcsname{21.7}
\expandafter\def\csname ev@intern/bio/Routing-F1/0M/routing_z_retain_layer_12\endcsname{23.6}
\expandafter\def\csname ev@intern/bio/Routing-F1/0M/routing_z_retain_layer_13\endcsname{23.5}
\expandafter\def\csname ev@intern/bio/Routing-F1/0M/routing_z_retain_layer_14\endcsname{23.7}
\expandafter\def\csname ev@intern/bio/Routing-F1/0M/routing_z_retain_layer_15\endcsname{23.4}
\expandafter\def\csname ev@intern/bio/Routing-F1/0M/routing_z_retain_layer_16\endcsname{22.7}
\expandafter\def\csname ev@intern/bio/Routing-F1/0M/routing_z_retain_layer_17\endcsname{22.7}
\expandafter\def\csname ev@intern/bio/Routing-F1/0M/routing_z_retain_layer_18\endcsname{24.1}
\expandafter\def\csname ev@intern/bio/Routing-F1/0M/routing_z_retain_layer_19\endcsname{23.6}
\expandafter\def\csname ev@intern/bio/Routing-F1/0M/routing_z_retain_layer_20\endcsname{24.1}
\expandafter\def\csname ev@intern/bio/Routing-F1/0M/routing_z_retain_layer_21\endcsname{23.6}
\expandafter\def\csname ev@intern/bio/Routing-F1/0M/routing_z_retain_layer_22\endcsname{22.8}
\expandafter\def\csname ev@intern/bio/Routing-F1/0M/routing_z_retain_layer_23\endcsname{22.5}
\expandafter\def\csname ev@intern/bio/Routing-F1/0M/routing_z_retain_layer_24\endcsname{23.4}
\expandafter\def\csname ev@intern/bio/Routing-F1/0M/routing_z_retain_layer_25\endcsname{23.2}
\expandafter\def\csname ev@intern/bio/Routing-F1/0M/routing_z_retain_layer_26\endcsname{22.4}
\expandafter\def\csname ev@intern/bio/Routing-F1/0M/routing_z_retain_layer_27\endcsname{21.7}
\expandafter\def\csname ev@intern/bio/Routing-F1/0M/routing_z_retain_layer_28\endcsname{21.0}
\expandafter\def\csname ev@intern/bio/Routing-F1/0M/routing_z_retain_layer_29\endcsname{20.2}
\expandafter\def\csname ev@intern/bio/Routing-F1/0M/routing_z_retain_layer_30\endcsname{21.2}
\expandafter\def\csname ev@intern/bio/Routing-F1/0M/routing_z_retain_layer_31\endcsname{20.3}
\expandafter\def\csname ev@intern/bio/Routing-F1/0M/routing_z_retain_layer_32\endcsname{17.4}
\expandafter\def\csname ev@intern/bio/Routing-F1/0M/routing_z_retain_layer_33\endcsname{16.4}
\expandafter\def\csname ev@intern/bio/Routing-F1/0M/routing_z_retain_layer_34\endcsname{16.4}
\expandafter\def\csname ev@intern/bio/Routing-F1/0M/routing_z_retain_layer_35\endcsname{18.0}
\expandafter\def\csname ev@intern/bio/Routing-F1/0M/routing_z_retain_layer_36\endcsname{16.3}
\expandafter\def\csname ev@intern/bio/Routing-F1/0M/routing_z_retain_layer_37\endcsname{15.3}
\expandafter\def\csname ev@intern/bio/Routing-F1/0M/routing_z_retain_layer_38\endcsname{13.4}
\expandafter\def\csname ev@intern/bio/Routing-F1/0M/routing_z_retain_layer_39\endcsname{6.8}
\expandafter\def\csname ev@qwen/bio/Routing-F0/0M/routing_z_forget\endcsname{17.1}
\expandafter\def\csname ev@qwen/bio/Routing-F0/0M/routing_z_forget_layer_0\endcsname{8.0}
\expandafter\def\csname ev@qwen/bio/Routing-F0/0M/routing_z_forget_layer_1\endcsname{7.4}
\expandafter\def\csname ev@qwen/bio/Routing-F0/0M/routing_z_forget_layer_2\endcsname{5.6}
\expandafter\def\csname ev@qwen/bio/Routing-F0/0M/routing_z_forget_layer_3\endcsname{12.4}
\expandafter\def\csname ev@qwen/bio/Routing-F0/0M/routing_z_forget_layer_4\endcsname{15.9}
\expandafter\def\csname ev@qwen/bio/Routing-F0/0M/routing_z_forget_layer_5\endcsname{16.0}
\expandafter\def\csname ev@qwen/bio/Routing-F0/0M/routing_z_forget_layer_6\endcsname{13.2}
\expandafter\def\csname ev@qwen/bio/Routing-F0/0M/routing_z_forget_layer_7\endcsname{17.0}
\expandafter\def\csname ev@qwen/bio/Routing-F0/0M/routing_z_forget_layer_8\endcsname{23.0}
\expandafter\def\csname ev@qwen/bio/Routing-F0/0M/routing_z_forget_layer_9\endcsname{21.6}
\expandafter\def\csname ev@qwen/bio/Routing-F0/0M/routing_z_forget_layer_10\endcsname{11.7}
\expandafter\def\csname ev@qwen/bio/Routing-F0/0M/routing_z_forget_layer_11\endcsname{11.6}
\expandafter\def\csname ev@qwen/bio/Routing-F0/0M/routing_z_forget_layer_12\endcsname{18.5}
\expandafter\def\csname ev@qwen/bio/Routing-F0/0M/routing_z_forget_layer_13\endcsname{19.1}
\expandafter\def\csname ev@qwen/bio/Routing-F0/0M/routing_z_forget_layer_14\endcsname{20.9}
\expandafter\def\csname ev@qwen/bio/Routing-F0/0M/routing_z_forget_layer_15\endcsname{20.8}
\expandafter\def\csname ev@qwen/bio/Routing-F0/0M/routing_z_forget_layer_16\endcsname{13.0}
\expandafter\def\csname ev@qwen/bio/Routing-F0/0M/routing_z_forget_layer_17\endcsname{20.2}
\expandafter\def\csname ev@qwen/bio/Routing-F0/0M/routing_z_forget_layer_18\endcsname{20.7}
\expandafter\def\csname ev@qwen/bio/Routing-F0/0M/routing_z_forget_layer_19\endcsname{19.0}
\expandafter\def\csname ev@qwen/bio/Routing-F0/0M/routing_z_forget_layer_20\endcsname{24.0}
\expandafter\def\csname ev@qwen/bio/Routing-F0/0M/routing_z_forget_layer_21\endcsname{23.5}
\expandafter\def\csname ev@qwen/bio/Routing-F0/0M/routing_z_forget_layer_22\endcsname{13.4}
\expandafter\def\csname ev@qwen/bio/Routing-F0/0M/routing_z_forget_layer_23\endcsname{12.1}
\expandafter\def\csname ev@qwen/bio/Routing-F0/0M/routing_z_forget_layer_24\endcsname{18.0}
\expandafter\def\csname ev@qwen/bio/Routing-F0/0M/routing_z_forget_layer_25\endcsname{18.6}
\expandafter\def\csname ev@qwen/bio/Routing-F0/0M/routing_z_forget_layer_26\endcsname{18.8}
\expandafter\def\csname ev@qwen/bio/Routing-F0/0M/routing_z_forget_layer_27\endcsname{20.1}
\expandafter\def\csname ev@qwen/bio/Routing-F0/0M/routing_z_forget_layer_28\endcsname{12.1}
\expandafter\def\csname ev@qwen/bio/Routing-F0/0M/routing_z_forget_layer_29\endcsname{20.1}
\expandafter\def\csname ev@qwen/bio/Routing-F0/0M/routing_z_forget_layer_30\endcsname{21.2}
\expandafter\def\csname ev@qwen/bio/Routing-F0/0M/routing_z_forget_layer_31\endcsname{18.6}
\expandafter\def\csname ev@qwen/bio/Routing-F0/0M/routing_z_forget_layer_32\endcsname{15.7}
\expandafter\def\csname ev@qwen/bio/Routing-F0/0M/routing_z_forget_layer_33\endcsname{19.0}
\expandafter\def\csname ev@qwen/bio/Routing-F0/0M/routing_z_forget_layer_34\endcsname{13.8}
\expandafter\def\csname ev@qwen/bio/Routing-F0/0M/routing_z_forget_layer_35\endcsname{23.3}
\expandafter\def\csname ev@qwen/bio/Routing-F0/0M/routing_z_forget_layer_36\endcsname{20.7}
\expandafter\def\csname ev@qwen/bio/Routing-F0/0M/routing_z_forget_layer_37\endcsname{19.8}
\expandafter\def\csname ev@qwen/bio/Routing-F0/0M/routing_z_forget_layer_38\endcsname{20.9}
\expandafter\def\csname ev@qwen/bio/Routing-F0/0M/routing_z_forget_layer_39\endcsname{16.1}
\expandafter\def\csname ev@qwen/bio/Routing-F0/0M/routing_z_retain\endcsname{3.1}
\expandafter\def\csname ev@qwen/bio/Routing-F0/0M/routing_z_retain_layer_0\endcsname{2.1}
\expandafter\def\csname ev@qwen/bio/Routing-F0/0M/routing_z_retain_layer_1\endcsname{1.6}
\expandafter\def\csname ev@qwen/bio/Routing-F0/0M/routing_z_retain_layer_2\endcsname{2.1}
\expandafter\def\csname ev@qwen/bio/Routing-F0/0M/routing_z_retain_layer_3\endcsname{2.6}
\expandafter\def\csname ev@qwen/bio/Routing-F0/0M/routing_z_retain_layer_4\endcsname{2.8}
\expandafter\def\csname ev@qwen/bio/Routing-F0/0M/routing_z_retain_layer_5\endcsname{2.3}
\expandafter\def\csname ev@qwen/bio/Routing-F0/0M/routing_z_retain_layer_6\endcsname{2.6}
\expandafter\def\csname ev@qwen/bio/Routing-F0/0M/routing_z_retain_layer_7\endcsname{3.3}
\expandafter\def\csname ev@qwen/bio/Routing-F0/0M/routing_z_retain_layer_8\endcsname{2.8}
\expandafter\def\csname ev@qwen/bio/Routing-F0/0M/routing_z_retain_layer_9\endcsname{3.7}
\expandafter\def\csname ev@qwen/bio/Routing-F0/0M/routing_z_retain_layer_10\endcsname{3.0}
\expandafter\def\csname ev@qwen/bio/Routing-F0/0M/routing_z_retain_layer_11\endcsname{2.4}
\expandafter\def\csname ev@qwen/bio/Routing-F0/0M/routing_z_retain_layer_12\endcsname{3.1}
\expandafter\def\csname ev@qwen/bio/Routing-F0/0M/routing_z_retain_layer_13\endcsname{3.1}
\expandafter\def\csname ev@qwen/bio/Routing-F0/0M/routing_z_retain_layer_14\endcsname{3.3}
\expandafter\def\csname ev@qwen/bio/Routing-F0/0M/routing_z_retain_layer_15\endcsname{3.4}
\expandafter\def\csname ev@qwen/bio/Routing-F0/0M/routing_z_retain_layer_16\endcsname{2.2}
\expandafter\def\csname ev@qwen/bio/Routing-F0/0M/routing_z_retain_layer_17\endcsname{3.2}
\expandafter\def\csname ev@qwen/bio/Routing-F0/0M/routing_z_retain_layer_18\endcsname{3.6}
\expandafter\def\csname ev@qwen/bio/Routing-F0/0M/routing_z_retain_layer_19\endcsname{4.5}
\expandafter\def\csname ev@qwen/bio/Routing-F0/0M/routing_z_retain_layer_20\endcsname{3.1}
\expandafter\def\csname ev@qwen/bio/Routing-F0/0M/routing_z_retain_layer_21\endcsname{3.3}
\expandafter\def\csname ev@qwen/bio/Routing-F0/0M/routing_z_retain_layer_22\endcsname{3.2}
\expandafter\def\csname ev@qwen/bio/Routing-F0/0M/routing_z_retain_layer_23\endcsname{2.4}
\expandafter\def\csname ev@qwen/bio/Routing-F0/0M/routing_z_retain_layer_24\endcsname{3.0}
\expandafter\def\csname ev@qwen/bio/Routing-F0/0M/routing_z_retain_layer_25\endcsname{3.8}
\expandafter\def\csname ev@qwen/bio/Routing-F0/0M/routing_z_retain_layer_26\endcsname{3.4}
\expandafter\def\csname ev@qwen/bio/Routing-F0/0M/routing_z_retain_layer_27\endcsname{3.8}
\expandafter\def\csname ev@qwen/bio/Routing-F0/0M/routing_z_retain_layer_28\endcsname{2.2}
\expandafter\def\csname ev@qwen/bio/Routing-F0/0M/routing_z_retain_layer_29\endcsname{3.7}
\expandafter\def\csname ev@qwen/bio/Routing-F0/0M/routing_z_retain_layer_30\endcsname{3.7}
\expandafter\def\csname ev@qwen/bio/Routing-F0/0M/routing_z_retain_layer_31\endcsname{4.2}
\expandafter\def\csname ev@qwen/bio/Routing-F0/0M/routing_z_retain_layer_32\endcsname{2.5}
\expandafter\def\csname ev@qwen/bio/Routing-F0/0M/routing_z_retain_layer_33\endcsname{3.2}
\expandafter\def\csname ev@qwen/bio/Routing-F0/0M/routing_z_retain_layer_34\endcsname{3.1}
\expandafter\def\csname ev@qwen/bio/Routing-F0/0M/routing_z_retain_layer_35\endcsname{3.9}
\expandafter\def\csname ev@qwen/bio/Routing-F0/0M/routing_z_retain_layer_36\endcsname{3.4}
\expandafter\def\csname ev@qwen/bio/Routing-F0/0M/routing_z_retain_layer_37\endcsname{3.0}
\expandafter\def\csname ev@qwen/bio/Routing-F0/0M/routing_z_retain_layer_38\endcsname{4.2}
\expandafter\def\csname ev@qwen/bio/Routing-F0/0M/routing_z_retain_layer_39\endcsname{3.3}
\expandafter\def\csname ev@qwen/bio/Routing-F1/0M/routing_z_forget\endcsname{72.1}
\expandafter\def\csname ev@qwen/bio/Routing-F1/0M/routing_z_forget_layer_0\endcsname{12.2}
\expandafter\def\csname ev@qwen/bio/Routing-F1/0M/routing_z_forget_layer_1\endcsname{27.3}
\expandafter\def\csname ev@qwen/bio/Routing-F1/0M/routing_z_forget_layer_2\endcsname{7.8}
\expandafter\def\csname ev@qwen/bio/Routing-F1/0M/routing_z_forget_layer_3\endcsname{48.5}
\expandafter\def\csname ev@qwen/bio/Routing-F1/0M/routing_z_forget_layer_4\endcsname{51.6}
\expandafter\def\csname ev@qwen/bio/Routing-F1/0M/routing_z_forget_layer_5\endcsname{52.1}
\expandafter\def\csname ev@qwen/bio/Routing-F1/0M/routing_z_forget_layer_6\endcsname{43.9}
\expandafter\def\csname ev@qwen/bio/Routing-F1/0M/routing_z_forget_layer_7\endcsname{62.2}
\expandafter\def\csname ev@qwen/bio/Routing-F1/0M/routing_z_forget_layer_8\endcsname{74.9}
\expandafter\def\csname ev@qwen/bio/Routing-F1/0M/routing_z_forget_layer_9\endcsname{70.9}
\expandafter\def\csname ev@qwen/bio/Routing-F1/0M/routing_z_forget_layer_10\endcsname{64.9}
\expandafter\def\csname ev@qwen/bio/Routing-F1/0M/routing_z_forget_layer_11\endcsname{58.9}
\expandafter\def\csname ev@qwen/bio/Routing-F1/0M/routing_z_forget_layer_12\endcsname{69.3}
\expandafter\def\csname ev@qwen/bio/Routing-F1/0M/routing_z_forget_layer_13\endcsname{77.4}
\expandafter\def\csname ev@qwen/bio/Routing-F1/0M/routing_z_forget_layer_14\endcsname{77.6}
\expandafter\def\csname ev@qwen/bio/Routing-F1/0M/routing_z_forget_layer_15\endcsname{80.8}
\expandafter\def\csname ev@qwen/bio/Routing-F1/0M/routing_z_forget_layer_16\endcsname{77.9}
\expandafter\def\csname ev@qwen/bio/Routing-F1/0M/routing_z_forget_layer_17\endcsname{81.3}
\expandafter\def\csname ev@qwen/bio/Routing-F1/0M/routing_z_forget_layer_18\endcsname{82.9}
\expandafter\def\csname ev@qwen/bio/Routing-F1/0M/routing_z_forget_layer_19\endcsname{81.7}
\expandafter\def\csname ev@qwen/bio/Routing-F1/0M/routing_z_forget_layer_20\endcsname{86.9}
\expandafter\def\csname ev@qwen/bio/Routing-F1/0M/routing_z_forget_layer_21\endcsname{85.0}
\expandafter\def\csname ev@qwen/bio/Routing-F1/0M/routing_z_forget_layer_22\endcsname{81.2}
\expandafter\def\csname ev@qwen/bio/Routing-F1/0M/routing_z_forget_layer_23\endcsname{81.4}
\expandafter\def\csname ev@qwen/bio/Routing-F1/0M/routing_z_forget_layer_24\endcsname{85.3}
\expandafter\def\csname ev@qwen/bio/Routing-F1/0M/routing_z_forget_layer_25\endcsname{85.9}
\expandafter\def\csname ev@qwen/bio/Routing-F1/0M/routing_z_forget_layer_26\endcsname{85.1}
\expandafter\def\csname ev@qwen/bio/Routing-F1/0M/routing_z_forget_layer_27\endcsname{84.4}
\expandafter\def\csname ev@qwen/bio/Routing-F1/0M/routing_z_forget_layer_28\endcsname{84.8}
\expandafter\def\csname ev@qwen/bio/Routing-F1/0M/routing_z_forget_layer_29\endcsname{84.6}
\expandafter\def\csname ev@qwen/bio/Routing-F1/0M/routing_z_forget_layer_30\endcsname{89.0}
\expandafter\def\csname ev@qwen/bio/Routing-F1/0M/routing_z_forget_layer_31\endcsname{87.5}
\expandafter\def\csname ev@qwen/bio/Routing-F1/0M/routing_z_forget_layer_32\endcsname{79.8}
\expandafter\def\csname ev@qwen/bio/Routing-F1/0M/routing_z_forget_layer_33\endcsname{77.2}
\expandafter\def\csname ev@qwen/bio/Routing-F1/0M/routing_z_forget_layer_34\endcsname{79.2}
\expandafter\def\csname ev@qwen/bio/Routing-F1/0M/routing_z_forget_layer_35\endcsname{88.8}
\expandafter\def\csname ev@qwen/bio/Routing-F1/0M/routing_z_forget_layer_36\endcsname{85.7}
\expandafter\def\csname ev@qwen/bio/Routing-F1/0M/routing_z_forget_layer_37\endcsname{86.4}
\expandafter\def\csname ev@qwen/bio/Routing-F1/0M/routing_z_forget_layer_38\endcsname{85.4}
\expandafter\def\csname ev@qwen/bio/Routing-F1/0M/routing_z_forget_layer_39\endcsname{76.6}
\expandafter\def\csname ev@qwen/bio/Routing-F1/0M/routing_z_retain\endcsname{1.8}
\expandafter\def\csname ev@qwen/bio/Routing-F1/0M/routing_z_retain_layer_0\endcsname{0.7}
\expandafter\def\csname ev@qwen/bio/Routing-F1/0M/routing_z_retain_layer_1\endcsname{1.0}
\expandafter\def\csname ev@qwen/bio/Routing-F1/0M/routing_z_retain_layer_2\endcsname{0.7}
\expandafter\def\csname ev@qwen/bio/Routing-F1/0M/routing_z_retain_layer_3\endcsname{2.0}
\expandafter\def\csname ev@qwen/bio/Routing-F1/0M/routing_z_retain_layer_4\endcsname{1.6}
\expandafter\def\csname ev@qwen/bio/Routing-F1/0M/routing_z_retain_layer_5\endcsname{1.5}
\expandafter\def\csname ev@qwen/bio/Routing-F1/0M/routing_z_retain_layer_6\endcsname{1.4}
\expandafter\def\csname ev@qwen/bio/Routing-F1/0M/routing_z_retain_layer_7\endcsname{2.0}
\expandafter\def\csname ev@qwen/bio/Routing-F1/0M/routing_z_retain_layer_8\endcsname{2.4}
\expandafter\def\csname ev@qwen/bio/Routing-F1/0M/routing_z_retain_layer_9\endcsname{1.9}
\expandafter\def\csname ev@qwen/bio/Routing-F1/0M/routing_z_retain_layer_10\endcsname{1.6}
\expandafter\def\csname ev@qwen/bio/Routing-F1/0M/routing_z_retain_layer_11\endcsname{1.3}
\expandafter\def\csname ev@qwen/bio/Routing-F1/0M/routing_z_retain_layer_12\endcsname{1.6}
\expandafter\def\csname ev@qwen/bio/Routing-F1/0M/routing_z_retain_layer_13\endcsname{2.0}
\expandafter\def\csname ev@qwen/bio/Routing-F1/0M/routing_z_retain_layer_14\endcsname{2.1}
\expandafter\def\csname ev@qwen/bio/Routing-F1/0M/routing_z_retain_layer_15\endcsname{2.0}
\expandafter\def\csname ev@qwen/bio/Routing-F1/0M/routing_z_retain_layer_16\endcsname{1.6}
\expandafter\def\csname ev@qwen/bio/Routing-F1/0M/routing_z_retain_layer_17\endcsname{1.9}
\expandafter\def\csname ev@qwen/bio/Routing-F1/0M/routing_z_retain_layer_18\endcsname{2.2}
\expandafter\def\csname ev@qwen/bio/Routing-F1/0M/routing_z_retain_layer_19\endcsname{2.1}
\expandafter\def\csname ev@qwen/bio/Routing-F1/0M/routing_z_retain_layer_20\endcsname{2.4}
\expandafter\def\csname ev@qwen/bio/Routing-F1/0M/routing_z_retain_layer_21\endcsname{2.2}
\expandafter\def\csname ev@qwen/bio/Routing-F1/0M/routing_z_retain_layer_22\endcsname{1.9}
\expandafter\def\csname ev@qwen/bio/Routing-F1/0M/routing_z_retain_layer_23\endcsname{1.8}
\expandafter\def\csname ev@qwen/bio/Routing-F1/0M/routing_z_retain_layer_24\endcsname{2.0}
\expandafter\def\csname ev@qwen/bio/Routing-F1/0M/routing_z_retain_layer_25\endcsname{2.1}
\expandafter\def\csname ev@qwen/bio/Routing-F1/0M/routing_z_retain_layer_26\endcsname{2.1}
\expandafter\def\csname ev@qwen/bio/Routing-F1/0M/routing_z_retain_layer_27\endcsname{2.0}
\expandafter\def\csname ev@qwen/bio/Routing-F1/0M/routing_z_retain_layer_28\endcsname{1.8}
\expandafter\def\csname ev@qwen/bio/Routing-F1/0M/routing_z_retain_layer_29\endcsname{2.0}
\expandafter\def\csname ev@qwen/bio/Routing-F1/0M/routing_z_retain_layer_30\endcsname{2.4}
\expandafter\def\csname ev@qwen/bio/Routing-F1/0M/routing_z_retain_layer_31\endcsname{2.2}
\expandafter\def\csname ev@qwen/bio/Routing-F1/0M/routing_z_retain_layer_32\endcsname{1.9}
\expandafter\def\csname ev@qwen/bio/Routing-F1/0M/routing_z_retain_layer_33\endcsname{1.7}
\expandafter\def\csname ev@qwen/bio/Routing-F1/0M/routing_z_retain_layer_34\endcsname{1.8}
\expandafter\def\csname ev@qwen/bio/Routing-F1/0M/routing_z_retain_layer_35\endcsname{2.2}
\expandafter\def\csname ev@qwen/bio/Routing-F1/0M/routing_z_retain_layer_36\endcsname{1.9}
\expandafter\def\csname ev@qwen/bio/Routing-F1/0M/routing_z_retain_layer_37\endcsname{1.9}
\expandafter\def\csname ev@qwen/bio/Routing-F1/0M/routing_z_retain_layer_38\endcsname{1.9}
\expandafter\def\csname ev@qwen/bio/Routing-F1/0M/routing_z_retain_layer_39\endcsname{1.6}
\expandafter\def\csname ev@gemma/bio/Routing-F0/0M/routing_z_forget\endcsname{26.2}
\expandafter\def\csname ev@gemma/bio/Routing-F0/0M/routing_z_forget_layer_0\endcsname{16.1}
\expandafter\def\csname ev@gemma/bio/Routing-F0/0M/routing_z_forget_layer_1\endcsname{15.9}
\expandafter\def\csname ev@gemma/bio/Routing-F0/0M/routing_z_forget_layer_2\endcsname{28.2}
\expandafter\def\csname ev@gemma/bio/Routing-F0/0M/routing_z_forget_layer_3\endcsname{29.5}
\expandafter\def\csname ev@gemma/bio/Routing-F0/0M/routing_z_forget_layer_4\endcsname{19.6}
\expandafter\def\csname ev@gemma/bio/Routing-F0/0M/routing_z_forget_layer_5\endcsname{33.1}
\expandafter\def\csname ev@gemma/bio/Routing-F0/0M/routing_z_forget_layer_6\endcsname{25.7}
\expandafter\def\csname ev@gemma/bio/Routing-F0/0M/routing_z_forget_layer_7\endcsname{20.8}
\expandafter\def\csname ev@gemma/bio/Routing-F0/0M/routing_z_forget_layer_8\endcsname{27.0}
\expandafter\def\csname ev@gemma/bio/Routing-F0/0M/routing_z_forget_layer_9\endcsname{20.1}
\expandafter\def\csname ev@gemma/bio/Routing-F0/0M/routing_z_forget_layer_10\endcsname{30.6}
\expandafter\def\csname ev@gemma/bio/Routing-F0/0M/routing_z_forget_layer_11\endcsname{30.9}
\expandafter\def\csname ev@gemma/bio/Routing-F0/0M/routing_z_forget_layer_12\endcsname{23.4}
\expandafter\def\csname ev@gemma/bio/Routing-F0/0M/routing_z_forget_layer_13\endcsname{26.5}
\expandafter\def\csname ev@gemma/bio/Routing-F0/0M/routing_z_forget_layer_14\endcsname{23.0}
\expandafter\def\csname ev@gemma/bio/Routing-F0/0M/routing_z_forget_layer_15\endcsname{23.1}
\expandafter\def\csname ev@gemma/bio/Routing-F0/0M/routing_z_forget_layer_16\endcsname{21.6}
\expandafter\def\csname ev@gemma/bio/Routing-F0/0M/routing_z_forget_layer_17\endcsname{28.5}
\expandafter\def\csname ev@gemma/bio/Routing-F0/0M/routing_z_forget_layer_18\endcsname{24.1}
\expandafter\def\csname ev@gemma/bio/Routing-F0/0M/routing_z_forget_layer_19\endcsname{23.6}
\expandafter\def\csname ev@gemma/bio/Routing-F0/0M/routing_z_forget_layer_20\endcsname{27.4}
\expandafter\def\csname ev@gemma/bio/Routing-F0/0M/routing_z_forget_layer_21\endcsname{29.7}
\expandafter\def\csname ev@gemma/bio/Routing-F0/0M/routing_z_forget_layer_22\endcsname{31.2}
\expandafter\def\csname ev@gemma/bio/Routing-F0/0M/routing_z_forget_layer_23\endcsname{29.7}
\expandafter\def\csname ev@gemma/bio/Routing-F0/0M/routing_z_forget_layer_24\endcsname{32.1}
\expandafter\def\csname ev@gemma/bio/Routing-F0/0M/routing_z_forget_layer_25\endcsname{32.1}
\expandafter\def\csname ev@gemma/bio/Routing-F0/0M/routing_z_forget_layer_26\endcsname{31.5}
\expandafter\def\csname ev@gemma/bio/Routing-F0/0M/routing_z_forget_layer_27\endcsname{28.3}
\expandafter\def\csname ev@gemma/bio/Routing-F0/0M/routing_z_forget_layer_28\endcsname{30.1}
\expandafter\def\csname ev@gemma/bio/Routing-F0/0M/routing_z_forget_layer_29\endcsname{21.2}
\expandafter\def\csname ev@gemma/bio/Routing-F0/0M/routing_z_retain\endcsname{8.9}
\expandafter\def\csname ev@gemma/bio/Routing-F0/0M/routing_z_retain_layer_0\endcsname{5.2}
\expandafter\def\csname ev@gemma/bio/Routing-F0/0M/routing_z_retain_layer_1\endcsname{5.2}
\expandafter\def\csname ev@gemma/bio/Routing-F0/0M/routing_z_retain_layer_2\endcsname{8.6}
\expandafter\def\csname ev@gemma/bio/Routing-F0/0M/routing_z_retain_layer_3\endcsname{9.7}
\expandafter\def\csname ev@gemma/bio/Routing-F0/0M/routing_z_retain_layer_4\endcsname{4.4}
\expandafter\def\csname ev@gemma/bio/Routing-F0/0M/routing_z_retain_layer_5\endcsname{10.1}
\expandafter\def\csname ev@gemma/bio/Routing-F0/0M/routing_z_retain_layer_6\endcsname{7.4}
\expandafter\def\csname ev@gemma/bio/Routing-F0/0M/routing_z_retain_layer_7\endcsname{8.3}
\expandafter\def\csname ev@gemma/bio/Routing-F0/0M/routing_z_retain_layer_8\endcsname{9.7}
\expandafter\def\csname ev@gemma/bio/Routing-F0/0M/routing_z_retain_layer_9\endcsname{6.1}
\expandafter\def\csname ev@gemma/bio/Routing-F0/0M/routing_z_retain_layer_10\endcsname{11.1}
\expandafter\def\csname ev@gemma/bio/Routing-F0/0M/routing_z_retain_layer_11\endcsname{12.0}
\expandafter\def\csname ev@gemma/bio/Routing-F0/0M/routing_z_retain_layer_12\endcsname{11.0}
\expandafter\def\csname ev@gemma/bio/Routing-F0/0M/routing_z_retain_layer_13\endcsname{8.1}
\expandafter\def\csname ev@gemma/bio/Routing-F0/0M/routing_z_retain_layer_14\endcsname{6.2}
\expandafter\def\csname ev@gemma/bio/Routing-F0/0M/routing_z_retain_layer_15\endcsname{14.0}
\expandafter\def\csname ev@gemma/bio/Routing-F0/0M/routing_z_retain_layer_16\endcsname{8.8}
\expandafter\def\csname ev@gemma/bio/Routing-F0/0M/routing_z_retain_layer_17\endcsname{15.0}
\expandafter\def\csname ev@gemma/bio/Routing-F0/0M/routing_z_retain_layer_18\endcsname{9.2}
\expandafter\def\csname ev@gemma/bio/Routing-F0/0M/routing_z_retain_layer_19\endcsname{9.9}
\expandafter\def\csname ev@gemma/bio/Routing-F0/0M/routing_z_retain_layer_20\endcsname{8.1}
\expandafter\def\csname ev@gemma/bio/Routing-F0/0M/routing_z_retain_layer_21\endcsname{8.9}
\expandafter\def\csname ev@gemma/bio/Routing-F0/0M/routing_z_retain_layer_22\endcsname{10.7}
\expandafter\def\csname ev@gemma/bio/Routing-F0/0M/routing_z_retain_layer_23\endcsname{5.8}
\expandafter\def\csname ev@gemma/bio/Routing-F0/0M/routing_z_retain_layer_24\endcsname{8.2}
\expandafter\def\csname ev@gemma/bio/Routing-F0/0M/routing_z_retain_layer_25\endcsname{9.7}
\expandafter\def\csname ev@gemma/bio/Routing-F0/0M/routing_z_retain_layer_26\endcsname{8.3}
\expandafter\def\csname ev@gemma/bio/Routing-F0/0M/routing_z_retain_layer_27\endcsname{8.9}
\expandafter\def\csname ev@gemma/bio/Routing-F0/0M/routing_z_retain_layer_28\endcsname{9.7}
\expandafter\def\csname ev@gemma/bio/Routing-F0/0M/routing_z_retain_layer_29\endcsname{7.6}
\expandafter\def\csname ev@gemma/bio/Routing-F1/0M/routing_z_forget\endcsname{65.9}
\expandafter\def\csname ev@gemma/bio/Routing-F1/0M/routing_z_forget_layer_0\endcsname{18.0}
\expandafter\def\csname ev@gemma/bio/Routing-F1/0M/routing_z_forget_layer_1\endcsname{18.2}
\expandafter\def\csname ev@gemma/bio/Routing-F1/0M/routing_z_forget_layer_2\endcsname{35.5}
\expandafter\def\csname ev@gemma/bio/Routing-F1/0M/routing_z_forget_layer_3\endcsname{36.3}
\expandafter\def\csname ev@gemma/bio/Routing-F1/0M/routing_z_forget_layer_4\endcsname{29.2}
\expandafter\def\csname ev@gemma/bio/Routing-F1/0M/routing_z_forget_layer_5\endcsname{48.1}
\expandafter\def\csname ev@gemma/bio/Routing-F1/0M/routing_z_forget_layer_6\endcsname{43.0}
\expandafter\def\csname ev@gemma/bio/Routing-F1/0M/routing_z_forget_layer_7\endcsname{42.2}
\expandafter\def\csname ev@gemma/bio/Routing-F1/0M/routing_z_forget_layer_8\endcsname{55.7}
\expandafter\def\csname ev@gemma/bio/Routing-F1/0M/routing_z_forget_layer_9\endcsname{59.0}
\expandafter\def\csname ev@gemma/bio/Routing-F1/0M/routing_z_forget_layer_10\endcsname{72.1}
\expandafter\def\csname ev@gemma/bio/Routing-F1/0M/routing_z_forget_layer_11\endcsname{73.1}
\expandafter\def\csname ev@gemma/bio/Routing-F1/0M/routing_z_forget_layer_12\endcsname{73.6}
\expandafter\def\csname ev@gemma/bio/Routing-F1/0M/routing_z_forget_layer_13\endcsname{75.8}
\expandafter\def\csname ev@gemma/bio/Routing-F1/0M/routing_z_forget_layer_14\endcsname{76.5}
\expandafter\def\csname ev@gemma/bio/Routing-F1/0M/routing_z_forget_layer_15\endcsname{72.1}
\expandafter\def\csname ev@gemma/bio/Routing-F1/0M/routing_z_forget_layer_16\endcsname{84.6}
\expandafter\def\csname ev@gemma/bio/Routing-F1/0M/routing_z_forget_layer_17\endcsname{82.1}
\expandafter\def\csname ev@gemma/bio/Routing-F1/0M/routing_z_forget_layer_18\endcsname{80.7}
\expandafter\def\csname ev@gemma/bio/Routing-F1/0M/routing_z_forget_layer_19\endcsname{81.3}
\expandafter\def\csname ev@gemma/bio/Routing-F1/0M/routing_z_forget_layer_20\endcsname{81.1}
\expandafter\def\csname ev@gemma/bio/Routing-F1/0M/routing_z_forget_layer_21\endcsname{82.3}
\expandafter\def\csname ev@gemma/bio/Routing-F1/0M/routing_z_forget_layer_22\endcsname{85.5}
\expandafter\def\csname ev@gemma/bio/Routing-F1/0M/routing_z_forget_layer_23\endcsname{82.3}
\expandafter\def\csname ev@gemma/bio/Routing-F1/0M/routing_z_forget_layer_24\endcsname{81.9}
\expandafter\def\csname ev@gemma/bio/Routing-F1/0M/routing_z_forget_layer_25\endcsname{83.4}
\expandafter\def\csname ev@gemma/bio/Routing-F1/0M/routing_z_forget_layer_26\endcsname{81.5}
\expandafter\def\csname ev@gemma/bio/Routing-F1/0M/routing_z_forget_layer_27\endcsname{80.9}
\expandafter\def\csname ev@gemma/bio/Routing-F1/0M/routing_z_forget_layer_28\endcsname{85.3}
\expandafter\def\csname ev@gemma/bio/Routing-F1/0M/routing_z_forget_layer_29\endcsname{75.7}
\expandafter\def\csname ev@gemma/bio/Routing-F1/0M/routing_z_retain\endcsname{8.5}
\expandafter\def\csname ev@gemma/bio/Routing-F1/0M/routing_z_retain_layer_0\endcsname{4.4}
\expandafter\def\csname ev@gemma/bio/Routing-F1/0M/routing_z_retain_layer_1\endcsname{4.0}
\expandafter\def\csname ev@gemma/bio/Routing-F1/0M/routing_z_retain_layer_2\endcsname{7.6}
\expandafter\def\csname ev@gemma/bio/Routing-F1/0M/routing_z_retain_layer_3\endcsname{8.4}
\expandafter\def\csname ev@gemma/bio/Routing-F1/0M/routing_z_retain_layer_4\endcsname{5.0}
\expandafter\def\csname ev@gemma/bio/Routing-F1/0M/routing_z_retain_layer_5\endcsname{8.5}
\expandafter\def\csname ev@gemma/bio/Routing-F1/0M/routing_z_retain_layer_6\endcsname{6.7}
\expandafter\def\csname ev@gemma/bio/Routing-F1/0M/routing_z_retain_layer_7\endcsname{6.2}
\expandafter\def\csname ev@gemma/bio/Routing-F1/0M/routing_z_retain_layer_8\endcsname{6.7}
\expandafter\def\csname ev@gemma/bio/Routing-F1/0M/routing_z_retain_layer_9\endcsname{6.5}
\expandafter\def\csname ev@gemma/bio/Routing-F1/0M/routing_z_retain_layer_10\endcsname{10.0}
\expandafter\def\csname ev@gemma/bio/Routing-F1/0M/routing_z_retain_layer_11\endcsname{10.6}
\expandafter\def\csname ev@gemma/bio/Routing-F1/0M/routing_z_retain_layer_12\endcsname{11.5}
\expandafter\def\csname ev@gemma/bio/Routing-F1/0M/routing_z_retain_layer_13\endcsname{8.2}
\expandafter\def\csname ev@gemma/bio/Routing-F1/0M/routing_z_retain_layer_14\endcsname{7.3}
\expandafter\def\csname ev@gemma/bio/Routing-F1/0M/routing_z_retain_layer_15\endcsname{15.3}
\expandafter\def\csname ev@gemma/bio/Routing-F1/0M/routing_z_retain_layer_16\endcsname{11.0}
\expandafter\def\csname ev@gemma/bio/Routing-F1/0M/routing_z_retain_layer_17\endcsname{15.3}
\expandafter\def\csname ev@gemma/bio/Routing-F1/0M/routing_z_retain_layer_18\endcsname{9.9}
\expandafter\def\csname ev@gemma/bio/Routing-F1/0M/routing_z_retain_layer_19\endcsname{10.3}
\expandafter\def\csname ev@gemma/bio/Routing-F1/0M/routing_z_retain_layer_20\endcsname{8.6}
\expandafter\def\csname ev@gemma/bio/Routing-F1/0M/routing_z_retain_layer_21\endcsname{9.4}
\expandafter\def\csname ev@gemma/bio/Routing-F1/0M/routing_z_retain_layer_22\endcsname{9.6}
\expandafter\def\csname ev@gemma/bio/Routing-F1/0M/routing_z_retain_layer_23\endcsname{7.4}
\expandafter\def\csname ev@gemma/bio/Routing-F1/0M/routing_z_retain_layer_24\endcsname{7.9}
\expandafter\def\csname ev@gemma/bio/Routing-F1/0M/routing_z_retain_layer_25\endcsname{8.7}
\expandafter\def\csname ev@gemma/bio/Routing-F1/0M/routing_z_retain_layer_26\endcsname{8.2}
\expandafter\def\csname ev@gemma/bio/Routing-F1/0M/routing_z_retain_layer_27\endcsname{7.0}
\expandafter\def\csname ev@gemma/bio/Routing-F1/0M/routing_z_retain_layer_28\endcsname{7.8}
\expandafter\def\csname ev@gemma/bio/Routing-F1/0M/routing_z_retain_layer_29\endcsname{5.7}
\expandafter\def\csname ev@intern/cyber/Routing-F0/0M/routing_z_forget\endcsname{19.5}
\expandafter\def\csname ev@intern/cyber/Routing-F0/0M/routing_z_forget_layer_0\endcsname{10.6}
\expandafter\def\csname ev@intern/cyber/Routing-F0/0M/routing_z_forget_layer_1\endcsname{14.0}
\expandafter\def\csname ev@intern/cyber/Routing-F0/0M/routing_z_forget_layer_2\endcsname{7.9}
\expandafter\def\csname ev@intern/cyber/Routing-F0/0M/routing_z_forget_layer_3\endcsname{17.4}
\expandafter\def\csname ev@intern/cyber/Routing-F0/0M/routing_z_forget_layer_4\endcsname{19.2}
\expandafter\def\csname ev@intern/cyber/Routing-F0/0M/routing_z_forget_layer_5\endcsname{23.1}
\expandafter\def\csname ev@intern/cyber/Routing-F0/0M/routing_z_forget_layer_6\endcsname{20.4}
\expandafter\def\csname ev@intern/cyber/Routing-F0/0M/routing_z_forget_layer_7\endcsname{26.6}
\expandafter\def\csname ev@intern/cyber/Routing-F0/0M/routing_z_forget_layer_8\endcsname{27.6}
\expandafter\def\csname ev@intern/cyber/Routing-F0/0M/routing_z_forget_layer_9\endcsname{25.8}
\expandafter\def\csname ev@intern/cyber/Routing-F0/0M/routing_z_forget_layer_10\endcsname{16.1}
\expandafter\def\csname ev@intern/cyber/Routing-F0/0M/routing_z_forget_layer_11\endcsname{16.6}
\expandafter\def\csname ev@intern/cyber/Routing-F0/0M/routing_z_forget_layer_12\endcsname{19.3}
\expandafter\def\csname ev@intern/cyber/Routing-F0/0M/routing_z_forget_layer_13\endcsname{21.5}
\expandafter\def\csname ev@intern/cyber/Routing-F0/0M/routing_z_forget_layer_14\endcsname{23.3}
\expandafter\def\csname ev@intern/cyber/Routing-F0/0M/routing_z_forget_layer_15\endcsname{19.5}
\expandafter\def\csname ev@intern/cyber/Routing-F0/0M/routing_z_forget_layer_16\endcsname{15.7}
\expandafter\def\csname ev@intern/cyber/Routing-F0/0M/routing_z_forget_layer_17\endcsname{18.4}
\expandafter\def\csname ev@intern/cyber/Routing-F0/0M/routing_z_forget_layer_18\endcsname{21.6}
\expandafter\def\csname ev@intern/cyber/Routing-F0/0M/routing_z_forget_layer_19\endcsname{19.9}
\expandafter\def\csname ev@intern/cyber/Routing-F0/0M/routing_z_forget_layer_20\endcsname{26.3}
\expandafter\def\csname ev@intern/cyber/Routing-F0/0M/routing_z_forget_layer_21\endcsname{25.6}
\expandafter\def\csname ev@intern/cyber/Routing-F0/0M/routing_z_forget_layer_22\endcsname{15.9}
\expandafter\def\csname ev@intern/cyber/Routing-F0/0M/routing_z_forget_layer_23\endcsname{18.5}
\expandafter\def\csname ev@intern/cyber/Routing-F0/0M/routing_z_forget_layer_24\endcsname{17.7}
\expandafter\def\csname ev@intern/cyber/Routing-F0/0M/routing_z_forget_layer_25\endcsname{19.7}
\expandafter\def\csname ev@intern/cyber/Routing-F0/0M/routing_z_forget_layer_26\endcsname{21.1}
\expandafter\def\csname ev@intern/cyber/Routing-F0/0M/routing_z_forget_layer_27\endcsname{17.3}
\expandafter\def\csname ev@intern/cyber/Routing-F0/0M/routing_z_forget_layer_28\endcsname{14.7}
\expandafter\def\csname ev@intern/cyber/Routing-F0/0M/routing_z_forget_layer_29\endcsname{16.9}
\expandafter\def\csname ev@intern/cyber/Routing-F0/0M/routing_z_forget_layer_30\endcsname{20.4}
\expandafter\def\csname ev@intern/cyber/Routing-F0/0M/routing_z_forget_layer_31\endcsname{20.2}
\expandafter\def\csname ev@intern/cyber/Routing-F0/0M/routing_z_forget_layer_32\endcsname{15.5}
\expandafter\def\csname ev@intern/cyber/Routing-F0/0M/routing_z_forget_layer_33\endcsname{20.2}
\expandafter\def\csname ev@intern/cyber/Routing-F0/0M/routing_z_forget_layer_34\endcsname{16.1}
\expandafter\def\csname ev@intern/cyber/Routing-F0/0M/routing_z_forget_layer_35\endcsname{23.3}
\expandafter\def\csname ev@intern/cyber/Routing-F0/0M/routing_z_forget_layer_36\endcsname{26.3}
\expandafter\def\csname ev@intern/cyber/Routing-F0/0M/routing_z_forget_layer_37\endcsname{21.3}
\expandafter\def\csname ev@intern/cyber/Routing-F0/0M/routing_z_forget_layer_38\endcsname{22.9}
\expandafter\def\csname ev@intern/cyber/Routing-F0/0M/routing_z_forget_layer_39\endcsname{15.6}
\expandafter\def\csname ev@intern/cyber/Routing-F0/0M/routing_z_retain\endcsname{0.7}
\expandafter\def\csname ev@intern/cyber/Routing-F0/0M/routing_z_retain_layer_0\endcsname{0.7}
\expandafter\def\csname ev@intern/cyber/Routing-F0/0M/routing_z_retain_layer_1\endcsname{0.9}
\expandafter\def\csname ev@intern/cyber/Routing-F0/0M/routing_z_retain_layer_2\endcsname{1.1}
\expandafter\def\csname ev@intern/cyber/Routing-F0/0M/routing_z_retain_layer_3\endcsname{0.7}
\expandafter\def\csname ev@intern/cyber/Routing-F0/0M/routing_z_retain_layer_4\endcsname{1.2}
\expandafter\def\csname ev@intern/cyber/Routing-F0/0M/routing_z_retain_layer_5\endcsname{0.5}
\expandafter\def\csname ev@intern/cyber/Routing-F0/0M/routing_z_retain_layer_6\endcsname{0.6}
\expandafter\def\csname ev@intern/cyber/Routing-F0/0M/routing_z_retain_layer_7\endcsname{0.6}
\expandafter\def\csname ev@intern/cyber/Routing-F0/0M/routing_z_retain_layer_8\endcsname{0.4}
\expandafter\def\csname ev@intern/cyber/Routing-F0/0M/routing_z_retain_layer_9\endcsname{0.6}
\expandafter\def\csname ev@intern/cyber/Routing-F0/0M/routing_z_retain_layer_10\endcsname{0.6}
\expandafter\def\csname ev@intern/cyber/Routing-F0/0M/routing_z_retain_layer_11\endcsname{0.6}
\expandafter\def\csname ev@intern/cyber/Routing-F0/0M/routing_z_retain_layer_12\endcsname{0.9}
\expandafter\def\csname ev@intern/cyber/Routing-F0/0M/routing_z_retain_layer_13\endcsname{0.8}
\expandafter\def\csname ev@intern/cyber/Routing-F0/0M/routing_z_retain_layer_14\endcsname{0.6}
\expandafter\def\csname ev@intern/cyber/Routing-F0/0M/routing_z_retain_layer_15\endcsname{0.5}
\expandafter\def\csname ev@intern/cyber/Routing-F0/0M/routing_z_retain_layer_16\endcsname{1.0}
\expandafter\def\csname ev@intern/cyber/Routing-F0/0M/routing_z_retain_layer_17\endcsname{1.6}
\expandafter\def\csname ev@intern/cyber/Routing-F0/0M/routing_z_retain_layer_18\endcsname{0.6}
\expandafter\def\csname ev@intern/cyber/Routing-F0/0M/routing_z_retain_layer_19\endcsname{0.6}
\expandafter\def\csname ev@intern/cyber/Routing-F0/0M/routing_z_retain_layer_20\endcsname{0.5}
\expandafter\def\csname ev@intern/cyber/Routing-F0/0M/routing_z_retain_layer_21\endcsname{0.7}
\expandafter\def\csname ev@intern/cyber/Routing-F0/0M/routing_z_retain_layer_22\endcsname{0.5}
\expandafter\def\csname ev@intern/cyber/Routing-F0/0M/routing_z_retain_layer_23\endcsname{1.0}
\expandafter\def\csname ev@intern/cyber/Routing-F0/0M/routing_z_retain_layer_24\endcsname{0.9}
\expandafter\def\csname ev@intern/cyber/Routing-F0/0M/routing_z_retain_layer_25\endcsname{0.7}
\expandafter\def\csname ev@intern/cyber/Routing-F0/0M/routing_z_retain_layer_26\endcsname{0.7}
\expandafter\def\csname ev@intern/cyber/Routing-F0/0M/routing_z_retain_layer_27\endcsname{0.6}
\expandafter\def\csname ev@intern/cyber/Routing-F0/0M/routing_z_retain_layer_28\endcsname{0.7}
\expandafter\def\csname ev@intern/cyber/Routing-F0/0M/routing_z_retain_layer_29\endcsname{0.7}
\expandafter\def\csname ev@intern/cyber/Routing-F0/0M/routing_z_retain_layer_30\endcsname{0.6}
\expandafter\def\csname ev@intern/cyber/Routing-F0/0M/routing_z_retain_layer_31\endcsname{0.5}
\expandafter\def\csname ev@intern/cyber/Routing-F0/0M/routing_z_retain_layer_32\endcsname{0.4}
\expandafter\def\csname ev@intern/cyber/Routing-F0/0M/routing_z_retain_layer_33\endcsname{0.5}
\expandafter\def\csname ev@intern/cyber/Routing-F0/0M/routing_z_retain_layer_34\endcsname{0.9}
\expandafter\def\csname ev@intern/cyber/Routing-F0/0M/routing_z_retain_layer_35\endcsname{0.8}
\expandafter\def\csname ev@intern/cyber/Routing-F0/0M/routing_z_retain_layer_36\endcsname{0.7}
\expandafter\def\csname ev@intern/cyber/Routing-F0/0M/routing_z_retain_layer_37\endcsname{0.7}
\expandafter\def\csname ev@intern/cyber/Routing-F0/0M/routing_z_retain_layer_38\endcsname{0.8}
\expandafter\def\csname ev@intern/cyber/Routing-F0/0M/routing_z_retain_layer_39\endcsname{0.8}
\expandafter\def\csname ev@intern/cyber/Routing-F1/0M/routing_z_forget\endcsname{46.8}
\expandafter\def\csname ev@intern/cyber/Routing-F1/0M/routing_z_forget_layer_0\endcsname{10.3}
\expandafter\def\csname ev@intern/cyber/Routing-F1/0M/routing_z_forget_layer_1\endcsname{15.1}
\expandafter\def\csname ev@intern/cyber/Routing-F1/0M/routing_z_forget_layer_2\endcsname{8.1}
\expandafter\def\csname ev@intern/cyber/Routing-F1/0M/routing_z_forget_layer_3\endcsname{20.7}
\expandafter\def\csname ev@intern/cyber/Routing-F1/0M/routing_z_forget_layer_4\endcsname{24.1}
\expandafter\def\csname ev@intern/cyber/Routing-F1/0M/routing_z_forget_layer_5\endcsname{33.9}
\expandafter\def\csname ev@intern/cyber/Routing-F1/0M/routing_z_forget_layer_6\endcsname{31.0}
\expandafter\def\csname ev@intern/cyber/Routing-F1/0M/routing_z_forget_layer_7\endcsname{43.1}
\expandafter\def\csname ev@intern/cyber/Routing-F1/0M/routing_z_forget_layer_8\endcsname{48.7}
\expandafter\def\csname ev@intern/cyber/Routing-F1/0M/routing_z_forget_layer_9\endcsname{50.8}
\expandafter\def\csname ev@intern/cyber/Routing-F1/0M/routing_z_forget_layer_10\endcsname{34.6}
\expandafter\def\csname ev@intern/cyber/Routing-F1/0M/routing_z_forget_layer_11\endcsname{35.8}
\expandafter\def\csname ev@intern/cyber/Routing-F1/0M/routing_z_forget_layer_12\endcsname{37.3}
\expandafter\def\csname ev@intern/cyber/Routing-F1/0M/routing_z_forget_layer_13\endcsname{44.5}
\expandafter\def\csname ev@intern/cyber/Routing-F1/0M/routing_z_forget_layer_14\endcsname{48.3}
\expandafter\def\csname ev@intern/cyber/Routing-F1/0M/routing_z_forget_layer_15\endcsname{49.1}
\expandafter\def\csname ev@intern/cyber/Routing-F1/0M/routing_z_forget_layer_16\endcsname{40.1}
\expandafter\def\csname ev@intern/cyber/Routing-F1/0M/routing_z_forget_layer_17\endcsname{44.8}
\expandafter\def\csname ev@intern/cyber/Routing-F1/0M/routing_z_forget_layer_18\endcsname{50.4}
\expandafter\def\csname ev@intern/cyber/Routing-F1/0M/routing_z_forget_layer_19\endcsname{53.2}
\expandafter\def\csname ev@intern/cyber/Routing-F1/0M/routing_z_forget_layer_20\endcsname{59.5}
\expandafter\def\csname ev@intern/cyber/Routing-F1/0M/routing_z_forget_layer_21\endcsname{57.4}
\expandafter\def\csname ev@intern/cyber/Routing-F1/0M/routing_z_forget_layer_22\endcsname{43.7}
\expandafter\def\csname ev@intern/cyber/Routing-F1/0M/routing_z_forget_layer_23\endcsname{43.9}
\expandafter\def\csname ev@intern/cyber/Routing-F1/0M/routing_z_forget_layer_24\endcsname{42.7}
\expandafter\def\csname ev@intern/cyber/Routing-F1/0M/routing_z_forget_layer_25\endcsname{48.0}
\expandafter\def\csname ev@intern/cyber/Routing-F1/0M/routing_z_forget_layer_26\endcsname{49.3}
\expandafter\def\csname ev@intern/cyber/Routing-F1/0M/routing_z_forget_layer_27\endcsname{49.9}
\expandafter\def\csname ev@intern/cyber/Routing-F1/0M/routing_z_forget_layer_28\endcsname{45.6}
\expandafter\def\csname ev@intern/cyber/Routing-F1/0M/routing_z_forget_layer_29\endcsname{53.1}
\expandafter\def\csname ev@intern/cyber/Routing-F1/0M/routing_z_forget_layer_30\endcsname{57.6}
\expandafter\def\csname ev@intern/cyber/Routing-F1/0M/routing_z_forget_layer_31\endcsname{66.3}
\expandafter\def\csname ev@intern/cyber/Routing-F1/0M/routing_z_forget_layer_32\endcsname{58.8}
\expandafter\def\csname ev@intern/cyber/Routing-F1/0M/routing_z_forget_layer_33\endcsname{59.2}
\expandafter\def\csname ev@intern/cyber/Routing-F1/0M/routing_z_forget_layer_34\endcsname{56.5}
\expandafter\def\csname ev@intern/cyber/Routing-F1/0M/routing_z_forget_layer_35\endcsname{71.0}
\expandafter\def\csname ev@intern/cyber/Routing-F1/0M/routing_z_forget_layer_36\endcsname{74.4}
\expandafter\def\csname ev@intern/cyber/Routing-F1/0M/routing_z_forget_layer_37\endcsname{75.0}
\expandafter\def\csname ev@intern/cyber/Routing-F1/0M/routing_z_forget_layer_38\endcsname{76.1}
\expandafter\def\csname ev@intern/cyber/Routing-F1/0M/routing_z_forget_layer_39\endcsname{61.9}
\expandafter\def\csname ev@intern/cyber/Routing-F1/0M/routing_z_retain\endcsname{0.3}
\expandafter\def\csname ev@intern/cyber/Routing-F1/0M/routing_z_retain_layer_0\endcsname{0.3}
\expandafter\def\csname ev@intern/cyber/Routing-F1/0M/routing_z_retain_layer_1\endcsname{0.3}
\expandafter\def\csname ev@intern/cyber/Routing-F1/0M/routing_z_retain_layer_2\endcsname{0.4}
\expandafter\def\csname ev@intern/cyber/Routing-F1/0M/routing_z_retain_layer_3\endcsname{0.3}
\expandafter\def\csname ev@intern/cyber/Routing-F1/0M/routing_z_retain_layer_4\endcsname{0.3}
\expandafter\def\csname ev@intern/cyber/Routing-F1/0M/routing_z_retain_layer_5\endcsname{0.2}
\expandafter\def\csname ev@intern/cyber/Routing-F1/0M/routing_z_retain_layer_6\endcsname{0.3}
\expandafter\def\csname ev@intern/cyber/Routing-F1/0M/routing_z_retain_layer_7\endcsname{0.3}
\expandafter\def\csname ev@intern/cyber/Routing-F1/0M/routing_z_retain_layer_8\endcsname{0.2}
\expandafter\def\csname ev@intern/cyber/Routing-F1/0M/routing_z_retain_layer_9\endcsname{0.3}
\expandafter\def\csname ev@intern/cyber/Routing-F1/0M/routing_z_retain_layer_10\endcsname{0.3}
\expandafter\def\csname ev@intern/cyber/Routing-F1/0M/routing_z_retain_layer_11\endcsname{0.2}
\expandafter\def\csname ev@intern/cyber/Routing-F1/0M/routing_z_retain_layer_12\endcsname{0.3}
\expandafter\def\csname ev@intern/cyber/Routing-F1/0M/routing_z_retain_layer_13\endcsname{0.3}
\expandafter\def\csname ev@intern/cyber/Routing-F1/0M/routing_z_retain_layer_14\endcsname{0.2}
\expandafter\def\csname ev@intern/cyber/Routing-F1/0M/routing_z_retain_layer_15\endcsname{0.2}
\expandafter\def\csname ev@intern/cyber/Routing-F1/0M/routing_z_retain_layer_16\endcsname{0.3}
\expandafter\def\csname ev@intern/cyber/Routing-F1/0M/routing_z_retain_layer_17\endcsname{0.3}
\expandafter\def\csname ev@intern/cyber/Routing-F1/0M/routing_z_retain_layer_18\endcsname{0.3}
\expandafter\def\csname ev@intern/cyber/Routing-F1/0M/routing_z_retain_layer_19\endcsname{0.2}
\expandafter\def\csname ev@intern/cyber/Routing-F1/0M/routing_z_retain_layer_20\endcsname{0.2}
\expandafter\def\csname ev@intern/cyber/Routing-F1/0M/routing_z_retain_layer_21\endcsname{0.3}
\expandafter\def\csname ev@intern/cyber/Routing-F1/0M/routing_z_retain_layer_22\endcsname{0.2}
\expandafter\def\csname ev@intern/cyber/Routing-F1/0M/routing_z_retain_layer_23\endcsname{0.3}
\expandafter\def\csname ev@intern/cyber/Routing-F1/0M/routing_z_retain_layer_24\endcsname{0.2}
\expandafter\def\csname ev@intern/cyber/Routing-F1/0M/routing_z_retain_layer_25\endcsname{0.2}
\expandafter\def\csname ev@intern/cyber/Routing-F1/0M/routing_z_retain_layer_26\endcsname{0.2}
\expandafter\def\csname ev@intern/cyber/Routing-F1/0M/routing_z_retain_layer_27\endcsname{0.2}
\expandafter\def\csname ev@intern/cyber/Routing-F1/0M/routing_z_retain_layer_28\endcsname{0.2}
\expandafter\def\csname ev@intern/cyber/Routing-F1/0M/routing_z_retain_layer_29\endcsname{0.3}
\expandafter\def\csname ev@intern/cyber/Routing-F1/0M/routing_z_retain_layer_30\endcsname{0.3}
\expandafter\def\csname ev@intern/cyber/Routing-F1/0M/routing_z_retain_layer_31\endcsname{0.3}
\expandafter\def\csname ev@intern/cyber/Routing-F1/0M/routing_z_retain_layer_32\endcsname{0.2}
\expandafter\def\csname ev@intern/cyber/Routing-F1/0M/routing_z_retain_layer_33\endcsname{0.2}
\expandafter\def\csname ev@intern/cyber/Routing-F1/0M/routing_z_retain_layer_34\endcsname{0.3}
\expandafter\def\csname ev@intern/cyber/Routing-F1/0M/routing_z_retain_layer_35\endcsname{0.3}
\expandafter\def\csname ev@intern/cyber/Routing-F1/0M/routing_z_retain_layer_36\endcsname{0.3}
\expandafter\def\csname ev@intern/cyber/Routing-F1/0M/routing_z_retain_layer_37\endcsname{0.3}
\expandafter\def\csname ev@intern/cyber/Routing-F1/0M/routing_z_retain_layer_38\endcsname{0.3}
\expandafter\def\csname ev@intern/cyber/Routing-F1/0M/routing_z_retain_layer_39\endcsname{0.2}
\expandafter\def\csname ev@qwen/cyber/Routing-F0/0M/routing_z_forget\endcsname{16.0}
\expandafter\def\csname ev@qwen/cyber/Routing-F0/0M/routing_z_forget_layer_0\endcsname{8.5}
\expandafter\def\csname ev@qwen/cyber/Routing-F0/0M/routing_z_forget_layer_1\endcsname{10.6}
\expandafter\def\csname ev@qwen/cyber/Routing-F0/0M/routing_z_forget_layer_2\endcsname{7.3}
\expandafter\def\csname ev@qwen/cyber/Routing-F0/0M/routing_z_forget_layer_3\endcsname{13.6}
\expandafter\def\csname ev@qwen/cyber/Routing-F0/0M/routing_z_forget_layer_4\endcsname{15.6}
\expandafter\def\csname ev@qwen/cyber/Routing-F0/0M/routing_z_forget_layer_5\endcsname{17.4}
\expandafter\def\csname ev@qwen/cyber/Routing-F0/0M/routing_z_forget_layer_6\endcsname{15.9}
\expandafter\def\csname ev@qwen/cyber/Routing-F0/0M/routing_z_forget_layer_7\endcsname{21.4}
\expandafter\def\csname ev@qwen/cyber/Routing-F0/0M/routing_z_forget_layer_8\endcsname{22.7}
\expandafter\def\csname ev@qwen/cyber/Routing-F0/0M/routing_z_forget_layer_9\endcsname{21.7}
\expandafter\def\csname ev@qwen/cyber/Routing-F0/0M/routing_z_forget_layer_10\endcsname{13.2}
\expandafter\def\csname ev@qwen/cyber/Routing-F0/0M/routing_z_forget_layer_11\endcsname{13.4}
\expandafter\def\csname ev@qwen/cyber/Routing-F0/0M/routing_z_forget_layer_12\endcsname{15.3}
\expandafter\def\csname ev@qwen/cyber/Routing-F0/0M/routing_z_forget_layer_13\endcsname{17.1}
\expandafter\def\csname ev@qwen/cyber/Routing-F0/0M/routing_z_forget_layer_14\endcsname{20.3}
\expandafter\def\csname ev@qwen/cyber/Routing-F0/0M/routing_z_forget_layer_15\endcsname{16.5}
\expandafter\def\csname ev@qwen/cyber/Routing-F0/0M/routing_z_forget_layer_16\endcsname{14.1}
\expandafter\def\csname ev@qwen/cyber/Routing-F0/0M/routing_z_forget_layer_17\endcsname{15.5}
\expandafter\def\csname ev@qwen/cyber/Routing-F0/0M/routing_z_forget_layer_18\endcsname{17.7}
\expandafter\def\csname ev@qwen/cyber/Routing-F0/0M/routing_z_forget_layer_19\endcsname{15.7}
\expandafter\def\csname ev@qwen/cyber/Routing-F0/0M/routing_z_forget_layer_20\endcsname{22.1}
\expandafter\def\csname ev@qwen/cyber/Routing-F0/0M/routing_z_forget_layer_21\endcsname{22.3}
\expandafter\def\csname ev@qwen/cyber/Routing-F0/0M/routing_z_forget_layer_22\endcsname{13.3}
\expandafter\def\csname ev@qwen/cyber/Routing-F0/0M/routing_z_forget_layer_23\endcsname{14.8}
\expandafter\def\csname ev@qwen/cyber/Routing-F0/0M/routing_z_forget_layer_24\endcsname{13.4}
\expandafter\def\csname ev@qwen/cyber/Routing-F0/0M/routing_z_forget_layer_25\endcsname{15.8}
\expandafter\def\csname ev@qwen/cyber/Routing-F0/0M/routing_z_forget_layer_26\endcsname{18.6}
\expandafter\def\csname ev@qwen/cyber/Routing-F0/0M/routing_z_forget_layer_27\endcsname{14.8}
\expandafter\def\csname ev@qwen/cyber/Routing-F0/0M/routing_z_forget_layer_28\endcsname{12.9}
\expandafter\def\csname ev@qwen/cyber/Routing-F0/0M/routing_z_forget_layer_29\endcsname{13.9}
\expandafter\def\csname ev@qwen/cyber/Routing-F0/0M/routing_z_forget_layer_30\endcsname{16.7}
\expandafter\def\csname ev@qwen/cyber/Routing-F0/0M/routing_z_forget_layer_31\endcsname{15.0}
\expandafter\def\csname ev@qwen/cyber/Routing-F0/0M/routing_z_forget_layer_32\endcsname{13.4}
\expandafter\def\csname ev@qwen/cyber/Routing-F0/0M/routing_z_forget_layer_33\endcsname{16.0}
\expandafter\def\csname ev@qwen/cyber/Routing-F0/0M/routing_z_forget_layer_34\endcsname{13.4}
\expandafter\def\csname ev@qwen/cyber/Routing-F0/0M/routing_z_forget_layer_35\endcsname{18.8}
\expandafter\def\csname ev@qwen/cyber/Routing-F0/0M/routing_z_forget_layer_36\endcsname{22.2}
\expandafter\def\csname ev@qwen/cyber/Routing-F0/0M/routing_z_forget_layer_37\endcsname{17.7}
\expandafter\def\csname ev@qwen/cyber/Routing-F0/0M/routing_z_forget_layer_38\endcsname{16.8}
\expandafter\def\csname ev@qwen/cyber/Routing-F0/0M/routing_z_forget_layer_39\endcsname{12.9}
\expandafter\def\csname ev@qwen/cyber/Routing-F0/0M/routing_z_retain\endcsname{1.1}
\expandafter\def\csname ev@qwen/cyber/Routing-F0/0M/routing_z_retain_layer_0\endcsname{1.6}
\expandafter\def\csname ev@qwen/cyber/Routing-F0/0M/routing_z_retain_layer_1\endcsname{1.3}
\expandafter\def\csname ev@qwen/cyber/Routing-F0/0M/routing_z_retain_layer_2\endcsname{1.7}
\expandafter\def\csname ev@qwen/cyber/Routing-F0/0M/routing_z_retain_layer_3\endcsname{1.0}
\expandafter\def\csname ev@qwen/cyber/Routing-F0/0M/routing_z_retain_layer_4\endcsname{1.6}
\expandafter\def\csname ev@qwen/cyber/Routing-F0/0M/routing_z_retain_layer_5\endcsname{0.8}
\expandafter\def\csname ev@qwen/cyber/Routing-F0/0M/routing_z_retain_layer_6\endcsname{0.9}
\expandafter\def\csname ev@qwen/cyber/Routing-F0/0M/routing_z_retain_layer_7\endcsname{0.9}
\expandafter\def\csname ev@qwen/cyber/Routing-F0/0M/routing_z_retain_layer_8\endcsname{1.0}
\expandafter\def\csname ev@qwen/cyber/Routing-F0/0M/routing_z_retain_layer_9\endcsname{1.0}
\expandafter\def\csname ev@qwen/cyber/Routing-F0/0M/routing_z_retain_layer_10\endcsname{1.0}
\expandafter\def\csname ev@qwen/cyber/Routing-F0/0M/routing_z_retain_layer_11\endcsname{1.2}
\expandafter\def\csname ev@qwen/cyber/Routing-F0/0M/routing_z_retain_layer_12\endcsname{1.1}
\expandafter\def\csname ev@qwen/cyber/Routing-F0/0M/routing_z_retain_layer_13\endcsname{0.9}
\expandafter\def\csname ev@qwen/cyber/Routing-F0/0M/routing_z_retain_layer_14\endcsname{0.9}
\expandafter\def\csname ev@qwen/cyber/Routing-F0/0M/routing_z_retain_layer_15\endcsname{1.1}
\expandafter\def\csname ev@qwen/cyber/Routing-F0/0M/routing_z_retain_layer_16\endcsname{1.3}
\expandafter\def\csname ev@qwen/cyber/Routing-F0/0M/routing_z_retain_layer_17\endcsname{1.5}
\expandafter\def\csname ev@qwen/cyber/Routing-F0/0M/routing_z_retain_layer_18\endcsname{1.0}
\expandafter\def\csname ev@qwen/cyber/Routing-F0/0M/routing_z_retain_layer_19\endcsname{0.8}
\expandafter\def\csname ev@qwen/cyber/Routing-F0/0M/routing_z_retain_layer_20\endcsname{0.9}
\expandafter\def\csname ev@qwen/cyber/Routing-F0/0M/routing_z_retain_layer_21\endcsname{1.0}
\expandafter\def\csname ev@qwen/cyber/Routing-F0/0M/routing_z_retain_layer_22\endcsname{1.2}
\expandafter\def\csname ev@qwen/cyber/Routing-F0/0M/routing_z_retain_layer_23\endcsname{1.3}
\expandafter\def\csname ev@qwen/cyber/Routing-F0/0M/routing_z_retain_layer_24\endcsname{1.1}
\expandafter\def\csname ev@qwen/cyber/Routing-F0/0M/routing_z_retain_layer_25\endcsname{1.0}
\expandafter\def\csname ev@qwen/cyber/Routing-F0/0M/routing_z_retain_layer_26\endcsname{1.0}
\expandafter\def\csname ev@qwen/cyber/Routing-F0/0M/routing_z_retain_layer_27\endcsname{1.1}
\expandafter\def\csname ev@qwen/cyber/Routing-F0/0M/routing_z_retain_layer_28\endcsname{1.2}
\expandafter\def\csname ev@qwen/cyber/Routing-F0/0M/routing_z_retain_layer_29\endcsname{0.9}
\expandafter\def\csname ev@qwen/cyber/Routing-F0/0M/routing_z_retain_layer_30\endcsname{1.1}
\expandafter\def\csname ev@qwen/cyber/Routing-F0/0M/routing_z_retain_layer_31\endcsname{0.7}
\expandafter\def\csname ev@qwen/cyber/Routing-F0/0M/routing_z_retain_layer_32\endcsname{0.8}
\expandafter\def\csname ev@qwen/cyber/Routing-F0/0M/routing_z_retain_layer_33\endcsname{0.6}
\expandafter\def\csname ev@qwen/cyber/Routing-F0/0M/routing_z_retain_layer_34\endcsname{1.2}
\expandafter\def\csname ev@qwen/cyber/Routing-F0/0M/routing_z_retain_layer_35\endcsname{1.0}
\expandafter\def\csname ev@qwen/cyber/Routing-F0/0M/routing_z_retain_layer_36\endcsname{1.0}
\expandafter\def\csname ev@qwen/cyber/Routing-F0/0M/routing_z_retain_layer_37\endcsname{1.1}
\expandafter\def\csname ev@qwen/cyber/Routing-F0/0M/routing_z_retain_layer_38\endcsname{1.1}
\expandafter\def\csname ev@qwen/cyber/Routing-F0/0M/routing_z_retain_layer_39\endcsname{1.3}
\expandafter\def\csname ev@qwen/cyber/Routing-F1/0M/routing_z_forget\endcsname{35.2}
\expandafter\def\csname ev@qwen/cyber/Routing-F1/0M/routing_z_forget_layer_0\endcsname{7.4}
\expandafter\def\csname ev@qwen/cyber/Routing-F1/0M/routing_z_forget_layer_1\endcsname{8.6}
\expandafter\def\csname ev@qwen/cyber/Routing-F1/0M/routing_z_forget_layer_2\endcsname{5.5}
\expandafter\def\csname ev@qwen/cyber/Routing-F1/0M/routing_z_forget_layer_3\endcsname{11.0}
\expandafter\def\csname ev@qwen/cyber/Routing-F1/0M/routing_z_forget_layer_4\endcsname{11.5}
\expandafter\def\csname ev@qwen/cyber/Routing-F1/0M/routing_z_forget_layer_5\endcsname{18.3}
\expandafter\def\csname ev@qwen/cyber/Routing-F1/0M/routing_z_forget_layer_6\endcsname{17.7}
\expandafter\def\csname ev@qwen/cyber/Routing-F1/0M/routing_z_forget_layer_7\endcsname{27.8}
\expandafter\def\csname ev@qwen/cyber/Routing-F1/0M/routing_z_forget_layer_8\endcsname{33.2}
\expandafter\def\csname ev@qwen/cyber/Routing-F1/0M/routing_z_forget_layer_9\endcsname{34.8}
\expandafter\def\csname ev@qwen/cyber/Routing-F1/0M/routing_z_forget_layer_10\endcsname{22.7}
\expandafter\def\csname ev@qwen/cyber/Routing-F1/0M/routing_z_forget_layer_11\endcsname{21.8}
\expandafter\def\csname ev@qwen/cyber/Routing-F1/0M/routing_z_forget_layer_12\endcsname{23.9}
\expandafter\def\csname ev@qwen/cyber/Routing-F1/0M/routing_z_forget_layer_13\endcsname{30.4}
\expandafter\def\csname ev@qwen/cyber/Routing-F1/0M/routing_z_forget_layer_14\endcsname{35.4}
\expandafter\def\csname ev@qwen/cyber/Routing-F1/0M/routing_z_forget_layer_15\endcsname{34.8}
\expandafter\def\csname ev@qwen/cyber/Routing-F1/0M/routing_z_forget_layer_16\endcsname{30.3}
\expandafter\def\csname ev@qwen/cyber/Routing-F1/0M/routing_z_forget_layer_17\endcsname{34.0}
\expandafter\def\csname ev@qwen/cyber/Routing-F1/0M/routing_z_forget_layer_18\endcsname{36.8}
\expandafter\def\csname ev@qwen/cyber/Routing-F1/0M/routing_z_forget_layer_19\endcsname{38.8}
\expandafter\def\csname ev@qwen/cyber/Routing-F1/0M/routing_z_forget_layer_20\endcsname{45.0}
\expandafter\def\csname ev@qwen/cyber/Routing-F1/0M/routing_z_forget_layer_21\endcsname{44.5}
\expandafter\def\csname ev@qwen/cyber/Routing-F1/0M/routing_z_forget_layer_22\endcsname{29.2}
\expandafter\def\csname ev@qwen/cyber/Routing-F1/0M/routing_z_forget_layer_23\endcsname{30.2}
\expandafter\def\csname ev@qwen/cyber/Routing-F1/0M/routing_z_forget_layer_24\endcsname{29.2}
\expandafter\def\csname ev@qwen/cyber/Routing-F1/0M/routing_z_forget_layer_25\endcsname{33.9}
\expandafter\def\csname ev@qwen/cyber/Routing-F1/0M/routing_z_forget_layer_26\endcsname{37.1}
\expandafter\def\csname ev@qwen/cyber/Routing-F1/0M/routing_z_forget_layer_27\endcsname{37.9}
\expandafter\def\csname ev@qwen/cyber/Routing-F1/0M/routing_z_forget_layer_28\endcsname{33.4}
\expandafter\def\csname ev@qwen/cyber/Routing-F1/0M/routing_z_forget_layer_29\endcsname{39.0}
\expandafter\def\csname ev@qwen/cyber/Routing-F1/0M/routing_z_forget_layer_30\endcsname{44.9}
\expandafter\def\csname ev@qwen/cyber/Routing-F1/0M/routing_z_forget_layer_31\endcsname{50.6}
\expandafter\def\csname ev@qwen/cyber/Routing-F1/0M/routing_z_forget_layer_32\endcsname{47.8}
\expandafter\def\csname ev@qwen/cyber/Routing-F1/0M/routing_z_forget_layer_33\endcsname{47.9}
\expandafter\def\csname ev@qwen/cyber/Routing-F1/0M/routing_z_forget_layer_34\endcsname{43.2}
\expandafter\def\csname ev@qwen/cyber/Routing-F1/0M/routing_z_forget_layer_35\endcsname{62.8}
\expandafter\def\csname ev@qwen/cyber/Routing-F1/0M/routing_z_forget_layer_36\endcsname{69.2}
\expandafter\def\csname ev@qwen/cyber/Routing-F1/0M/routing_z_forget_layer_37\endcsname{68.2}
\expandafter\def\csname ev@qwen/cyber/Routing-F1/0M/routing_z_forget_layer_38\endcsname{73.5}
\expandafter\def\csname ev@qwen/cyber/Routing-F1/0M/routing_z_forget_layer_39\endcsname{55.9}
\expandafter\def\csname ev@qwen/cyber/Routing-F1/0M/routing_z_retain\endcsname{0.1}
\expandafter\def\csname ev@qwen/cyber/Routing-F1/0M/routing_z_retain_layer_0\endcsname{0.6}
\expandafter\def\csname ev@qwen/cyber/Routing-F1/0M/routing_z_retain_layer_1\endcsname{0.1}
\expandafter\def\csname ev@qwen/cyber/Routing-F1/0M/routing_z_retain_layer_2\endcsname{0.2}
\expandafter\def\csname ev@qwen/cyber/Routing-F1/0M/routing_z_retain_layer_3\endcsname{0.1}
\expandafter\def\csname ev@qwen/cyber/Routing-F1/0M/routing_z_retain_layer_4\endcsname{0.1}
\expandafter\def\csname ev@qwen/cyber/Routing-F1/0M/routing_z_retain_layer_5\endcsname{0.1}
\expandafter\def\csname ev@qwen/cyber/Routing-F1/0M/routing_z_retain_layer_6\endcsname{0.1}
\expandafter\def\csname ev@qwen/cyber/Routing-F1/0M/routing_z_retain_layer_7\endcsname{0.1}
\expandafter\def\csname ev@qwen/cyber/Routing-F1/0M/routing_z_retain_layer_8\endcsname{0.0}
\expandafter\def\csname ev@qwen/cyber/Routing-F1/0M/routing_z_retain_layer_9\endcsname{0.1}
\expandafter\def\csname ev@qwen/cyber/Routing-F1/0M/routing_z_retain_layer_10\endcsname{0.1}
\expandafter\def\csname ev@qwen/cyber/Routing-F1/0M/routing_z_retain_layer_11\endcsname{0.1}
\expandafter\def\csname ev@qwen/cyber/Routing-F1/0M/routing_z_retain_layer_12\endcsname{0.1}
\expandafter\def\csname ev@qwen/cyber/Routing-F1/0M/routing_z_retain_layer_13\endcsname{0.1}
\expandafter\def\csname ev@qwen/cyber/Routing-F1/0M/routing_z_retain_layer_14\endcsname{0.1}
\expandafter\def\csname ev@qwen/cyber/Routing-F1/0M/routing_z_retain_layer_15\endcsname{0.1}
\expandafter\def\csname ev@qwen/cyber/Routing-F1/0M/routing_z_retain_layer_16\endcsname{0.1}
\expandafter\def\csname ev@qwen/cyber/Routing-F1/0M/routing_z_retain_layer_17\endcsname{0.1}
\expandafter\def\csname ev@qwen/cyber/Routing-F1/0M/routing_z_retain_layer_18\endcsname{0.1}
\expandafter\def\csname ev@qwen/cyber/Routing-F1/0M/routing_z_retain_layer_19\endcsname{0.1}
\expandafter\def\csname ev@qwen/cyber/Routing-F1/0M/routing_z_retain_layer_20\endcsname{0.1}
\expandafter\def\csname ev@qwen/cyber/Routing-F1/0M/routing_z_retain_layer_21\endcsname{0.1}
\expandafter\def\csname ev@qwen/cyber/Routing-F1/0M/routing_z_retain_layer_22\endcsname{0.1}
\expandafter\def\csname ev@qwen/cyber/Routing-F1/0M/routing_z_retain_layer_23\endcsname{0.1}
\expandafter\def\csname ev@qwen/cyber/Routing-F1/0M/routing_z_retain_layer_24\endcsname{0.0}
\expandafter\def\csname ev@qwen/cyber/Routing-F1/0M/routing_z_retain_layer_25\endcsname{0.1}
\expandafter\def\csname ev@qwen/cyber/Routing-F1/0M/routing_z_retain_layer_26\endcsname{0.0}
\expandafter\def\csname ev@qwen/cyber/Routing-F1/0M/routing_z_retain_layer_27\endcsname{0.1}
\expandafter\def\csname ev@qwen/cyber/Routing-F1/0M/routing_z_retain_layer_28\endcsname{0.1}
\expandafter\def\csname ev@qwen/cyber/Routing-F1/0M/routing_z_retain_layer_29\endcsname{0.1}
\expandafter\def\csname ev@qwen/cyber/Routing-F1/0M/routing_z_retain_layer_30\endcsname{0.1}
\expandafter\def\csname ev@qwen/cyber/Routing-F1/0M/routing_z_retain_layer_31\endcsname{0.1}
\expandafter\def\csname ev@qwen/cyber/Routing-F1/0M/routing_z_retain_layer_32\endcsname{0.1}
\expandafter\def\csname ev@qwen/cyber/Routing-F1/0M/routing_z_retain_layer_33\endcsname{0.1}
\expandafter\def\csname ev@qwen/cyber/Routing-F1/0M/routing_z_retain_layer_34\endcsname{0.2}
\expandafter\def\csname ev@qwen/cyber/Routing-F1/0M/routing_z_retain_layer_35\endcsname{0.2}
\expandafter\def\csname ev@qwen/cyber/Routing-F1/0M/routing_z_retain_layer_36\endcsname{0.2}
\expandafter\def\csname ev@qwen/cyber/Routing-F1/0M/routing_z_retain_layer_37\endcsname{0.2}
\expandafter\def\csname ev@qwen/cyber/Routing-F1/0M/routing_z_retain_layer_38\endcsname{0.2}
\expandafter\def\csname ev@qwen/cyber/Routing-F1/0M/routing_z_retain_layer_39\endcsname{0.1}
\expandafter\def\csname ev@gemma/cyber/Routing-F0/0M/routing_z_forget\endcsname{21.1}
\expandafter\def\csname ev@gemma/cyber/Routing-F0/0M/routing_z_forget_layer_0\endcsname{10.8}
\expandafter\def\csname ev@gemma/cyber/Routing-F0/0M/routing_z_forget_layer_1\endcsname{8.8}
\expandafter\def\csname ev@gemma/cyber/Routing-F0/0M/routing_z_forget_layer_2\endcsname{19.3}
\expandafter\def\csname ev@gemma/cyber/Routing-F0/0M/routing_z_forget_layer_3\endcsname{21.1}
\expandafter\def\csname ev@gemma/cyber/Routing-F0/0M/routing_z_forget_layer_4\endcsname{16.8}
\expandafter\def\csname ev@gemma/cyber/Routing-F0/0M/routing_z_forget_layer_5\endcsname{24.6}
\expandafter\def\csname ev@gemma/cyber/Routing-F0/0M/routing_z_forget_layer_6\endcsname{22.7}
\expandafter\def\csname ev@gemma/cyber/Routing-F0/0M/routing_z_forget_layer_7\endcsname{19.2}
\expandafter\def\csname ev@gemma/cyber/Routing-F0/0M/routing_z_forget_layer_8\endcsname{20.5}
\expandafter\def\csname ev@gemma/cyber/Routing-F0/0M/routing_z_forget_layer_9\endcsname{18.5}
\expandafter\def\csname ev@gemma/cyber/Routing-F0/0M/routing_z_forget_layer_10\endcsname{22.1}
\expandafter\def\csname ev@gemma/cyber/Routing-F0/0M/routing_z_forget_layer_11\endcsname{23.0}
\expandafter\def\csname ev@gemma/cyber/Routing-F0/0M/routing_z_forget_layer_12\endcsname{18.7}
\expandafter\def\csname ev@gemma/cyber/Routing-F0/0M/routing_z_forget_layer_13\endcsname{18.9}
\expandafter\def\csname ev@gemma/cyber/Routing-F0/0M/routing_z_forget_layer_14\endcsname{20.4}
\expandafter\def\csname ev@gemma/cyber/Routing-F0/0M/routing_z_forget_layer_15\endcsname{17.2}
\expandafter\def\csname ev@gemma/cyber/Routing-F0/0M/routing_z_forget_layer_16\endcsname{18.2}
\expandafter\def\csname ev@gemma/cyber/Routing-F0/0M/routing_z_forget_layer_17\endcsname{18.9}
\expandafter\def\csname ev@gemma/cyber/Routing-F0/0M/routing_z_forget_layer_18\endcsname{21.8}
\expandafter\def\csname ev@gemma/cyber/Routing-F0/0M/routing_z_forget_layer_19\endcsname{19.3}
\expandafter\def\csname ev@gemma/cyber/Routing-F0/0M/routing_z_forget_layer_20\endcsname{25.4}
\expandafter\def\csname ev@gemma/cyber/Routing-F0/0M/routing_z_forget_layer_21\endcsname{22.2}
\expandafter\def\csname ev@gemma/cyber/Routing-F0/0M/routing_z_forget_layer_22\endcsname{29.5}
\expandafter\def\csname ev@gemma/cyber/Routing-F0/0M/routing_z_forget_layer_23\endcsname{25.1}
\expandafter\def\csname ev@gemma/cyber/Routing-F0/0M/routing_z_forget_layer_24\endcsname{28.5}
\expandafter\def\csname ev@gemma/cyber/Routing-F0/0M/routing_z_forget_layer_25\endcsname{27.6}
\expandafter\def\csname ev@gemma/cyber/Routing-F0/0M/routing_z_forget_layer_26\endcsname{23.0}
\expandafter\def\csname ev@gemma/cyber/Routing-F0/0M/routing_z_forget_layer_27\endcsname{22.5}
\expandafter\def\csname ev@gemma/cyber/Routing-F0/0M/routing_z_forget_layer_28\endcsname{28.6}
\expandafter\def\csname ev@gemma/cyber/Routing-F0/0M/routing_z_forget_layer_29\endcsname{20.5}
\expandafter\def\csname ev@gemma/cyber/Routing-F0/0M/routing_z_retain\endcsname{2.2}
\expandafter\def\csname ev@gemma/cyber/Routing-F0/0M/routing_z_retain_layer_0\endcsname{2.0}
\expandafter\def\csname ev@gemma/cyber/Routing-F0/0M/routing_z_retain_layer_1\endcsname{1.7}
\expandafter\def\csname ev@gemma/cyber/Routing-F0/0M/routing_z_retain_layer_2\endcsname{0.9}
\expandafter\def\csname ev@gemma/cyber/Routing-F0/0M/routing_z_retain_layer_3\endcsname{2.4}
\expandafter\def\csname ev@gemma/cyber/Routing-F0/0M/routing_z_retain_layer_4\endcsname{1.3}
\expandafter\def\csname ev@gemma/cyber/Routing-F0/0M/routing_z_retain_layer_5\endcsname{0.9}
\expandafter\def\csname ev@gemma/cyber/Routing-F0/0M/routing_z_retain_layer_6\endcsname{1.1}
\expandafter\def\csname ev@gemma/cyber/Routing-F0/0M/routing_z_retain_layer_7\endcsname{1.3}
\expandafter\def\csname ev@gemma/cyber/Routing-F0/0M/routing_z_retain_layer_8\endcsname{1.3}
\expandafter\def\csname ev@gemma/cyber/Routing-F0/0M/routing_z_retain_layer_9\endcsname{2.2}
\expandafter\def\csname ev@gemma/cyber/Routing-F0/0M/routing_z_retain_layer_10\endcsname{0.6}
\expandafter\def\csname ev@gemma/cyber/Routing-F0/0M/routing_z_retain_layer_11\endcsname{2.4}
\expandafter\def\csname ev@gemma/cyber/Routing-F0/0M/routing_z_retain_layer_12\endcsname{5.0}
\expandafter\def\csname ev@gemma/cyber/Routing-F0/0M/routing_z_retain_layer_13\endcsname{1.4}
\expandafter\def\csname ev@gemma/cyber/Routing-F0/0M/routing_z_retain_layer_14\endcsname{7.0}
\expandafter\def\csname ev@gemma/cyber/Routing-F0/0M/routing_z_retain_layer_15\endcsname{2.3}
\expandafter\def\csname ev@gemma/cyber/Routing-F0/0M/routing_z_retain_layer_16\endcsname{2.9}
\expandafter\def\csname ev@gemma/cyber/Routing-F0/0M/routing_z_retain_layer_17\endcsname{2.3}
\expandafter\def\csname ev@gemma/cyber/Routing-F0/0M/routing_z_retain_layer_18\endcsname{4.2}
\expandafter\def\csname ev@gemma/cyber/Routing-F0/0M/routing_z_retain_layer_19\endcsname{0.8}
\expandafter\def\csname ev@gemma/cyber/Routing-F0/0M/routing_z_retain_layer_20\endcsname{1.4}
\expandafter\def\csname ev@gemma/cyber/Routing-F0/0M/routing_z_retain_layer_21\endcsname{1.2}
\expandafter\def\csname ev@gemma/cyber/Routing-F0/0M/routing_z_retain_layer_22\endcsname{1.5}
\expandafter\def\csname ev@gemma/cyber/Routing-F0/0M/routing_z_retain_layer_23\endcsname{1.0}
\expandafter\def\csname ev@gemma/cyber/Routing-F0/0M/routing_z_retain_layer_24\endcsname{3.3}
\expandafter\def\csname ev@gemma/cyber/Routing-F0/0M/routing_z_retain_layer_25\endcsname{1.7}
\expandafter\def\csname ev@gemma/cyber/Routing-F0/0M/routing_z_retain_layer_26\endcsname{2.0}
\expandafter\def\csname ev@gemma/cyber/Routing-F0/0M/routing_z_retain_layer_27\endcsname{3.4}
\expandafter\def\csname ev@gemma/cyber/Routing-F0/0M/routing_z_retain_layer_28\endcsname{2.8}
\expandafter\def\csname ev@gemma/cyber/Routing-F0/0M/routing_z_retain_layer_29\endcsname{4.0}
\expandafter\def\csname ev@gemma/cyber/Routing-F1/0M/routing_z_forget\endcsname{35.7}
\expandafter\def\csname ev@gemma/cyber/Routing-F1/0M/routing_z_forget_layer_0\endcsname{12.9}
\expandafter\def\csname ev@gemma/cyber/Routing-F1/0M/routing_z_forget_layer_1\endcsname{10.3}
\expandafter\def\csname ev@gemma/cyber/Routing-F1/0M/routing_z_forget_layer_2\endcsname{22.6}
\expandafter\def\csname ev@gemma/cyber/Routing-F1/0M/routing_z_forget_layer_3\endcsname{24.7}
\expandafter\def\csname ev@gemma/cyber/Routing-F1/0M/routing_z_forget_layer_4\endcsname{22.6}
\expandafter\def\csname ev@gemma/cyber/Routing-F1/0M/routing_z_forget_layer_5\endcsname{32.9}
\expandafter\def\csname ev@gemma/cyber/Routing-F1/0M/routing_z_forget_layer_6\endcsname{33.8}
\expandafter\def\csname ev@gemma/cyber/Routing-F1/0M/routing_z_forget_layer_7\endcsname{29.3}
\expandafter\def\csname ev@gemma/cyber/Routing-F1/0M/routing_z_forget_layer_8\endcsname{31.9}
\expandafter\def\csname ev@gemma/cyber/Routing-F1/0M/routing_z_forget_layer_9\endcsname{33.1}
\expandafter\def\csname ev@gemma/cyber/Routing-F1/0M/routing_z_forget_layer_10\endcsname{39.5}
\expandafter\def\csname ev@gemma/cyber/Routing-F1/0M/routing_z_forget_layer_11\endcsname{42.1}
\expandafter\def\csname ev@gemma/cyber/Routing-F1/0M/routing_z_forget_layer_12\endcsname{30.5}
\expandafter\def\csname ev@gemma/cyber/Routing-F1/0M/routing_z_forget_layer_13\endcsname{31.7}
\expandafter\def\csname ev@gemma/cyber/Routing-F1/0M/routing_z_forget_layer_14\endcsname{33.0}
\expandafter\def\csname ev@gemma/cyber/Routing-F1/0M/routing_z_forget_layer_15\endcsname{27.5}
\expandafter\def\csname ev@gemma/cyber/Routing-F1/0M/routing_z_forget_layer_16\endcsname{30.9}
\expandafter\def\csname ev@gemma/cyber/Routing-F1/0M/routing_z_forget_layer_17\endcsname{34.8}
\expandafter\def\csname ev@gemma/cyber/Routing-F1/0M/routing_z_forget_layer_18\endcsname{34.8}
\expandafter\def\csname ev@gemma/cyber/Routing-F1/0M/routing_z_forget_layer_19\endcsname{39.4}
\expandafter\def\csname ev@gemma/cyber/Routing-F1/0M/routing_z_forget_layer_20\endcsname{43.0}
\expandafter\def\csname ev@gemma/cyber/Routing-F1/0M/routing_z_forget_layer_21\endcsname{39.9}
\expandafter\def\csname ev@gemma/cyber/Routing-F1/0M/routing_z_forget_layer_22\endcsname{48.6}
\expandafter\def\csname ev@gemma/cyber/Routing-F1/0M/routing_z_forget_layer_23\endcsname{48.2}
\expandafter\def\csname ev@gemma/cyber/Routing-F1/0M/routing_z_forget_layer_24\endcsname{48.8}
\expandafter\def\csname ev@gemma/cyber/Routing-F1/0M/routing_z_forget_layer_25\endcsname{50.0}
\expandafter\def\csname ev@gemma/cyber/Routing-F1/0M/routing_z_forget_layer_26\endcsname{46.6}
\expandafter\def\csname ev@gemma/cyber/Routing-F1/0M/routing_z_forget_layer_27\endcsname{48.1}
\expandafter\def\csname ev@gemma/cyber/Routing-F1/0M/routing_z_forget_layer_28\endcsname{54.8}
\expandafter\def\csname ev@gemma/cyber/Routing-F1/0M/routing_z_forget_layer_29\endcsname{46.1}
\expandafter\def\csname ev@gemma/cyber/Routing-F1/0M/routing_z_retain\endcsname{2.5}
\expandafter\def\csname ev@gemma/cyber/Routing-F1/0M/routing_z_retain_layer_0\endcsname{2.0}
\expandafter\def\csname ev@gemma/cyber/Routing-F1/0M/routing_z_retain_layer_1\endcsname{1.6}
\expandafter\def\csname ev@gemma/cyber/Routing-F1/0M/routing_z_retain_layer_2\endcsname{0.9}
\expandafter\def\csname ev@gemma/cyber/Routing-F1/0M/routing_z_retain_layer_3\endcsname{2.4}
\expandafter\def\csname ev@gemma/cyber/Routing-F1/0M/routing_z_retain_layer_4\endcsname{1.4}
\expandafter\def\csname ev@gemma/cyber/Routing-F1/0M/routing_z_retain_layer_5\endcsname{1.2}
\expandafter\def\csname ev@gemma/cyber/Routing-F1/0M/routing_z_retain_layer_6\endcsname{1.4}
\expandafter\def\csname ev@gemma/cyber/Routing-F1/0M/routing_z_retain_layer_7\endcsname{1.6}
\expandafter\def\csname ev@gemma/cyber/Routing-F1/0M/routing_z_retain_layer_8\endcsname{1.5}
\expandafter\def\csname ev@gemma/cyber/Routing-F1/0M/routing_z_retain_layer_9\endcsname{2.4}
\expandafter\def\csname ev@gemma/cyber/Routing-F1/0M/routing_z_retain_layer_10\endcsname{1.0}
\expandafter\def\csname ev@gemma/cyber/Routing-F1/0M/routing_z_retain_layer_11\endcsname{3.1}
\expandafter\def\csname ev@gemma/cyber/Routing-F1/0M/routing_z_retain_layer_12\endcsname{4.5}
\expandafter\def\csname ev@gemma/cyber/Routing-F1/0M/routing_z_retain_layer_13\endcsname{1.8}
\expandafter\def\csname ev@gemma/cyber/Routing-F1/0M/routing_z_retain_layer_14\endcsname{7.4}
\expandafter\def\csname ev@gemma/cyber/Routing-F1/0M/routing_z_retain_layer_15\endcsname{2.8}
\expandafter\def\csname ev@gemma/cyber/Routing-F1/0M/routing_z_retain_layer_16\endcsname{3.3}
\expandafter\def\csname ev@gemma/cyber/Routing-F1/0M/routing_z_retain_layer_17\endcsname{2.8}
\expandafter\def\csname ev@gemma/cyber/Routing-F1/0M/routing_z_retain_layer_18\endcsname{4.3}
\expandafter\def\csname ev@gemma/cyber/Routing-F1/0M/routing_z_retain_layer_19\endcsname{1.2}
\expandafter\def\csname ev@gemma/cyber/Routing-F1/0M/routing_z_retain_layer_20\endcsname{1.8}
\expandafter\def\csname ev@gemma/cyber/Routing-F1/0M/routing_z_retain_layer_21\endcsname{1.5}
\expandafter\def\csname ev@gemma/cyber/Routing-F1/0M/routing_z_retain_layer_22\endcsname{2.0}
\expandafter\def\csname ev@gemma/cyber/Routing-F1/0M/routing_z_retain_layer_23\endcsname{1.7}
\expandafter\def\csname ev@gemma/cyber/Routing-F1/0M/routing_z_retain_layer_24\endcsname{3.2}
\expandafter\def\csname ev@gemma/cyber/Routing-F1/0M/routing_z_retain_layer_25\endcsname{2.3}
\expandafter\def\csname ev@gemma/cyber/Routing-F1/0M/routing_z_retain_layer_26\endcsname{2.5}
\expandafter\def\csname ev@gemma/cyber/Routing-F1/0M/routing_z_retain_layer_27\endcsname{3.5}
\expandafter\def\csname ev@gemma/cyber/Routing-F1/0M/routing_z_retain_layer_28\endcsname{3.4}
\expandafter\def\csname ev@gemma/cyber/Routing-F1/0M/routing_z_retain_layer_29\endcsname{3.7}
\expandafter\def\csname ev@historical:qwen/bio/F2-A2/0M/wmdp\endcsname{48.1}
\expandafter\def\csname ev@historical:qwen/bio/F2-A2/0M/mmlu\endcsname{2.8}
\expandafter\def\csname ev@historical:qwen/bio/F3-A3/250M/wmdp\endcsname{38.7}
\expandafter\def\csname ev@historical:qwen/bio/F3-A3/250M/mmlu\endcsname{2.3}
\expandafter\def\csname ev@historical:qwen/bio/B-A2/0M/wmdp\endcsname{50.1}
\expandafter\def\csname ev@historical:qwen/bio/B-A2/0M/mmlu\endcsname{5.3}
\expandafter\def\csname ev@historical:qwen/bio/A3-direct-A3/250M/wmdp\endcsname{19.2}
\expandafter\def\csname ev@historical:qwen/bio/A3-direct-A3/250M/mmlu\endcsname{0.9}
\expandafter\def\csname ev@historical:qwen/bio/A3-rise/250M/wmdp\endcsname{10.5}
\expandafter\def\csname ev@historical:qwen/bio/F3-rise/250M/wmdp\endcsname{1.1}
\expandafter\def\csname ev@historical:qwen/bio/CEU-SFT-rise/250M/wmdp\endcsname{1.8}
\expandafter\def\csname ev@historical:qwen/bio/GA-SFT-rise/250M/wmdp\endcsname{57.1}
\expandafter\def\csname ev@historical:qwen/bio/GD-SFT-rise/250M/wmdp\endcsname{1.1}
\expandafter\def\csname ev@historical:qwen/bio/PureNPO-SFT-rise/250M/wmdp\endcsname{-0.6}
\expandafter\def\csname ev@historical:qwen/bio/PDU-SFT-rise/250M/wmdp\endcsname{31.7}
\expandafter\def\csname ev@historical:qwen/bio/ExternalRMU-SFT-rise/250M/wmdp\endcsname{33.7}
\expandafter\def\csname ev@historical:qwen/bio/SatImp-SFT-rise/250M/wmdp\endcsname{2.7}
\expandafter\def\csname ev@historical:qwen/bio/SimNPO-SFT-rise/250M/wmdp\endcsname{0.4}
\expandafter\def\csname ev@historical:qwen/bio/WGA-SFT-rise/250M/wmdp\endcsname{1.6}
\expandafter\def\csname ev@historical:qwen/bio/best-baseline/0M/wmdp\endcsname{24.7}
\expandafter\def\csname ev@historical:qwen/bio/A2-best/0M/wmdp\endcsname{8.2}
\expandafter\def\csname ev@historical:qwen/bio/best-A2/0M/wmdp\endcsname{-8.2}
\expandafter\def\csname ev@historical:qwen/bio/best-baseline/250M/wmdp\endcsname{75.7}
\expandafter\def\csname ev@historical:qwen/bio/best-A3/250M/wmdp\endcsname{32.3}
\expandafter\def\csname ev@historical:intern/bio/F2-A2/0M/wmdp\endcsname{32.5}
\expandafter\def\csname ev@historical:intern/bio/F2-A2/0M/mmlu\endcsname{1.2}
\expandafter\def\csname ev@historical:intern/bio/F3-A3/250M/wmdp\endcsname{23.6}
\expandafter\def\csname ev@historical:intern/bio/F3-A3/250M/mmlu\endcsname{1.4}
\expandafter\def\csname ev@historical:intern/bio/B-A2/0M/wmdp\endcsname{54.6}
\expandafter\def\csname ev@historical:intern/bio/B-A2/0M/mmlu\endcsname{4.1}
\expandafter\def\csname ev@historical:intern/bio/A3-direct-A3/250M/wmdp\endcsname{9.1}
\expandafter\def\csname ev@historical:intern/bio/A3-direct-A3/250M/mmlu\endcsname{0.2}
\expandafter\def\csname ev@historical:intern/bio/A3-rise/250M/wmdp\endcsname{25.6}
\expandafter\def\csname ev@historical:intern/bio/F3-rise/250M/wmdp\endcsname{16.6}
\expandafter\def\csname ev@historical:intern/bio/CEU-SFT-rise/250M/wmdp\endcsname{13.8}
\expandafter\def\csname ev@historical:intern/bio/GA-SFT-rise/250M/wmdp\endcsname{23.8}
\expandafter\def\csname ev@historical:intern/bio/GD-SFT-rise/250M/wmdp\endcsname{16.0}
\expandafter\def\csname ev@historical:intern/bio/PureNPO-SFT-rise/250M/wmdp\endcsname{1.5}
\expandafter\def\csname ev@historical:intern/bio/PDU-SFT-rise/250M/wmdp\endcsname{29.5}
\expandafter\def\csname ev@historical:intern/bio/ExternalRMU-SFT-rise/250M/wmdp\endcsname{46.2}
\expandafter\def\csname ev@historical:intern/bio/SatImp-SFT-rise/250M/wmdp\endcsname{9.4}
\expandafter\def\csname ev@historical:intern/bio/SimNPO-SFT-rise/250M/wmdp\endcsname{0.7}
\expandafter\def\csname ev@historical:intern/bio/WGA-SFT-rise/250M/wmdp\endcsname{10.6}
\expandafter\def\csname ev@historical:intern/bio/best-baseline/0M/wmdp\endcsname{24.7}
\expandafter\def\csname ev@historical:intern/bio/A2-best/0M/wmdp\endcsname{4.8}
\expandafter\def\csname ev@historical:intern/bio/best-A2/0M/wmdp\endcsname{-4.8}
\expandafter\def\csname ev@historical:intern/bio/best-baseline/250M/wmdp\endcsname{75.7}
\expandafter\def\csname ev@historical:intern/bio/best-A3/250M/wmdp\endcsname{20.6}
\expandafter\def\csname ev@historical:gemma/bio/F2-A2/0M/wmdp\endcsname{38.4}
\expandafter\def\csname ev@historical:gemma/bio/F2-A2/0M/mmlu\endcsname{3.9}
\expandafter\def\csname ev@historical:gemma/bio/F3-A3/250M/wmdp\endcsname{36.1}
\expandafter\def\csname ev@historical:gemma/bio/F3-A3/250M/mmlu\endcsname{3.4}
\expandafter\def\csname ev@historical:gemma/bio/B-A2/0M/wmdp\endcsname{46.7}
\expandafter\def\csname ev@historical:gemma/bio/B-A2/0M/mmlu\endcsname{5.5}
\expandafter\def\csname ev@historical:gemma/bio/A3-direct-A3/250M/wmdp\endcsname{14.8}
\expandafter\def\csname ev@historical:gemma/bio/A3-direct-A3/250M/mmlu\endcsname{1.0}
\expandafter\def\csname ev@historical:gemma/bio/A3-rise/250M/wmdp\endcsname{7.9}
\expandafter\def\csname ev@historical:gemma/bio/F3-rise/250M/wmdp\endcsname{5.7}
\expandafter\def\csname ev@historical:gemma/bio/CEU-SFT-rise/250M/wmdp\endcsname{3.4}
\expandafter\def\csname ev@historical:gemma/bio/GA-SFT-rise/250M/wmdp\endcsname{-0.7}
\expandafter\def\csname ev@historical:gemma/bio/GD-SFT-rise/250M/wmdp\endcsname{49.6}
\expandafter\def\csname ev@historical:gemma/bio/PureNPO-SFT-rise/250M/wmdp\endcsname{2.9}
\expandafter\def\csname ev@historical:gemma/bio/PDU-SFT-rise/250M/wmdp\endcsname{9.1}
\expandafter\def\csname ev@historical:gemma/bio/ExternalRMU-SFT-rise/250M/wmdp\endcsname{31.4}
\expandafter\def\csname ev@historical:gemma/bio/SatImp-SFT-rise/250M/wmdp\endcsname{16.0}
\expandafter\def\csname ev@historical:gemma/bio/SimNPO-SFT-rise/250M/wmdp\endcsname{1.7}
\expandafter\def\csname ev@historical:gemma/bio/WGA-SFT-rise/250M/wmdp\endcsname{12.6}
\expandafter\def\csname ev@historical:gemma/bio/best-baseline/0M/wmdp\endcsname{24.7}
\expandafter\def\csname ev@historical:gemma/bio/A2-best/0M/wmdp\endcsname{7.8}
\expandafter\def\csname ev@historical:gemma/bio/best-A2/0M/wmdp\endcsname{-7.8}
\expandafter\def\csname ev@historical:gemma/bio/best-baseline/250M/wmdp\endcsname{60.9}
\expandafter\def\csname ev@historical:gemma/bio/best-A3/250M/wmdp\endcsname{20.5}
\expandafter\def\csname ev@historical:qwen/cyber/F2-A2/0M/wmdp\endcsname{18.6}
\expandafter\def\csname ev@historical:qwen/cyber/F2-A2/0M/mmlu\endcsname{0.7}
\expandafter\def\csname ev@historical:qwen/cyber/F3-A3/250M/wmdp\endcsname{23.2}
\expandafter\def\csname ev@historical:qwen/cyber/F3-A3/250M/mmlu\endcsname{0.5}
\expandafter\def\csname ev@historical:qwen/cyber/B-A2/0M/wmdp\endcsname{28.7}
\expandafter\def\csname ev@historical:qwen/cyber/B-A2/0M/mmlu\endcsname{1.9}
\expandafter\def\csname ev@historical:qwen/cyber/A3-direct-A3/250M/wmdp\endcsname{8.3}
\expandafter\def\csname ev@historical:qwen/cyber/A3-direct-A3/250M/mmlu\endcsname{0.1}
\expandafter\def\csname ev@historical:qwen/cyber/A3-rise/250M/wmdp\endcsname{-0.8}
\expandafter\def\csname ev@historical:qwen/cyber/F3-rise/250M/wmdp\endcsname{3.7}
\expandafter\def\csname ev@historical:qwen/cyber/CEU-SFT-rise/250M/wmdp\endcsname{11.7}
\expandafter\def\csname ev@historical:qwen/cyber/GA-SFT-rise/250M/wmdp\endcsname{9.8}
\expandafter\def\csname ev@historical:qwen/cyber/GD-SFT-rise/250M/wmdp\endcsname{9.8}
\expandafter\def\csname ev@historical:qwen/cyber/PureNPO-SFT-rise/250M/wmdp\endcsname{6.2}
\expandafter\def\csname ev@historical:qwen/cyber/PDU-SFT-rise/250M/wmdp\endcsname{3.2}
\expandafter\def\csname ev@historical:qwen/cyber/ExternalRMU-SFT-rise/250M/wmdp\endcsname{12.8}
\expandafter\def\csname ev@historical:qwen/cyber/SatImp-SFT-rise/250M/wmdp\endcsname{11.4}
\expandafter\def\csname ev@historical:qwen/cyber/SimNPO-SFT-rise/250M/wmdp\endcsname{9.6}
\expandafter\def\csname ev@historical:qwen/cyber/WGA-SFT-rise/250M/wmdp\endcsname{11.9}
\expandafter\def\csname ev@historical:qwen/cyber/best-baseline/0M/wmdp\endcsname{27.7}
\expandafter\def\csname ev@historical:qwen/cyber/A2-best/0M/wmdp\endcsname{0.1}
\expandafter\def\csname ev@historical:qwen/cyber/best-A2/0M/wmdp\endcsname{-0.1}
\expandafter\def\csname ev@historical:qwen/cyber/best-baseline/250M/wmdp\endcsname{40.6}
\expandafter\def\csname ev@historical:qwen/cyber/best-A3/250M/wmdp\endcsname{13.6}
\expandafter\def\csname ev@historical:intern/cyber/F2-A2/0M/wmdp\endcsname{27.0}
\expandafter\def\csname ev@historical:intern/cyber/F2-A2/0M/mmlu\endcsname{0.4}
\expandafter\def\csname ev@historical:intern/cyber/F3-A3/250M/wmdp\endcsname{26.4}
\expandafter\def\csname ev@historical:intern/cyber/F3-A3/250M/mmlu\endcsname{0.2}
\expandafter\def\csname ev@historical:intern/cyber/B-A2/0M/wmdp\endcsname{33.5}
\expandafter\def\csname ev@historical:intern/cyber/B-A2/0M/mmlu\endcsname{1.0}
\expandafter\def\csname ev@historical:intern/cyber/A3-direct-A3/250M/wmdp\endcsname{7.0}
\expandafter\def\csname ev@historical:intern/cyber/A3-direct-A3/250M/mmlu\endcsname{-0.1}
\expandafter\def\csname ev@historical:intern/cyber/A3-rise/250M/wmdp\endcsname{3.1}
\expandafter\def\csname ev@historical:intern/cyber/F3-rise/250M/wmdp\endcsname{2.5}
\expandafter\def\csname ev@historical:intern/cyber/CEU-SFT-rise/250M/wmdp\endcsname{7.8}
\expandafter\def\csname ev@historical:intern/cyber/GA-SFT-rise/250M/wmdp\endcsname{22.9}
\expandafter\def\csname ev@historical:intern/cyber/GD-SFT-rise/250M/wmdp\endcsname{17.2}
\expandafter\def\csname ev@historical:intern/cyber/PureNPO-SFT-rise/250M/wmdp\endcsname{9.5}
\expandafter\def\csname ev@historical:intern/cyber/PDU-SFT-rise/250M/wmdp\endcsname{11.8}
\expandafter\def\csname ev@historical:intern/cyber/ExternalRMU-SFT-rise/250M/wmdp\endcsname{14.5}
\expandafter\def\csname ev@historical:intern/cyber/SatImp-SFT-rise/250M/wmdp\endcsname{13.7}
\expandafter\def\csname ev@historical:intern/cyber/SimNPO-SFT-rise/250M/wmdp\endcsname{5.9}
\expandafter\def\csname ev@historical:intern/cyber/WGA-SFT-rise/250M/wmdp\endcsname{13.2}
\expandafter\def\csname ev@historical:intern/cyber/best-baseline/0M/wmdp\endcsname{24.6}
\expandafter\def\csname ev@historical:intern/cyber/A2-best/0M/wmdp\endcsname{1.5}
\expandafter\def\csname ev@historical:intern/cyber/best-A2/0M/wmdp\endcsname{-1.5}
\expandafter\def\csname ev@historical:intern/cyber/best-baseline/250M/wmdp\endcsname{39.5}
\expandafter\def\csname ev@historical:intern/cyber/best-A3/250M/wmdp\endcsname{10.4}
\expandafter\def\csname ev@historical:gemma/cyber/F2-A2/0M/wmdp\endcsname{8.8}
\expandafter\def\csname ev@historical:gemma/cyber/F2-A2/0M/mmlu\endcsname{0.4}
\expandafter\def\csname ev@historical:gemma/cyber/F3-A3/250M/wmdp\endcsname{11.2}
\expandafter\def\csname ev@historical:gemma/cyber/F3-A3/250M/mmlu\endcsname{0.9}
\expandafter\def\csname ev@historical:gemma/cyber/B-A2/0M/wmdp\endcsname{19.9}
\expandafter\def\csname ev@historical:gemma/cyber/B-A2/0M/mmlu\endcsname{2.2}
\expandafter\def\csname ev@historical:gemma/cyber/A3-direct-A3/250M/wmdp\endcsname{2.8}
\expandafter\def\csname ev@historical:gemma/cyber/A3-direct-A3/250M/mmlu\endcsname{0.0}
\expandafter\def\csname ev@historical:gemma/cyber/A3-rise/250M/wmdp\endcsname{3.2}
\expandafter\def\csname ev@historical:gemma/cyber/F3-rise/250M/wmdp\endcsname{5.5}
\expandafter\def\csname ev@historical:gemma/cyber/CEU-SFT-rise/250M/wmdp\endcsname{6.0}
\expandafter\def\csname ev@historical:gemma/cyber/GA-SFT-rise/250M/wmdp\endcsname{5.3}
\expandafter\def\csname ev@historical:gemma/cyber/GD-SFT-rise/250M/wmdp\endcsname{17.5}
\expandafter\def\csname ev@historical:gemma/cyber/PureNPO-SFT-rise/250M/wmdp\endcsname{2.8}
\expandafter\def\csname ev@historical:gemma/cyber/PDU-SFT-rise/250M/wmdp\endcsname{10.1}
\expandafter\def\csname ev@historical:gemma/cyber/ExternalRMU-SFT-rise/250M/wmdp\endcsname{9.6}
\expandafter\def\csname ev@historical:gemma/cyber/SatImp-SFT-rise/250M/wmdp\endcsname{14.5}
\expandafter\def\csname ev@historical:gemma/cyber/SimNPO-SFT-rise/250M/wmdp\endcsname{10.3}
\expandafter\def\csname ev@historical:gemma/cyber/WGA-SFT-rise/250M/wmdp\endcsname{15.1}
\expandafter\def\csname ev@historical:gemma/cyber/best-baseline/0M/wmdp\endcsname{24.8}
\expandafter\def\csname ev@historical:gemma/cyber/A2-best/0M/wmdp\endcsname{5.5}
\expandafter\def\csname ev@historical:gemma/cyber/best-A2/0M/wmdp\endcsname{-5.5}
\expandafter\def\csname ev@historical:gemma/cyber/best-baseline/250M/wmdp\endcsname{31.9}
\expandafter\def\csname ev@historical:gemma/cyber/best-A3/250M/wmdp\endcsname{-1.6}
\expandafter\def\csname ev@historical:all/all/A2-inband/0M/count\endcsname{2}
\expandafter\def\csname ev@historical:all/all/A2-below-best/0M/count\endcsname{0}
\expandafter\def\csname ev@historical:all/all/A2-best-max/0M/wmdp\endcsname{8.2}
\expandafter\def\csname ev@historical:all/all/usable-spread-min/0M/points\endcsname{14}
\expandafter\def\csname ev@historical:all/all/A3-lowest/250M/count\endcsname{5}
\expandafter\def\csname ev@historical:all/all/late-collapse/250M/count\endcsname{2}
\expandafter\def\csname ev@historical:all/bio/B-min/0M/wmdp\endcsname{79.1}
\expandafter\def\csname ev@historical:all/bio/B-max/0M/wmdp\endcsname{84.0}
\expandafter\def\csname ev@historical:all/bio/A2-min/0M/wmdp\endcsname{29.5}
\expandafter\def\csname ev@historical:all/bio/A2-max/0M/wmdp\endcsname{32.9}
\expandafter\def\csname ev@historical:all/cyber/B-min/0M/wmdp\endcsname{50.2}
\expandafter\def\csname ev@historical:all/cyber/B-max/0M/wmdp\endcsname{59.5}
\expandafter\def\csname ev@historical:all/cyber/A2-min/0M/wmdp\endcsname{26.0}
\expandafter\def\csname ev@historical:all/cyber/A2-max/0M/wmdp\endcsname{30.3}
\expandafter\def\csname ev@historical:all/all/sep-mmlu-max/0M/mmlu\endcsname{3.9}
\expandafter\def\csname ev@historical:all/bio/A3-kept-min/250M/pct\endcsname{53}
\expandafter\def\csname ev@historical:all/cyber/A3-kept-min/250M/pct\endcsname{84}
\expandafter\def\csname ev@historical:all/bio/top2-kept-max/250M/pct\endcsname{102}
\expandafter\def\csname ev@historical:all/cyber/top2-kept-max/250M/pct\endcsname{76}
\expandafter\def\csname ev@historical:all/bio/best-A3-min/250M/wmdp\endcsname{20.5}
\expandafter\def\csname ev@historical:all/bio/best-A3-max/250M/wmdp\endcsname{32.3}
\expandafter\def\csname ev@historical:all/cyber/best-A3-min/250M/wmdp\endcsname{-1.6}
\expandafter\def\csname ev@historical:all/cyber/best-A3-max/250M/wmdp\endcsname{13.6}
\expandafter\def\csname ev@historical:all/bio/A3-mmlu-rise-min/250M/mmlu\endcsname{1.3}
\expandafter\def\csname ev@historical:all/bio/A3-mmlu-rise-max/250M/mmlu\endcsname{1.9}
\expandafter\def\csname ev@historical:all/bio/R-SFT-A3-direct-min/250M/wmdp\endcsname{15.6}
\expandafter\def\csname ev@historical:all/bio/R-SFT-A3-direct-max/250M/wmdp\endcsname{16.4}
\expandafter\def\csname ev@historical:all/bio/R-SFT-A3-direct-positive/250M/count\endcsname{3}
\expandafter\def\csname ev@historical:all/cyber/R-SFT-A3-direct-min/250M/wmdp\endcsname{1.2}
\expandafter\def\csname ev@historical:all/cyber/R-SFT-A3-direct-max/250M/wmdp\endcsname{8.9}
\expandafter\def\csname ev@historical:all/cyber/R-SFT-A3-direct-positive/250M/count\endcsname{3}
\expandafter\def\csname ev@historical:all/bio/A3-direct-A3-min/250M/wmdp\endcsname{9.1}
\expandafter\def\csname ev@historical:all/bio/A3-direct-A3-max/250M/wmdp\endcsname{19.2}
\expandafter\def\csname ev@historical:all/cyber/A3-direct-A3-min/250M/wmdp\endcsname{2.8}
\expandafter\def\csname ev@historical:all/cyber/A3-direct-A3-max/250M/wmdp\endcsname{8.3}
\expandafter\def\csname ev@historical:all/all/F3-ref-gap-min/250M/wmdp\endcsname{1.8}
\expandafter\def\csname ev@historical:all/all/F3-ref-gap-max/250M/wmdp\endcsname{6.2}
\expandafter\def\csname ev@historical:all/all/combos/0M/count\endcsname{6}
\expandafter\def\csname ev@historical:all/all/band/0M/points\endcsname{3}
\expandafter\def\csname ev@historical:all/all/collapse/0M/points\endcsname{10}
\expandafter\def\csname ev@variant:intern/bio/Random-excision-mean/0M/wmdp\endcsname{78.9}
\expandafter\def\csname ev@variant:intern/bio/Random-excision-sd/0M/wmdp\endcsname{0.9}
\expandafter\def\csname ev@variant:intern/bio/F1-Random-excision/0M/wmdp\endcsname{0.5}
\expandafter\def\csname ev@variant:intern/bio/Random-excision-mean/0M/mmlu\endcsname{82.4}
\expandafter\def\csname ev@variant:intern/bio/Random-excision-sd/0M/mmlu\endcsname{0.2}
\expandafter\def\csname ev@variant:intern/bio/F1-Random-excision/0M/mmlu\endcsname{0.8}
\expandafter\def\csname ev@qwen/bio/F0-A0/0M/wmdp\endcsname{13.1}
\expandafter\def\csname ev@qwen/bio/F0-A0/0M/mmlu\endcsname{1.4}
\expandafter\def\csname ev@qwen/bio/F1-A1/0M/wmdp\endcsname{18.9}
\expandafter\def\csname ev@qwen/bio/F1-A1/0M/mmlu\endcsname{1.8}
\expandafter\def\csname ev@qwen/bio/F2-A2/0M/wmdp\endcsname{48.4}
\expandafter\def\csname ev@qwen/bio/F2-A2/0M/mmlu\endcsname{2.9}
\expandafter\def\csname ev@qwen/bio/F3-A3/250M/wmdp\endcsname{38.9}
\expandafter\def\csname ev@qwen/bio/F3-A3/250M/mmlu\endcsname{2.0}
\expandafter\def\csname ev@qwen/bio/R-A2/0M/wmdp\endcsname{9.1}
\expandafter\def\csname ev@qwen/bio/R-A2/0M/mmlu\endcsname{4.6}
\expandafter\def\csname ev@qwen/bio/A1-A2/0M/wmdp\endcsname{30.5}
\expandafter\def\csname ev@qwen/bio/A1-A2/0M/mmlu\endcsname{1.4}
\expandafter\def\csname ev@qwen/bio/B-A2/0M/wmdp\endcsname{50.8}
\expandafter\def\csname ev@qwen/bio/B-A2/0M/mmlu\endcsname{5.2}
\expandafter\def\csname ev@qwen/bio/B-F1/0M/wmdp\endcsname{1.3}
\expandafter\def\csname ev@qwen/bio/B-F1/0M/mmlu\endcsname{2.0}
\expandafter\def\csname ev@qwen/bio/A3-direct-A3/250M/wmdp\endcsname{20.2}
\expandafter\def\csname ev@qwen/bio/A3-direct-A3/250M/mmlu\endcsname{0.7}
\expandafter\def\csname ev@qwen/bio/A3-rise/250M/wmdp\endcsname{10.5}
\expandafter\def\csname ev@qwen/bio/F3-rise/250M/wmdp\endcsname{1.1}
\expandafter\def\csname ev@qwen/bio/CEU-SFT-rise/250M/wmdp\endcsname{1.8}
\expandafter\def\csname ev@qwen/bio/GA-SFT-rise/250M/wmdp\endcsname{57.1}
\expandafter\def\csname ev@qwen/bio/GD-SFT-rise/250M/wmdp\endcsname{1.1}
\expandafter\def\csname ev@qwen/bio/PureNPO-SFT-rise/250M/wmdp\endcsname{-0.6}
\expandafter\def\csname ev@qwen/bio/PDU-SFT-rise/250M/wmdp\endcsname{31.7}
\expandafter\def\csname ev@qwen/bio/ExternalRMU-SFT-rise/250M/wmdp\endcsname{33.7}
\expandafter\def\csname ev@qwen/bio/SatImp-SFT-rise/250M/wmdp\endcsname{2.7}
\expandafter\def\csname ev@qwen/bio/SimNPO-SFT-rise/250M/wmdp\endcsname{0.4}
\expandafter\def\csname ev@qwen/bio/WGA-SFT-rise/250M/wmdp\endcsname{1.6}
\expandafter\def\csname ev@qwen/bio/best-baseline/0M/wmdp\endcsname{44.1}
\expandafter\def\csname ev@qwen/bio/A2-best/0M/wmdp\endcsname{-11.6}
\expandafter\def\csname ev@qwen/bio/best-A2/0M/wmdp\endcsname{11.6}
\expandafter\def\csname ev@qwen/bio/best-baseline/250M/wmdp\endcsname{76.0}
\expandafter\def\csname ev@qwen/bio/best-A3/250M/wmdp\endcsname{33.2}
\expandafter\def\csname ev@intern/bio/F0-A0/0M/wmdp\endcsname{13.7}
\expandafter\def\csname ev@intern/bio/F0-A0/0M/mmlu\endcsname{1.8}
\expandafter\def\csname ev@intern/bio/F1-A1/0M/wmdp\endcsname{15.1}
\expandafter\def\csname ev@intern/bio/F1-A1/0M/mmlu\endcsname{1.0}
\expandafter\def\csname ev@intern/bio/F2-A2/0M/wmdp\endcsname{32.3}
\expandafter\def\csname ev@intern/bio/F2-A2/0M/mmlu\endcsname{1.4}
\expandafter\def\csname ev@intern/bio/F3-A3/250M/wmdp\endcsname{23.6}
\expandafter\def\csname ev@intern/bio/F3-A3/250M/mmlu\endcsname{1.5}
\expandafter\def\csname ev@intern/bio/R-A2/0M/wmdp\endcsname{3.7}
\expandafter\def\csname ev@intern/bio/R-A2/0M/mmlu\endcsname{3.5}
\expandafter\def\csname ev@intern/bio/A1-A2/0M/wmdp\endcsname{34.7}
\expandafter\def\csname ev@intern/bio/A1-A2/0M/mmlu\endcsname{1.3}
\expandafter\def\csname ev@intern/bio/B-A2/0M/wmdp\endcsname{53.8}
\expandafter\def\csname ev@intern/bio/B-A2/0M/mmlu\endcsname{4.1}
\expandafter\def\csname ev@intern/bio/B-F1/0M/wmdp\endcsname{4.0}
\expandafter\def\csname ev@intern/bio/B-F1/0M/mmlu\endcsname{1.8}
\expandafter\def\csname ev@intern/bio/A3-direct-A3/250M/wmdp\endcsname{9.6}
\expandafter\def\csname ev@intern/bio/A3-direct-A3/250M/mmlu\endcsname{0.4}
\expandafter\def\csname ev@intern/bio/A3-rise/250M/wmdp\endcsname{25.6}
\expandafter\def\csname ev@intern/bio/F3-rise/250M/wmdp\endcsname{16.6}
\expandafter\def\csname ev@intern/bio/CEU-SFT-rise/250M/wmdp\endcsname{13.8}
\expandafter\def\csname ev@intern/bio/GA-SFT-rise/250M/wmdp\endcsname{23.8}
\expandafter\def\csname ev@intern/bio/GD-SFT-rise/250M/wmdp\endcsname{16.0}
\expandafter\def\csname ev@intern/bio/PureNPO-SFT-rise/250M/wmdp\endcsname{1.5}
\expandafter\def\csname ev@intern/bio/PDU-SFT-rise/250M/wmdp\endcsname{29.5}
\expandafter\def\csname ev@intern/bio/ExternalRMU-SFT-rise/250M/wmdp\endcsname{46.2}
\expandafter\def\csname ev@intern/bio/SatImp-SFT-rise/250M/wmdp\endcsname{9.4}
\expandafter\def\csname ev@intern/bio/SimNPO-SFT-rise/250M/wmdp\endcsname{0.7}
\expandafter\def\csname ev@intern/bio/WGA-SFT-rise/250M/wmdp\endcsname{10.6}
\expandafter\def\csname ev@intern/bio/best-baseline/0M/wmdp\endcsname{37.4}
\expandafter\def\csname ev@intern/bio/A2-best/0M/wmdp\endcsname{-7.8}
\expandafter\def\csname ev@intern/bio/best-A2/0M/wmdp\endcsname{7.8}
\expandafter\def\csname ev@intern/bio/best-baseline/250M/wmdp\endcsname{75.6}
\expandafter\def\csname ev@intern/bio/best-A3/250M/wmdp\endcsname{20.6}
\expandafter\def\csname ev@gemma/bio/F0-A0/0M/wmdp\endcsname{5.7}
\expandafter\def\csname ev@gemma/bio/F0-A0/0M/mmlu\endcsname{1.4}
\expandafter\def\csname ev@gemma/bio/F1-A1/0M/wmdp\endcsname{6.0}
\expandafter\def\csname ev@gemma/bio/F1-A1/0M/mmlu\endcsname{1.9}
\expandafter\def\csname ev@gemma/bio/F2-A2/0M/wmdp\endcsname{37.4}
\expandafter\def\csname ev@gemma/bio/F2-A2/0M/mmlu\endcsname{3.9}
\expandafter\def\csname ev@gemma/bio/F3-A3/250M/wmdp\endcsname{35.1}
\expandafter\def\csname ev@gemma/bio/F3-A3/250M/mmlu\endcsname{3.4}
\expandafter\def\csname ev@gemma/bio/R-A2/0M/wmdp\endcsname{14.9}
\expandafter\def\csname ev@gemma/bio/R-A2/0M/mmlu\endcsname{3.5}
\expandafter\def\csname ev@gemma/bio/A1-A2/0M/wmdp\endcsname{36.6}
\expandafter\def\csname ev@gemma/bio/A1-A2/0M/mmlu\endcsname{3.3}
\expandafter\def\csname ev@gemma/bio/B-A2/0M/wmdp\endcsname{45.3}
\expandafter\def\csname ev@gemma/bio/B-A2/0M/mmlu\endcsname{5.6}
\expandafter\def\csname ev@gemma/bio/B-F1/0M/wmdp\endcsname{2.7}
\expandafter\def\csname ev@gemma/bio/B-F1/0M/mmlu\endcsname{0.4}
\expandafter\def\csname ev@gemma/bio/A3-direct-A3/250M/wmdp\endcsname{14.4}
\expandafter\def\csname ev@gemma/bio/A3-direct-A3/250M/mmlu\endcsname{1.2}
\expandafter\def\csname ev@gemma/bio/A3-rise/250M/wmdp\endcsname{7.9}
\expandafter\def\csname ev@gemma/bio/F3-rise/250M/wmdp\endcsname{5.7}
\expandafter\def\csname ev@gemma/bio/CEU-SFT-rise/250M/wmdp\endcsname{3.4}
\expandafter\def\csname ev@gemma/bio/GA-SFT-rise/250M/wmdp\endcsname{-0.7}
\expandafter\def\csname ev@gemma/bio/GD-SFT-rise/250M/wmdp\endcsname{49.6}
\expandafter\def\csname ev@gemma/bio/PureNPO-SFT-rise/250M/wmdp\endcsname{2.9}
\expandafter\def\csname ev@gemma/bio/PDU-SFT-rise/250M/wmdp\endcsname{9.1}
\expandafter\def\csname ev@gemma/bio/ExternalRMU-SFT-rise/250M/wmdp\endcsname{31.4}
\expandafter\def\csname ev@gemma/bio/SatImp-SFT-rise/250M/wmdp\endcsname{16.0}
\expandafter\def\csname ev@gemma/bio/SimNPO-SFT-rise/250M/wmdp\endcsname{1.7}
\expandafter\def\csname ev@gemma/bio/WGA-SFT-rise/250M/wmdp\endcsname{12.6}
\expandafter\def\csname ev@gemma/bio/best-baseline/0M/wmdp\endcsname{28.6}
\expandafter\def\csname ev@gemma/bio/A2-best/0M/wmdp\endcsname{5.0}
\expandafter\def\csname ev@gemma/bio/best-A2/0M/wmdp\endcsname{-5.0}
\expandafter\def\csname ev@gemma/bio/best-baseline/250M/wmdp\endcsname{60.8}
\expandafter\def\csname ev@gemma/bio/best-A3/250M/wmdp\endcsname{19.5}
\expandafter\def\csname ev@qwen/cyber/F0-A0/0M/wmdp\endcsname{7.9}
\expandafter\def\csname ev@qwen/cyber/F0-A0/0M/mmlu\endcsname{0.6}
\expandafter\def\csname ev@qwen/cyber/F1-A1/0M/wmdp\endcsname{14.4}
\expandafter\def\csname ev@qwen/cyber/F1-A1/0M/mmlu\endcsname{0.4}
\expandafter\def\csname ev@qwen/cyber/F2-A2/0M/wmdp\endcsname{18.4}
\expandafter\def\csname ev@qwen/cyber/F2-A2/0M/mmlu\endcsname{0.7}
\expandafter\def\csname ev@qwen/cyber/F3-A3/250M/wmdp\endcsname{23.0}
\expandafter\def\csname ev@qwen/cyber/F3-A3/250M/mmlu\endcsname{0.8}
\expandafter\def\csname ev@qwen/cyber/R-A2/0M/wmdp\endcsname{3.6}
\expandafter\def\csname ev@qwen/cyber/R-A2/0M/mmlu\endcsname{1.6}
\expandafter\def\csname ev@qwen/cyber/A1-A2/0M/wmdp\endcsname{15.0}
\expandafter\def\csname ev@qwen/cyber/A1-A2/0M/mmlu\endcsname{0.3}
\expandafter\def\csname ev@qwen/cyber/B-A2/0M/wmdp\endcsname{29.4}
\expandafter\def\csname ev@qwen/cyber/B-A2/0M/mmlu\endcsname{1.7}
\expandafter\def\csname ev@qwen/cyber/B-F1/0M/wmdp\endcsname{-0.0}
\expandafter\def\csname ev@qwen/cyber/B-F1/0M/mmlu\endcsname{1.0}
\expandafter\def\csname ev@qwen/cyber/A3-direct-A3/250M/wmdp\endcsname{8.6}
\expandafter\def\csname ev@qwen/cyber/A3-direct-A3/250M/mmlu\endcsname{0.3}
\expandafter\def\csname ev@qwen/cyber/A3-rise/250M/wmdp\endcsname{-0.8}
\expandafter\def\csname ev@qwen/cyber/F3-rise/250M/wmdp\endcsname{3.7}
\expandafter\def\csname ev@qwen/cyber/CEU-SFT-rise/250M/wmdp\endcsname{11.7}
\expandafter\def\csname ev@qwen/cyber/GA-SFT-rise/250M/wmdp\endcsname{9.8}
\expandafter\def\csname ev@qwen/cyber/GD-SFT-rise/250M/wmdp\endcsname{9.8}
\expandafter\def\csname ev@qwen/cyber/PureNPO-SFT-rise/250M/wmdp\endcsname{6.2}
\expandafter\def\csname ev@qwen/cyber/PDU-SFT-rise/250M/wmdp\endcsname{3.2}
\expandafter\def\csname ev@qwen/cyber/ExternalRMU-SFT-rise/250M/wmdp\endcsname{12.8}
\expandafter\def\csname ev@qwen/cyber/SatImp-SFT-rise/250M/wmdp\endcsname{11.4}
\expandafter\def\csname ev@qwen/cyber/SimNPO-SFT-rise/250M/wmdp\endcsname{9.6}
\expandafter\def\csname ev@qwen/cyber/WGA-SFT-rise/250M/wmdp\endcsname{11.9}
\expandafter\def\csname ev@qwen/cyber/best-baseline/0M/wmdp\endcsname{27.5}
\expandafter\def\csname ev@qwen/cyber/A2-best/0M/wmdp\endcsname{0.6}
\expandafter\def\csname ev@qwen/cyber/best-A2/0M/wmdp\endcsname{-0.6}
\expandafter\def\csname ev@qwen/cyber/best-baseline/250M/wmdp\endcsname{40.6}
\expandafter\def\csname ev@qwen/cyber/best-A3/250M/wmdp\endcsname{13.3}
\expandafter\def\csname ev@intern/cyber/F0-A0/0M/wmdp\endcsname{7.7}
\expandafter\def\csname ev@intern/cyber/F0-A0/0M/mmlu\endcsname{0.4}
\expandafter\def\csname ev@intern/cyber/F1-A1/0M/wmdp\endcsname{12.8}
\expandafter\def\csname ev@intern/cyber/F1-A1/0M/mmlu\endcsname{0.2}
\expandafter\def\csname ev@intern/cyber/F2-A2/0M/wmdp\endcsname{26.6}
\expandafter\def\csname ev@intern/cyber/F2-A2/0M/mmlu\endcsname{0.3}
\expandafter\def\csname ev@intern/cyber/F3-A3/250M/wmdp\endcsname{26.0}
\expandafter\def\csname ev@intern/cyber/F3-A3/250M/mmlu\endcsname{0.0}
\expandafter\def\csname ev@intern/cyber/R-A2/0M/wmdp\endcsname{4.7}
\expandafter\def\csname ev@intern/cyber/R-A2/0M/mmlu\endcsname{0.8}
\expandafter\def\csname ev@intern/cyber/A1-A2/0M/wmdp\endcsname{18.8}
\expandafter\def\csname ev@intern/cyber/A1-A2/0M/mmlu\endcsname{0.2}
\expandafter\def\csname ev@intern/cyber/B-A2/0M/wmdp\endcsname{32.7}
\expandafter\def\csname ev@intern/cyber/B-A2/0M/mmlu\endcsname{1.0}
\expandafter\def\csname ev@intern/cyber/B-F1/0M/wmdp\endcsname{1.0}
\expandafter\def\csname ev@intern/cyber/B-F1/0M/mmlu\endcsname{0.6}
\expandafter\def\csname ev@intern/cyber/A3-direct-A3/250M/wmdp\endcsname{6.1}
\expandafter\def\csname ev@intern/cyber/A3-direct-A3/250M/mmlu\endcsname{0.2}
\expandafter\def\csname ev@intern/cyber/A3-rise/250M/wmdp\endcsname{3.1}
\expandafter\def\csname ev@intern/cyber/F3-rise/250M/wmdp\endcsname{2.5}
\expandafter\def\csname ev@intern/cyber/CEU-SFT-rise/250M/wmdp\endcsname{7.8}
\expandafter\def\csname ev@intern/cyber/GA-SFT-rise/250M/wmdp\endcsname{22.9}
\expandafter\def\csname ev@intern/cyber/GD-SFT-rise/250M/wmdp\endcsname{17.2}
\expandafter\def\csname ev@intern/cyber/PureNPO-SFT-rise/250M/wmdp\endcsname{9.5}
\expandafter\def\csname ev@intern/cyber/PDU-SFT-rise/250M/wmdp\endcsname{11.8}
\expandafter\def\csname ev@intern/cyber/ExternalRMU-SFT-rise/250M/wmdp\endcsname{14.5}
\expandafter\def\csname ev@intern/cyber/SatImp-SFT-rise/250M/wmdp\endcsname{13.7}
\expandafter\def\csname ev@intern/cyber/SimNPO-SFT-rise/250M/wmdp\endcsname{5.9}
\expandafter\def\csname ev@intern/cyber/WGA-SFT-rise/250M/wmdp\endcsname{13.2}
\expandafter\def\csname ev@intern/cyber/best-baseline/0M/wmdp\endcsname{27.8}
\expandafter\def\csname ev@intern/cyber/A2-best/0M/wmdp\endcsname{-1.6}
\expandafter\def\csname ev@intern/cyber/best-A2/0M/wmdp\endcsname{1.6}
\expandafter\def\csname ev@intern/cyber/best-baseline/250M/wmdp\endcsname{40.1}
\expandafter\def\csname ev@intern/cyber/best-A3/250M/wmdp\endcsname{10.3}
\expandafter\def\csname ev@gemma/cyber/F0-A0/0M/wmdp\endcsname{1.5}
\expandafter\def\csname ev@gemma/cyber/F0-A0/0M/mmlu\endcsname{0.3}
\expandafter\def\csname ev@gemma/cyber/F1-A1/0M/wmdp\endcsname{1.5}
\expandafter\def\csname ev@gemma/cyber/F1-A1/0M/mmlu\endcsname{0.4}
\expandafter\def\csname ev@gemma/cyber/F2-A2/0M/wmdp\endcsname{9.5}
\expandafter\def\csname ev@gemma/cyber/F2-A2/0M/mmlu\endcsname{0.6}
\expandafter\def\csname ev@gemma/cyber/F3-A3/250M/wmdp\endcsname{11.1}
\expandafter\def\csname ev@gemma/cyber/F3-A3/250M/mmlu\endcsname{0.6}
\expandafter\def\csname ev@gemma/cyber/R-A2/0M/wmdp\endcsname{1.5}
\expandafter\def\csname ev@gemma/cyber/R-A2/0M/mmlu\endcsname{0.2}
\expandafter\def\csname ev@gemma/cyber/A1-A2/0M/wmdp\endcsname{18.7}
\expandafter\def\csname ev@gemma/cyber/A1-A2/0M/mmlu\endcsname{1.6}
\expandafter\def\csname ev@gemma/cyber/B-A2/0M/wmdp\endcsname{21.4}
\expandafter\def\csname ev@gemma/cyber/B-A2/0M/mmlu\endcsname{2.4}
\expandafter\def\csname ev@gemma/cyber/B-F1/0M/wmdp\endcsname{1.2}
\expandafter\def\csname ev@gemma/cyber/B-F1/0M/mmlu\endcsname{0.4}
\expandafter\def\csname ev@gemma/cyber/A3-direct-A3/250M/wmdp\endcsname{1.7}
\expandafter\def\csname ev@gemma/cyber/A3-direct-A3/250M/mmlu\endcsname{-0.1}
\expandafter\def\csname ev@gemma/cyber/A3-rise/250M/wmdp\endcsname{3.2}
\expandafter\def\csname ev@gemma/cyber/F3-rise/250M/wmdp\endcsname{5.5}
\expandafter\def\csname ev@gemma/cyber/CEU-SFT-rise/250M/wmdp\endcsname{6.0}
\expandafter\def\csname ev@gemma/cyber/GA-SFT-rise/250M/wmdp\endcsname{5.3}
\expandafter\def\csname ev@gemma/cyber/GD-SFT-rise/250M/wmdp\endcsname{17.5}
\expandafter\def\csname ev@gemma/cyber/PureNPO-SFT-rise/250M/wmdp\endcsname{2.8}
\expandafter\def\csname ev@gemma/cyber/PDU-SFT-rise/250M/wmdp\endcsname{10.1}
\expandafter\def\csname ev@gemma/cyber/ExternalRMU-SFT-rise/250M/wmdp\endcsname{9.6}
\expandafter\def\csname ev@gemma/cyber/SatImp-SFT-rise/250M/wmdp\endcsname{14.5}
\expandafter\def\csname ev@gemma/cyber/SimNPO-SFT-rise/250M/wmdp\endcsname{10.3}
\expandafter\def\csname ev@gemma/cyber/WGA-SFT-rise/250M/wmdp\endcsname{15.1}
\expandafter\def\csname ev@gemma/cyber/best-baseline/0M/wmdp\endcsname{25.1}
\expandafter\def\csname ev@gemma/cyber/A2-best/0M/wmdp\endcsname{4.4}
\expandafter\def\csname ev@gemma/cyber/best-A2/0M/wmdp\endcsname{-4.4}
\expandafter\def\csname ev@gemma/cyber/best-baseline/250M/wmdp\endcsname{35.5}
\expandafter\def\csname ev@gemma/cyber/best-A3/250M/wmdp\endcsname{1.5}
\expandafter\def\csname ev@all/all/A2-inband/0M/count\endcsname{4}
\expandafter\def\csname ev@all/all/A2-below-best/0M/count\endcsname{3}
\expandafter\def\csname ev@all/all/A2-best-max/0M/wmdp\endcsname{5.0}
\expandafter\def\csname ev@all/all/usable-spread-min/0M/points\endcsname{15}
\expandafter\def\csname ev@all/all/A3-lowest/250M/count\endcsname{6}
\expandafter\def\csname ev@all/all/late-collapse/250M/count\endcsname{0}
\expandafter\def\csname ev@all/all/A2-retention-min/0M/pct\endcsname{91}
\expandafter\def\csname ev@all/bio/B-min/0M/wmdp\endcsname{78.9}
\expandafter\def\csname ev@all/bio/B-max/0M/wmdp\endcsname{83.4}
\expandafter\def\csname ev@all/bio/A2-min/0M/wmdp\endcsname{29.6}
\expandafter\def\csname ev@all/bio/A2-max/0M/wmdp\endcsname{33.6}
\expandafter\def\csname ev@all/cyber/B-min/0M/wmdp\endcsname{50.8}
\expandafter\def\csname ev@all/cyber/B-max/0M/wmdp\endcsname{58.9}
\expandafter\def\csname ev@all/cyber/A2-min/0M/wmdp\endcsname{26.3}
\expandafter\def\csname ev@all/cyber/A2-max/0M/wmdp\endcsname{29.4}
\expandafter\def\csname ev@all/all/sep-drop-min/0M/wmdp\endcsname{9.5}
\expandafter\def\csname ev@all/all/sep-drop-max/0M/wmdp\endcsname{48.4}
\expandafter\def\csname ev@all/all/sep-drop-min/250M/wmdp\endcsname{11.1}
\expandafter\def\csname ev@all/all/sep-drop-max/250M/wmdp\endcsname{38.9}
\expandafter\def\csname ev@all/all/sep-general-median/0M/acc\endcsname{0.6}
\expandafter\def\csname ev@all/all/sep-general-max/0M/acc\endcsname{9.5}
\expandafter\def\csname ev@all/all/sep-general-second/0M/acc\endcsname{4.3}
\expandafter\def\csname ev@all/all/sep-general-within1/0M/count\endcsname{34}
\expandafter\def\csname ev@all/all/sep-general-within2/0M/count\endcsname{39}
\expandafter\def\csname ev@all/all/sep-general-cells/0M/count\endcsname{48}
\expandafter\def\csname ev@all/all/sep-mmlu-max/0M/mmlu\endcsname{3.9}
\expandafter\def\csname ev@all/bio/A3-kept-min/250M/pct\endcsname{52}
\expandafter\def\csname ev@all/cyber/A3-kept-min/250M/pct\endcsname{78}
\expandafter\def\csname ev@all/bio/top2-kept-max/250M/pct\endcsname{37}
\expandafter\def\csname ev@all/cyber/top2-kept-max/250M/pct\endcsname{63}
\expandafter\def\csname ev@all/bio/best-A3-min/250M/wmdp\endcsname{19.5}
\expandafter\def\csname ev@all/bio/best-A3-max/250M/wmdp\endcsname{33.2}
\expandafter\def\csname ev@all/cyber/best-A3-min/250M/wmdp\endcsname{1.5}
\expandafter\def\csname ev@all/cyber/best-A3-max/250M/wmdp\endcsname{13.3}
\expandafter\def\csname ev@all/bio/A3-mmlu-rise-min/250M/mmlu\endcsname{1.5}
\expandafter\def\csname ev@all/bio/A3-mmlu-rise-max/250M/mmlu\endcsname{1.8}
\expandafter\def\csname ev@all/bio/A3-ifeval-gap-max/250M/ifeval\endcsname{0.4}
\expandafter\def\csname ev@all/bio/R-A2-min/0M/wmdp\endcsname{3.7}
\expandafter\def\csname ev@all/bio/R-A2-max/0M/wmdp\endcsname{14.9}
\expandafter\def\csname ev@all/bio/R-A2-positive/0M/count\endcsname{3}
\expandafter\def\csname ev@all/cyber/R-A2-min/0M/wmdp\endcsname{1.5}
\expandafter\def\csname ev@all/cyber/R-A2-max/0M/wmdp\endcsname{4.7}
\expandafter\def\csname ev@all/cyber/R-A2-positive/0M/count\endcsname{3}
\expandafter\def\csname ev@all/bio/A1-A2-min/0M/wmdp\endcsname{30.5}
\expandafter\def\csname ev@all/bio/A1-A2-max/0M/wmdp\endcsname{36.6}
\expandafter\def\csname ev@all/bio/A1-A2-positive/0M/count\endcsname{3}
\expandafter\def\csname ev@all/cyber/A1-A2-min/0M/wmdp\endcsname{15.0}
\expandafter\def\csname ev@all/cyber/A1-A2-max/0M/wmdp\endcsname{18.8}
\expandafter\def\csname ev@all/cyber/A1-A2-positive/0M/count\endcsname{3}
\expandafter\def\csname ev@all/bio/R-SFT-A3-direct-min/250M/wmdp\endcsname{15.8}
\expandafter\def\csname ev@all/bio/R-SFT-A3-direct-max/250M/wmdp\endcsname{16.4}
\expandafter\def\csname ev@all/bio/R-SFT-A3-direct-positive/250M/count\endcsname{3}
\expandafter\def\csname ev@all/cyber/R-SFT-A3-direct-min/250M/wmdp\endcsname{1.9}
\expandafter\def\csname ev@all/cyber/R-SFT-A3-direct-max/250M/wmdp\endcsname{8.8}
\expandafter\def\csname ev@all/cyber/R-SFT-A3-direct-positive/250M/count\endcsname{3}
\expandafter\def\csname ev@all/bio/A3-direct-A3-min/250M/wmdp\endcsname{9.6}
\expandafter\def\csname ev@all/bio/A3-direct-A3-max/250M/wmdp\endcsname{20.2}
\expandafter\def\csname ev@all/cyber/A3-direct-A3-min/250M/wmdp\endcsname{1.7}
\expandafter\def\csname ev@all/cyber/A3-direct-A3-max/250M/wmdp\endcsname{8.6}
\expandafter\def\csname ev@variant:intern/bio/F1-A1-without_std_activation_mask/0M/wmdp\endcsname{14.8}
\expandafter\def\csname ev@variant:intern/bio/F1-A1-without_routing_loss/0M/wmdp\endcsname{13.8}
\expandafter\def\csname ev@variant:intern/bio/F1-A1-gradient_mask_only/0M/wmdp\endcsname{15.8}
\expandafter\def\csname ev@all/all/F3-ref-gap-min/250M/wmdp\endcsname{2.3}
\expandafter\def\csname ev@all/all/F3-ref-gap-max/250M/wmdp\endcsname{6.0}
\expandafter\def\csname ev@all/bio/adj-share-min/0M/pct\endcsname{57}
\expandafter\def\csname ev@all/bio/adj-share-max/0M/pct\endcsname{75}
\expandafter\def\csname ev@all/bio/rest-drop-min/0M/mmlu\endcsname{1.2}
\expandafter\def\csname ev@all/bio/rest-drop-max/0M/mmlu\endcsname{2.8}
\expandafter\def\csname ev@all/all/combos/0M/count\endcsname{6}
\expandafter\def\csname ev@all/all/band/0M/points\endcsname{3}
\expandafter\def\csname ev@all/all/collapse/0M/points\endcsname{10}
\expandafter\def\csname ev@all/bio/Routing-F0-forget-min/0M/pct\endcsname{17}
\expandafter\def\csname ev@all/bio/Routing-F0-forget-max/0M/pct\endcsname{27}
\expandafter\def\csname ev@all/bio/Routing-F1-forget-min/0M/pct\endcsname{65}
\expandafter\def\csname ev@all/bio/Routing-F1-forget-max/0M/pct\endcsname{76}
\expandafter\def\csname ev@all/cyber/Routing-F0-forget-min/0M/pct\endcsname{15}
\expandafter\def\csname ev@all/cyber/Routing-F0-forget-max/0M/pct\endcsname{22}
\expandafter\def\csname ev@all/cyber/Routing-F1-forget-min/0M/pct\endcsname{35}
\expandafter\def\csname ev@all/cyber/Routing-F1-forget-max/0M/pct\endcsname{47}
\expandafter\def\csname ev@all/all/Routing-F1-retain-below3/0M/count\endcsname{4}

\newcommand{\ev}[1]{\ifcsname ev@#1\endcsname\csname ev@#1\endcsname\else\todo{missing: \detokenize{#1}}\fi}

\newcommand{\todo}[1]{\textcolor{red}{[TODO: #1]}}

\usepackage{multirow}
\usepackage{tabularx}
\usepackage{placeins}
\newcolumntype{Y}{>{\centering\arraybackslash}X}
\definecolor{phcolor}{HTML}{C2187A}

\definecolor{band}{HTML}{DCEAF8}
\definecolor{collapsed}{HTML}{9A9A94}
\definecolor{sepdropbg}{HTML}{DCE8F3}
\definecolor{sepmuted}{HTML}{7A8187}
\definecolor{oursrowmain}{HTML}{EEEEEA}
\definecolor{removalmax}{HTML}{5A93D0}
\definecolor{removal2}{HTML}{D3E3F3}

\newcommand{\shorttitle}{Relocating Hazardous Knowledge into Removable Experts in Mixture-of-Experts Models}
\setheadertext{\shorttitle}
\setheaderlogos{%
  \includegraphics[height=2em]{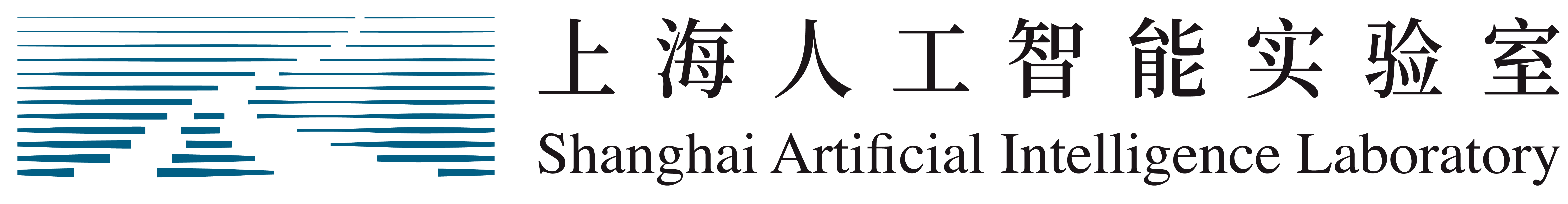}
}

\title{Towards Architectural-Level Safety Control:\\Relocating Hazardous Knowledge into Removable Experts in Mixture-of-Experts Models}
\author{%
Cheng Jin$^*$, Zhongtian Ma$^*$, Xiao Wang, Han Qi, Xiaoyu Wen, Xiang Wang, Youbang Sun, Xia Hu, \quad Bowen Zhou, Chaochao Lu$^{\dagger \sharp}$, Qiaosheng Zhang$^{\ddagger \sharp}$
    \vspace{1mm} \\
    Shanghai AI Laboratory\\
    \textbf{$^*$ Equal Contribution.}~~
    \textbf{$^\dagger$ Project Lead.}~~
    \textbf{$^\ddagger$ Technical Lead.}~~
    \textbf{$^\sharp$ Corresponding Authors.}
    \vspace{1mm} \\

\projectlink{\githuburl}{\faGithub\hspace{4pt}GitHub}%
\hspace{5pt}%
\projectlink{\huggingfaceurl}{%
    \raisebox{-2pt}{\includegraphics[height=10pt]{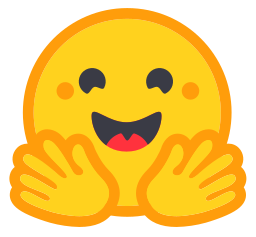}}%
    \hspace{3pt}Hugging Face}
}
\reportnumber{}
\begin{abstract}
  Large language models acquire broad knowledge from web-scale corpora during pretraining, including knowledge that could be hazardous if misused. Removing such knowledge from the model itself without degrading general capability is desirable but difficult, as it is deeply entangled with general capability within the same parameters. Existing safety alignment and unlearning methods typically adjust these shared parameters, tending to suppress the expression of hazardous knowledge rather than remove it. As a result, the knowledge remains in the model and is readily recoverable by subsequent fine-tuning. To address this limitation, we propose \textbf{\sysname{}} (\textbf{R}elocating into Removable \textbf{EX}perts), a knowledge relocation method that separates hazardous knowledge from general capability at the architectural level. Exploiting the modular structure of Mixture-of-Experts models, \sysname{} expands the model with a set of new, physically removable experts designated to hold hazardous knowledge. 
Through expert expansion, selective gradient masking and routing differentiation, it relocates hazardous knowledge out of the original experts and concentrates it in the new ones. Excising these new experts yields a model suitable for public release. This removal holds consistently across Intern-S2, Qwen3.5-35B-A3B and Gemma-4-26B-A4B; on Qwen3.5-35B-A3B, for instance, excision lowers WMDP-Bio accuracy by
\ev{qwen/bio/F2-A2/0M/wmdp} points while MMLU changes by less than 3 points.
The excised model matches the strongest of nine unlearning baselines in removing hazardous knowledge, and it is the only one that preserves
this removal under subsequent fine-tuning. After 250M tokens of standard fine-tuning on Qwen3.5, the best baseline rebounds on WMDP-Bio from
\ev{qwen/bio/ExternalRMU/0M/wmdp}\% to \ev{qwen/bio/ExternalRMU-SFT/250M/wmdp}\%, whereas \sysname{} rises only from \ev{qwen/bio/A2/0M/wmdp}\% to \ev{qwen/bio/A3/250M/wmdp}\%.

\end{abstract}

\renewcommand{\absfont}{\linespread{1.1}\fontsize{10}{12}\selectfont}

\begin{document}
\begin{tcolorbox}[
    colback=AbstractBgColor,
    colframe=AbstractBgColor,
    arc=5pt,
    auto outer arc,
    boxrule=0pt,
    left=8pt, right=8pt,
    top=4pt, bottom=4pt,
    parbox=false,
    width=\textwidth,
    before skip=-800pt,
    after skip=0pt,
    enlarge top by=-12pt
]
\setlength{\parskip}{3pt}
\maketitle
\end{tcolorbox}

\vspace{2pt}
\noindent\begin{minipage}{\textwidth}
  \centering
  \begin{minipage}[c]{0.64\textwidth}
    \includegraphics[width=\linewidth]{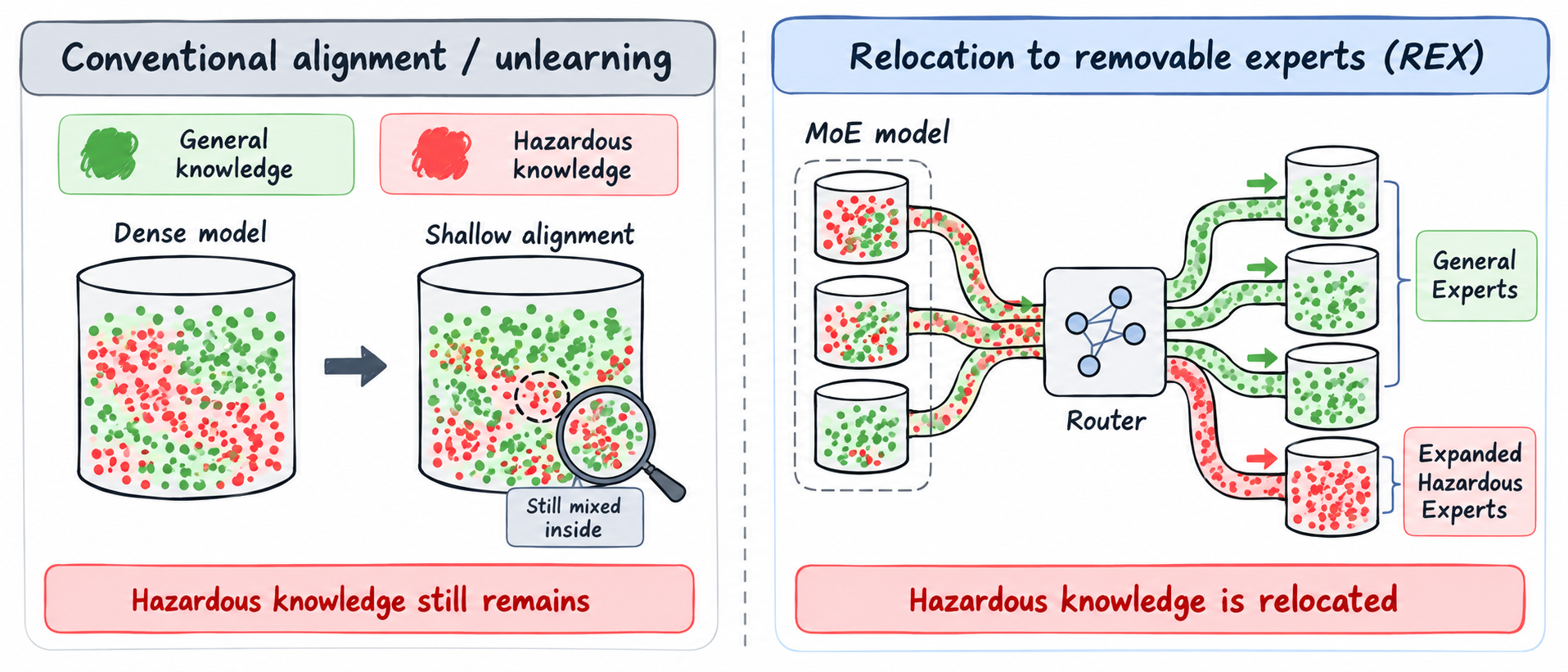}
  \end{minipage}\hfill
  \begin{minipage}[c]{0.35\textwidth}
    \includegraphics[width=\linewidth]{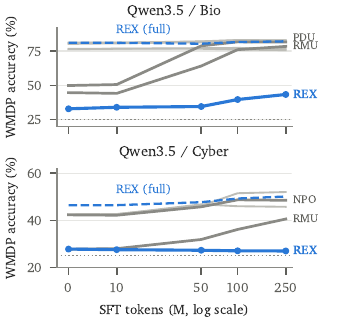}
  \end{minipage}
  \captionof{figure}{\small Left: Conventional alignment and unlearning versus \sysname{}. In a dense model, general (green) and hazardous (red) knowledge are mixed, and alignment or unlearning changes only the surface behavior. \sysname{} routes hazardous knowledge into removable experts, which are excised before public release. Right: Qwen3.5 on WMDP-Bio and WMDP-Cyber during standard post-training (lower is better). Blue solid and dashed curves show the excised and full model. Gray curves show baselines.}
  \label{fig:motivation}
\end{minipage}

\newpage
\begingroup
\setlength{\parskip}{2pt}
\tableofcontents
\endgroup

\newpage
\section{Introduction}
\label{sec:intro}

Hazardous knowledge and general knowledge are deeply entangled in pretrained large language models. Pretraining inevitably teaches a model dual-use knowledge, such as synthesis routes in biology or attack techniques in cybersecurity, and this knowledge shares parameters with general knowledge, so the same weights encode legitimate academic competence and potentially dangerous capability.
For open-weight models, external safeguards can be removed once the weights are released, so the hazardous knowledge itself should be addressed properly. An ideal removal approach should satisfy the following requirements:
\begin{enumerate}
    \item Hazardous capability needs to be removed  as thoroughly as possible, while doing minimal damage to general capability.
    \item Removal should not come at the cost of destruction: developers still need to serve authorized research and defense users with the full capability, while releasing a version with controlled hazardous capability to the public.
    \item The released model should remain safe after standard post-training, since open-weight models are routinely fine-tuned after release. Yet when existing methods only adjust original parameter values, standard supervised fine-tuning can restore or strengthen the removed knowledge, even on data unrelated to the hazardous domain~\citep{doshi2024recovery}.
\end{enumerate}

Most existing approaches share a common limitation. They modify the original parameter set without changing the model architecture. They therefore fall short of the three requirements above in three related ways. First, behavioral interventions such as refusal training and inference-time filtering leave the hazardous knowledge encoded in the model unchanged and provide only shallow alignment~\citep{qi2025shallow}. Unlearning methods such as Representation Misdirection for Unlearning (RMU)~\citep{li2024wmdp}, NPO~\citep{zhang2024npo} and gradient ascent~\citep{jang2022knowledge} act directly on the original parameters that also support general capability, so removing hazardous capability also affects general capability. Second, none of these approaches provides a single model architecture from which both a full model and a public model with a separately removable hazardous component can be delivered. Third, because current alignment and unlearning methods suppress or overwrite hazardous knowledge rather than removing it from the model, later post-training can make the same knowledge visible again (even when the new data is unrelated to the hazardous domain~\citep{doshi2024recovery}). The released model therefore cannot be expected to remain safe after post-training.
These limitations motivate a change at the architectural level. If we can augment a pretrained model with additional parameters, concentrate hazardous knowledge in them while preserving general capability in the original parameters, and remove these additional parameters before release, this could satisfy the three requirements above.

To realize this idea, we turn to Mixture-of-Experts (MoE) models~\citep{shazeer2017moe,fedus2022switch} and propose \sysname{} (Relocating into Removable EXperts), a method for relocating hazardous knowledge into newly added experts that can be removed before public release. MoE models are well suited to this approach because their feed-forward parameters are partitioned into independently addressable experts, the expert set can be expanded, a trainable router determines which experts process each token, and selected experts can be removed at delivery. \sysname{} expands a pretrained MoE model with dedicated target experts and relocates hazardous knowledge from the original experts into them, retaining general capability in the original experts and thereby separating hazardous capability from general capability at the architectural level.

The main challenge is to achieve this capability separation when hazardous and general knowledge are already mixed across the original experts. \sysname{} addresses this challenge through three components: expert expansion, selective gradient masking and routing differentiation. Expert expansion initializes the target experts from the original experts that hazardous data relies on beyond general data, so relocation starts from experts already aligned with the target domain. Selective gradient masking lets target data update only the target experts, and on general data it excludes the target experts before top-$K$ selection, so the original experts learn to maintain general capability without them while the hazardous knowledge they hold receives no training signal and fades. Routing differentiation directs target-related tokens toward the target experts and keeps general tokens away from them. After the target experts are excised, we use a light RMU to clean up the hazardous knowledge left in the retained parameters. Through these three components, \sysname{} separates hazardous capability from general capability at the architectural level. This separation allows general capability to improve during post-training without a substantial rebound in hazardous capability. Retaining or removing the target experts yields a full model for authorized users or a controlled model for public release.

An intuitive comparison is shown in \cref{fig:motivation}. The pretrained model can be viewed as a glass of water containing general knowledge (green) and hazardous knowledge (red) mixed together. Conventional alignment and unlearning operate directly inside this glass, trying to make the water appear green by suppressing hazardous outputs, while the underlying mixture remains unchanged. Our method uses the routing mechanism of an MoE model as an additional layer of separation. In this analogy, the router acts like a filter that directs the (red) hazardous signal to newly added target experts, while the (green) general signal remains in the original experts. As training continues on both data sources, routing and parameter updates reinforce this division. The model can then be separated into a full form that retains hazardous capability and an excised form with the target experts removed, providing a safer and more controllable model for release.

Our contributions are threefold.
\begin{itemize}
  \item We propose \sysname{}, the first method to separate and remove hazardous capability in MoE models at the architectural level. \sysname{} moves the hazardous knowledge already present in a pretrained MoE model into expanded target experts while keeping general capability in the original experts. We measure this separation by comparing the full model with the excised model, obtained by removing the expanded experts, on hazardous and general benchmarks. In both domains and on three open-weight MoE models, the excised model scores substantially lower on hazardous capability while its general capability remains largely unchanged (\cref{tab:separation}).

  \item Although \sysname{} is designed for knowledge removal and capability separation rather than for benchmark forgetting alone, when applied to the base model it removes hazardous knowledge on the standard WMDP-Bio and WMDP-Cyber unlearning benchmarks on par with the strongest of 9 parameter-level unlearning baselines (\cref{sec:exp-base,sec:exp-ablation}).

  \item  We show that \sysname{} is the only method that preserves forgetting under post-training. In contrast to the baselines that reach a low starting point and then climb back towards the original model under the same standard fine-tuning, the delivered model of \sysname{} remains well below them in hazardous knowledge. On Qwen3.5, RMU rises from \ev{qwen/bio/ExternalRMU/0M/wmdp}\% to \ev{qwen/bio/ExternalRMU-SFT/250M/wmdp}\% on WMDP-Bio, close to the \ev{qwen/bio/B/0M/wmdp}\% of the original model, whereas \sysname{} ends at \ev{qwen/bio/A3/250M/wmdp}\% and rises by only \ev{qwen/bio/A3-rise/250M/wmdp} points over the same trajectory (\cref{sec:exp-trajectory}).
\end{itemize}

\section{Related Work}
\label{sec:related}

Machine unlearning removes hazardous knowledge by changing parameter values in place, including in recent methods designed for Mixture-of-Experts models, and a growing body of work shows that such removal does not survive later training. RMU~\citep{li2024wmdp} perturbs intermediate representations of forget data toward a random direction while preserving those of retain data, and it was introduced together with the WMDP benchmark that we use to measure hazardous knowledge. Gradient ascent~\citep{jang2022knowledge} maximizes the loss on forget data, gradient difference~\citep{maini2024tofu} adds a retain term to it, and NPO~\citep{zhang2024npo} replaces its unbounded objective with a preference-style loss that avoids collapse. All of these methods act on parameters that general capability also uses. As a result, \citet{doshi2024recovery} show that ordinary fine-tuning on unrelated data restores the knowledge they remove, and \citet{qi2024durability} find that safeguards on open-weight models are routinely undone by later training. Tamper-resistant methods try to make the removal harder to reverse by shaping the loss landscape around the released weights, through adversarial meta-learning against simulated fine-tuning attacks in TAR~\citep{tamirisa2024tar}, an invariance objective in ILU~\citep{wang2025ilu}, or sharpness-aware minimization~\citep{fan2025sam}, yet the suppressed knowledge still resides in parameters that remain in the model. For MoE models, recent work uses the expert structure to make unlearning more targeted. SEUF~\citep{zhuang2024seuf} attributes the target knowledge to the experts most engaged with it, confines unlearning to those experts and anchors the router so that they stay selected. Such methods use experts to decide where to unlearn, but they still overwrite parameter values inside experts that the delivered model keeps.

Knowledge localization takes a different approach and confines a capability to designated parameters so that it can be removed as a whole. Gradient routing~\citep{cloud2024gradient} masks gradients during training so that chosen data updates only designated subregions of a network, which can then be ablated. SGTM~\citep{shilov2025sgtm} builds on it and applies selective gradient masking during pretraining so that target data updates only a designated subset of parameters, which are then removed, and GRAM~\citep{roland2026gram} uses a modular architecture to enable access control over capabilities learned during pretraining. Because these methods prevent knowledge from entering the rest of the model while it is first learned, they require training from scratch and do not apply to a pretrained model in which the knowledge is already spread across the parameters.

\section{Method}
\label{sec:method}

\begin{figure}[t]
  \centering
  \includegraphics[width=\textwidth]{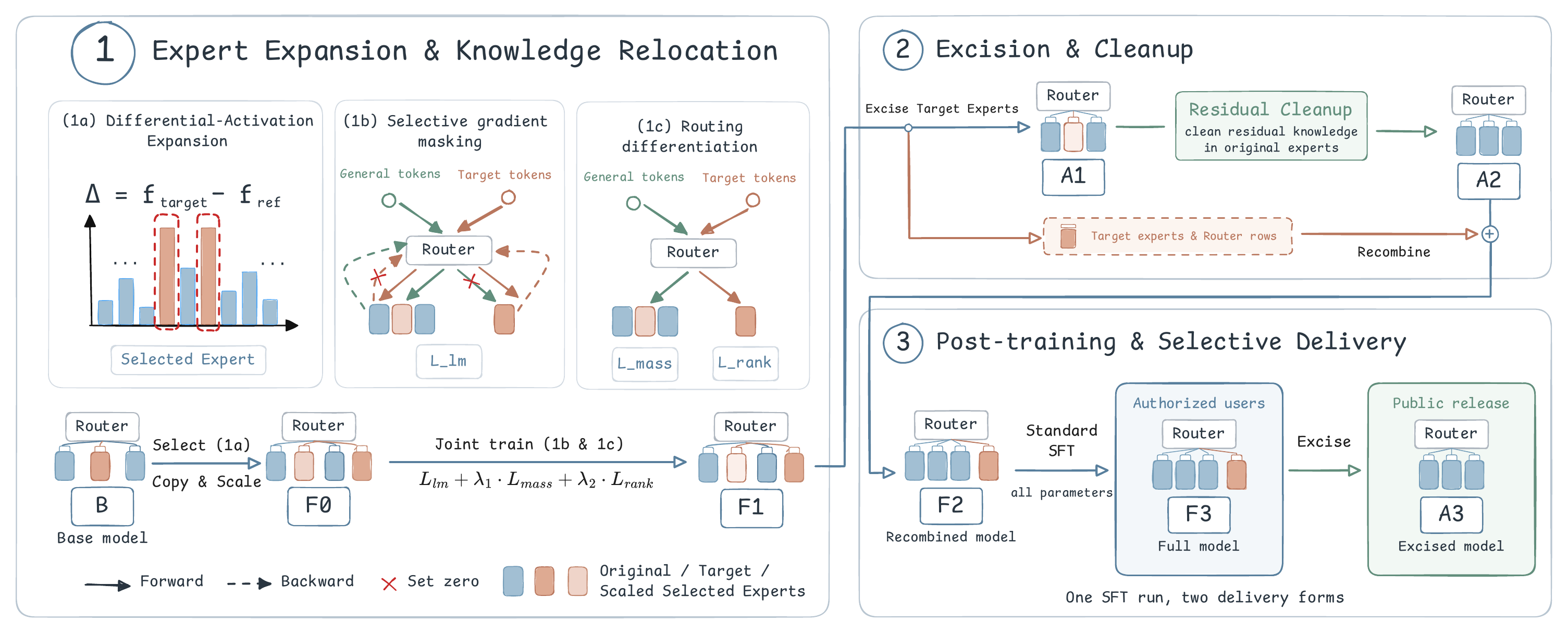}
  \caption{Overview of the \sysname{} pipeline. The method first expands the pretrained MoE model and relocates hazardous knowledge into newly added experts through expert expansion, selective gradient masking and routing differentiation. A light RMU cleanup then cleans residual hazardous knowledge from the original parameters. Post-training runs on the recombined full model, and the target experts are excised at delivery to obtain the public model.}
  \label{fig:overview}
\end{figure}

\sysname{} starts from a pretrained MoE model $B$ and relocates the hazardous knowledge spread across its original experts into a set of new, physically removable experts, which we call target experts. The pipeline has three training stages and is illustrated in \cref{fig:overview}. The first stage expands $B$ into $F_0$ with the target experts and relocates the hazardous knowledge into them. Three components perform the relocation, namely expert expansion (\cref{sec:method-expand}), selective gradient masking (\cref{sec:method-mask}) and routing differentiation (\cref{sec:method-route}). They turn the catastrophic forgetting of continued training into directed forgetting, and together they realize knowledge relocation and capability separation in the full model $F_1$. The second stage excises the target experts of $F_1$ and their router rows, which we keep as $H_1$, and applies a light RMU cleanup, which turns the excised model $A_1$ into the delivered base model $A_2$ (\cref{sec:method-excise}). The third stage recombines $A_2$ with $H_1$ into the full structure $F_2$, runs standard post-training on $F_2$ to obtain the full model for authorized users $F_3$, and excises the target experts at delivery to obtain the public model $A_3$, which bounds the growth of hazardous knowledge during post-training (\cref{sec:method-sft}). We first fix the notation (\cref{sec:method-setup}).

\subsection{Preliminaries and notation}
\label{sec:method-setup}

We use two kinds of training data. Target data $\mathcal{D}_{\mathrm{harm}}$ carries the target knowledge, such as a biosecurity or cybersecurity corpus, and general data $\mathcal{D}_{\mathrm{std}}$ covers general capability. In an MoE layer, the router maps the hidden state $x$ of a token to expert logits and routing probabilities
\begin{equation}
  r(x) = W_r x, \qquad P(x) = \mathrm{softmax}\big(r(x)\big).
  \label{eq:router}
\end{equation}
The layer output is the weighted sum of the top-$K$ experts,
\begin{equation}
  y = \sum_{i \in \mathrm{TopK}(r(x))} g_i(x)\, E_i(x),
  \label{eq:moe}
\end{equation}
where $\mathrm{TopK}(r(x))$ is the set of the $K$ experts with the largest logits, $E_i$ is the $i$-th expert and $g_i$ are the normalized gate weights. We write $\theta_{\mathrm{harm}}$ for the target experts and their router rows, $\theta_{\mathrm{std}}$ for the original experts and their router rows, and $\theta_{\mathrm{shared}}$ for the remaining parameters, including attention, embeddings, norms, the language-model head and any router parameters not tied to a single expert. Shared-expert MLPs belong to neither group of experts.

\subsection{Expert expansion}
\label{sec:method-expand}

Expert expansion adds the target experts and initializes them by differential-activation expansion. On the training corpora $\mathcal{D}_{\mathrm{std}}$ and
$\mathcal{D}_{\mathrm{harm}}$, we count how often each expert $i$ is selected into the top-$K$, which gives the frequencies $f_i^{\mathrm{std}}$ and $f_i^{\mathrm{harm}}$,
and compute $\Delta_i = f_i^{\mathrm{harm}} - f_i^{\mathrm{std}}$. In each layer, the $M$ experts with the largest $\Delta_i$ form the source set $S$, where $M$ is the number of target experts per layer. Selecting by the
frequency on target data alone tends to pick general-purpose experts that are active on all data, and their copies would carry nothing specific to the target domain.
Subtracting the frequency on general data keeps the experts that target data relies on beyond general data, so the target experts match the computation of the target domain
from the start. The selection uses only routing statistics on the training data and never reads evaluation data.

For each expert in $S$, we copy its parameters and router row as a new target expert. We write $H$ for the set of new target experts. Expanding the original model $B$ in this way gives the expanded model $F_0$. All original experts are kept, and
the source experts are scaled by a fixed factor $s$ to weaken the hazardous knowledge they hold before training. If the source experts were left unchanged, they would
compete with the target experts for target tokens, making routing differentiation harder and greatly increasing the training budget that relocation needs. If they were set to zero, general capability would suffer severe damage. Scaling weakens the hazardous knowledge in the source experts while keeping most of their general computation.

\subsection{Selective gradient masking}
\label{sec:method-mask}

Selective gradient masking decides where the updates of each data source are allowed to be written. During relocation the shared parameters stay frozen, including attention, embeddings, norms and the language-model head, together with the shared-expert MLPs, and only the routed experts and the routers are trained. For the expert parameters, gradients from target tokens update only the target experts, and gradients from general tokens update only the original experts. The routers are updated by both. In terms of parameter sets,
\begin{equation}
  \text{updates} =
  \begin{cases}
    \theta_{\mathrm{harm}} \cup \theta_{\mathrm{router}} & \text{on } \mathcal{D}_{\mathrm{harm}},\\[2pt]
    \theta_{\mathrm{std}} \cup \theta_{\mathrm{router}} & \text{on } \mathcal{D}_{\mathrm{std}},
  \end{cases}
  \label{eq:grad-mask}
\end{equation}
where $\theta_{\mathrm{router}}$ collects all router rows. Forward computation is masked on general data only. A pre-top-$K$ activation mask sets the logits of the target experts to $-\infty$ before the selection,
\begin{equation}
  \tilde r_i(x) =
  \begin{cases}
    r_i(x) & i \notin H,\\[2pt]
    -\infty & i \in H,
  \end{cases}
  \qquad x \in \mathcal{D}_{\mathrm{std}},
  \label{eq:activation-mask}
\end{equation}
so the top-$K$ is taken over the original experts and the gate weights are renormalized. Each token of general data is therefore dispatched without the target experts, and the model must perform well with the original experts alone.

The two rules reinforce each other. The gradient mask on target data makes the target experts the only place where new hazardous knowledge is written. The pre-top-$K$ activation mask on general data trains the original experts to hold general capability without the target experts, so the hazardous knowledge they already carry loses its support and decays.

\subsection{Routing differentiation}
\label{sec:method-route}

Routing differentiation steers target tokens toward the target experts, and together with selective gradient masking it accelerates capability separation. It adds two losses. The mass loss decides how much routing mass the target experts receive, and the rank loss decides where they sit in the top-$K$ order. For token $t$ in a given layer, let $z_t = \sum_{i \in H \cap \mathrm{TopK}(r(x_t))} g_i(x_t)$ be the normalized top-$K$ gate weight assigned to the target experts. $z_t$ is computed before masking, while the actual dispatch uses the top-$K$ selected again after masking. Writing $\bar z$ for the mean of $z_t$ over the tokens of a batch and over MoE layers, the mass loss requires $\bar z$ to be no lower than a floor $\tau_h$ on target data and no higher than a ceiling $\tau_s$ on general data,
\begin{equation}
  \mathcal{L}_{\mathrm{mass}} =
  \begin{cases}
    \lambda_h \max\left(0,\, \tau_h - \bar z\right) & \text{on } \mathcal{D}_{\mathrm{harm}},\\[2pt]
    \lambda_s \max\left(0,\, \bar z - \tau_s\right) & \text{on } \mathcal{D}_{\mathrm{std}}
  \end{cases}\,,
  \label{eq:mass}
\end{equation}
so it raises the mass on target tokens and holds it down on general tokens. The hinge form produces gradients only when routing violates this split.

The rank loss constrains the order of the top-$K$ selection instead of the mass. A bound on the mass does not decide which experts are activated, because a target expert enters the activated set as soon as its logit crosses the top-$K$ boundary, even when it then receives little mass. On $\mathcal{D}_{\mathrm{std}}$ we therefore write $h_{\max} = \max_{i \in H} r_i(x_t)$ for the largest target-expert logit and $s_K$ for the $K$-th largest logit among the original experts, which is the top-$K$ boundary, and require $h_{\max}$ to stay below $s_K$ by a margin $\delta$,
\begin{equation}
  \mathcal{L}_{\mathrm{rank}} = \mathbb{E}_{t \in \mathcal{D}_{\mathrm{std}}}\left[\max\left(0,\, \delta - (s_K - h_{\max})\right)\right],
  \label{eq:rank}
\end{equation}
so the target experts stay outside the activated set on general tokens. The pre-top-$K$ activation mask keeps the target experts out of dispatch on general data only during relocation. At inference the router scores all experts, so the rank loss trains the unmasked logits themselves to keep the target experts below the top-$K$ boundary on general tokens.

The two losses complement each other. The mass loss shapes how much routing mass the target experts receive, but it bounds only the total mass, so a satisfied bound does not prevent an expert from being wrongly selected. The rank loss acts on the top-$K$ selection itself and keeps the target experts out of the activated set on general tokens. Because it constrains the selection so directly, however, its gradient can pull against the language-modeling loss. We therefore use the two losses together, with the rank loss correcting the selection and the mass loss shaping routing more gently.
Thus, the final relocation objective for the first stage of \sysname{}, which turns $F_0$ into the full model $F_1$, is
\begin{equation}
    \mathcal{L} = \mathcal{L}_{\mathrm{LM}} + \mathcal{L}_{\mathrm{mass}} + \lambda_r \mathcal{L}_{\mathrm{rank}},
    \label{eq:objective}
\end{equation}
where the language-modeling loss is computed and back-propagated under the masking rules of \cref{sec:method-mask}.

\subsection{Excision and residual cleanup}
\label{sec:method-excise}

Not all hazardous knowledge is held in the target experts. Part of it stays in the original experts, where directed forgetting is incomplete, and part of it is encoded in parameters that relocation never updates, such as attention and the shared-expert MLPs. Excision cannot remove this part, so the second stage of \sysname{} adds a parameter-level cleanup after the structural removal.

The second stage has two steps. It first excises the target experts of $F_1$ and their router rows, which we keep as $H_1$ for the third stage, and obtains the model $A_1$. The residual cleanup then applies RMU~\citep{li2024wmdp} to the expert MLPs that remain in $A_1$. At a chosen monitor layer, the objective pushes the hidden states of target data toward a fixed random direction and keeps the hidden states of general data close to those of $A_1$ before cleanup,
\begin{equation}
  \mathcal{L}_{\mathrm{RMU}} = \mathbb{E}_{t \sim \mathcal{D}_{\mathrm{harm}}} \big\| \phi_\theta(t) - c\,u \big\|_2^2
  + \alpha\, \mathbb{E}_{t \sim \mathcal{D}_{\mathrm{std}}} \big\| \phi_\theta(t) - \phi_{\theta_{\mathrm{ref}}}(t) \big\|_2^2,
  \label{eq:rmu}
\end{equation}
where $\phi_\theta(t)$ is the hidden state of token $t$ at the monitor layer, $u$ a fixed random unit vector, $c$ the steering coefficient, $\alpha$ the weight of the retain term, and $\theta_{\mathrm{ref}}$ the parameters of $A_1$ before cleanup. The trainable parameters are the routed-expert MLPs and the shared-expert MLPs of a few early layers, while the routers, attention, embeddings, norms and the language-model head stay frozen. The cleanup gives the delivered base model $A_2$, which the third stage recombines with $H_1$ into $F_2$.

\subsection{Post-training and delivery}
\label{sec:method-sft}

The third stage takes the full structure $F_2$ built in the previous stage and runs ordinary full-parameter fine-tuning $\Sstd$ on it, which gives the full instruction model $F_3$. Hazardous knowledge does not rebound during post-training because the router and the target experts work together. The router keeps sending target-related tokens to the target experts, so the updates those tokens produce are written mainly into the target experts, while the growth in general capability is written into the parameters that stay. 

At delivery we excise the target experts, and the hazardous capability leaves with them, which gives the public model $A_3$. The delivered model therefore improves on general capability without a matching rise in hazardous capability. The releasing party holds the target experts, so it can decide which artifact to expose. $F_3$ serves authorized users and keeps the hazardous capability, while $A_3$ is the public model without it.

\section{Experiments}
\label{sec:experiments}

Our experiments follow the order of the three contributions. \Cref{sec:exp-excision} measures the capability separation that \sysname{} builds, comparing the full model with the excised model both directly after \sysname{} and after standard post-training. \Cref{sec:exp-base} compares the excised base model with nine parameter-level unlearning baselines on WMDP before any post-training, and \cref{sec:exp-trajectory} compares the same routes after one shared standard fine-tuning procedure. \Cref{sec:exp-ablation} attributes each result to the step of the pipeline that produces it.

\subsection{Setup}
\label{sec:exp-setup}

We study three MoE backbones with different routing designs and two hazardous domains. Intern-S2~\citep{bai2026interns2previewscientificagenticfoundation} and Qwen3.5-35B-A3B~\citep{qwen2026qwen35} have 40 layers with 256 routed experts, top-8 routing and a shared expert. Gemma-4-26B-A4B~\citep{gemma2026gemma4} has 30 layers with 128 routed experts, top-8 routing, per-expert router scales and an always-on dense MLP. The target domains are biosecurity and cybersecurity, with the WMDP forget corpora~\citep{li2024wmdp} as $\mathcal{D}_{\mathrm{harm}}$ and FineWeb-Edu~\citep{penedo2024fineweb} as $\mathcal{D}_{\mathrm{std}}$. Expansion adds $M = 8$ target experts to every MoE layer, and the full recipe of relocation and residual cleanup is given in \cref{app:setup}. Model names follow \cref{sec:method}. $B$ is the original model and $F_0$ its expansion, $F_1$ and $A_1$ are the full and excised models after relocation, $A_2$ is the delivered base model after residual cleanup and $F_2$ its recombination with the target experts, and post-training turns $F_2$ into the full model for authorized users $F_3$ and, after excision, the public model $A_3$.

We compare against nine parameter-level unlearning baselines: RMU~\citep{li2024wmdp}, gradient ascent (GA)~\citep{jang2022knowledge}, gradient difference (GD)~\citep{maini2024tofu}, NPO~\citep{zhang2024npo}, SimNPO~\citep{fan2024simnpo}, CEU~\citep{yang2025ceu}, PDU~\citep{entesari2025pdu}, SatImp~\citep{yang2025satimp} and WGA~\citep{wang2025wga}. Every baseline is trained on the same forget and retain data as \sysname{}, tuned with a hyperparameter search that follows its original paper (\cref{app:impl-baselines}) and evaluated in its standard form without excision, so RMU in the tables denotes this external baseline rather than the matched control $R$ of \cref{sec:exp-ablation}. A method has collapsed if it scores below half of the original model on any of the four general-capability benchmarks, or, along the post-training trajectory, more than \ev{all/all/collapse/0M/points} MMLU points below the original model. Collapsed methods are still reported but take no part in the comparisons of \cref{sec:exp-base,sec:exp-trajectory}, and since no baseline usable on the base model collapses during post-training, the comparison set is the same at both stages.

Post-training uses one fixed standard fine-tuning procedure $\Sstd$ for every route (\cref{app:impl-sft}). It is full-parameter supervised fine-tuning on a text-only instruction mixture with the life-science subsets removed, run along one nominal 250M-token cosine trajectory at learning rate $2\times10^{-6}$ with the Muon optimizer~\citep{jordan2024muon} and the same data order and seed, and WMDP and MMLU are evaluated at 0, 10M, 50M, 100M and 250M tokens. Hazardous knowledge is accuracy on WMDP-Bio or WMDP-Cyber~\citep{li2024wmdp}, a four-way multiple-choice test with chance accuracy at 25\%. General capability is measured by MMLU~\citep{hendrycks2021mmlu}, ARC-C~\citep{clark2018arc}, HellaSwag~\citep{zellers2019hellaswag} and GSM8K~\citep{cobbe2021gsm8k}, and instruction following by IFEval prompt-level strict accuracy~\citep{zhou2023ifeval}, all run with lm-eval~\citep{biderman2024lmeval} and vLLM~\citep{kwon2023vllm} under the protocol of \cref{app:impl-eval}. Along the trajectory, the excised side of \sysname{} is evaluated with the pre-top-$K$ activation mask on the target experts, which agrees with physical excision within one standard error where both were measured (\cref{tab:mask-physical}).

\subsection{Separating hazardous from general capability}
\label{sec:exp-excision}

Excising the target experts takes away the target knowledge and leaves general capability almost unchanged, so the same model takes two distinct capability profiles depending on whether these experts are kept (\cref{tab:separation}). For every backbone and domain, \cref{tab:separation} compares $F_2$ with $A_2$ at the base-model stage and $F_3$ with $A_3$ after standard fine-tuning. Within each pair the two models share every parameter except the target experts, and \cref{tab:excision-full} lists the pair at every stage of the pipeline. At the base-model stage, excision lowers WMDP by \ev{all/all/sep-drop-min/0M/wmdp}--\ev{all/all/sep-drop-max/0M/wmdp} points, while across the \ev{all/all/sep-general-cells/0M/count} general-capability comparisons of the two stages the median change is \ev{all/all/sep-general-median/0M/acc} points and \ev{all/all/sep-general-within1/0M/count} of them stay below one point. The pattern holds after post-training as well. After 250M tokens of full-parameter fine-tuning, the target experts still carry the target knowledge, and excising them removes \ev{all/all/sep-drop-min/250M/wmdp}--\ev{all/all/sep-drop-max/250M/wmdp} points of WMDP at the same mild cost to general capability.

The full model retains the hazardous capability, so the removal performed by \sysname{} is separation rather than destruction. After post-training, the full model for authorized users $F_3$ is \ev{all/all/F3-ref-gap-min/250M/wmdp}--\ev{all/all/F3-ref-gap-max/250M/wmdp} points of WMDP below the original model fine-tuned with the same procedure, $\Sstd(B)$, across the six settings, while $A_3$, excised from the same run, is the public model. The base-model pair gives more direct evidence. The parameters of $F_2$ outside the target experts have already been cleaned and are identical to those of $A_2$, yet adding back the target experts alone raises WMDP by up to \ev{qwen/bio/F2-A2/0M/wmdp} points, from \ev{qwen/bio/A2/0M/wmdp}\% to \ev{qwen/bio/F2/0M/wmdp}\% on Qwen3.5 / Bio. The target knowledge is therefore stored in these experts rather than wiped out beyond recovery, and both forms of the model come from one fine-tuning run.

The drop in target knowledge comes from the target experts, not from excising any parameters of the same size (\cref{tab:random-excision}). On the Intern-S2 / Bio model $F_1$, we remove the same number of original experts as there are target experts in every layer, keep the added experts, and repeat with five random seeds. After random excision WMDP-Bio is $\ev{variant:intern/bio/Random-excision-mean/0M/wmdp} \pm \ev{variant:intern/bio/Random-excision-sd/0M/wmdp}$\%, essentially the \ev{intern/bio/F1/0M/wmdp}\% of $F_1$, whereas excising the target experts lowers it to \ev{intern/bio/A1/0M/wmdp}\%. MMLU stays at $\ev{variant:intern/bio/Random-excision-mean/0M/mmlu} \pm \ev{variant:intern/bio/Random-excision-sd/0M/mmlu}$\% and \ev{intern/bio/A1/0M/mmlu}\% under the two excisions. Removing structure of the same size does not by itself take the target knowledge away. What takes it away is the set of experts that \sysname{} selected and relocated.

\begin{table*}[p]
\centering
\fontsize{7.5}{9}\selectfont
\captionsetup{font=small}
\setlength{\tabcolsep}{2.0pt}
\renewcommand{\arraystretch}{1.08}
\caption{Capability separation in the same model (\%). Full is the model with the target experts and Excised the same model with them removed; $\Delta$ excision (shaded row) is Full minus Excised, i.e.\ the accuracy taken away by the excision, with the target-knowledge entries in blue. Grey text marks the references: the original model $B$ for the base model, and the same standard SFT (250M tokens) applied to it, $S_{\mathrm{std}}(B)$, after SFT.}
\label{tab:separation}
\begin{tabularx}{\textwidth}{@{}l@{\hspace{4pt}}l@{\hspace{4pt}}YYYYY@{\hspace{6pt}}YYYYY@{\hspace{6pt}}YYYYY@{}}
\toprule
 &  & \multicolumn{5}{@{}c@{\hspace{6pt}}}{Qwen3.5} & \multicolumn{5}{@{}c@{\hspace{6pt}}}{Intern-S2} & \multicolumn{5}{@{}c@{\hspace{0pt}}}{Gemma 4} \\
\cmidrule(l{0pt}r{6pt}){3-7}\cmidrule(l{0pt}r{6pt}){8-12}\cmidrule(l{0pt}r{0pt}){13-17}
 &  & WMDP$\!\downarrow$ & MMLU & ARC-C & Hella. & GSM8K & WMDP$\!\downarrow$ & MMLU & ARC-C & Hella. & GSM8K & WMDP$\!\downarrow$ & MMLU & ARC-C & Hella. & GSM8K \\
\midrule
\multicolumn{17}{c}{\rule{0pt}{9pt}\fontsize{8.5}{10}\selectfont\textsc{Bio}} \\[1pt]
\arrayrulecolor{black!25}\cmidrule(lr){1-17}\arrayrulecolor{black}
\multirow{4}{*}{Base} & \textcolor{sepmuted}{Original $B$} & \textcolor{sepmuted}{83.2} & \textcolor{sepmuted}{82.4} & \textcolor{sepmuted}{73.5} & \textcolor{sepmuted}{85.6} & \textcolor{sepmuted}{85.8} & \textcolor{sepmuted}{83.4} & \textcolor{sepmuted}{85.0} & \textcolor{sepmuted}{74.3} & \textcolor{sepmuted}{85.8} & \textcolor{sepmuted}{87.1} & \textcolor{sepmuted}{78.9} & \textcolor{sepmuted}{74.3} & \textcolor{sepmuted}{73.2} & \textcolor{sepmuted}{85.6} & \textcolor{sepmuted}{71.4} \\
 & Full $F_2$ & 80.8 & 80.1 & 70.1 & 83.7 & 83.4 & 61.9 & 82.4 & 70.7 & 84.8 & 84.8 & 71.0 & 72.6 & 71.0 & 85.0 & 69.1 \\
 & Excised $A_2$ & 32.4 & 77.2 & 68.8 & 83.6 & 83.3 & 29.6 & 81.0 & 70.7 & 84.9 & 83.8 & 33.6 & 68.7 & 66.7 & 84.6 & 66.0 \\
 & \cellcolor{oursrowmain}$\Delta$ excision & \cellcolor{removal2}\textbf{$48.4$} & \cellcolor{oursrowmain}$2.9$ & \cellcolor{oursrowmain}$1.4$ & \cellcolor{oursrowmain}$0.1$ & \cellcolor{oursrowmain}$0.1$ & \cellcolor{removal2}\textbf{$32.3$} & \cellcolor{oursrowmain}$1.4$ & \cellcolor{oursrowmain}$0.0$ & \cellcolor{oursrowmain}$-0.1$ & \cellcolor{oursrowmain}$1.0$ & \cellcolor{removal2}\textbf{$37.4$} & \cellcolor{oursrowmain}$3.9$ & \cellcolor{oursrowmain}$4.3$ & \cellcolor{oursrowmain}$0.4$ & \cellcolor{oursrowmain}$3.1$ \\
\cmidrule(l){2-17}
\multirow{4}{*}{SFT} & \textcolor{sepmuted}{$S_{\mathrm{std}}(B)$} & \textcolor{sepmuted}{84.0} & \textcolor{sepmuted}{82.2} & \textcolor{sepmuted}{73.6} & \textcolor{sepmuted}{85.4} & \textcolor{sepmuted}{85.1} & \textcolor{sepmuted}{84.5} & \textcolor{sepmuted}{84.8} & \textcolor{sepmuted}{74.4} & \textcolor{sepmuted}{85.8} & \textcolor{sepmuted}{86.9} & \textcolor{sepmuted}{78.8} & \textcolor{sepmuted}{74.9} & \textcolor{sepmuted}{73.7} & \textcolor{sepmuted}{85.5} & \textcolor{sepmuted}{72.2} \\
 & Full $F_3$ & 81.7 & 80.7 & 70.7 & 84.2 & 85.9 & 78.6 & 84.1 & 72.0 & 85.3 & 85.2 & 76.4 & 73.9 & 71.8 & 85.2 & 69.2 \\
 & Excised $A_3$ & 42.8 & 78.7 & 70.3 & 84.1 & 85.7 & 55.0 & 82.5 & 71.3 & 85.1 & 86.2 & 41.3 & 70.5 & 67.9 & 84.8 & 68.2 \\
 & \cellcolor{oursrowmain}$\Delta$ excision & \cellcolor{removal2}\textbf{$38.9$} & \cellcolor{oursrowmain}$2.0$ & \cellcolor{oursrowmain}$0.4$ & \cellcolor{oursrowmain}$0.1$ & \cellcolor{oursrowmain}$0.2$ & \cellcolor{removal2}\textbf{$23.6$} & \cellcolor{oursrowmain}$1.5$ & \cellcolor{oursrowmain}$0.7$ & \cellcolor{oursrowmain}$0.1$ & \cellcolor{oursrowmain}$-1.0$ & \cellcolor{removal2}\textbf{$35.1$} & \cellcolor{oursrowmain}$3.4$ & \cellcolor{oursrowmain}$3.9$ & \cellcolor{oursrowmain}$0.4$ & \cellcolor{oursrowmain}$1.1$ \\
\midrule
\multicolumn{17}{c}{\rule{0pt}{9pt}\fontsize{8.5}{10}\selectfont\textsc{Cyber}} \\[1pt]
\arrayrulecolor{black!25}\cmidrule(lr){1-17}\arrayrulecolor{black}
\multirow{4}{*}{Base} & \textcolor{sepmuted}{Original $B$} & \textcolor{sepmuted}{57.6} & \textcolor{sepmuted}{82.4} & \textcolor{sepmuted}{73.5} & \textcolor{sepmuted}{85.6} & \textcolor{sepmuted}{85.8} & \textcolor{sepmuted}{58.9} & \textcolor{sepmuted}{85.0} & \textcolor{sepmuted}{74.3} & \textcolor{sepmuted}{85.8} & \textcolor{sepmuted}{87.1} & \textcolor{sepmuted}{50.8} & \textcolor{sepmuted}{74.3} & \textcolor{sepmuted}{73.2} & \textcolor{sepmuted}{85.6} & \textcolor{sepmuted}{71.4} \\
 & Full $F_2$ & 46.5 & 81.4 & 72.9 & 82.9 & 84.4 & 52.8 & 84.4 & 73.3 & 82.3 & 84.0 & 38.9 & 72.5 & 72.6 & 84.6 & 63.1 \\
 & Excised $A_2$ & 28.2 & 80.7 & 72.3 & 82.4 & 83.2 & 26.3 & 84.1 & 73.6 & 81.7 & 83.9 & 29.4 & 71.9 & 72.0 & 83.9 & 53.6 \\
 & \cellcolor{oursrowmain}$\Delta$ excision & \cellcolor{removal2}\textbf{$18.4$} & \cellcolor{oursrowmain}$0.7$ & \cellcolor{oursrowmain}$0.6$ & \cellcolor{oursrowmain}$0.4$ & \cellcolor{oursrowmain}$1.1$ & \cellcolor{removal2}\textbf{$26.6$} & \cellcolor{oursrowmain}$0.3$ & \cellcolor{oursrowmain}$-0.3$ & \cellcolor{oursrowmain}$0.7$ & \cellcolor{oursrowmain}$0.1$ & \cellcolor{removal2}\textbf{$9.5$} & \cellcolor{oursrowmain}$0.6$ & \cellcolor{oursrowmain}$0.6$ & \cellcolor{oursrowmain}$0.7$ & \cellcolor{oursrowmain}$9.5$ \\
\cmidrule(l){2-17}
\multirow{4}{*}{SFT} & \textcolor{sepmuted}{$S_{\mathrm{std}}(B)$} & \textcolor{sepmuted}{55.3} & \textcolor{sepmuted}{82.2} & \textcolor{sepmuted}{73.6} & \textcolor{sepmuted}{85.4} & \textcolor{sepmuted}{85.1} & \textcolor{sepmuted}{59.1} & \textcolor{sepmuted}{84.8} & \textcolor{sepmuted}{74.4} & \textcolor{sepmuted}{85.8} & \textcolor{sepmuted}{86.9} & \textcolor{sepmuted}{51.1} & \textcolor{sepmuted}{74.9} & \textcolor{sepmuted}{73.7} & \textcolor{sepmuted}{85.5} & \textcolor{sepmuted}{72.2} \\
 & Full $F_3$ & 50.3 & 82.0 & 72.5 & 84.3 & 85.6 & 55.8 & 84.0 & 73.7 & 84.7 & 86.5 & 45.1 & 74.0 & 72.8 & 85.4 & 72.1 \\
 & Excised $A_3$ & 27.3 & 81.3 & 73.0 & 83.9 & 85.4 & 29.8 & 84.0 & 73.7 & 84.3 & 85.6 & 34.0 & 73.3 & 72.7 & 85.2 & 69.9 \\
 & \cellcolor{oursrowmain}$\Delta$ excision & \cellcolor{removal2}\textbf{$23.0$} & \cellcolor{oursrowmain}$0.8$ & \cellcolor{oursrowmain}$-0.4$ & \cellcolor{oursrowmain}$0.4$ & \cellcolor{oursrowmain}$0.2$ & \cellcolor{removal2}\textbf{$26.0$} & \cellcolor{oursrowmain}$0.0$ & \cellcolor{oursrowmain}$0.0$ & \cellcolor{oursrowmain}$0.4$ & \cellcolor{oursrowmain}$0.9$ & \cellcolor{removal2}\textbf{$11.1$} & \cellcolor{oursrowmain}$0.6$ & \cellcolor{oursrowmain}$0.1$ & \cellcolor{oursrowmain}$0.2$ & \cellcolor{oursrowmain}$2.2$ \\
\bottomrule
\end{tabularx}
\end{table*}

\begin{figure}[p]
  \centering
  \includegraphics[width=\textwidth]{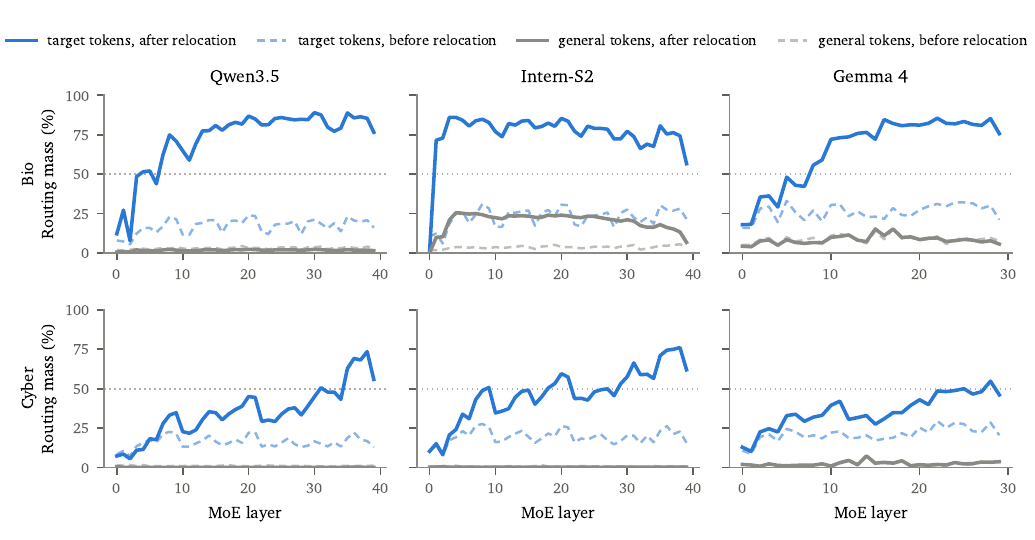}
  \caption{Routing mass on the target experts at every MoE layer. Each curve is the mean normalized top-$K$ mass that valid tokens assign to the target experts, so values above the dashed 50\% line mean the target experts take most of the selected mass. Solid lines are after relocation ($F_1$) and faint dashed lines before it ($F_0$). Target and general tokens use separate fixed samples shared between $F_0$ and $F_1$, and all panels share one vertical axis.}
  \label{fig:routing-summary}
\end{figure}

Relocation makes the target experts the inference path for target tokens, while general tokens stay on the original experts (\cref{fig:routing-summary}). On target tokens, the routing mass assigned to the target experts rises in all six settings, from \ev{all/bio/Routing-F0-forget-min/0M/pct}--\ev{all/bio/Routing-F0-forget-max/0M/pct}\% to \ev{all/bio/Routing-F1-forget-min/0M/pct}--\ev{all/bio/Routing-F1-forget-max/0M/pct}\% on Bio and from \ev{all/cyber/Routing-F0-forget-min/0M/pct}--\ev{all/cyber/Routing-F0-forget-max/0M/pct}\% to \ev{all/cyber/Routing-F1-forget-min/0M/pct}--\ev{all/cyber/Routing-F1-forget-max/0M/pct}\% on Cyber, starting from a share inherited from the source experts that expansion copies. On Bio the target experts hold more than half of the selected mass at most layers of all three backbones, and on Cyber, whose smaller forget corpus gives a shorter relocation budget, the share approaches or passes half in the last layers. On general tokens the mass stays far below this level in every setting and is highest on Intern-S2 / Bio at \ev{intern/bio/Routing-F1/0M/routing_z_retain}\%, where excision still lowers MMLU only from \ev{intern/bio/F1/0M/mmlu}\% to \ev{intern/bio/A1/0M/mmlu}\%. Because target tokens take the target experts at inference time, excision carries the target knowledge away, and post-training gradients on target-related data flow into these experts (\cref{sec:exp-trajectory}).

\FloatBarrier
\subsection{Removing target knowledge from the base model}
\label{sec:exp-base}

\begin{table*}[!t]
\centering
\fontsize{7.5}{9}\selectfont
\captionsetup{font=small}
\setlength{\tabcolsep}{2.0pt}
\renewcommand{\arraystretch}{1.08}
\caption{Base-model unlearning before post-training (\%). \Ours{} (shaded row) is the delivered base $A_2$. Deeper blue marks a larger share of the above-chance WMDP accuracy of $B$ removed, $(W_B - W)/(W_B - 25)$; general capability is a constraint and is not ranked. A dagger marks a collapsed method, with a score below 50\% of $B$ on any general-capability benchmark; it is muted and excluded from the comparison.}
\label{tab:main-results}
\begin{tabularx}{\textwidth}{@{}l@{\hspace{4pt}}YYYYY@{\hspace{8pt}}YYYYY@{\hspace{8pt}}YYYYY@{}}
\toprule
 & \multicolumn{5}{@{}c@{\hspace{8pt}}}{Qwen3.5} & \multicolumn{5}{@{}c@{\hspace{8pt}}}{Intern-S2} & \multicolumn{5}{@{}c@{\hspace{0pt}}}{Gemma 4} \\
\cmidrule(l{0pt}r{8pt}){2-6}\cmidrule(l{0pt}r{8pt}){7-11}\cmidrule(l{0pt}r{0pt}){12-16}
 & WMDP$\!\downarrow$ & MMLU & ARC-C & Hella. & GSM8K & WMDP$\!\downarrow$ & MMLU & ARC-C & Hella. & GSM8K & WMDP$\!\downarrow$ & MMLU & ARC-C & Hella. & GSM8K \\
\midrule
\multicolumn{16}{c}{\rule{0pt}{9pt}\fontsize{8.5}{10}\selectfont\textsc{Bio}} \\[1pt]
\arrayrulecolor{black!25}\cmidrule(lr){1-16}\arrayrulecolor{black}
\textcolor{sepmuted}{Original ($B$)} & \textcolor{sepmuted}{83.2} & \textcolor{sepmuted}{82.4} & \textcolor{sepmuted}{73.5} & \textcolor{sepmuted}{85.6} & \textcolor{sepmuted}{85.8} & \textcolor{sepmuted}{83.4} & \textcolor{sepmuted}{85.0} & \textcolor{sepmuted}{74.3} & \textcolor{sepmuted}{85.8} & \textcolor{sepmuted}{87.1} & \textcolor{sepmuted}{78.9} & \textcolor{sepmuted}{74.3} & \textcolor{sepmuted}{73.2} & \textcolor{sepmuted}{85.6} & \textcolor{sepmuted}{71.4} \\
\rowcolor{oursrowmain}\textbf{\Ours{} (ours)} & \cellcolor{removalmax!53}32.4 & 77.2 & 68.8 & 83.6 & 83.3 & \cellcolor{removalmax!59}29.6 & 81.0 & 70.7 & 84.9 & 83.8 & \cellcolor{removalmax!49}33.6 & 68.7 & 66.7 & 84.6 & 66.0 \\
\addlinespace[2pt]
\textcolor{collapsed}{CEU} & \textcolor{collapsed}{81.2}$^{\dag}$ & \textcolor{collapsed}{81.0}$^{\dag}$ & \textcolor{collapsed}{40.1}$^{\dag}$ & \textcolor{collapsed}{39.3}$^{\dag}$ & \textcolor{collapsed}{3.8}$^{\dag}$ & \textcolor{collapsed}{70.2}$^{\dag}$ & \textcolor{collapsed}{75.7}$^{\dag}$ & \textcolor{collapsed}{29.0}$^{\dag}$ & \textcolor{collapsed}{30.6}$^{\dag}$ & \textcolor{collapsed}{0.0}$^{\dag}$ & \textcolor{collapsed}{75.1}$^{\dag}$ & \textcolor{collapsed}{72.0}$^{\dag}$ & \textcolor{collapsed}{34.5}$^{\dag}$ & \textcolor{collapsed}{33.5}$^{\dag}$ & \textcolor{collapsed}{0.0}$^{\dag}$ \\
\textcolor{collapsed}{GA} & \textcolor{collapsed}{25.0}$^{\dag}$ & \textcolor{collapsed}{28.1}$^{\dag}$ & \textcolor{collapsed}{26.0}$^{\dag}$ & \textcolor{collapsed}{26.1}$^{\dag}$ & \textcolor{collapsed}{0.0}$^{\dag}$ & \textcolor{collapsed}{24.7}$^{\dag}$ & \textcolor{collapsed}{23.0}$^{\dag}$ & \textcolor{collapsed}{25.6}$^{\dag}$ & \textcolor{collapsed}{25.5}$^{\dag}$ & \textcolor{collapsed}{0.0}$^{\dag}$ & \textcolor{collapsed}{24.7}$^{\dag}$ & \textcolor{collapsed}{22.9}$^{\dag}$ & \textcolor{collapsed}{27.2}$^{\dag}$ & \textcolor{collapsed}{26.6}$^{\dag}$ & \textcolor{collapsed}{0.0}$^{\dag}$ \\
GD & \cellcolor{removalmax!0}81.0 & 81.2 & 72.0 & 85.2 & 74.7 & \cellcolor{removalmax!6}66.9 & 82.5 & 73.5 & 85.0 & 85.1 & \cellcolor{removalmax!61}28.6 & 70.7 & 65.5 & 79.2 & 40.5 \\
NPO & \cellcolor{removalmax!1}76.4 & 77.4 & 71.5 & 83.4 & 83.6 & \cellcolor{removalmax!1}76.9 & 78.7 & 67.4 & 83.2 & 82.2 & \cellcolor{removalmax!5}65.0 & 69.2 & 68.4 & 84.1 & 57.1 \\
PDU & \cellcolor{removalmax!26}47.8 & 78.2 & 72.5 & 84.7 & 67.7 & \cellcolor{removalmax!17}54.8 & 83.8 & 75.3 & 86.2 & 87.1 & \cellcolor{removalmax!2}69.4 & 73.3 & 71.9 & 84.4 & 68.5 \\
RMU & \cellcolor{removalmax!32}44.1 & 82.4 & 73.9 & 85.4 & 84.7 & \cellcolor{removalmax!43}37.4 & 84.3 & 74.6 & 85.4 & 86.1 & \cellcolor{removalmax!59}29.2 & 70.0 & 68.0 & 78.6 & 68.4 \\
SatImp & \cellcolor{removalmax!0}80.4 & 80.6 & 43.3 & 44.1 & 70.3 & \textcolor{collapsed}{73.8}$^{\dag}$ & \textcolor{collapsed}{77.6}$^{\dag}$ & \textcolor{collapsed}{35.1}$^{\dag}$ & \textcolor{collapsed}{38.0}$^{\dag}$ & \textcolor{collapsed}{0.3}$^{\dag}$ & \textcolor{collapsed}{61.6}$^{\dag}$ & \textcolor{collapsed}{57.1}$^{\dag}$ & \textcolor{collapsed}{35.1}$^{\dag}$ & \textcolor{collapsed}{33.4}$^{\dag}$ & \textcolor{collapsed}{0.0}$^{\dag}$ \\
SimNPO & \cellcolor{removalmax!1}76.4 & 77.2 & 71.0 & 83.7 & 82.6 & \cellcolor{removalmax!2}74.6 & 77.8 & 69.4 & 83.5 & 81.5 & \cellcolor{removalmax!5}64.3 & 68.4 & 69.2 & 84.1 & 58.5 \\
WGA & \cellcolor{removalmax!0}80.9 & 80.7 & 43.5 & 44.2 & 71.2 & \textcolor{collapsed}{74.0}$^{\dag}$ & \textcolor{collapsed}{76.9}$^{\dag}$ & \textcolor{collapsed}{35.0}$^{\dag}$ & \textcolor{collapsed}{37.0}$^{\dag}$ & \textcolor{collapsed}{0.5}$^{\dag}$ & \textcolor{collapsed}{66.3}$^{\dag}$ & \textcolor{collapsed}{61.0}$^{\dag}$ & \textcolor{collapsed}{35.5}$^{\dag}$ & \textcolor{collapsed}{34.3}$^{\dag}$ & \textcolor{collapsed}{0.0}$^{\dag}$ \\
\midrule
\multicolumn{16}{c}{\rule{0pt}{9pt}\fontsize{8.5}{10}\selectfont\textsc{Cyber}} \\[1pt]
\arrayrulecolor{black!25}\cmidrule(lr){1-16}\arrayrulecolor{black}
\textcolor{sepmuted}{Original ($B$)} & \textcolor{sepmuted}{57.6} & \textcolor{sepmuted}{82.4} & \textcolor{sepmuted}{73.5} & \textcolor{sepmuted}{85.6} & \textcolor{sepmuted}{85.8} & \textcolor{sepmuted}{58.9} & \textcolor{sepmuted}{85.0} & \textcolor{sepmuted}{74.3} & \textcolor{sepmuted}{85.8} & \textcolor{sepmuted}{87.1} & \textcolor{sepmuted}{50.8} & \textcolor{sepmuted}{74.3} & \textcolor{sepmuted}{73.2} & \textcolor{sepmuted}{85.6} & \textcolor{sepmuted}{71.4} \\
\rowcolor{oursrowmain}\textbf{\Ours{} (ours)} & \cellcolor{removalmax!57}28.2 & 80.7 & 72.3 & 82.4 & 83.2 & \cellcolor{removalmax!65}26.3 & 84.1 & 73.6 & 81.7 & 83.9 & \cellcolor{removalmax!48}29.4 & 71.9 & 72.0 & 83.9 & 53.6 \\
\addlinespace[2pt]
\textcolor{collapsed}{CEU} & \textcolor{collapsed}{37.8}$^{\dag}$ & \textcolor{collapsed}{75.4}$^{\dag}$ & \textcolor{collapsed}{35.5}$^{\dag}$ & \textcolor{collapsed}{33.4}$^{\dag}$ & \textcolor{collapsed}{0.0}$^{\dag}$ & \textcolor{collapsed}{44.8}$^{\dag}$ & \textcolor{collapsed}{77.8}$^{\dag}$ & \textcolor{collapsed}{34.1}$^{\dag}$ & \textcolor{collapsed}{36.0}$^{\dag}$ & \textcolor{collapsed}{0.0}$^{\dag}$ & \textcolor{collapsed}{26.6}$^{\dag}$ & \textcolor{collapsed}{32.3}$^{\dag}$ & \textcolor{collapsed}{30.4}$^{\dag}$ & \textcolor{collapsed}{30.1}$^{\dag}$ & \textcolor{collapsed}{0.0}$^{\dag}$ \\
\textcolor{collapsed}{GA} & \textcolor{collapsed}{34.7}$^{\dag}$ & \textcolor{collapsed}{74.4}$^{\dag}$ & \textcolor{collapsed}{30.4}$^{\dag}$ & \textcolor{collapsed}{26.9}$^{\dag}$ & \textcolor{collapsed}{0.0}$^{\dag}$ & \textcolor{collapsed}{24.6}$^{\dag}$ & \textcolor{collapsed}{35.5}$^{\dag}$ & \textcolor{collapsed}{26.3}$^{\dag}$ & \textcolor{collapsed}{26.7}$^{\dag}$ & \textcolor{collapsed}{0.0}$^{\dag}$ & \textcolor{collapsed}{43.6}$^{\dag}$ & \textcolor{collapsed}{72.4}$^{\dag}$ & \textcolor{collapsed}{28.2}$^{\dag}$ & \textcolor{collapsed}{29.0}$^{\dag}$ & \textcolor{collapsed}{0.0}$^{\dag}$ \\
\textcolor{collapsed}{GD} & \textcolor{collapsed}{36.9}$^{\dag}$ & \textcolor{collapsed}{78.6}$^{\dag}$ & \textcolor{collapsed}{74.7}$^{\dag}$ & \textcolor{collapsed}{84.3}$^{\dag}$ & \textcolor{collapsed}{0.0}$^{\dag}$ & \textcolor{collapsed}{31.0}$^{\dag}$ & \textcolor{collapsed}{81.8}$^{\dag}$ & \textcolor{collapsed}{74.3}$^{\dag}$ & \textcolor{collapsed}{84.2}$^{\dag}$ & \textcolor{collapsed}{0.4}$^{\dag}$ & \textcolor{collapsed}{27.3}$^{\dag}$ & \textcolor{collapsed}{66.6}$^{\dag}$ & \textcolor{collapsed}{60.2}$^{\dag}$ & \textcolor{collapsed}{73.9}$^{\dag}$ & \textcolor{collapsed}{0.0}$^{\dag}$ \\
NPO & \cellcolor{removalmax!16}41.9 & 81.3 & 74.2 & 87.6 & 85.3 & \cellcolor{removalmax!19}41.4 & 83.2 & 72.3 & 86.5 & 87.6 & \cellcolor{removalmax!5}44.1 & 70.9 & 67.9 & 84.1 & 56.2 \\
PDU & \cellcolor{removalmax!15}42.4 & 81.9 & 73.2 & 85.4 & 76.0 & \cellcolor{removalmax!59}27.8 & 84.7 & 74.5 & 86.4 & 85.8 & \cellcolor{removalmax!70}25.1 & 73.7 & 71.3 & 84.5 & 69.8 \\
RMU & \cellcolor{removalmax!60}27.5 & 81.8 & 72.6 & 84.9 & 76.9 & \cellcolor{removalmax!33}35.7 & 84.8 & 74.3 & 83.4 & 85.0 & \cellcolor{removalmax!60}26.8 & 72.6 & 67.9 & 78.7 & 65.4 \\
\textcolor{collapsed}{SatImp} & \textcolor{collapsed}{32.8}$^{\dag}$ & \textcolor{collapsed}{74.4}$^{\dag}$ & \textcolor{collapsed}{39.1}$^{\dag}$ & \textcolor{collapsed}{37.2}$^{\dag}$ & \textcolor{collapsed}{0.0}$^{\dag}$ & \textcolor{collapsed}{37.6}$^{\dag}$ & \textcolor{collapsed}{78.8}$^{\dag}$ & \textcolor{collapsed}{48.5}$^{\dag}$ & \textcolor{collapsed}{54.2}$^{\dag}$ & \textcolor{collapsed}{0.0}$^{\dag}$ & \textcolor{collapsed}{28.2}$^{\dag}$ & \textcolor{collapsed}{40.4}$^{\dag}$ & \textcolor{collapsed}{30.5}$^{\dag}$ & \textcolor{collapsed}{34.1}$^{\dag}$ & \textcolor{collapsed}{0.0}$^{\dag}$ \\
SimNPO & \cellcolor{removalmax!15}42.5 & 81.1 & 74.9 & 88.5 & 74.3 & \cellcolor{removalmax!8}47.5 & 81.6 & 72.7 & 82.4 & 86.0 & \cellcolor{removalmax!60}26.8 & 67.9 & 66.1 & 82.2 & 48.7 \\
\textcolor{collapsed}{WGA} & \textcolor{collapsed}{32.2}$^{\dag}$ & \textcolor{collapsed}{72.9}$^{\dag}$ & \textcolor{collapsed}{37.5}$^{\dag}$ & \textcolor{collapsed}{35.7}$^{\dag}$ & \textcolor{collapsed}{0.0}$^{\dag}$ & \textcolor{collapsed}{37.6}$^{\dag}$ & \textcolor{collapsed}{78.7}$^{\dag}$ & \textcolor{collapsed}{48.4}$^{\dag}$ & \textcolor{collapsed}{53.4}$^{\dag}$ & \textcolor{collapsed}{0.0}$^{\dag}$ & \textcolor{collapsed}{26.3}$^{\dag}$ & \textcolor{collapsed}{34.1}$^{\dag}$ & \textcolor{collapsed}{27.5}$^{\dag}$ & \textcolor{collapsed}{30.6}$^{\dag}$ & \textcolor{collapsed}{0.0}$^{\dag}$ \\
\bottomrule
\end{tabularx}
\end{table*}

Before any post-training, the base model $A_2$ delivered by \sysname{} removes target knowledge to a degree comparable with the strongest external parameter-level unlearning methods (\cref{tab:main-results}). Once collapsed methods are set aside, the usable baselines in every setting still span at least \ev{all/all/usable-spread-min/0M/points} points of WMDP accuracy. Most of them leave target accuracy far above chance, and only a few reach the low region. $A_2$ lies in this low region in all six settings and is never more than \ev{all/all/A2-best-max/0M/wmdp} points above the lowest usable baseline. In three settings $A_2$ is itself the lowest, by \ev{qwen/bio/best-A2/0M/wmdp} points below RMU on Qwen3.5 / Bio, \ev{intern/bio/best-A2/0M/wmdp} points below RMU on Intern-S2 / Bio and \ev{intern/cyber/best-A2/0M/wmdp} points below PDU on Intern-S2 / Cyber, and on Qwen3.5 / Cyber it is within \ev{qwen/cyber/A2-best/0M/wmdp} points of RMU. The strongest baseline is not the same method across settings, alternating among RMU, GD and PDU, whereas $A_2$ reaches this group in every setting with one pipeline. This level is reached with a light RMU cleanup because excision has already taken away the knowledge carried by the target experts, so the cleanup acts only on what remains in the retained parameters. \Cref{sec:exp-ablation} shows that the same cleanup applied to the original model leaves more target knowledge in every setting. On Gemma~4 / Bio, GD and RMU start below $A_2$ at the base-model stage. \Cref{sec:exp-trajectory} shows that this starting point does not hold under the same standard post-training.

This removal does not come at the expense of general capability, which $A_2$ retains at the same level as the usable external methods. In all six settings, the MMLU, ARC-C and HellaSwag scores of $A_2$ stay above \ev{all/all/A2-retention-min/0M/pct}\% of those of the original model $B$, the same level of retention as usable methods such as RMU and PDU (\cref{tab:main-results}). The more visible MMLU decrease on Bio is concentrated in the subjects adjacent to the target domain: the ten biology and health subjects make up 15\% of MMLU questions but account for \ev{all/bio/adj-share-min/0M/pct}--\ev{all/bio/adj-share-max/0M/pct}\% of the decrease, while the remaining subjects stay within \ev{all/bio/rest-drop-min/0M/mmlu}--\ev{all/bio/rest-drop-max/0M/mmlu} points of $B$ (\cref{tab:mmlu-split}). What leaves together with the target domain is therefore mainly knowledge in the adjacent subjects rather than general capability at large. On Gemma~4 / Bio, RMU, which removes target knowledge to a similar depth, falls to the same level on these subjects (\ev{gemma/bio/ExternalRMU/0M/mmlu_adj}\% against \ev{gemma/bio/A2/0M/mmlu_adj}\% for $A_2$). \Cref{sec:exp-trajectory} further shows that standard post-training raises the MMLU of the delivered model.

\FloatBarrier
\subsection{Persistence of hazardous knowledge removal under standard post-training}
\label{sec:exp-trajectory}

\begin{table*}[!t]
\centering
\fontsize{7.5}{9}\selectfont
\captionsetup{font=small}
\setlength{\tabcolsep}{2.0pt}
\renewcommand{\arraystretch}{1.08}
\caption{Hazardous knowledge under the same standard post-training (\%). Every route runs the same 250M-token $\Sstd$; Before and After give WMDP at its input and output, and MMLU is measured after. \Ours{} (shaded row) is the delivered model $A_3$, fine-tuned as the recombined full model and excised at delivery. Deeper blue marks a larger share of the above-chance WMDP accuracy of $B$ still removed after fine-tuning. Only baselines that did not collapse on the base model (\cref{tab:main-results}) are reported; a dash marks a collapsed setting.}
\label{tab:sft-endpoints}
\begin{tabularx}{\textwidth}{@{}l@{\hspace{4pt}}YYY@{\hspace{8pt}}YYY@{\hspace{8pt}}YYY@{}}
\toprule
 & \multicolumn{3}{@{}c@{\hspace{8pt}}}{Qwen3.5} & \multicolumn{3}{@{}c@{\hspace{8pt}}}{Intern-S2} & \multicolumn{3}{@{}c@{\hspace{0pt}}}{Gemma 4} \\
\cmidrule(l{0pt}r{8pt}){2-4}\cmidrule(l{0pt}r{8pt}){5-7}\cmidrule(l{0pt}r{0pt}){8-10}
 & Before & After$\downarrow$ & MMLU & Before & After$\downarrow$ & MMLU & Before & After$\downarrow$ & MMLU \\
\midrule
\multicolumn{10}{c}{\rule{0pt}{9pt}\fontsize{8.5}{10}\selectfont\textsc{Bio}} \\[1pt]
\arrayrulecolor{black!25}\cmidrule(lr){1-10}\arrayrulecolor{black}
\textcolor{sepmuted}{Original + $\Sstd$} & \textcolor{sepmuted}{83.2} & \textcolor{sepmuted}{84.0} & \textcolor{sepmuted}{82.2} & \textcolor{sepmuted}{83.4} & \textcolor{sepmuted}{84.5} & \textcolor{sepmuted}{84.8} & \textcolor{sepmuted}{78.9} & \textcolor{sepmuted}{78.8} & \textcolor{sepmuted}{74.9} \\
\rowcolor{oursrowmain}\textbf{\Ours{} (ours)} & 32.4 & \cellcolor{removalmax!34}42.8 & 78.7 & 29.6 & \cellcolor{removalmax!17}55.0 & 82.5 & 33.6 & \cellcolor{removalmax!34}41.3 & 70.5 \\
\addlinespace[2pt]
GD & 81.0 & \cellcolor{removalmax!0}82.2 & 81.5 & 66.9 & \cellcolor{removalmax!0}83.0 & 84.4 & 28.6 & \cellcolor{removalmax!0}77.4 & 74.8 \\
NPO & 76.4 & \cellcolor{removalmax!1}76.0 & 78.3 & 76.9 & \cellcolor{removalmax!1}77.6 & 80.7 & 65.0 & \cellcolor{removalmax!3}67.8 & 70.3 \\
PDU & 47.8 & \cellcolor{removalmax!0}81.8 & 81.4 & 54.8 & \cellcolor{removalmax!0}82.7 & 84.8 & 69.4 & \cellcolor{removalmax!0}78.5 & 74.7 \\
RMU & 44.1 & \cellcolor{removalmax!0}78.3 & 82.0 & 37.4 & \cellcolor{removalmax!0}82.0 & 84.6 & 29.2 & \cellcolor{removalmax!8}60.8 & 73.2 \\
SatImp & 80.4 & \cellcolor{removalmax!0}82.7 & 82.1 & -- & -- & -- & -- & -- & -- \\
SimNPO & 76.4 & \cellcolor{removalmax!1}77.5 & 78.3 & 74.6 & \cellcolor{removalmax!1}75.6 & 79.9 & 64.3 & \cellcolor{removalmax!3}66.9 & 69.0 \\
WGA & 80.9 & \cellcolor{removalmax!0}82.7 & 82.0 & -- & -- & -- & -- & -- & -- \\
\midrule
\multicolumn{10}{c}{\rule{0pt}{9pt}\fontsize{8.5}{10}\selectfont\textsc{Cyber}} \\[1pt]
\arrayrulecolor{black!25}\cmidrule(lr){1-10}\arrayrulecolor{black}
\textcolor{sepmuted}{Original + $\Sstd$} & \textcolor{sepmuted}{57.6} & \textcolor{sepmuted}{55.3} & \textcolor{sepmuted}{82.2} & \textcolor{sepmuted}{58.9} & \textcolor{sepmuted}{59.1} & \textcolor{sepmuted}{84.8} & \textcolor{sepmuted}{50.8} & \textcolor{sepmuted}{51.1} & \textcolor{sepmuted}{74.9} \\
\rowcolor{oursrowmain}\textbf{\Ours{} (ours)} & 28.2 & \cellcolor{removalmax!61}27.3 & 81.3 & 26.3 & \cellcolor{removalmax!52}29.8 & 84.0 & 29.4 & \cellcolor{removalmax!30}34.0 & 73.3 \\
\addlinespace[2pt]
NPO & 41.9 & \cellcolor{removalmax!4}49.4 & 81.9 & 41.4 & \cellcolor{removalmax!4}50.6 & 84.5 & 44.1 & \cellcolor{removalmax!2}46.5 & 72.6 \\
PDU & 42.4 & \cellcolor{removalmax!10}45.5 & 82.3 & 27.8 & \cellcolor{removalmax!22}40.1 & 84.8 & 25.1 & \cellcolor{removalmax!25}35.5 & 74.5 \\
RMU & 27.5 & \cellcolor{removalmax!19}40.6 & 82.2 & 35.7 & \cellcolor{removalmax!4}51.2 & 84.3 & 26.8 & \cellcolor{removalmax!24}35.7 & 73.8 \\
SimNPO & 42.5 & \cellcolor{removalmax!2}52.5 & 81.9 & 47.5 & \cellcolor{removalmax!3}52.1 & 82.8 & 26.8 & \cellcolor{removalmax!17}38.0 & 71.4 \\
\bottomrule
\end{tabularx}
\end{table*}

\begin{figure}[!t]
  \centering
  \includegraphics[width=\textwidth]{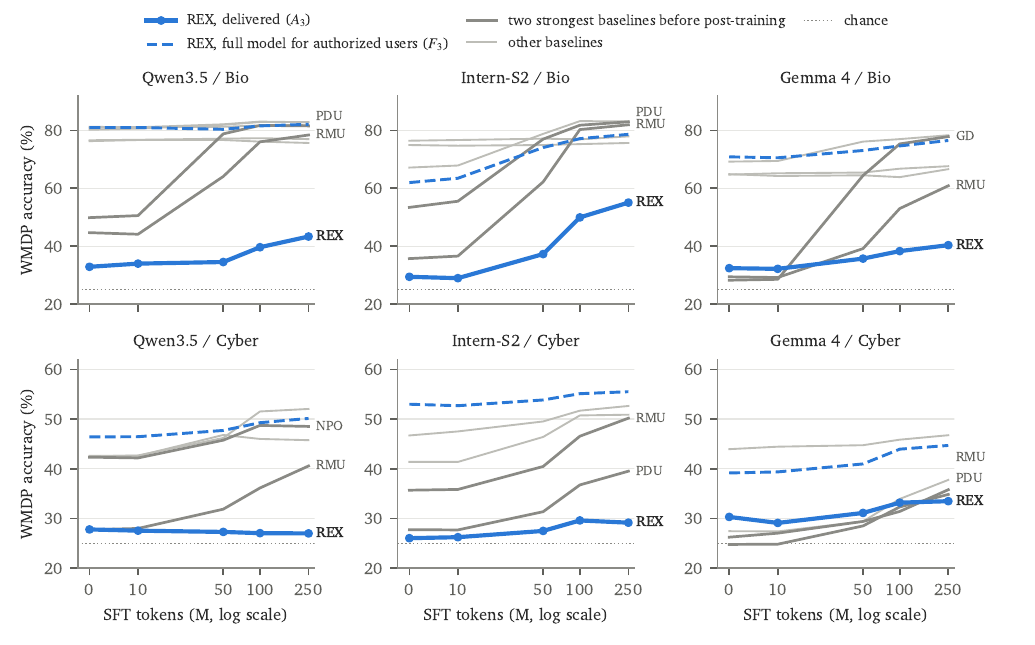}
  \caption{Target knowledge along one shared 250M-token standard SFT trajectory (same data, order and seed for every route, with the 0 point the input checkpoint). Blue curves are \sysname{}, solid for the delivered model $A_3$ and dashed for the full model for authorized users $F_3$ from the same run. Grey curves are the external unlearning baselines that did not collapse on the base model, with the two lowest in WMDP before post-training drawn darker and labelled.}
  \label{fig:sft-grid}
\end{figure}

Under the same standard fine-tuning, the model delivered by \sysname{} retains the least target knowledge among all methods that are usable on the base model (\cref{tab:sft-endpoints,fig:sft-grid}). After 250M tokens, $A_3$ has the lowest WMDP in \ev{all/all/A3-lowest/250M/count} of \ev{all/all/combos/0M/count} settings, \ev{all/bio/best-A3-min/250M/wmdp}--\ev{all/bio/best-A3-max/250M/wmdp} points below the lowest baseline on Bio and \ev{all/cyber/best-A3-min/250M/wmdp}--\ev{all/cyber/best-A3-max/250M/wmdp} points below it on Cyber. The baselines that matched $A_2$ in \cref{sec:exp-base} return close to the fine-tuned original model. On Intern-S2 / Bio, RMU climbs from \ev{intern/bio/ExternalRMU/0M/wmdp}\% to \ev{intern/bio/ExternalRMU-SFT/250M/wmdp}\%, against \ev{intern/bio/S_std(B)/250M/wmdp}\% for the original model under the same $\Sstd$, while $A_3$ ends at \ev{intern/bio/A3/250M/wmdp}\%. On Gemma~4 / Bio, GD and RMU start below $A_2$ (\ev{gemma/bio/GD/0M/wmdp}\% and \ev{gemma/bio/ExternalRMU/0M/wmdp}\% against \ev{gemma/bio/A2/0M/wmdp}\%) yet end at \ev{gemma/bio/GD-SFT/250M/wmdp}\% and \ev{gemma/bio/ExternalRMU-SFT/250M/wmdp}\%, above the \ev{gemma/bio/A3/250M/wmdp}\% of $A_3$. A low starting point before post-training therefore does not bound what post-training restores.

The advantage of $A_3$ is not inherited from a lower starting point. It keeps most of its base-model removal, whereas the baselines lose most of theirs. We define the kept share of a route as $(W_B - W_{250\mathrm{M}})/(W_B - W_0)$, where $W_B$, $W_0$ and $W_{250\mathrm{M}}$ are the WMDP accuracy of $B$, of the route before post-training and of the route after 250M tokens, so it measures how much of the removal before post-training is still present. On Bio, $A_3$ keeps at least \ev{all/bio/A3-kept-min/250M/pct}\%, while the two strongest baselines of each setting keep at most \ev{all/bio/top2-kept-max/250M/pct}\%. On Cyber the figures are at least \ev{all/cyber/A3-kept-min/250M/pct}\% against at most \ev{all/cyber/top2-kept-max/250M/pct}\%. The removal of every baseline resides in parameter values, and standard fine-tuning modifies those same values. \sysname{} instead fine-tunes the recombined full model. The router sends target-related tokens to the target experts, the fine-tuning gradients related to target knowledge follow them there, and the resulting updates leave with those experts at delivery. $F_3$ from the same run serves as the full model for authorized users. The direct-SFT control in \cref{sec:exp-ablation} starts from the same $A_2$ without recombination and ends \ev{all/bio/A3-direct-A3-min/250M/wmdp}--\ev{all/bio/A3-direct-A3-max/250M/wmdp} points higher on Bio, which attributes the smaller rise of $A_3$ to fine-tuning on the full structure and excising at delivery rather than to the low starting point of $A_2$.

The general-capability gains of post-training still reach the delivered model. On Bio, the MMLU of $A_3$ rises by \ev{all/bio/A3-mmlu-rise-min/250M/mmlu}--\ev{all/bio/A3-mmlu-rise-max/250M/mmlu} points over fine-tuning, and its IFEval is comparable or even better than the original model under the same $\Sstd$ (\cref{tab:endpoints-general}). Standard post-training thus improves the delivered model as it improves the original one, while the hazardous capability of the delivered model stays low. In \sysname{}, capability growth and risk growth are decoupled.

\FloatBarrier
\subsection{Ablations}
\label{sec:exp-ablation}

The two mechanisms that \sysname{} adds to selective gradient masking together leave the least target knowledge in the retained parameters (\cref{tab:mechanism-ablation}). Each variant is measured by the WMDP of its excised model $A_1$ before residual cleanup. Gradient masking alone, the localization rule of SGTM~\citep{shilov2025sgtm} applied inside our expanded model, leaves $A_1$ at \ev{variant:intern/bio/A1-gradient_mask_only/0M/wmdp}\%. The rule keeps target data from being written into the original experts, but it does not act on the target knowledge they already hold, which is the gap between localization during pretraining and relocation in a pretrained model (\cref{sec:related}). With the pre-top-$K$ activation mask on general data and the routing-differentiation losses together, the full method lowers $A_1$ to \ev{intern/bio/A1/0M/wmdp}\%. Removing either mechanism leaves more target knowledge, \ev{variant:intern/bio/A1-without_std_activation_mask/0M/wmdp}\% without the pre-top-$K$ activation mask and \ev{variant:intern/bio/A1-without_routing_loss/0M/wmdp}\% without routing differentiation, so directed forgetting in the original experts and the routing path into the target experts both contribute to moving existing knowledge out.

\begin{table}[!t]
\centering
\fontsize{7.5}{9}\selectfont
\captionsetup{font=small}
\setlength{\tabcolsep}{2.0pt}
\renewcommand{\arraystretch}{1.08}
\caption{Relocation-mechanism ablation on Intern-S2 / Bio (\%). Each row enables the marked mechanisms: selective gradient masking on target data (Grad.), the pre-top-$K$ activation mask on general data (Act.) and the routing-differentiation losses (Route). The Grad. only row keeps the masking rule that \sysname{} shares with SGTM~\citep{shilov2025sgtm}, applied to the same expanded model and recipe. $A_1$ is the relocated full model $F_1$ with the target experts excised, before residual cleanup, and its WMDP is the target knowledge that stays in the retained parameters; $\Delta$WMDP $= F_1 - A_1$ is the part that the excision takes away. All rows share one starting model, token budget, data order and relocation recipe, so the \Ours{} row is the model in the other tables.}
\label{tab:mechanism-ablation}
\begin{tabular}{@{}lccc@{\hspace{10pt}}ccc@{}}
\toprule
 & Grad. & Act. & Route & $A_1$ WMDP$\downarrow$ & $A_1$ MMLU & $\Delta$WMDP \\
\midrule
\rowcolor{oursrowmain}\textbf{\Ours{}} & \checkmark & \checkmark & \checkmark & \textbf{64.3} & 82.3 & 15.1 \\
w/o Act. & \checkmark & \textcolor{sepmuted}{--} & \checkmark & 65.9 & 82.2 & 14.8 \\
w/o Route & \checkmark & \checkmark & \textcolor{sepmuted}{--} & 70.2 & 83.6 & 13.8 \\
Grad. only & \checkmark & \textcolor{sepmuted}{--} & \textcolor{sepmuted}{--} & 68.4 & 83.3 & 15.8 \\
\bottomrule
\end{tabular}
\end{table}

\begin{table*}[!t]
\centering
\fontsize{7.5}{9}\selectfont
\captionsetup{font=small}
\setlength{\tabcolsep}{2.0pt}
\renewcommand{\arraystretch}{1.08}
\caption{Pipeline ablation (\%). Base: removing either step of base-model delivery; $R$ applies the RMU operator, hyperparameters and forget budget of the residual cleanup directly to $B$. SFT: the same 250M-token $\Sstd$ applied to $R$, to the excised base $A_2$, and to the recombined full model $F_2$ with the target experts excised at delivery. Shaded rows are the full pipeline; deeper blue marks a larger share of the above-chance WMDP accuracy of $B$ removed, as in \cref{tab:main-results}.}
\label{tab:pipeline-ablation}
\begin{tabularx}{\textwidth}{@{}l@{\hspace{4pt}}l@{\hspace{4pt}}YY@{\hspace{8pt}}YY@{\hspace{8pt}}YY@{}}
\toprule
 &  & \multicolumn{2}{@{}c@{\hspace{8pt}}}{Qwen3.5} & \multicolumn{2}{@{}c@{\hspace{8pt}}}{Intern-S2} & \multicolumn{2}{@{}c@{\hspace{0pt}}}{Gemma 4} \\
\cmidrule(l{0pt}r{8pt}){3-4}\cmidrule(l{0pt}r{8pt}){5-6}\cmidrule(l{0pt}r{0pt}){7-8}
 &  & WMDP$\downarrow$ & MMLU & WMDP$\downarrow$ & MMLU & WMDP$\downarrow$ & MMLU \\
\midrule
\multicolumn{8}{c}{\rule{0pt}{9pt}\fontsize{8.5}{10}\selectfont\textsc{Bio}} \\[1pt]
\arrayrulecolor{black!25}\cmidrule(lr){1-8}\arrayrulecolor{black}
\multirow{3}{*}{Base} & Excision only ($A_1$) & \cellcolor{removalmax!8}62.9 & 78.6 & \cellcolor{removalmax!7}64.3 & 82.3 & \cellcolor{removalmax!2}70.2 & 72.1 \\
 & RMU only ($R$) & \cellcolor{removalmax!36}41.6 & 81.8 & \cellcolor{removalmax!52}33.3 & 84.5 & \cellcolor{removalmax!22}48.5 & 72.2 \\
 & \cellcolor{oursrowmain}\textbf{Excision + RMU ($A_2$)} & \cellcolor{removalmax!53}32.4 & \cellcolor{oursrowmain}77.2 & \cellcolor{removalmax!59}29.6 & \cellcolor{oursrowmain}81.0 & \cellcolor{removalmax!49}33.6 & \cellcolor{oursrowmain}68.7 \\
\cmidrule(l){2-8}
\multirow{3}{*}{SFT} & SFT on $R$ ($R$-SFT) & \cellcolor{removalmax!0}78.8 & 82.0 & \cellcolor{removalmax!0}81.0 & 84.6 & \cellcolor{removalmax!1}71.5 & 74.1 \\
 & SFT on $A_2$ ($A_3$-direct) & \cellcolor{removalmax!8}63.0 & 79.4 & \cellcolor{removalmax!7}64.6 & 83.0 & \cellcolor{removalmax!13}55.7 & 71.7 \\
 & \cellcolor{oursrowmain}\textbf{SFT on $F_2$, excise ($A_3$)} & \cellcolor{removalmax!34}42.8 & \cellcolor{oursrowmain}78.7 & \cellcolor{removalmax!17}55.0 & \cellcolor{oursrowmain}82.5 & \cellcolor{removalmax!34}41.3 & \cellcolor{oursrowmain}70.5 \\
\midrule
\multicolumn{8}{c}{\rule{0pt}{9pt}\fontsize{8.5}{10}\selectfont\textsc{Cyber}} \\[1pt]
\arrayrulecolor{black!25}\cmidrule(lr){1-8}\arrayrulecolor{black}
\multirow{3}{*}{Base} & Excision only ($A_1$) & \cellcolor{removalmax!14}43.2 & 81.0 & \cellcolor{removalmax!12}45.1 & 84.2 & \cellcolor{removalmax!1}48.1 & 73.6 \\
 & RMU only ($R$) & \cellcolor{removalmax!44}31.8 & 82.3 & \cellcolor{removalmax!48}30.9 & 84.9 & \cellcolor{removalmax!42}30.9 & 72.1 \\
 & \cellcolor{oursrowmain}\textbf{Excision + RMU ($A_2$)} & \cellcolor{removalmax!57}28.2 & \cellcolor{oursrowmain}80.7 & \cellcolor{removalmax!65}26.3 & \cellcolor{oursrowmain}84.1 & \cellcolor{removalmax!48}29.4 & \cellcolor{oursrowmain}71.9 \\
\cmidrule(l){2-8}
\multirow{3}{*}{SFT} & SFT on $R$ ($R$-SFT) & \cellcolor{removalmax!13}43.5 & 82.3 & \cellcolor{removalmax!12}44.7 & 84.2 & \cellcolor{removalmax!18}37.6 & 73.7 \\
 & SFT on $A_2$ ($A_3$-direct) & \cellcolor{removalmax!31}35.9 & 81.6 & \cellcolor{removalmax!32}35.9 & 84.2 & \cellcolor{removalmax!24}35.7 & 73.3 \\
 & \cellcolor{oursrowmain}\textbf{SFT on $F_2$, excise ($A_3$)} & \cellcolor{removalmax!61}27.3 & \cellcolor{oursrowmain}81.3 & \cellcolor{removalmax!52}29.8 & \cellcolor{oursrowmain}84.0 & \cellcolor{removalmax!30}34.0 & \cellcolor{oursrowmain}73.3 \\
\bottomrule
\end{tabularx}
\end{table*}

On the base model, excision and residual cleanup are complementary. In all six settings, $A_2$ has lower WMDP than both excision alone ($A_1$) and the same cleanup applied without excision ($R$) (\cref{tab:pipeline-ablation}). Excision without cleanup leaves \ev{all/bio/A1-A2-min/0M/wmdp}--\ev{all/bio/A1-A2-max/0M/wmdp} points of WMDP-Bio in the retained parameters, which the residual cleanup then removes. $R$ applies the same RMU operator, hyperparameters and forget budget directly to the original model, so its only difference from $A_2$ is the starting point of the cleanup. It stays \ev{all/bio/R-A2-min/0M/wmdp}--\ev{all/bio/R-A2-max/0M/wmdp} points above $A_2$ on Bio and \ev{all/cyber/R-A2-min/0M/wmdp}--\ev{all/cyber/R-A2-max/0M/wmdp} points above it on Cyber, and this gap is what relocation and excision contribute to the base model. The higher MMLU of $R$ lies almost entirely in the subjects adjacent to the target domain: on Intern-S2 / Bio, $R$ and $A_2$ score \ev{intern/bio/R/0M/mmlu_adj}\% and \ev{intern/bio/A2/0M/mmlu_adj}\% on the adjacent subjects but \ev{intern/bio/R/0M/mmlu_rest}\% and \ev{intern/bio/A2/0M/mmlu_rest}\% on all others (\cref{tab:mmlu-split}). Cleanup alone thus suppresses answers to the target benchmark while knowledge in the adjacent subjects stays, whereas excision removes it together with the target domain, as discussed in \cref{sec:exp-base}.

Under the same standard post-training, the three routes form the same ordering in every setting, $R$-SFT above $A_3$-direct above $A_3$, and each gap in this ordering isolates one part of the pipeline (\cref{tab:pipeline-ablation}, SFT rows). Excision is the difference between $R$-SFT and $A_3$-direct, which fine-tune the cleaned original model $R$ and the excised and cleaned base $A_2$, respectively. It accounts for \ev{all/bio/R-SFT-A3-direct-min/250M/wmdp}--\ev{all/bio/R-SFT-A3-direct-max/250M/wmdp} points on Bio and \ev{all/cyber/R-SFT-A3-direct-min/250M/wmdp}--\ev{all/cyber/R-SFT-A3-direct-max/250M/wmdp} points on Cyber. Fine-tuning the recombined full model and excising at delivery is the difference between $A_3$-direct and $A_3$, which start from the same $A_2$. It accounts for a further \ev{all/bio/A3-direct-A3-min/250M/wmdp}--\ev{all/bio/A3-direct-A3-max/250M/wmdp} points on Bio and \ev{all/cyber/A3-direct-A3-min/250M/wmdp}--\ev{all/cyber/A3-direct-A3-max/250M/wmdp} points on Cyber.

\FloatBarrier

\section{Discussion}
\label{sec:discussion}

Our results suggest that methods for removing hazardous knowledge should be compared on the model as it is delivered, not only on the score right after unlearning. Open-weight models are post-trained both before and after release, and prior work found that unlearning implemented as in-place modification of parameter values is undone by ordinary fine-tuning~\citep{doshi2024recovery}. A score taken right after unlearning therefore reflects how deeply the knowledge has been suppressed, but not whether it will return under later training, and the risk of a public model is decided by its state at delivery, after post-training. Judged by this state, the comparison between methods differs from the one on the base model: methods that are comparable on the base model separate clearly after the same standard fine-tuning (\cref{sec:exp-base,sec:exp-trajectory}). Whether the public model of \sysname{} carries the hazardous capability depends on which parameters are delivered, a decision made only after post-training, so its effect corresponds to this evaluation point by construction.

Taken together, the results of \cref{sec:experiments} show that a change at the architectural level can relocate the hazardous knowledge already present in a pretrained model into removable experts and decouple hazardous from general capability in the structure of the model, which opens a direction for controlling hazardous capability beyond alignment and unlearning. Behavioral alignment changes how the model responds and parameter-level unlearning changes parameter values, but both work within a fixed set of parameters, so the hazardous knowledge always remains in the delivered model. \sysname{} instead changes the structure of the model itself. The hazardous capability is concentrated in a set of experts that can be removed as a whole, the public model no longer contains these experts, and it keeps most of its removal under standard post-training (\cref{sec:exp-excision,sec:exp-trajectory}). This direction is aligned with knowledge localization during pretraining~\citep{cloud2024gradient,shilov2025sgtm}, but it does not require training from scratch and applies to the models that releasing parties already hold, whose pretraining is complete. Its only premise is that the model contains units of parameters that can be removed as a whole, and the experts and router that MoE models introduce for computational efficiency provide exactly such units. For MoE models, controlling which capabilities are released thus becomes an architectural choice rather than a matter of repeatedly adjusting parameter values.

\section{Conclusion}
\label{sec:conclusion}

Relocating hazardous knowledge into removable target experts changes where that knowledge is stored in an MoE model, and thereby turns its removal from an in-place modification of parameter values into a structural operation. Through expert expansion, selective gradient masking and routing differentiation, one model takes two forms, a full model that keeps the hazardous capability for authorized users and an excised model for public release, with almost the same general capability. On the base model, excision followed by a light RMU cleanup removes target knowledge to a degree comparable with the strongest parameter-level unlearning methods. Post-training on the full structure and excising at delivery gives the delivered model the lowest WMDP among usable methods after standard fine-tuning, while its general capability keeps improving with post-training. What to release thus becomes a choice of which parameters to deliver from a single trained model.

\bibliography{main}

\clearpage
\appendix
\makeatletter
\setlength{\@fptop}{0pt}
\setlength{\@fpsep}{18pt}
\setlength{\@fpbot}{0pt plus 1fil}
\makeatother

\section{Implementation details}
\label{app:setup}

This section gives the implementation details of \sysname{} in the order of the pipeline, followed by the training of the external baselines (\cref{app:impl-baselines}). \Cref{tab:shared-hparams} lists the hyperparameters of \sysname{} shared by all six settings and \cref{tab:recipes} those that vary across settings. Both tables are generated directly from the training configuration of each setting. Every training stage uses the same random seed, which fixes the data order and the remaining sources of randomness.

\begin{table}[!htbp]
\centering
\footnotesize
\setlength{\tabcolsep}{4pt}
\caption{Hyperparameters shared by all six settings. The learning rates, routing losses and the schedule and budget of residual cleanup depend on the setting and are listed in \cref{tab:recipes}.}
\label{tab:shared-hparams}
\begin{minipage}[t]{0.45\textwidth}
\begin{tabularx}{\linewidth}[t]{@{}l>{\raggedright\arraybackslash}X@{}}
\toprule
\multicolumn{2}{@{}l}{\textit{Expert expansion}} \\
\quad Target experts per layer $M$ & 8 \\
\quad Noise on cloned experts & $10^{-5}$ \\
\quad Statistics chunk & 2048 tokens \\
\quad General rows for statistics & first 300,000 \\
\addlinespace
\multicolumn{2}{@{}l}{\textit{Relocation}} \\
\quad Optimizer ($\beta_1, \beta_2$) & AdamW (0.9, 0.999) \\
\quad Weight decay / clipping & 0 / 1 \\
\quad Schedule & no warmup, cosine to 0 \\
\quad Packed length & 4096 \\
\addlinespace
\multicolumn{2}{@{}l}{\textit{Residual cleanup}} \\
\quad Optimizer ($\beta_1, \beta_2$) & AdamW (0.9, 0.999) \\
\quad Weight decay / clipping & 0 / 1 \\
\quad Packed length & 4096 \\
& \\
\bottomrule
\end{tabularx}
\end{minipage}
\hfill
\begin{minipage}[t]{0.52\textwidth}
\begin{tabularx}{\linewidth}[t]{@{}l>{\raggedright\arraybackslash}X@{}}
\toprule
\multicolumn{2}{@{}l}{\textit{Standard post-training $\Sstd$}} \\
\quad Tokens / per step & 250M / 500K \\
\quad Packed length & 31,250 (Gemma 4: 15,625) \\
\quad Learning rate & $2{\times}10^{-6}$ \\
\quad Schedule & 10\% warmup, cosine to $10^{-7}$ \\
\quad Weight decay / clipping & 0.05 / 1 \\
\quad Muon momentum & 0.95, Nesterov \\
\quad Muon update scale & spectral, factor 0.2 \\
\quad Non-matrix parameters & Adam \\
\quad Evaluation points & 0M, 10M, 50M, 100M, 250M \\
\quad Seed & 42 \\
\addlinespace
\multicolumn{2}{@{}l}{\textit{Evaluation}} \\
\quad Libraries & lm-eval 0.4.12, vLLM 0.19.1 \\
\quad Precision / max length & BF16 / 8,192 \\
\quad Chat template & IFEval only, no thinking \\
\addlinespace
\bottomrule
\end{tabularx}
\end{minipage}
\end{table}

\begin{table*}[!htbp]
\centering
\scriptsize
\setlength{\tabcolsep}{2pt}
\caption{Hyperparameters that differ across settings. Token pairs give target / general data, multiplier pairs give original / target experts, and the BOS pair gives relocation / residual cleanup. Multipliers scale the relocation learning rate; a target-expert multiplier of 0 keeps the cloned experts fixed. The cleanup budget is one pass over the forget corpus or a fixed number of forget tokens. Hyperparameters shared by all settings are listed in \cref{tab:shared-hparams}.}
\label{tab:recipes}
\begin{tabular}{lcccccc}
\toprule
 & Qwen3.5 & Intern-S2 & Gemma 4 & Qwen3.5 & Intern-S2 & Gemma 4 \\
 & Bio & Bio & Bio & Cyber & Cyber & Cyber \\
\midrule
Source-expert scale $s$ & 0.5 & 0.5 & 0.7 & 0.3 & 0.5 & 0.7 \\
Relocation tokens & 164M / 109M & 164M / 109M & 157M / 52M & 7M / 5M & 7M / 5M & 7M / 2M \\
Learning rate & $10^{-6}$ & $10^{-6}$ & $10^{-6}$ & $10^{-6}$ & $10^{-6}$ & $10^{-6}$ \\
Router multipliers & 40 / 60 & 40 / 60 & 40 / 60 & 40 / 60 & 40 / 60 & 40 / 60 \\
Expert multipliers & 1 / 0 & 1 / 40 & 1 / 20 & 20 / 0 & 10 / 0 & 1 / 0 \\
Mass loss $\tau_h$ (weight) & 0.75 (6) & 0.75 (6) & 0.65 (4) & 0.75 (6) & 0.75 (6) & 0.65 (4) \\
Mass loss $\tau_s$ (weight) & 0.02 (1) & 0.02 (1) & 0.02 (1) & 0.02 (1) & 0.02 (1) & 0.02 (1) \\
Rank loss $\delta$ (weight) & 0.1 (30) & 0.1 (30) & 0.1 (30) & 0.1 (30) & 0.1 (30) & 0.1 (30) \\
RMU monitor / updated layers & 9 / 7--9 & 9 / 7--9 & 7 / 5--7 & 9 / 7--9 & 9 / 7--9 & 9 / 7--9 \\
RMU steering coefficient & 300 & 300 & 1200 & 300 & 300 & 1200 \\
RMU retain weight & 350 & 350 & 4.45 & 350 & 350 & 4.45 \\
RMU learning rate & $10^{-5}$ & $10^{-5}$ & $10^{-6}$ & $10^{-5}$ & $10^{-5}$ & $2{\times}10^{-4}$ \\
RMU schedule & constant & constant & warmup + cosine & constant & constant & constant \\
RMU forget budget & 1 pass & 1 pass & 1 pass & 13.6M tokens & 13.6M tokens & 1 pass \\
RMU master weights & BF16 & BF16 & FP32 & BF16 & BF16 & BF16 \\
RMU parameter scope (MLPs) & routed + shared & routed + shared & routed + dense & routed + shared & routed + shared & routed down \\
BOS per packed sequence & no / no & no / no & yes / yes & no / no & no / no & yes / no \\
\bottomrule
\end{tabular}
\end{table*}

\FloatBarrier
\subsection{Expansion and relocation}
\label{app:impl-relocation}

The routing statistics for expansion are collected on the original model $B$ using training corpora only. The target side uses the full forget corpus and the general side a prefix of the shuffled FineWeb-Edu corpus, with long documents split into fixed-length chunks that are counted separately. Differential-activation expansion selects $M = 8$ source experts in every MoE layer, the cloned target experts receive small Gaussian noise, and the source experts are scaled by the factor $s$ of each setting.

Relocation trains the routed experts and the routers with AdamW, while $\theta_{\mathrm{shared}}$ and the shared-expert MLPs stay frozen. The learning rate has no warmup and decays to zero on a cosine schedule, and the original experts, the target experts and the router rows of each bank each carry their own learning-rate multiplier. Training text is packed into fixed-length sequences, with EOS appended to every document and no BOS, except that on Gemma~4 each packed sequence starts with BOS. Target and general data are mixed under the token budgets of \cref{tab:recipes}. The Cyber forget corpus is much smaller than the Bio corpus, so the Cyber relocation budgets are correspondingly smaller.

\subsection{Residual cleanup}
\label{app:impl-cleanup}

Residual cleanup applies RMU to the expert MLPs that remain in $A_1$. It monitors one early layer and updates the routed-expert MLPs and the shared-expert MLPs (the always-on dense MLP on Gemma~4) of the three layers up to it, and on Gemma~4 / Cyber it updates only the down projections of the routed experts. Routers, attention, embeddings, norms and the language-model head remain frozen. The random direction $u$ is drawn from a fixed seed and kept for the whole cleanup, and both the forget and the retain terms are computed on labelled tokens only and averaged per sequence. The learning rate, schedule and forget budget vary across settings and are listed in \cref{tab:recipes}. The matched control $R$ of the pipeline ablation uses exactly the same configuration, and its only difference is that it starts from the original model $B$ instead of $A_1$.

\subsection{Standard post-training}
\label{app:impl-sft}

$\Sstd$ is full-parameter supervised fine-tuning on the Megatron backend of ms-swift. The data is a text-only instruction mixture with the life-science subsets removed, converted with the chat template and thinking format of each backbone and packed into fixed-length sequences, and samples longer than the packed length are dropped. Matrix parameters are trained with Muon and all other parameters with Adam. The same recipe produces every post-training trajectory in the paper, namely $F_2 \to F_3$, $A_2 \to A_3$-direct, $R \to R$-SFT and the post-training of every baseline. These trajectories share the data, data order and seed, and differ only in their starting checkpoint.

\subsection{Unlearning baselines}
\label{app:impl-baselines}

The nine baselines are trained in all six settings under one shared configuration, with the same forget corpus $\mathcal{D}_{\mathrm{harm}}$ and the same FineWeb-Edu retain corpus $\mathcal{D}_{\mathrm{std}}$ as \sysname{}. Every run uses a Megatron-Core backend on 8 GPUs with expert parallelism 8, tensor, pipeline and context parallelism 1, BF16 precision and seed 42. The corpora are tokenized for each backbone and packed into 4096-token sequences, and on Gemma~4 each packed sequence starts with BOS, as in relocation. All methods use AdamW ($\beta_1 = 0.9$, $\beta_2 = 0.999$, $\epsilon = 10^{-8}$) without weight decay, gradient clipping at 1, a global batch of 32 sequences and a cosine schedule with a linear warmup over the first 5\% of steps. They train all parameters, including attention, experts and routers, except RMU, which updates only layers 7--9. The forget term carries the weight $\gamma = 1$ in every method, and the retain term, where present, the weight $\alpha = 1$ unless stated otherwise. \Cref{tab:baseline-hparams} lists the remaining hyperparameters.

\begin{table}[!t]
\centering
\footnotesize
\setlength{\tabcolsep}{4pt}
\caption{Hyperparameters of the nine unlearning baselines. Left: method-specific hyperparameters and base learning rate on Bio / Cyber; PDU on Intern-S2 / Bio uses $10^{-6}$. All methods share the configuration described in the text, including the forget weight $\gamma = 1$. Right: values that depend on the backbone, with the RMU values shared by both domains, and the Cyber epoch (23 steps each) selected by the largest $\mathrm{Acc}_{\mathrm{MMLU}} - \mathrm{Acc}_{\mathrm{WMDP}}$; Bio uses the final checkpoint.}
\label{tab:baseline-hparams}
\begin{minipage}[t]{0.55\textwidth}
\begin{tabularx}{\linewidth}[t]{@{}l>{\raggedright\arraybackslash}Xl@{}}
\toprule
Method & Hyperparameters & LR (Bio / Cyber) \\
\midrule
CEU & -- & $10^{-6}$ / $5{\times}10^{-6}$ \\
GA & -- & $10^{-6}$ / $5{\times}10^{-6}$ \\
GD & $\alpha = 1$ & $10^{-6}$ / $5{\times}10^{-6}$ \\
NPO & $\beta = 0.1$, $\alpha = 1$, reference $B$ & $5{\times}10^{-5}$ / $5{\times}10^{-5}$ \\
PDU & $\lambda_0 = 1$, $\eta = 0.1$, $\varepsilon$ per backbone & $4{\times}10^{-6}$ / $5{\times}10^{-6}$ \\
RMU & layers 7--9 (monitor 9), $c$, $\alpha$ per backbone & $10^{-6}$ / $5{\times}10^{-6}$ \\
SatImp & $\beta_1 = 2$, $\beta_2 = 0.1$, $\alpha = 0$ & $10^{-6}$ / $5{\times}10^{-6}$ \\
SimNPO & $\beta = 0.1$, $\delta = 0$, $\alpha = 1$ & $5{\times}10^{-5}$ / $5{\times}10^{-5}$ \\
WGA & $\beta = 2$, $\alpha = 0$ & $10^{-6}$ / $5{\times}10^{-6}$ \\
\bottomrule
\end{tabularx}
\end{minipage}
\hfill
\begin{minipage}[t]{0.42\textwidth}
\begin{tabularx}{\linewidth}[t]{@{}>{\raggedright\arraybackslash}Xccc@{}}
\toprule
 & Qwen3.5 & Intern-S2 & Gemma 4 \\
\midrule
\multicolumn{4}{@{}l}{\textit{Backbone-specific}} \\
\quad PDU $\varepsilon$, Bio & 2.51 & 1.98 & 2.6 \\
\quad PDU $\varepsilon$, Cyber & 2.51 & 2.05 & 2.6 \\
\quad RMU $c$ & 300 & 300 & 1200 \\
\quad RMU $\alpha$ & 634 & 350 & 4.45 \\
\addlinespace
\multicolumn{4}{@{}l}{\textit{Selected Cyber epoch}} \\
\quad CEU & 8 & 4 & 10 \\
\quad GA & 2 & 2 & 1 \\
\quad GD & 3 & 6 & 7 \\
\quad NPO & 3 & 3 & 3 \\
\quad PDU & 2 & 3 & 9 \\
\quad RMU & 4 & 2 & 4 \\
\quad SatImp & 2 & 1 & 9 \\
\quad SimNPO & 2 & 1 & 1 \\
\quad WGA & 8 & 1 & 3 \\
\bottomrule
\end{tabularx}
\end{minipage}
\end{table}

The objectives below share one notation, and every formula is a loss to be minimized. For token $t$ of a sequence $x$, let $o_t \in \mathbb{R}^{|V|}$ be the vocabulary logits of the trained model $\pi_\theta$, $\ell_t = -\log \pi_\theta(x_t \mid x_{<t})$ the token negative log-likelihood and $p_t = \exp(-\ell_t)$ the probability of the correct token. We write $\bar\ell(x) = \frac{1}{|x|}\sum_t \ell_t$ for the length-normalized NLL of a sequence, $\mathbb{E}_{\mathrm{harm}}[\cdot]$ for the token average over $\mathcal{D}_{\mathrm{harm}}$, and $\mathcal{L}_{\mathrm{std}} = \mathbb{E}_{\mathrm{std}}[\ell_t]$ for the mean NLL on $\mathcal{D}_{\mathrm{std}}$. The reference model $\pi_{\mathrm{ref}}$ is the frozen original model $B$. The symbols $\alpha$, $\beta$, $\delta$, $\lambda$ and $\eta$ are local to this subsection.

Five baselines act on the token-level predictions of the target data. Four of them operate directly on the token NLL and differ in how they weight it:
\begin{equation*}
\begin{aligned}
\mathcal{L}_{\mathrm{GA}} &= -\gamma\,\mathbb{E}_{\mathrm{harm}}[\ell_t], \\
\mathcal{L}_{\mathrm{GD}} &= -\gamma\,\mathbb{E}_{\mathrm{harm}}[\ell_t] + \alpha\,\mathcal{L}_{\mathrm{std}}, \\
\mathcal{L}_{\mathrm{WGA}} &= -\mathbb{E}_{\mathrm{harm}}\big[\operatorname{stopgrad}(p_t^{\beta})\,\ell_t\big], \\
\mathcal{L}_{\mathrm{SatImp}} &= -\mathbb{E}_{\mathrm{harm}}\big[\operatorname{stopgrad}(p_t^{\beta_1}(1-p_t)^{\beta_2})\,\ell_t\big].
\end{aligned}
\end{equation*}
Here $\operatorname{stopgrad}$ preserves the forward value but blocks its gradient, so the probability weights are recomputed at each step and treated as constants during backpropagation. GA~\citep{jang2022knowledge} raises the NLL of the target data, and GD~\citep{maini2024tofu} adds standard language modelling on the retain data. WGA~\citep{wang2025wga} weights each token by a power of its current probability, so the tokens that the model still predicts confidently dominate the update and the tokens that are already forgotten contribute little. SatImp~\citep{yang2025satimp} multiplies this saturation weight by the importance factor $(1-p_t)^{\beta_2}$. WGA and SatImp are run without the retain term ($\alpha = 0$), the form in which they produce effective forgetting in our setting. The fifth, CEU~\citep{yang2025ceu}, replaces ascent with cross-entropy towards a target distribution that excludes the correct token. At each position it renormalizes the model's own distribution over the remaining vocabulary,
\begin{equation*}
q_t(v) = \frac{\exp(o_t(v))}{\sum_{v' \neq x_t} \exp(o_t(v'))} \quad (v \neq x_t), \qquad q_t(x_t) = 0,
\end{equation*}
treats $q_t$ as a fixed target and minimizes $\mathcal{L}_{\mathrm{CEU}} = -\mathbb{E}_{\mathrm{harm}}\big[\sum_v q_t(v) \log \pi_\theta(v \mid x_{<t})\big]$, skipping the first token of each sequence.

NPO~\citep{zhang2024npo} and SimNPO cast forgetting as a preference against the target sequences and operate on sequence-level likelihoods,
\begin{equation*}
\mathcal{L}_{\mathrm{NPO}} = -\frac{2\gamma}{\beta}\,\mathbb{E}_{x \sim \mathcal{D}_{\mathrm{harm}}}\Big[\log\sigma\big(\beta(\bar\ell_\theta(x) - \bar\ell_{\mathrm{ref}}(x))\big)\Big] + \alpha\,\mathcal{L}_{\mathrm{std}},
\end{equation*}
where $\sigma$ is the sigmoid. SimNPO~\citep{fan2024simnpo} is the reference-free variant. It replaces $\bar\ell_{\mathrm{ref}}(x)$ with a constant margin $\delta$, set to 0, and therefore needs no forward pass through a frozen reference.

PDU~\citep{entesari2025pdu} flattens the output distribution on target data and enforces retain performance as an explicit constraint. Its forget loss $\mathcal{L}_{\mathrm{flat}} = \mathbb{E}_{\mathrm{harm}}\big[(\max_v o_t(v) - \bar o_t)^2\big]$, with $\bar o_t$ the mean logit over the vocabulary, drives the output towards uniform. The constraint $\mathcal{L}_{\mathrm{std}} \le \varepsilon$ enters through a Lagrangian, $\mathcal{L}_{\mathrm{PDU}} = \gamma\,\mathcal{L}_{\mathrm{flat}} + \lambda(\mathcal{L}_{\mathrm{std}} - \varepsilon)$, whose dual variable starts at $\lambda_0$ and is updated after every step as $\lambda \leftarrow \max\big(0, \lambda + \eta(\mathcal{L}_{\mathrm{std}} - \varepsilon)\big)$, so the balance between forgetting and retention is adjusted automatically. The threshold $\varepsilon$ is set per backbone.

The external RMU baseline~\citep{li2024wmdp} uses the objective of \cref{eq:rmu} with the original model $B$ as reference, monitor layer 9 and all parameters of layers 7--9 trainable. It thus differs from the residual cleanup of \sysname{} in both its starting point and its parameter scope, since the cleanup starts from $A_1$ and updates only expert MLPs. Its steering coefficient $c$ and retain weight $\alpha$ are set per backbone and shared by both domains.

The two domains differ in schedule and checkpoint selection because the Bio forget corpus is more than an order of magnitude larger than the Cyber one. Bio runs train for 3 epochs (2142 steps) with the learning rate decaying to 0 and are evaluated at the final checkpoint. Cyber runs train for 10 epochs (230 steps) with the learning rate decaying to a floor of $10^{-6}$, and are evaluated after every epoch with WMDP-Cyber and 0-shot MMLU. The epoch with the largest $\mathrm{Acc}_{\mathrm{MMLU}} - \mathrm{Acc}_{\mathrm{WMDP}}$ is kept, which gives each baseline its most favourable trade-off between forgetting and retention. The selected checkpoint is the one reported in \cref{tab:main-results} and the starting point of the baseline's standard post-training (\cref{app:impl-sft}).

\FloatBarrier

\subsection{Evaluation}
\label{app:impl-eval}

Endpoint evaluation uses lm-eval with vLLM at BF16 precision and without a chat template. WMDP and MMLU report 0-shot accuracy, ARC-C 25-shot normalized accuracy, HellaSwag 10-shot normalized accuracy and GSM8K 5-shot strict-match exact match. IFEval reports prompt-level strict accuracy with the chat template applied and thinking disabled. The Gemma~4 checkpoints carry no chat template, so for evaluation only a copy of the tokenizer receives a single-turn, text-only, non-thinking template equivalent to the ms-swift template used in training, and the checkpoint itself is unchanged. The endpoint evaluation of $A_3$ reloads the endpoint checkpoint and excises the target experts physically.

Evaluation along the post-training trajectory runs on the resident training model, without export or reloading. At each evaluation point the trajectory of \sysname{} is scored once with the full model, which gives $F_3$, and once with the pre-top-$K$ activation mask on the target experts, which gives $A_3$. \Cref{tab:mask-physical} compares this mode with physical excision of the same delivered base under the same evaluator, in the two settings that were evaluated both ways, and the two agree to within one standard error on WMDP and MMLU.

\begin{table}[!t]
\centering
\footnotesize
\caption{Pre-top-$K$ activation mask versus physical excision of the same delivered base $A_2$ (\%). Both columns use the evaluator of the resident trajectory evaluation, which differs from the endpoint evaluator of \cref{tab:main-results}, so the physical values differ slightly from those reported there. The two settings shown are those in which $A_2$ was evaluated both ways under this evaluator.}
\label{tab:mask-physical}
\begin{tabular}{lcccccc}
\toprule
\multirow{2}{*}{Setting} & \multicolumn{3}{c}{WMDP} & \multicolumn{3}{c}{MMLU} \\
\cmidrule(lr){2-4}\cmidrule(lr){5-7}
 & Masked & Physical & $\Delta$ & Masked & Physical & $\Delta$ \\
\midrule
Qwen3.5 / Cyber & 27.8 & 27.9 & -0.1 & 80.6 & 80.7 & -0.1 \\
Intern-S2 / Cyber & 26.0 & 26.5 & -0.4 & 84.0 & 84.1 & -0.0 \\
\bottomrule
\end{tabular}
\end{table}

\FloatBarrier
\subsection{Routing diagnostic}
\label{app:routing}

The routing diagnostic uses only training text. With seed 42, it samples up to 512 nonoverlapping windows of at most 2048 tokens per corpus by reservoir sampling, over the full forget corpus and over the first 4096 retain documents in sorted file order. Text is encoded without BOS and with EOS appended, and each $F_0/F_1$ pair is measured on the same windows. The measurement hooks read the routing logits without changing them, and the model parameters are verified to be unchanged by the collection. The mass is computed from the normalized top-$K$ gate weights without masking and before the per-expert scaling of Gemma~4, and each layer mean weights every valid token equally.

\FloatBarrier
\section{Supplementary results}
\label{app:results}

\subsection{Separation at every stage}
\label{app:separation}

\Cref{tab:excision-full} lists the full-versus-excised pair at every stage. Each pair shares every parameter except the target experts, so the difference between the two members of a pair is the target knowledge that those experts carry at that point in the pipeline.

\begin{table}[!t]
\centering
\footnotesize
\caption{Full model versus removal of the target experts at every stage (\%). $\Delta$ is full minus excised. $B$ is the original model: WMDP Qwen3.5 / Bio 83.2, Intern-S2 / Bio 83.4, Gemma 4 / Bio 78.9, Qwen3.5 / Cyber 57.6, Intern-S2 / Cyber 58.9, Gemma 4 / Cyber 50.8.}
\label{tab:excision-full}
\begin{tabular}{llcccccc}
\toprule
\multirow{2}{*}{Setting} & \multirow{2}{*}{Stage} & \multicolumn{3}{c}{WMDP} & \multicolumn{3}{c}{MMLU} \\
\cmidrule(lr){3-5}\cmidrule(lr){6-8}
 & & Full & Excised & $\Delta\uparrow$ & Full & Excised & $\Delta\downarrow$ \\
\midrule
\multirow{3}{*}{Qwen3.5 / Bio} & Relocated ($F_1$ / $A_1$) & 81.8 & 62.9 & +18.9 & 80.4 & 78.6 & +1.8 \\
 & Delivered base ($F_2$ / $A_2$) & 80.8 & 32.4 & +48.4 & 80.1 & 77.2 & +2.9 \\
 & After SFT ($F_3$ / $A_3$, 250M) & 81.7 & 42.8 & +38.9 & 80.7 & 78.7 & +2.0 \\
\midrule
\multirow{3}{*}{Intern-S2 / Bio} & Relocated ($F_1$ / $A_1$) & 79.4 & 64.3 & +15.1 & 83.2 & 82.3 & +1.0 \\
 & Delivered base ($F_2$ / $A_2$) & 61.9 & 29.6 & +32.3 & 82.4 & 81.0 & +1.4 \\
 & After SFT ($F_3$ / $A_3$, 250M) & 78.6 & 55.0 & +23.6 & 84.1 & 82.5 & +1.5 \\
\midrule
\multirow{3}{*}{Gemma 4 / Bio} & Relocated ($F_1$ / $A_1$) & 76.2 & 70.2 & +6.0 & 74.0 & 72.1 & +1.9 \\
 & Delivered base ($F_2$ / $A_2$) & 71.0 & 33.6 & +37.4 & 72.6 & 68.7 & +3.9 \\
 & After SFT ($F_3$ / $A_3$, 250M) & 76.4 & 41.3 & +35.1 & 73.9 & 70.5 & +3.4 \\
\midrule
\multirow{3}{*}{Qwen3.5 / Cyber} & Relocated ($F_1$ / $A_1$) & 57.6 & 43.2 & +14.4 & 81.4 & 81.0 & +0.4 \\
 & Delivered base ($F_2$ / $A_2$) & 46.5 & 28.2 & +18.4 & 81.4 & 80.7 & +0.7 \\
 & After SFT ($F_3$ / $A_3$, 250M) & 50.3 & 27.3 & +23.0 & 82.0 & 81.3 & +0.8 \\
\midrule
\multirow{3}{*}{Intern-S2 / Cyber} & Relocated ($F_1$ / $A_1$) & 57.9 & 45.1 & +12.8 & 84.5 & 84.2 & +0.2 \\
 & Delivered base ($F_2$ / $A_2$) & 52.8 & 26.3 & +26.6 & 84.4 & 84.1 & +0.3 \\
 & After SFT ($F_3$ / $A_3$, 250M) & 55.8 & 29.8 & +26.0 & 84.0 & 84.0 & +0.0 \\
\midrule
\multirow{3}{*}{Gemma 4 / Cyber} & Relocated ($F_1$ / $A_1$) & 49.6 & 48.1 & +1.5 & 74.0 & 73.6 & +0.4 \\
 & Delivered base ($F_2$ / $A_2$) & 38.9 & 29.4 & +9.5 & 72.5 & 71.9 & +0.6 \\
 & After SFT ($F_3$ / $A_3$, 250M) & 45.1 & 34.0 & +11.1 & 74.0 & 73.3 & +0.6 \\
\bottomrule
\end{tabular}
\end{table}

\subsection{Random-expert excision}
\label{app:random}

\Cref{tab:random-excision} gives the per-seed results behind the five-seed summary reported in \cref{sec:exp-excision}.

\begin{table}[!t]
\centering
\footnotesize
\caption{Random-expert excision control on Intern-S2 / Bio (\%). Each seed removes the same number of original experts in each layer of $F_1$ as target-expert excision, while retaining the added experts. $\Delta$ is $F_1$ minus the excised model; SD is the sample standard deviation across five seeds, not an evaluation standard error.}
\label{tab:random-excision}
\begin{tabular}{lcccc}
\toprule
Excision & WMDP & MMLU & $\Delta$WMDP & $\Delta$MMLU \\
\midrule
Seed 0 & 80.1 & 82.6 & -0.7 & +0.7 \\
Seed 1 & 78.2 & 82.7 & +1.3 & +0.6 \\
Seed 2 & 79.3 & 82.2 & +0.1 & +1.0 \\
Seed 3 & 77.9 & 82.4 & +1.5 & +0.9 \\
Seed 4 & 79.2 & 82.3 & +0.2 & +1.0 \\
\midrule
Mean $\pm$ seed SD & $78.9 \pm 0.9$ & $82.4 \pm 0.2$ & +0.5 & +0.8 \\
\midrule
Target experts ($A_1$) & 64.3 & 82.3 & +15.1 & +1.0 \\
\bottomrule
\end{tabular}
\end{table}

\subsection{General capability}
\label{app:general}

\Cref{tab:endpoints-general} lists the general capability of every route at the 250M-token endpoint. The IFEval level differs across backbones under the shared $\Sstd$ for all routes alike, so it is compared within a backbone rather than across backbones. \Cref{tab:mmlu-split} splits MMLU into the subjects adjacent to the target domain and all other subjects.

\begin{table*}[!t]
\centering
\scriptsize
\setlength{\tabcolsep}{2.4pt}
\caption{General capability at the 250M-token SFT endpoint (\%): MMLU and IFEval (prompt-level strict, chat template, thinking disabled). For \Ours{} $A_3$, IFEval reloads the endpoint and physically excises the target experts. A dagger marks a baseline that collapsed on the base model (\cref{tab:main-results}); it is muted and takes no part in the comparisons.}
\label{tab:endpoints-general}
\begin{tabular}{lcccccccccccc}
\toprule
\multirow{3}{*}{Method} & \multicolumn{6}{c}{Bio} & \multicolumn{6}{c}{Cyber} \\
\cmidrule(lr){2-7}\cmidrule(lr){8-13}
 & \multicolumn{2}{c}{Qwen3.5} & \multicolumn{2}{c}{Intern-S2} & \multicolumn{2}{c}{Gemma 4} & \multicolumn{2}{c}{Qwen3.5} & \multicolumn{2}{c}{Intern-S2} & \multicolumn{2}{c}{Gemma 4} \\
\cmidrule(lr){2-3}\cmidrule(lr){4-5}\cmidrule(lr){6-7}\cmidrule(lr){8-9}\cmidrule(lr){10-11}\cmidrule(lr){12-13}
 & MMLU & IFEval & MMLU & IFEval & MMLU & IFEval & MMLU & IFEval & MMLU & IFEval & MMLU & IFEval \\
\midrule
Original + $\Sstd$ & 82.2 & 43.1 & 84.8 & 78.9 & 74.9 & 26.8 & 82.2 & 43.1 & 84.8 & 78.9 & 74.9 & 26.8 \\
\midrule
CEU & \textcolor{collapsed}{82.1}$^{\dag}$ & \textcolor{collapsed}{50.1}$^{\dag}$ & \textcolor{collapsed}{84.7}$^{\dag}$ & \textcolor{collapsed}{79.7}$^{\dag}$ & \textcolor{collapsed}{74.5}$^{\dag}$ & \textcolor{collapsed}{26.2}$^{\dag}$ & \textcolor{collapsed}{81.5}$^{\dag}$ & \textcolor{collapsed}{54.2}$^{\dag}$ & \textcolor{collapsed}{83.4}$^{\dag}$ & \textcolor{collapsed}{78.9}$^{\dag}$ & \textcolor{collapsed}{66.3}$^{\dag}$ & \textcolor{collapsed}{10.5}$^{\dag}$ \\
GA & \textcolor{collapsed}{80.5}$^{\dag}$ & \textcolor{collapsed}{51.9}$^{\dag}$ & \textcolor{collapsed}{46.9}$^{\dag}$ & \textcolor{collapsed}{79.3}$^{\dag}$ & \textcolor{collapsed}{27.3}$^{\dag}$ & \textcolor{collapsed}{16.1}$^{\dag}$ & \textcolor{collapsed}{80.8}$^{\dag}$ & \textcolor{collapsed}{46.6}$^{\dag}$ & \textcolor{collapsed}{81.1}$^{\dag}$ & \textcolor{collapsed}{81.9}$^{\dag}$ & \textcolor{collapsed}{73.7}$^{\dag}$ & \textcolor{collapsed}{17.2}$^{\dag}$ \\
GD & 81.5 & 38.8 & 84.4 & 78.7 & 74.8 & 27.5 & \textcolor{collapsed}{81.3}$^{\dag}$ & \textcolor{collapsed}{43.8}$^{\dag}$ & \textcolor{collapsed}{83.2}$^{\dag}$ & \textcolor{collapsed}{78.4}$^{\dag}$ & \textcolor{collapsed}{73.1}$^{\dag}$ & \textcolor{collapsed}{26.6}$^{\dag}$ \\
NPO & 78.3 & 37.9 & 80.7 & 70.4 & 70.3 & 22.4 & 81.9 & 37.7 & 84.5 & 75.8 & 72.6 & 23.8 \\
PDU & 81.4 & 50.3 & 84.8 & 78.4 & 74.7 & 26.2 & 82.3 & 48.2 & 84.8 & 77.8 & 74.5 & 25.7 \\
RMU & 82.0 & 40.7 & 84.6 & 79.3 & 73.2 & 27.7 & 82.2 & 38.5 & 84.3 & 77.5 & 73.8 & 27.4 \\
SatImp & 82.1 & 48.2 & \textcolor{collapsed}{84.4}$^{\dag}$ & \textcolor{collapsed}{79.5}$^{\dag}$ & \textcolor{collapsed}{74.2}$^{\dag}$ & \textcolor{collapsed}{25.7}$^{\dag}$ & \textcolor{collapsed}{81.1}$^{\dag}$ & \textcolor{collapsed}{50.5}$^{\dag}$ & \textcolor{collapsed}{83.9}$^{\dag}$ & \textcolor{collapsed}{80.8}$^{\dag}$ & \textcolor{collapsed}{71.0}$^{\dag}$ & \textcolor{collapsed}{15.5}$^{\dag}$ \\
SimNPO & 78.3 & 42.5 & 79.9 & 65.8 & 69.0 & 24.4 & 81.9 & 40.7 & 82.8 & 75.8 & 71.4 & 20.1 \\
WGA & 82.0 & 49.2 & \textcolor{collapsed}{84.6}$^{\dag}$ & \textcolor{collapsed}{78.9}$^{\dag}$ & \textcolor{collapsed}{74.3}$^{\dag}$ & \textcolor{collapsed}{26.4}$^{\dag}$ & \textcolor{collapsed}{81.0}$^{\dag}$ & \textcolor{collapsed}{51.8}$^{\dag}$ & \textcolor{collapsed}{83.7}$^{\dag}$ & \textcolor{collapsed}{82.6}$^{\dag}$ & \textcolor{collapsed}{66.9}$^{\dag}$ & \textcolor{collapsed}{13.3}$^{\dag}$ \\
Matched $R$-SFT & 82.0 & 45.7 & 84.6 & 78.6 & 74.1 & 28.8 & 82.3 & 39.6 & 84.2 & 78.0 & 73.7 & 27.0 \\
$A_3$-direct & 79.4 & 45.3 & 83.0 & 79.3 & 71.7 & 27.7 & 81.6 & 41.0 & 84.2 & 78.0 & 73.3 & 26.8 \\
\midrule
\Ours{} full ($F_3$) & 80.7 & 43.2 & 84.1 & 78.6 & 73.9 & 26.1 & 82.0 & 39.2 & 84.0 & 79.1 & 74.0 & 27.5 \\
\Ours{} ($A_3$) & 78.7 & 44.7 & 82.5 & 78.6 & 70.5 & 28.1 & 81.3 & 38.1 & 84.0 & 76.2 & 73.3 & 26.2 \\
\bottomrule
\end{tabular}
\end{table*}

\begin{table}[!t]
\centering
\footnotesize
\caption{MMLU split into target-adjacent subjects and all other subjects (accuracy, \%, weighted by question count). Bio-adjacent: the ten biology and health subjects; Cyber-adjacent: the four computer-science subjects. \textit{Share}: fraction of the questions that are adjacent. \textit{Adj.\ share of drop}: fraction of the $B \to A_2$ MMLU drop that comes from adjacent subjects.}
\label{tab:mmlu-split}
\begin{tabular}{lccccccccc}
\toprule
\multirow{2}{*}{Setting} & \multirow{2}{*}{Share} & \multicolumn{3}{c}{Adjacent} & \multicolumn{3}{c}{Other subjects} & \multirow{2}{*}{\shortstack{Adj.\ share\\of drop}} \\
\cmidrule(lr){3-5}\cmidrule(lr){6-8}
 & & $B$ & $A_2$ & $A_3$ & $B$ & $A_2$ & $A_3$ & \\
\midrule
Qwen3.5 / Bio & 15 & 86.9 & 66.9 & 72.8 & 81.7 & 79.0 & 79.7 & 57.1 \\
Intern-S2 / Bio & 15 & 88.7 & 68.2 & 76.8 & 84.4 & 83.2 & 83.5 & 75.0 \\
Gemma 4 / Bio & 15 & 81.5 & 59.8 & 66.5 & 73.1 & 70.3 & 71.2 & 57.6 \\
Qwen3.5 / Cyber & 3 & 81.8 & 58.5 & 65.3 & 82.5 & 81.4 & 81.8 & 39.5 \\
Intern-S2 / Cyber & 3 & 87.4 & 72.1 & 77.4 & 85.0 & 84.4 & 84.2 & 45.7 \\
Gemma 4 / Cyber & 3 & 69.7 & 47.3 & 59.2 & 74.5 & 72.7 & 73.8 & 27.3 \\
\bottomrule
\end{tabular}
\par\vspace{3pt}{\footnotesize $A_3$ is the 250M SFT endpoint.\par}
\end{table}

\end{document}